\documentclass{article}

\usepackage{times}
\usepackage{geometry}
\usepackage{tcolorbox}
\usepackage[utf8]{inputenc}
\usepackage[T1]{fontenc}
\usepackage{hyperref}
\usepackage{url}
\usepackage{booktabs}
\usepackage{subcaption}
\usepackage{xcolor}
\usepackage{multirow}
\usepackage{multicol}
\usepackage{wrapfig}
\usepackage{adjustbox}
\usepackage{pdfx}
\usepackage{xspace}
\usepackage{amsmath}
\usepackage{amsfonts}
\usepackage{microtype}

\newcommand{\myparatight}[1]{\smallskip\noindent{\bf {#1}:}~}
\newcommand{\mysubparatight}[1]{{\bf \textit{#1}:}}

\newcommand{\datasetname}{VideoMarkData\xspace}
\newcommand{\benchmarkname}{VideoMarkBench\xspace}

\newtcolorbox{custombox}[1][]{
    colback=white!90!gray!10,
    colframe=black!60,
    fonttitle=\bfseries,
    coltitle=black,
    colbacktitle=black!10!white,
    title={#1}
}

\begin{document}
\begin{center}
{\Large{\bf{\benchmarkname: Benchmarking Robustness of Video Watermarking}}}

\vspace{1cm}

\begin{tabular}{cccc}
    Zhengyuan Jiang$^1$ & Moyang Guo$^1$ & Kecen Li$^2$ & Yuepeng Hu$^1$ \\
    Yupu Wang$^1$ & Zhicong Huang$^2$ & Cheng Hong$^2$ & Neil Zhenqiang Gong$^1$ \\
    \multicolumn{4}{c}{$^1$Duke University \quad \quad $^2$Ant Group} \\
    \multicolumn{4}{c}{\{zhengyuan.jiang, moyang.guo, yuepeng.hu, yupu.wang, neil.gong\}@duke.edu} \\
    \multicolumn{4}{c}{likecen2023@ia.ac.cn, zhicong303@gmail.com, vince.hc@antgroup.com} \\
\end{tabular}

\end{center}

\begin{abstract}
    The rapid development of video generative models has led to a surge in highly realistic synthetic videos, raising ethical concerns related to disinformation and copyright infringement. Recently, video watermarking has been proposed as a mitigation strategy by embedding invisible marks into AI-generated videos to enable subsequent detection. However, the robustness of existing video watermarking methods against both common and adversarial perturbations remains underexplored. In this work, we introduce VideoMarkBench, the first systematic benchmark designed to evaluate the robustness of video watermarks under \textit{watermark removal} and \textit{watermark forgery} attacks. Our study encompasses a unified dataset generated by three state-of-the-art video generative models, across three video styles, incorporating four watermarking methods and seven aggregation strategies used during detection. We comprehensively evaluate 12 types of perturbations under white-box, black-box, and no-box threat models. Our findings reveal significant vulnerabilities in current watermarking approaches and highlight the urgent need for more robust solutions.

    \begin{flushleft}
    \raisebox{-0.5ex}{\includegraphics[height=1.2em]{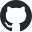}} \textbf{Code:} \url{https://github.com/zhengyuan-jiang/VideoMarkBench}
    \end{flushleft}
    
    \begin{flushleft}
    \raisebox{-0.5ex}{\includegraphics[height=1.2em]{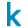}} \textbf{Data:} \url{https://www.kaggle.com/datasets/zhengyuanjiang/videomarkbench}
    \end{flushleft}

\end{abstract}

\section{Introduction}
Recent advancements in video generative models have enabled the creation of highly realistic synthetic videos that are nearly indistinguishable from authentic videos of real individuals. Despite their remarkable technological achievements, these generative capabilities introduce significant risks, including the spread of misinformation and potential copyright violations~\cite{christodorescu2024securing}. For instance, video generative models were used to create convincing deepfake footage of Ukrainian President Volodymyr Zelenskyy surrendering during the ongoing conflict, illustrating how synthetic videos can be weaponized to spread political misinformation and undermine public trust~\cite{fakevideo}.

Thus, it is important to detect whether a video containing sensitive information is AI-generated. Watermarks can be employed as a detection mechanism~\cite{jiang2024watermark}. Specifically, a watermarking method consists of two stages: watermark \emph{insertion} and \emph{detection}. In the insertion stage, the watermark is embedded into the AI-generated video during or after the generation process, producing a watermarked video. In the detection stage, a decoder extracts the watermark from the video and compares it with the ground-truth watermark. The video is detected as watermarked--and therefore AI-generated--if the similarity exceeds a predefined detection threshold.

Current video watermarking methods~\cite{zhang2023novel,tancik2020stegastamp,fernandez2024video,hu2025videoshield} are capable of embedding a watermark into a video and accurately decoding it in the absence of perturbations. However, videos often undergo common editing operations, such as MPEG-4 compression and cropping. Moreover, in adversarial settings, an attacker may deliberately introduce perturbations to remove or forge the watermark~\cite{jiang2023evading,lukas2023leveraging,saberi2023robustness,an2024waves,zhao2023invisible,nie2022diffusion,hu2024transfer,hu2024stable}, thereby evading detection. Despite this, the robustness of existing video watermarking methods against those perturbations has been largely underexplored.

\begin{figure}[t!]
    \centering
    \includegraphics[width=\linewidth]{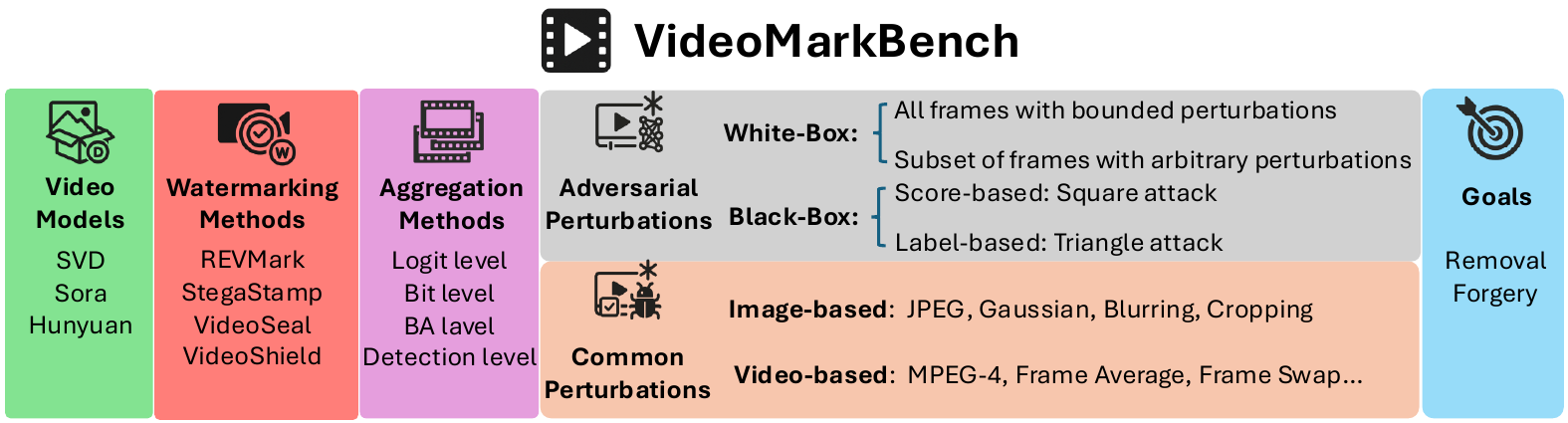}
    \caption{\label{fig:framework}Summary of our VideoMarkBench.}
\end{figure}

\myparatight{Our work}
In this work, we aim to bridge this gap by introducing \textbf{\benchmarkname} (\textbf{Video} Water\textbf{mark}ing \textbf{Bench}mark), the first systematic study that evaluates the effectiveness, utility, efficiency, and robustness of existing video watermarking methods. Figure~\ref{fig:framework} summarizes \benchmarkname{}. We conduct a comprehensive evaluation of watermark robustness against both \emph{removal} and \emph{forgery} perturbations, where perturbations are added to cause a watermarked video to be misclassified as unwatermarked, or an unwatermarked video to be falsely detected as watermarked, respectively.

\mysubparatight{- \quad Dataset} In addition to the real-world video dataset Kinetics-400~\cite{kay2017kinetics}, we construct a new AI-generated dataset, \datasetname, using three state-of-the-art video generative models. The video samples in \datasetname vary in style, length, and content, providing a diverse testbed for future research to explore the unique characteristics of AI-generated videos.

\mysubparatight{- \quad Systematic benchmarking} We introduce the \textit{first} systematic benchmark for evaluating the robustness of four state-of-the-art video watermarking methods against 12 types of perturbations used in watermark removal and forgery across different threat models. Our benchmark includes four adversarial perturbations in the white-box and black-box settings and eight common video perturbations in the no-box setting. Furthermore, we extend image watermarking methods to the video domain by treating each frame as an individual image, and we propose seven aggregation strategies to combine detection results across frames.

\mysubparatight{- \quad Observations} We summarize several key takeaways. First, current video watermarking methods perform accurately in the absence of perturbations. Second, existing video watermarking methods are broken against both watermark removal and forgery attacks in the white-box setting. Third, while these methods are relatively robust against forgery perturbations, they are vulnerable to adversarial removal perturbations in the black-box setting with a sufficient number of queries to the detection API and certain common removal perturbations in the no-box setting. Fourth, logit-level aggregation generally outperforms other aggregation strategies, and aggregation strategies based on median are more robust than those based on mean.

\section{Video Watermarking Methods}
Existing video watermarking methods can be broadly categorized into two types: \emph{post-generation} and \emph{pre-generation}. Post-generation methods~\cite{zhang2023novel,tancik2020stegastamp,fernandez2024video} embed a ground-truth watermark $w_g$ (a bitstring) into a video $x$ using a watermark encoder $E$, resulting in a watermarked video $x_w$, i.e., $x_w = E(x, w_g)$. These methods then employ a watermark detector $D$ to detect whether a test video $x_t$ has the watermark $w_g$. In contrast, pre-generation methods~\cite{wen2023tree,hu2025videoshield} do not use a dedicated watermark encoder. Instead, watermark insertion is integrated into the generative model itself, and the watermark is embedded during the video generation process. For detection, these methods use techniques such as DDIM Inversion~\cite{dhariwal2021diffusion} to extract the embedded watermark.

\myparatight{REVMark} REVMark~\cite{zhang2023novel} treats the video as a whole during both watermark insertion and detection. Specifically, the watermark encoder takes 8 cropped frames (the first 8 frames, each of size $128 \times 128$) as input and outputs their watermarked versions. Given a test video $x_t$, REVMark crops its first 8 frames into 8 consecutive frames of size $128 \times 128$, and then uses a watermark decoder $Dec$ to extract watermark logits from these frames, which are subsequently rounded to produce the decoded watermark $w$. If the bitwise accuracy (BA)—defined as the fraction of matching bits between the decoded watermark $w$ and the ground-truth watermark $w_g$—is no less than a predefined detection threshold $\tau$, i.e., $BA(w, w_g) \geq \tau$, the video is detected as watermarked; otherwise, it is considered unwatermarked. To enable fair comparison with other frame-level methods, we extend REVMark to operate across all frames of the video. We apply the decoder to each consecutive group of 8 frames and take the BA average for those decoded watermarks to obtain the final decision.

\myparatight{StegaStamp} StegaStamp~\cite{tancik2020stegastamp} is a state-of-the-art image watermarking method, which we extend to video watermarking by treating each video frame as an individual image. Specifically, the watermark encoder $E$ embeds the watermark $w_g$ into each frame of the video. During detection, given a test video $x_t$, the watermark decoder $Dec$ extracts watermark logits from each frame in $x_t$, and these logits are then aggregated to produce the final detection result. We discuss various aggregation strategies in Section~\ref{sec:aggregation}.

\myparatight{VideoSeal} VideoSeal~\cite{fernandez2024video} is a state-of-the-art video watermarking method. Unlike approaches that embed the watermark into every frame, VideoSeal uses the watermark encoder $E$ to embed the watermark into selected frames at a fixed interval. The perturbations introduced during watermark insertion are then propagated to neighboring frames. During detection, the watermark decoder extracts a watermark from each frame and computes the bitwise accuracy (BA) for each. These BA scores are then aggregated by taking their average to produce the final detection result.

\myparatight{VideoShield} VideoShield~\cite{hu2025videoshield} is a state-of-the-art video watermarking method designed specifically for videos generated by diffusion models. It embeds the watermark into the Gaussian noise image used during generation by modifying its sign. During detection, VideoShield applies DDIM Inversion~\cite{dhariwal2021diffusion} to estimate the original Gaussian noise image from the input video and then extracts the watermark from the sign of the estimated noise.

\subsection{\label{sec:aggregation}Aggregation Strategies for Frame-level Watermark Extraction}
For frame-level watermark extraction methods--such as StegaStamp and VideoSeal--the outputs consist of logits decoded from each individual video frame. To derive a per-video detection result, we propose seven aggregation strategies that combine these per-frame outputs, as detailed below:

\textbf{(1) Logit-mean:} We compute the average of the decoded logits across all frames to obtain aggregated logits. These aggregated logits are then rounded to a bitstring and compared with the ground-truth watermark to determine the final detection result. \textbf{(2) Logit-median:} Given $F$ frames and their corresponding $F$ vectors of decoded logits, we compute the geometric median of these $F$ vectors using the Powell method~\cite{powell1964efficient}. The resulting median vector is treated as the aggregated logits. \textbf{(3) Bit-median:} We first round the decoded logits from each frame to bitstrings, and then take a majority vote (0 or 1) across frames for each bit position to form the aggregated decoded watermark. \textbf{(4) BA-mean:} We compute the bitwise accuracy (BA) between the decoded watermark and the ground-truth for each frame, and then take the average BA across all frames. The final detection decision is made by comparing this average with the detection threshold $\tau$. Note that BA-mean aggregation was originally adopted by VideoSeal. \textbf{(5) BA-median:} Similar to BA-mean, we compute BA for each frame, but take the median BA across all frames and compare it with the threshold $\tau$ for detection. \textbf{(6) Detection-median:} For each frame, we compute BA and compare it with the detection threshold $\tau$ to obtain a binary detection result (watermarked or not). The final video-level decision is then obtained by taking the majority vote across all frame-level decisions. \textbf{(7) Detection-threshold:} We compute the detection result for each frame as in Detection-median. If the number of frames detected as watermarked is no less than a predefined threshold, the video is detected as watermarked. A detailed explanation is provided in the Appendix~\ref{sec:aggregation strategies}
\section{Perturbations for Video Watermarking}
\emph{Watermark removal} adds a perturbation $\delta$ to a watermarked video $x_w$ such that the perturbed version $x_w + \delta$ is falsely classified as unwatermarked. In contrast, \emph{watermark forgery} adds a perturbation $\delta$ to an unwatermarked video $x_u$ such that the detector falsely detects $x_u + \delta$ as watermarked.

\subsection{White-box Perturbations}
In the white-box setting, we assume an attacker has full access to the watermark detector, including its parameters. Perturbations are strategically crafted by solving an optimization problem to evade detection. Depending on the attacker's capabilities, we consider two scenarios, as described below.

\myparatight{Attacking each frame with bounded perturbations} In this scenario, we assume the attacker can add perturbations to all frames, but the perturbation size is bounded to preserve the visual quality of each frame. Specifically, the attacker crafts an adversarial perturbation $\delta$~\cite{szegedy2013intriguing} to remove the watermark $w_g$ by solving the following optimization problem via Projected Gradient Descent (PGD)~\cite{madry2018towards}:
\begin{align}
\min_\delta \ l(Dec(I + \delta), w_g), \quad \quad \text{s.t.} ||\delta||_\infty \leq \epsilon, \label{eq:wbox scenario1}
\end{align}
where $l$ is a loss function that measures the distance between two vectors, $Dec$ is the watermark decoder, $I$ is a video frame, and $\epsilon$ is the perturbation bound. For REVMark, which does not operate on a single frame during detection, $I$ corresponds to a stack of 8 frames of size $128 \times 128$, and $\delta$ represents the optimized video-level perturbation. To perform a watermark forgery attack, the objective is reversed to maximize the loss on an unwatermarked video.

\myparatight{Attacking a subset of frames with arbitrary perturbations} In this scenario, we assume that certain frames in the video are critical and must be preserved without perturbation, while the attacker is allowed to apply arbitrarily large perturbations to the remaining non-critical frames. Such an attack can be strategically designed to break logit-mean aggregation, as this strategy can be dominated by logits with large absolute values. Specifically, if some frames are perturbed so that their decoded logits attain extremely large values, the aggregated result may be skewed, making it easier to evade video-level detection. Our optimization objective is to reduce the decoded logit values as much as possible for bits where the ground-truth watermark $w_g$ is 1, and to increase them as much as possible where $w_g$ is 0. To achieve this, we formulate the following optimization problem over the decoded logits $Dec(I + \delta)$ to remove the watermark: $\min_\delta -\sum_{i=1}^{n}(\text{sign}(w_g - 0.5) * Dec(I + \delta))_i,$ where $n$ is the watermark length, $\text{sign}(\cdot)$ extracts the sign of each element, $*$ denotes element-wise multiplication, and $(\cdot)_i$ indicates the $i$-th element of the vector. To forge a watermark, we instead maximize this loss on an unwatermarked video.

\subsection{Black-box Perturbations}
In the black-box setting, the watermark detector is treated as an API: the attacker submits a video and observes the detection result without access to the internal workings of the detector. Specifically, the attacker iteratively refines the perturbation by repeatedly querying the detection API based on the feedback received. Black-box attacks can be categorized as either score-based or label-based, depending on the type of information available to the attacker from the detection API.

\myparatight{Score-based (Square Attack~\cite{andriushchenko2020square})} For score-based black-box perturbations, each query to the detection API returns a score indicating the likelihood that the input video contains a watermark. Square Attack~\cite{andriushchenko2020square} is a representative score-based method for images, and we extend it to videos by aggregating detection results across individual frames; implementation details are provided in the Appendix~\ref{sec:blackbox aggregation}. Specifically, Square Attack searches for a perturbation $\delta$ that removes or forges a watermark by strategically decreasing or increasing the score.

\myparatight{Label-based (Triangle Attack~\cite{wang2022triangle})} For label-based black-box perturbations, the detection API returns only a binary label (watermarked or unwatermarked) for each query. We extend Triangle Attack~\cite{wang2022triangle}—a label-based attack originally designed for images—to videos by flattening the video frames and treating it as a large image. Specifically, Triangle Attack begins with an initial sample that has the desired label but may contain a large perturbation relative to the target test video, and then iteratively searches for a smaller perturbation that maintains evasion by querying the detection API. The implementation details are provided in the Appendix~\ref{sec:blackbox aggregation}.

\subsection{Common Perturbations}
We consider both image-based and video-based common perturbations, which correspond to common image/video editing operations. Note that these perturbations can be applied by attackers or regular users. We apply the image-based perturbations to each frame of the video to perturb the entire video.

\myparatight{Image-based perturbations}
(1) \emph{JPEG}: a widely used image compression standard that reduces image size with a quality factor of $Q$. (2) \emph{Gaussian Noise}: adding random noise to the image, following a Gaussian distribution with a mean of 0 and a standard deviation of $\sigma$. (3) \emph{Gaussian Blur}: blurring the image with the  gaussian kernel with a standard deviation of $\sigma$. (4) \emph{Cropping}: cropping the image with a  proportion of $c$ and then resize the cropped image to the original size.

\myparatight{Video-based perturbations}
(1) \emph{MPEG-4}: a widely used video compression standard that reduces video size with a quality factor of $Q$. (2) \emph{Frame Average}: for each frame, computing the mean of its adjacent $N$ frames in the temporal dimension, with $N=1$ indicating no change. (3) \emph{Frame Swap}: for each frame, a random exchange with an adjacent frame (either the previous or the next frame) is conducted with a probability $p$. (4) \emph{Frame Removal}: removing each frame from the video with a probability $p$.

\section{\label{sec:dataset}Collecting Datasets}
\myparatight{AI-generated, watermarked videos}
To conduct a comprehensive evaluation of video watermarking methods across diverse visual styles and temporal dynamics, we construct a balanced benchmark dataset, \textbf{\datasetname}. It consists of videos generated by three state-of-the-art models: Stable Video Diffusion (SVD)~\cite{svd}, Sora~\cite{sora}, and Hunyuan Video~\cite{kong2024hunyuanvideo}. And we embed watermarks into those AI-generated videos. For each model, we generate videos in three styles—\emph{realistic}, \emph{cartoon}, and \emph{sci-fi}—capturing a broad range of visual characteristics. Temporal variation is explicitly controlled by specifying either slow or fast frame transitions within each style to modulate motion complexity. To ensure content consistency, a shared set of prompts is used across all models and styles. We use GPT-4~\cite{gpt} to generate those base prompts for us and turn them into different styles. Example prompts are shown in Table~\ref{tab:exmaple prompt} in the Appendix. Each prompt is annotated with its intended style, scene content, and motion type (i.e., speed of frame transitions), allowing us to evaluate watermark robustness across different generative models, contents, and styles.

Due to OpenAI's API query limitations, we collect 50 videos per style for Sora. For both SVD and Hunyuan Video, we collect 200 videos per style. In all cases, we maintain a 1:1 ratio of fast to slow motion videos, ensuring balanced temporal coverage. Table~\ref{attributeofdata} shows details of \datasetname.

\myparatight{Non-AI-generated, unwatermarked videos} We use the Kinetics-400 dataset~\cite{kay2017kinetics} for non-AI-generated videos—a widely used benchmark for video understanding. It contains approximately 240,000 YouTube clips across 400 diverse human actions, with variations in background, lighting, camera angle, and motion. Videos average 10 seconds in length and range from 240p to 1080p, offering a comprehensive reflection of real-world video diversity.

\begin{table}[!t]
\centering
\caption{Details of our \datasetname.}
\fontsize{8pt}{10pt}\selectfont
\begin{tabular}{|c|c|c|c|c|}
\hline
Video Generative Model & \#Frames & Resolution (H$\times$W) & Style & \#Samples per Style \\ \hline
Stable Video Diffusion (SVD) & 14 & 576$\times$1024 & Realistic, Cartoon, Sci-Fi & 200 \\
\hline
Sora & 150 & 720$\times$1280 & Realistic, Cartoon, Sci-Fi & 50 \\
\hline
Hunyuan Video & 61 & 576$\times$1024 & Realistic, Cartoon, Sci-Fi & 200 \\
\hline
\end{tabular}
\label{attributeofdata}
\end{table}
\section{\label{sec:evaluation}Benchmark Results}

\myparatight{Evaluation metrics}
We evaluate the robustness of video watermarking methods against watermark removal and forgery perturbations using \emph{False Negative Rate (FNR)} and \emph{False Positive Rate (FPR)}. FNR is defined as the proportion of (perturbed) watermarked videos that are falsely classified as unwatermarked, while FPR is the proportion of (perturbed) unwatermarked videos falsely detected as watermarked. Lower FNR and FPR indicate better robustness against removal and forgery perturbations, respectively. 

To assess the visual quality of watermarked videos, we report the average \emph{Peak Signal-to-Noise Ratio (PSNR)}~\cite{hore2010image} and \emph{Structural Similarity Index Measure (SSIM)}~\cite{wang2004image}, where higher values denote better visual similarity to the original (non-watermarked) videos. We also include the \emph{temporal LPIPS (tLP)}~\cite{chu2020learning}, which quantifies perceptual consistency across consecutive video frames. Lower tLP values suggest smoother temporal transitions and better preservation of temporal coherence.

\myparatight{Selection of detection threshold $\tau$}
REVMark~\cite{zhang2023novel} and VideoSeal~\cite{fernandez2024video} use a 96-bit watermark. The detection threshold $\tau$ is set to $\frac{67}{96}$, which guarantees a theoretical FPR of less than 0.01\%~\cite{jiang2023evading} (detailed in the Appendix~\ref{sec:select tau}). StegaStamp~\cite{tancik2020stegastamp} employs a 32-bit watermark, with the detection threshold $\tau$ set to $\frac{27}{32}$. VideoShield~\cite{hu2025videoshield} employs a 448-bit watermark, with the detection threshold $\tau$ set to the maximum detection score of 1,000 unwatermarked videos.

\subsection{Results under No Perturbation}
\begin{table}[]
\centering
\caption{Visual quality of watermarked video.}
\small
\begin{adjustbox}{max width=0.48\textwidth}
\begin{tabular}{lcccc}
    \toprule
     & REVMark & StegaStamp & VideoSeal & VideoShield \\
    \midrule
    PSNR $\uparrow$ & 37.13 & 37.91 & 37.85 & 7.945 \\
    \midrule
    SSIM $\uparrow$ & 0.948 & 0.945 & 0.942 &  0.264\\
    \midrule
    tLP $\downarrow$ & 2.762 & 0.198 & 0.145 & 6.674 \\
    \bottomrule
\end{tabular}
\end{adjustbox}
\label{tab:utility-no-perturbation}
\end{table}

Table~\ref{tab:fnr-no-perturbation} and Table~\ref{tab:fpr-no-perturbation} in the Appendix present the FNRs and FPRs of different video watermarking methods and aggregation strategies on the three AI-generated video datasets and real video dataset, under the setting where no perturbations are added to remove or forge the watermarks. We highlight two key observations from the results: First, the FNRs and FPRs of existing video watermarking methods are consistently near zero, demonstrating their effectiveness in distinguishing watermarked from non-watermarked videos in the absence of perturbations. Second, although certain aggregation strategies—such as BA-mean and BA-median—occasionally yield non-zero FNRs, the performance across different aggregation strategies remains comparable.

\begin{wraptable}{r}{0.5\textwidth}
\centering
\caption{Average time cost (ms) per video.}
\small
\begin{adjustbox}{max width=0.48\textwidth}
\begin{tabular}{lcccc}
    \toprule
     & REVMark & StegaStamp & VideoSeal & VideoShield\\
    \midrule
    Encoding & 26.66 & 14.99 & 157.6 & 1.598\\
    \midrule
    Decoding & 20.88 & 1.460 & 45.68 & 1.089$\times 10^4$\\
    \midrule
    Total & 47.54 & 16.45 & 203.3 & 1.090$\times 10^4$\\
    \bottomrule
\end{tabular}
\end{adjustbox}
\label{tab:efficiency-no-perturbation}
\end{wraptable}

Table~\ref{tab:utility-no-perturbation} reports the visual quality of watermarked videos for four video watermarking methods. Overall, post-generation watermarking methods generally preserve high visual quality. VideoShield--the only in-generation watermarking method--exhibits lower PSNR and SSIM values, likely due to the watermark being inserted during the video generation process, which can lead to more perceptible alterations in the video content. Table~\ref{tab:efficiency-no-perturbation} presents the time costs associated with watermark embedding and extraction. Among all methods, StegaStamp is the most efficient, requiring the least time for both encoding and decoding. In contrast, VideoShield incurs the highest time cost, primarily because its detection process involves DDIM inversion, which is computationally intensive.
\subsection{Robustness against White-box Video Perturbations}
Note that the inverse DDIM process used in VideoShield leads to gradient accumulation, resulting in excessive GPU memory consumption during white-box attacks. Due to our limited computational resources, we exclude VideoShield from our evaluation in the white-box setting.

\subsubsection{First Scenario: Attacking Each Frame with Bounded Perturbations}
In the first scenario, an attacker adds perturbations to each frame to remove or forge the watermark. To preserve the video's visual quality, the perturbations are constrained by an $\ell_{\infty}$-norm bound. Unless otherwise specified, comparisons across watermarking methods use the best-performing aggregation strategy for each watermarking method (StegaStamp or VideoSeal) where aggregation strategy is applicable, with results averaged over different generative models and video styles. When comparing aggregation strategies, we average the results across generative models and styles for StegaStamp or VideoSeal. For comparisons across generative models, we average results over all watermarking methods using various aggregation strategies and video styles. Similarly, when comparing across video styles, we average results over all watermarking methods with different aggregation strategies and generative models.

\begin{figure}[t!]
    \centering
    \begin{subfigure}{.23\linewidth}
        \centering
        \includegraphics[width=\linewidth]{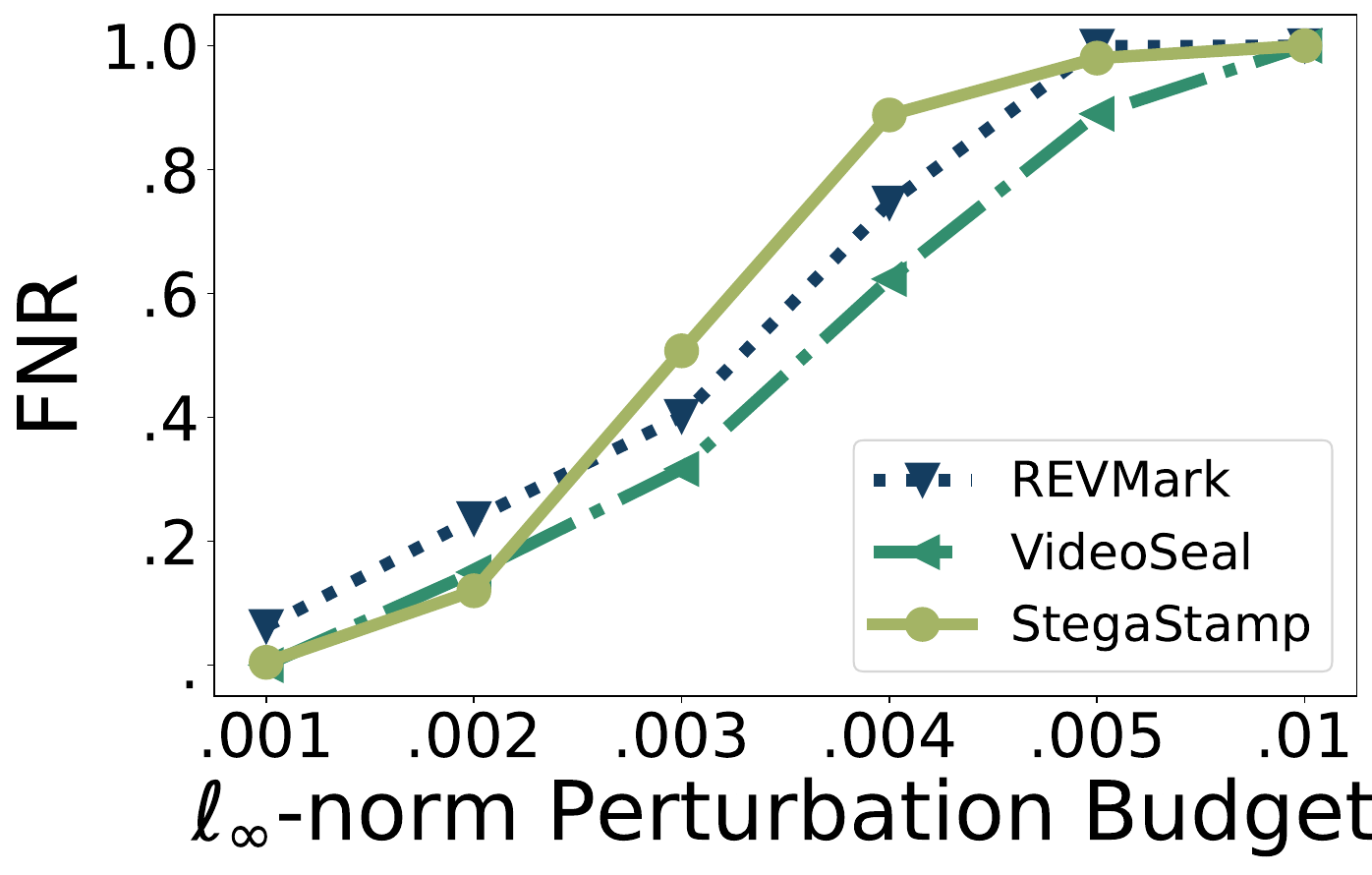}
        \caption{Watermarking}
        \label{subfig:wbox1watermark}
    \end{subfigure}
    \begin{subfigure}{.23\linewidth}
        \centering
        \includegraphics[width=\linewidth]{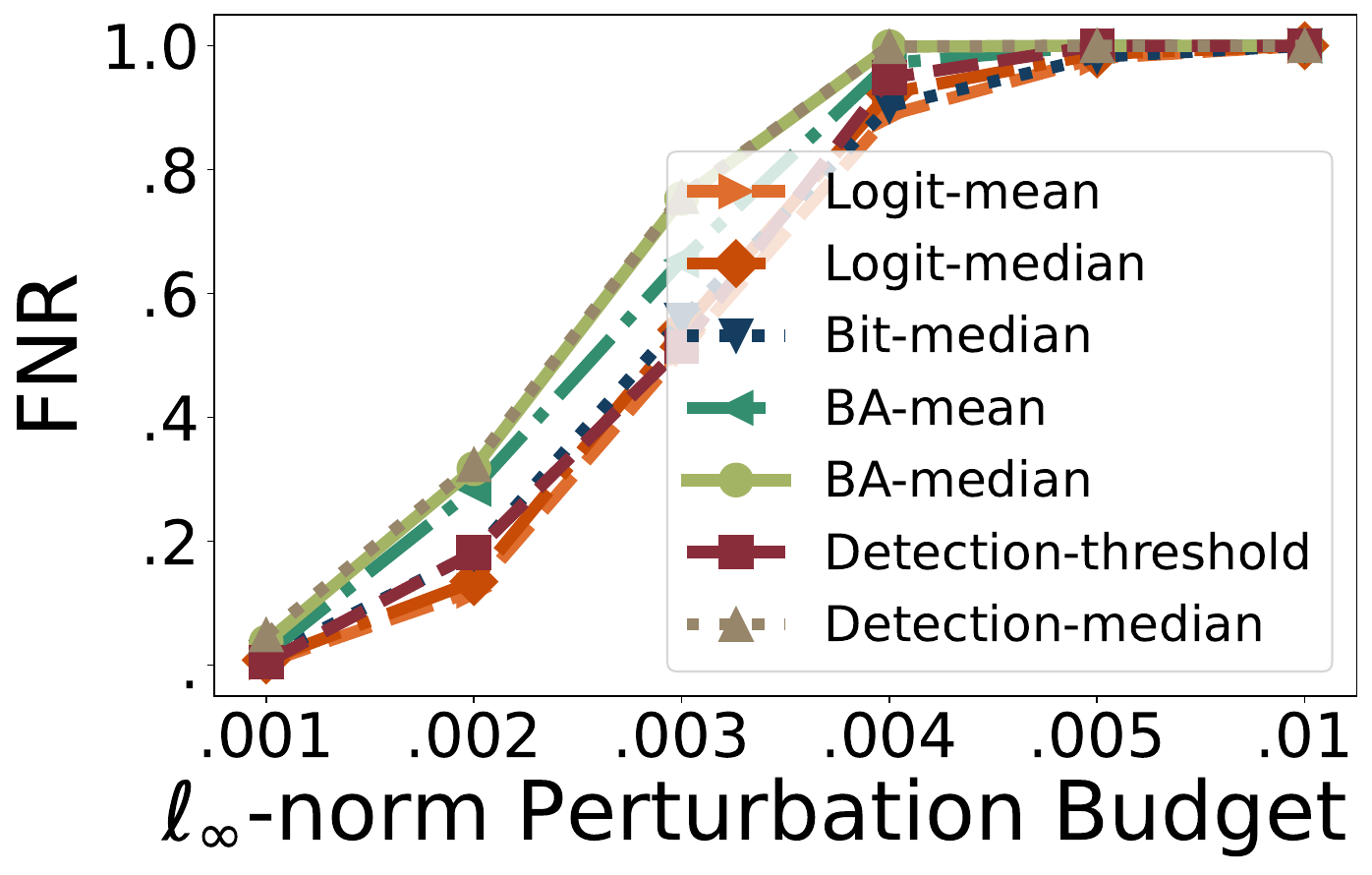}
        \caption{Aggregation}
        \label{subfig:wbox1aggregate}
    \end{subfigure}
    \begin{subfigure}{.23\linewidth}
        \centering
        \includegraphics[width=\linewidth]{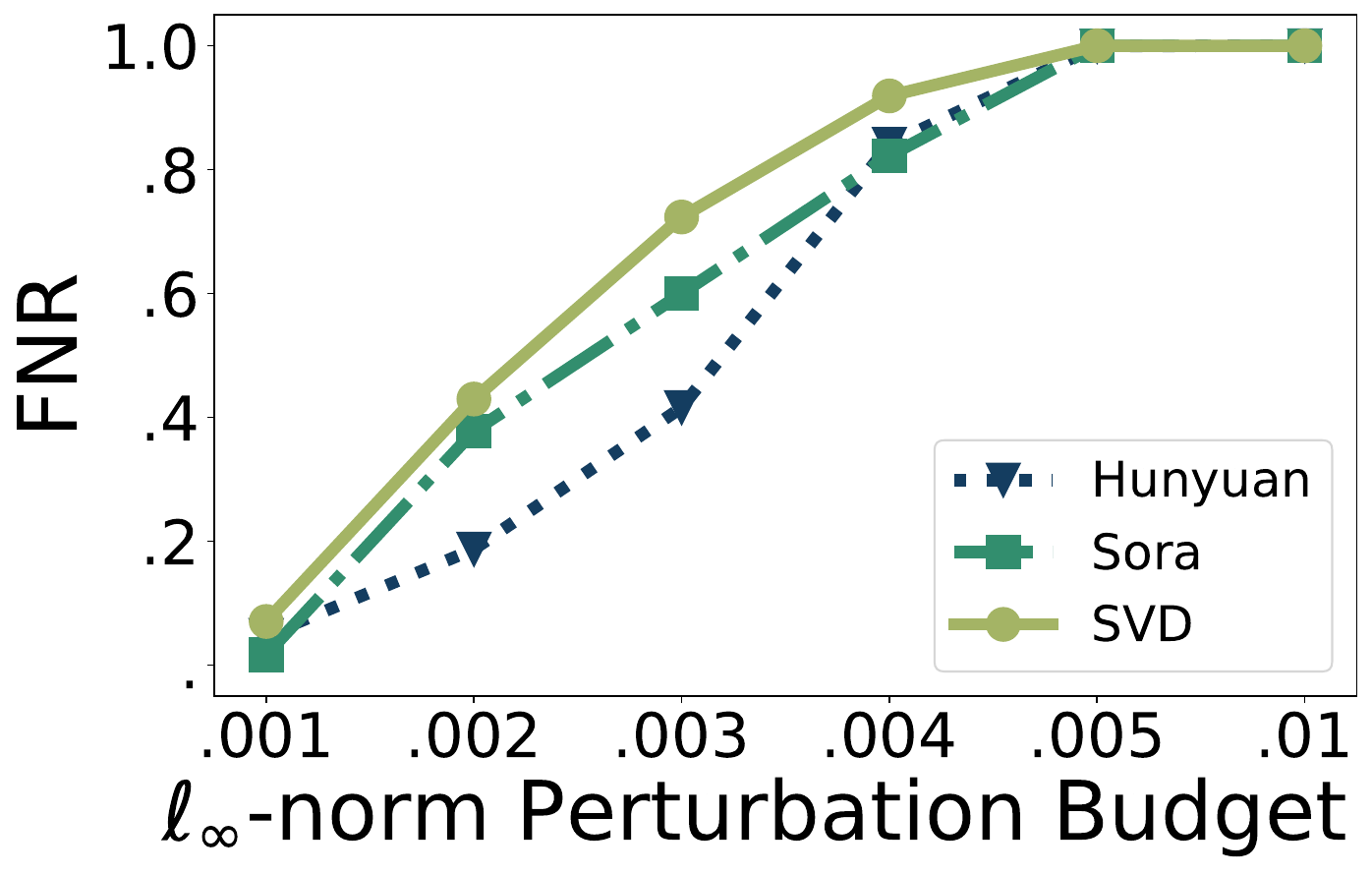}
        \caption{Model}
        \label{subfig:wbox1model}
    \end{subfigure}
    \begin{subfigure}{.23\linewidth}
        \centering
        \includegraphics[width=\linewidth]{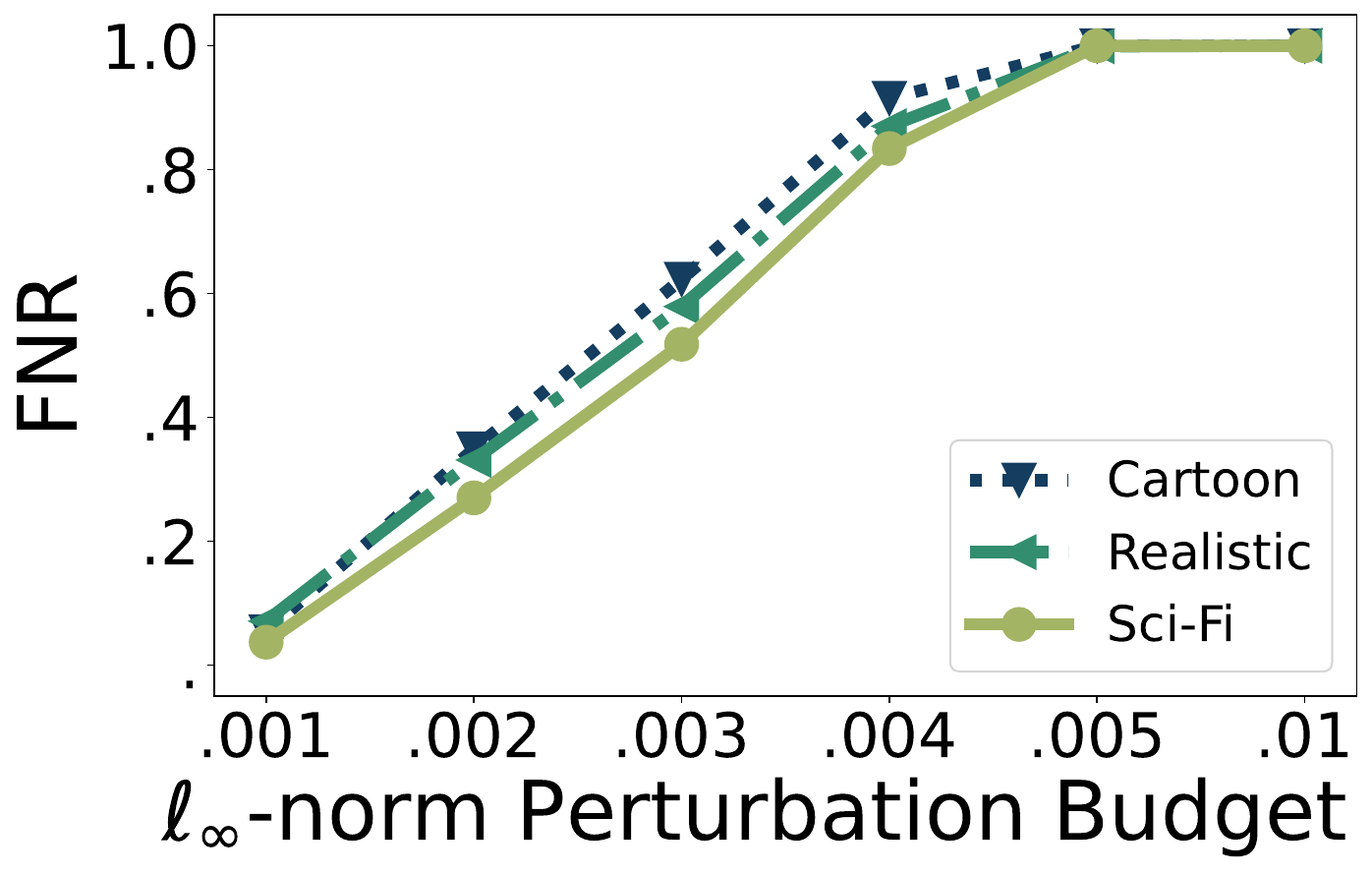}
        \caption{Style}
        \label{subfig:wbox1style}
    \end{subfigure}
    \caption{White-box watermark removal results in the first scenario.}
\end{figure}

\begin{wrapfigure}{r}{0.5\textwidth}
    \centering
    \begin{subfigure}{0.45\linewidth}
        \centering
        \includegraphics[width=\linewidth]{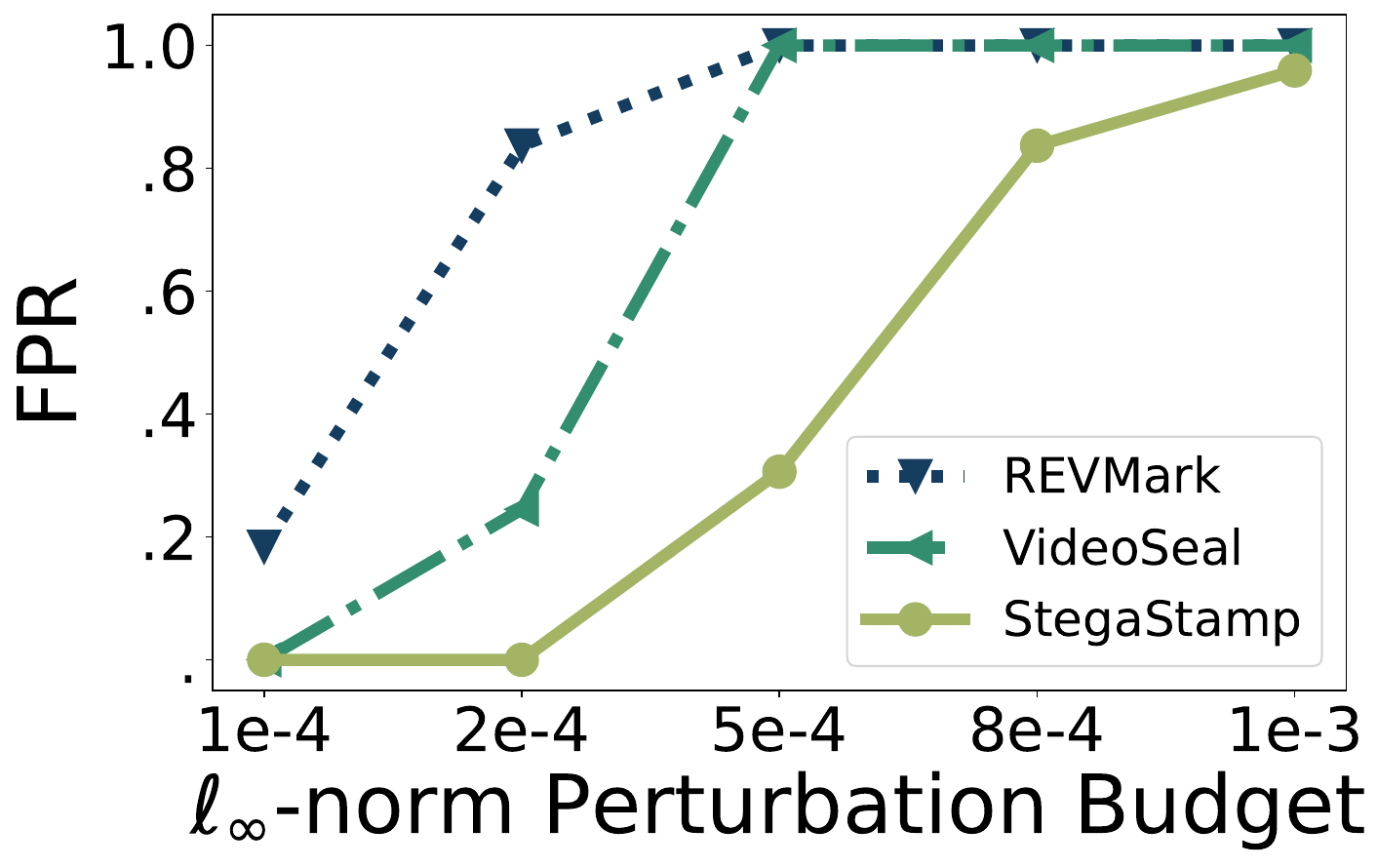}
        \caption{Watermarking}
        \label{subfig:wbox1forgerywatermark}
    \end{subfigure}
    \hfill
    \begin{subfigure}{0.45\linewidth}
        \centering
        \includegraphics[width=\linewidth]{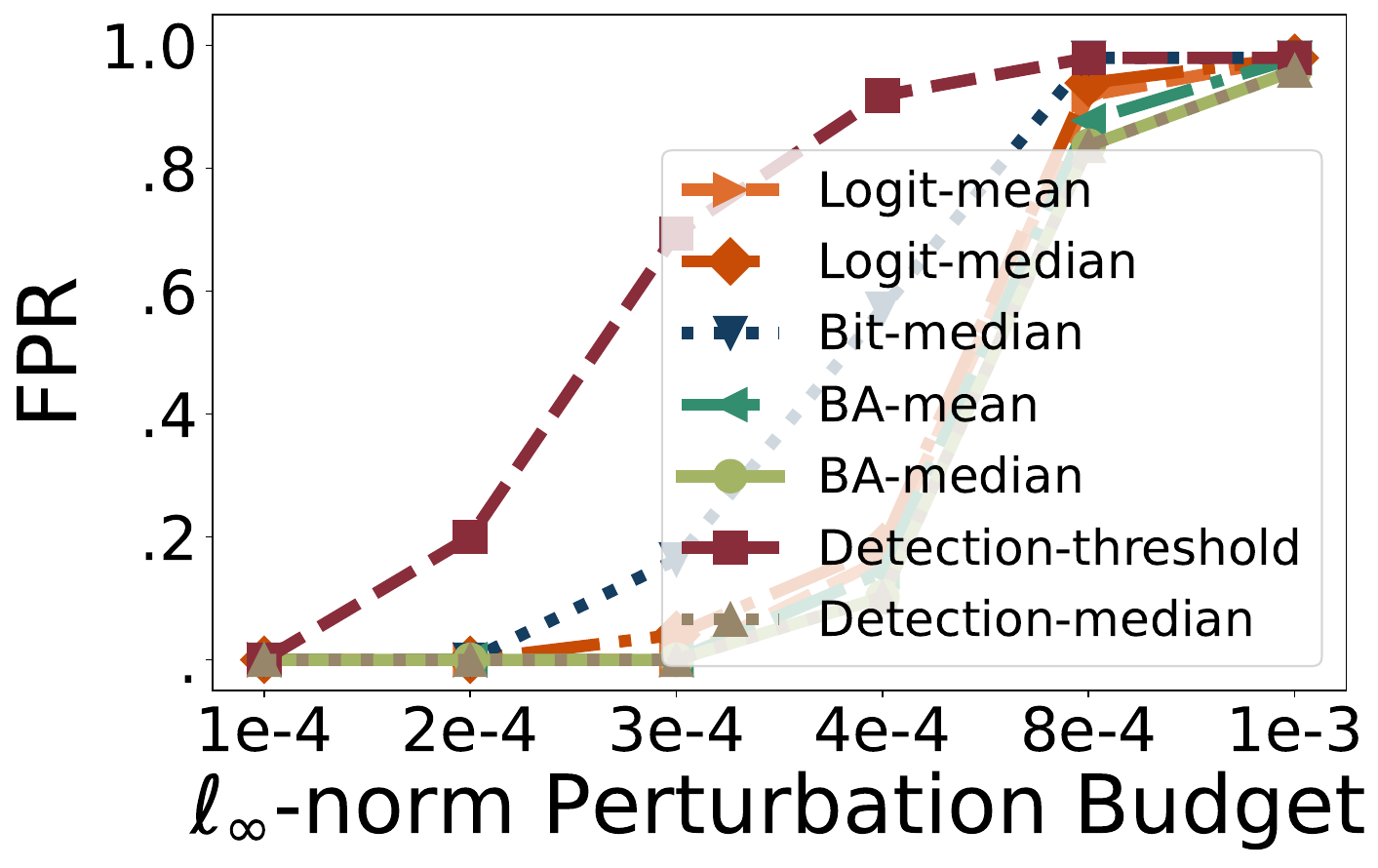}
        \caption{Aggregation}
        \label{subfig:wbox1forgeryaggregate}
    \end{subfigure}
    \caption{White-box watermark forgery results in the first scenario.}
\end{wrapfigure}

\myparatight{Comparison across watermarking methods} Figure~\ref{subfig:wbox1watermark} and~\ref{subfig:wbox1forgerywatermark} present the results of both watermark removal and forgery attacks across three watermarking methods. We have several observations. First, all existing video watermarking methods fail under the white-box setting—both FNR and FPR reach 1 even with small perturbations. This indicates that an attacker can effectively remove or forge a watermark while maintaining the video's visual quality. Second, among the three watermarking methods, VideoSeal has better robustness against watermark removal attacks, while StegaStamp is consistently more robust against forgery attacks. Third, the perturbations required for forgery attacks are significantly smaller than those needed for removal attacks, suggesting that watermark forgery is easier in the white-box setting. This is primarily because the watermark encoder and decoder are adversarially trained to resist removal perturbations, but forgery perturbations are largely ignored during training.

\myparatight{Comparison across aggregation strategies} 
We evaluate seven aggregation strategies on StegaStamp and VideoSeal, whose watermark decoder work on frame-level. Figure~\ref{subfig:wbox1aggregate} and~\ref{subfig:wbox1forgeryaggregate} present the results for StegaStamp. Results for VideoSeal are shown in Figure~\ref{fig:scenario1 aggregation VideoSeal} in the Appendix. We highlight several key observations. First, logit-level aggregation strategy consistently outperforms BA-level aggregation. Second, the detection-threshold aggregation strategy is the most robust against removal attacks, but it is the least robust against forgery attacks. This is because this strategy detects a video as watermarked as long as a predefined number of frames are detected as such. Therefore, a successful removal attack must target most frames in the video, whereas a successful forgery attack requires only a few frames to be falsely detected as watermarked. Third, detection-median aggregation is the most robust strategy against forgery attacks, as an attacker must successfully alter about half of the frames to influence the median-based detection result.

\myparatight{Comparison across generative models and styles}
Figures~\ref{subfig:wbox1model} and~\ref{subfig:wbox1style} present the results of watermark removal attacks across videos generated by different models and different video styles, respectively. Forgery results are not applicable in this case, as our real-world dataset is not generated by models and does not include style labels. We observe notable robustness gaps against watermark removal attacks both across models and across video styles. To statistically validate these differences, we conduct two-tailed t-tests under the null hypothesis that there is no difference in FNRs. We use a significance level of $\alpha = 0.05$. The calculated $p$-value for differences among models $\approx0.038<\alpha$, and the $p$-value for differences among video styles $\approx0.029<\alpha$. These results indicate that the observed robustness gaps across both models and video styles are statistically significant.

\subsubsection{Second Scenario: Attacking a Subset of Frames with Arbitrary Perturbations}
In the second scenario, an attacker adds arbitrary perturbations to a fraction of frames in the video to remove or forge the watermark. The attack objective is to manipulate the decoded logits of the perturbed frames to be extremely large or small, thereby dominating the final detection result. 
Since both REVMark and StegaStamp use a sigmoid activation in the logit layer—constraining their output logits to the range [0,1]—we only evaluate VideoSeal in this scenario.

\begin{wrapfigure}{r}{0.6\textwidth}
    \centering
    \begin{subfigure}{0.48\linewidth}
        \centering
        \includegraphics[width=\linewidth]{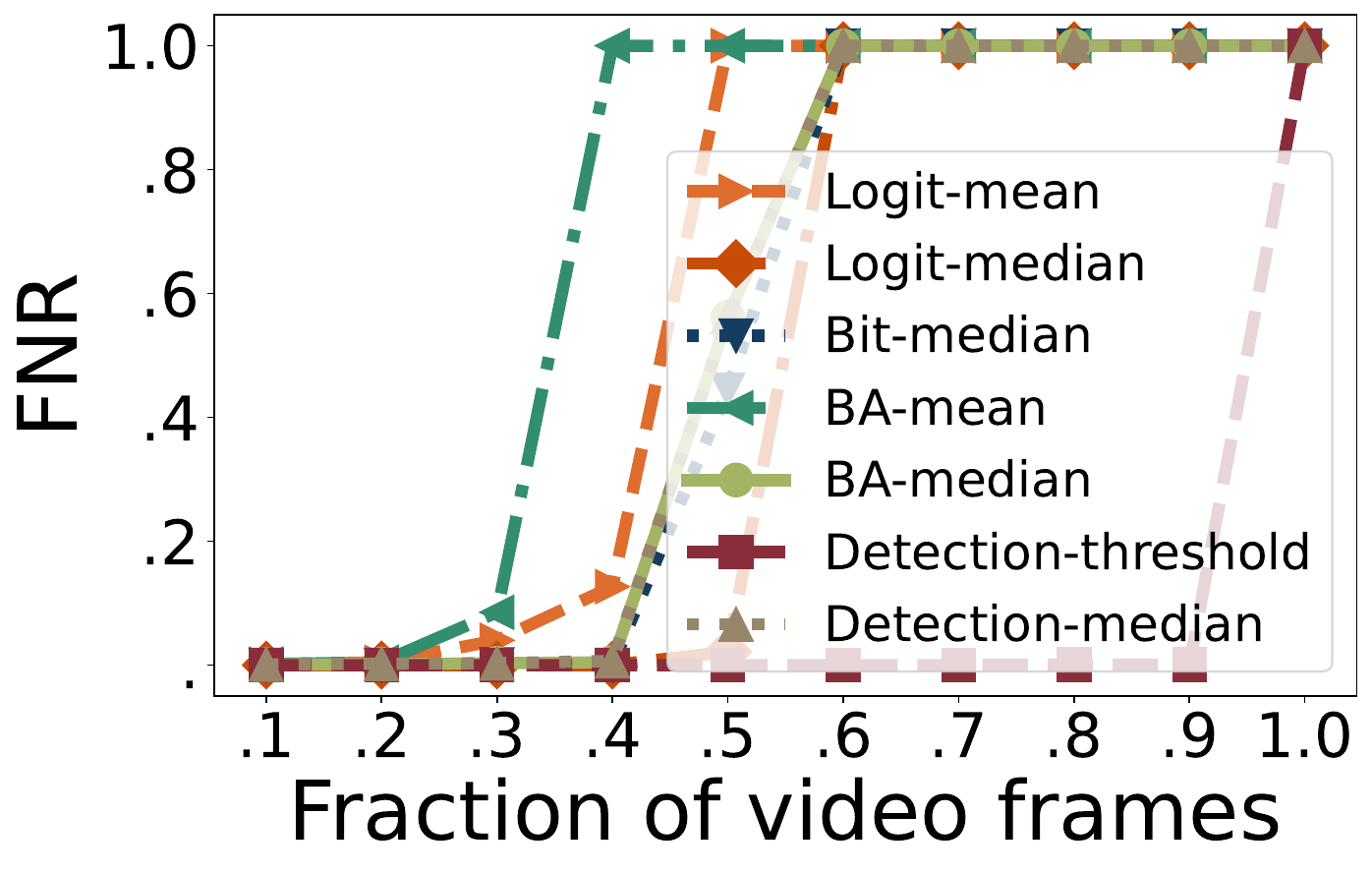}
        \caption{Removal}
    \end{subfigure}
    \hfill
    \begin{subfigure}{0.48\linewidth}
        \centering
        \includegraphics[width=\linewidth]{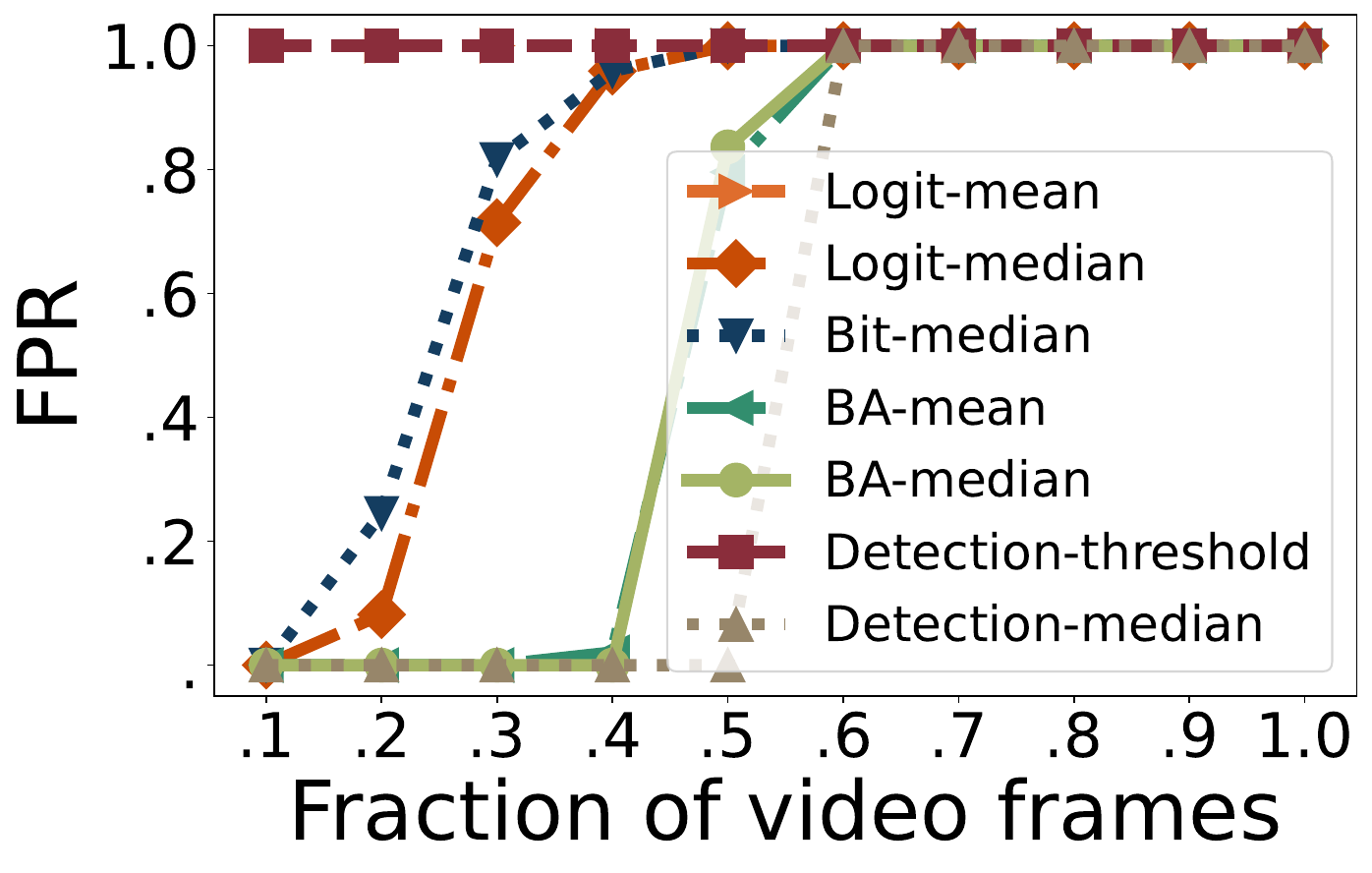}
        \caption{Forgery}
    \end{subfigure}
    \caption{\label{fig:scenario2}White-box attack results in the second scenario with different aggregation strategies.}
\end{wrapfigure}

\myparatight{Comparison across aggregation strategies for VideoSeal} Figure~\ref{fig:scenario2} presents VideoSeal's performance under white-box attacks in the second scenario. The x-axis represents the fraction of frames perturbed by the attacker. We have several observations. First, the logit-mean and BA-mean aggregation strategies are the least robust against watermark removal attacks. This vulnerability arises because the attacker optimizes the logits to have signs opposite to $w_g - 0.5$, which results in low bitwise accuracy for the perturbed frames. Second, logit-mean and detection-threshold aggregation strategies are the most vulnerable to watermark forgery attacks. In these cases, the attacker only needs to successfully perturb a small number of frames—exceeding the detection threshold—to forge a watermark. Third, BA-median and detection-median aggregation strategies demonstrate relatively strong and stable performance. This robustness comes from the fact that perturbing a subset of frames does not significantly affect the overall median, making these aggregation strategies based on median more robust.
\subsection{Robustness against Black-box Video Perturbations}
In the black-box setting, the watermark detection API is queried multiple times with perturbed video to iteratively find an adversarial perturbation based on the feedback. VideoShield is excluded from this evaluation due to the inefficiency of its detection process, which relies on time-consuming DDIM inversion. Since black-box attacks are computationally expensive, we use a subset of videos to conduct experiments (40 videos per model and style). For removal attacks in this setting, by default, we only evaluate on videos generated by SVD and realistic style, using BA-mean aggregation.

\begin{figure}[t!]
    \centering
    \begin{subfigure}{.23\linewidth}
        \centering
        \includegraphics[width=\linewidth]{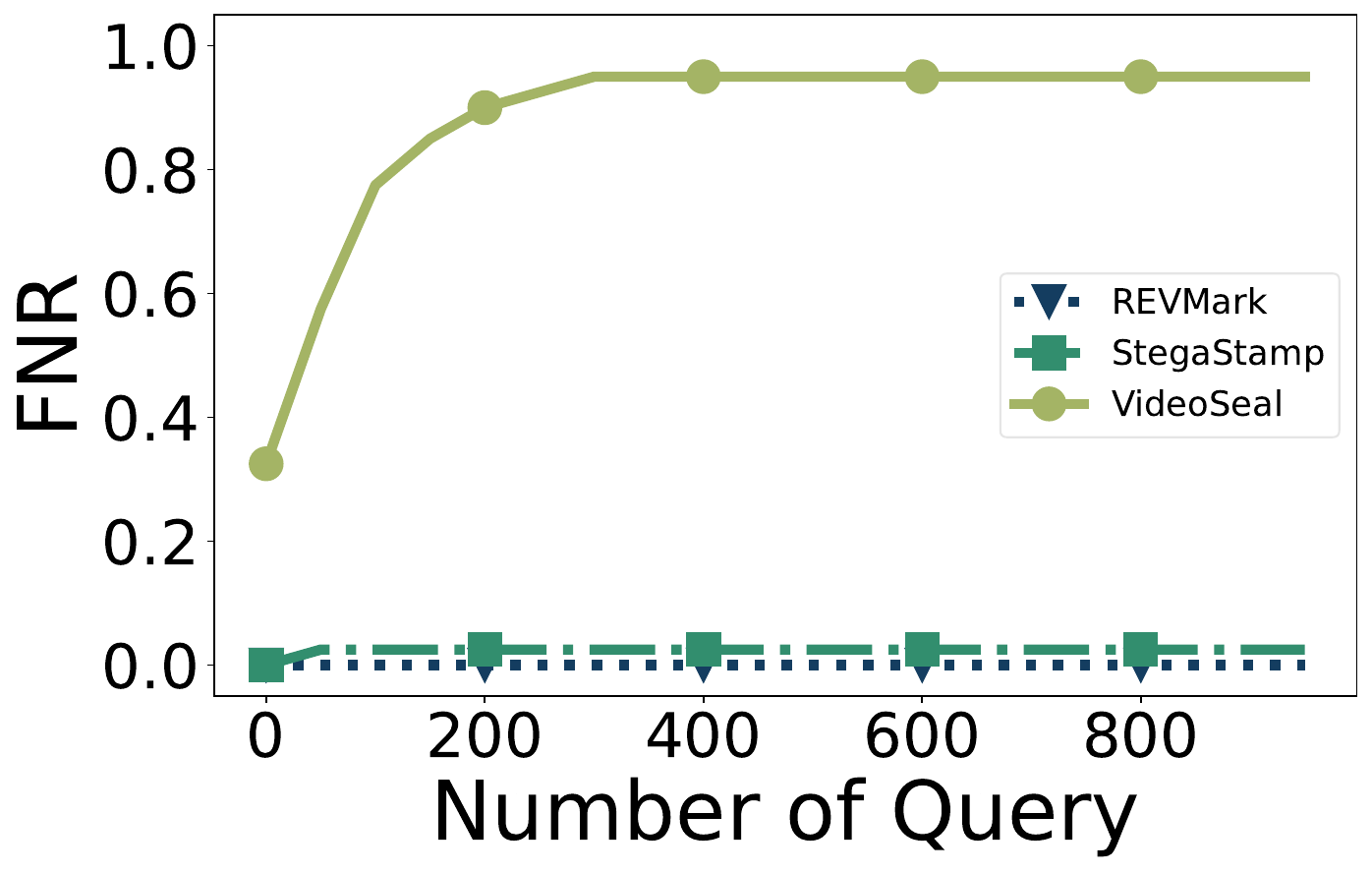}
        \caption{Watermarking}
    \end{subfigure}
    \begin{subfigure}{.23\linewidth}
        \centering
        \includegraphics[width=\linewidth]{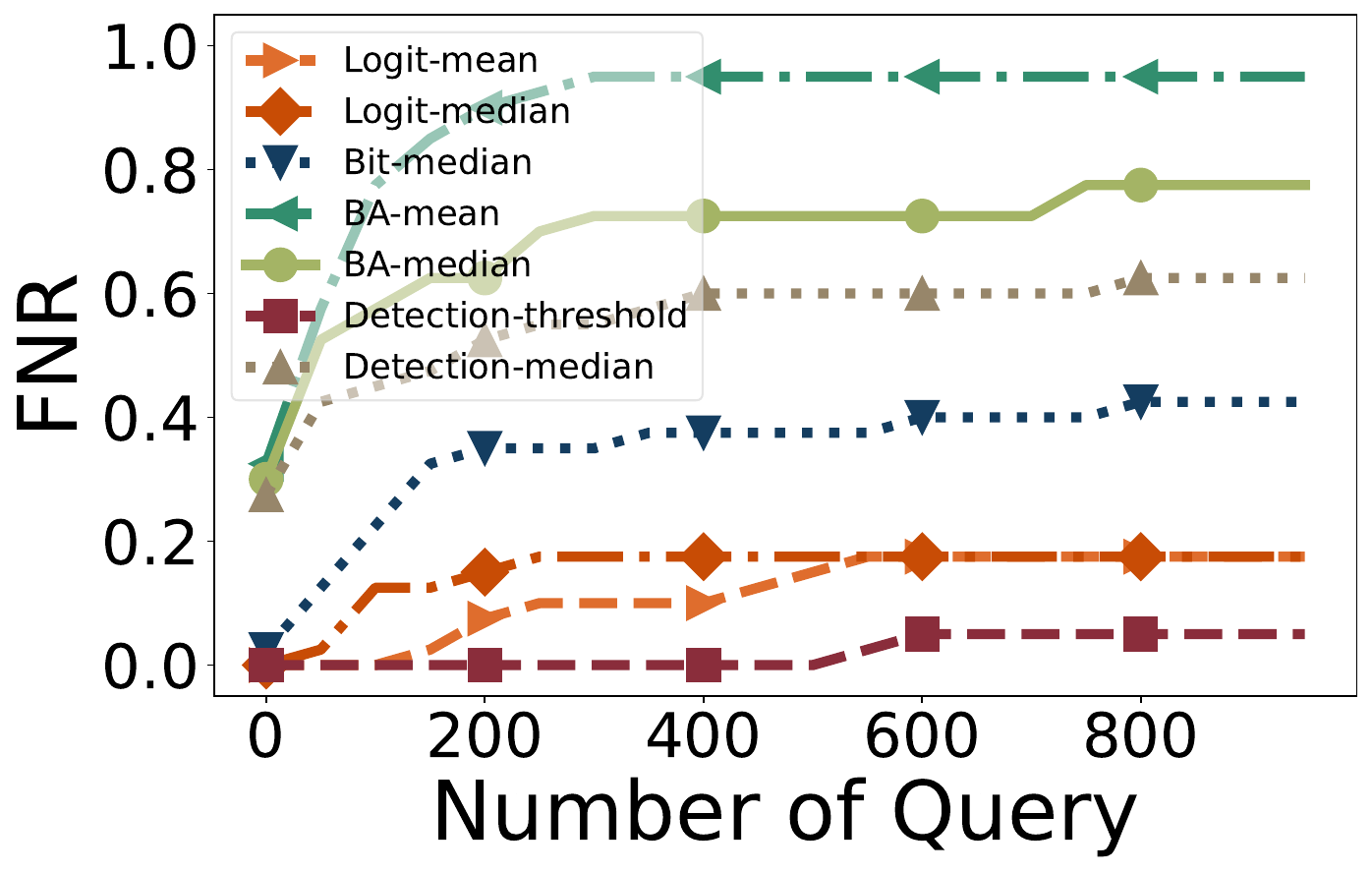}
        \caption{Aggregation}
    \end{subfigure}
    \begin{subfigure}{.23\linewidth}
        \centering
        \includegraphics[width=\linewidth]{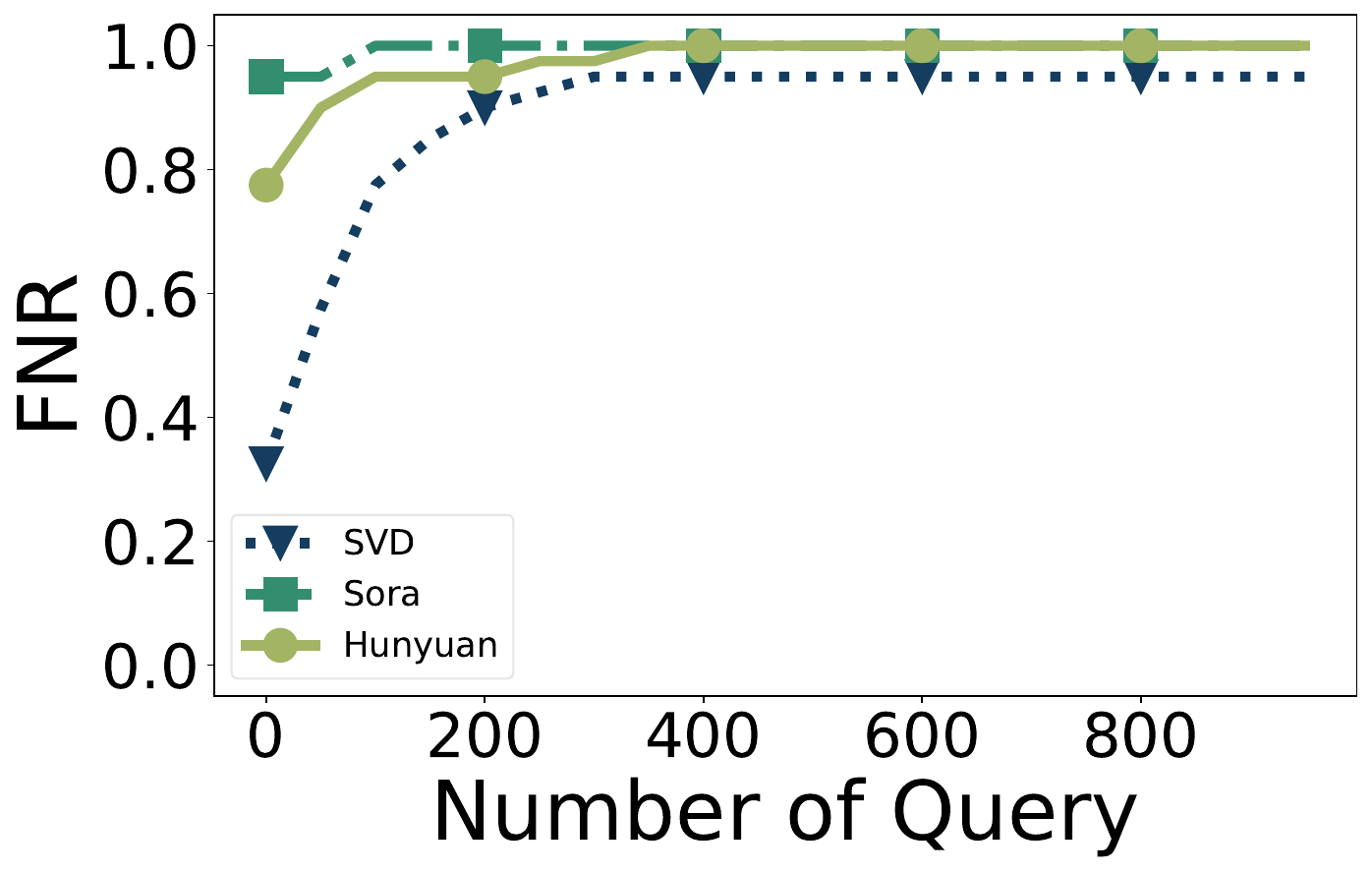}
        \caption{Model}
    \end{subfigure}
    \begin{subfigure}{.23\linewidth}
        \centering
        \includegraphics[width=\linewidth]{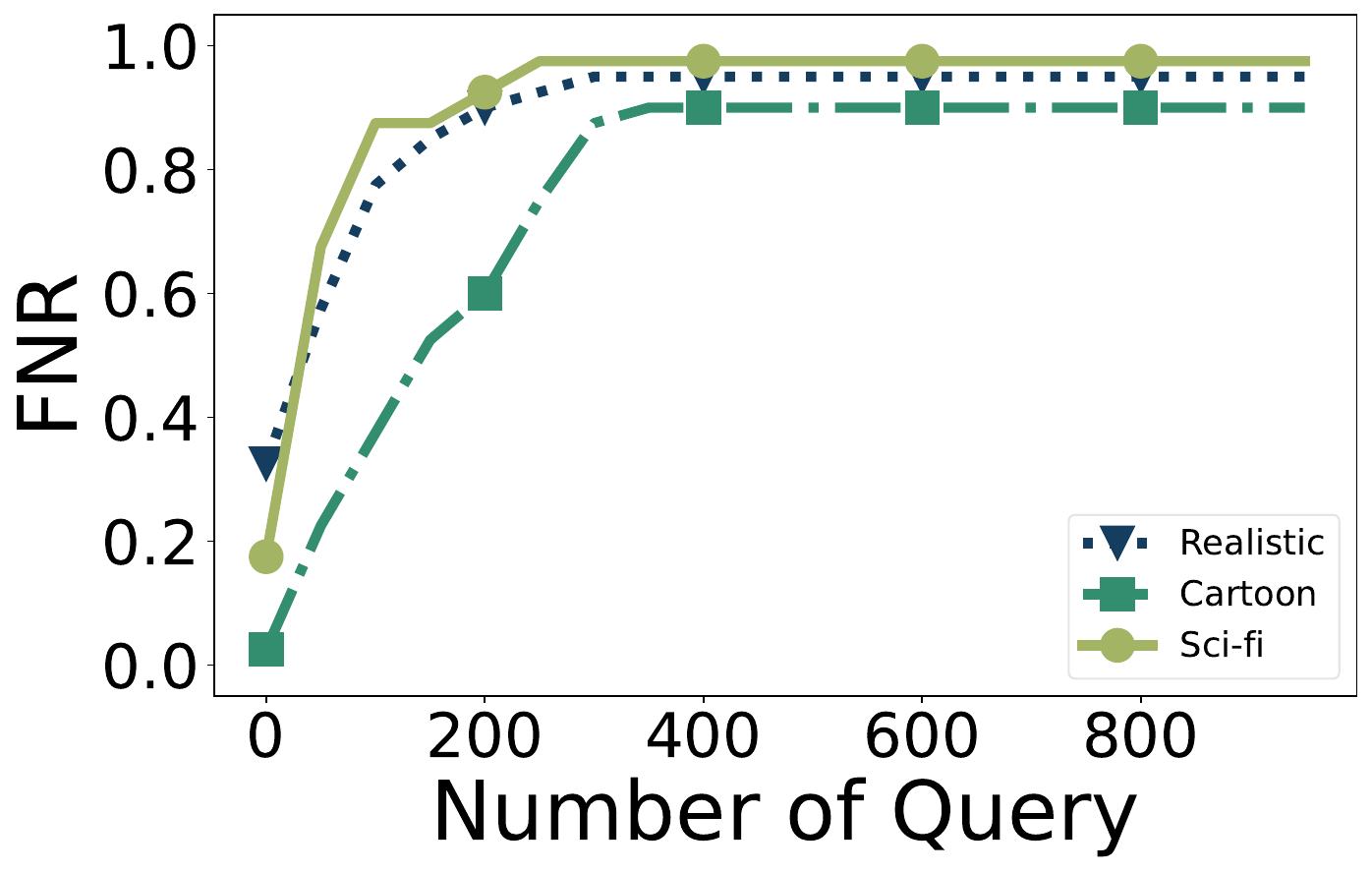}
        \caption{Style}
    \end{subfigure}
    \caption{\label{fig:square attack removal}Square Attack watermark removal results. Perturbations are $l_{\infty}$ bounded by 0.05.}
    \vspace{-2mm}
\end{figure}

\myparatight{Square Attack (score-based)} In our experiments, we follow the default settings of Square Attack~\cite{andriushchenko2020square}, with perturbations constrained by an $l_{\infty}$ bound of 0.05. Figure~\ref{fig:square attack removal} presents the results of Square Attack for watermark removal; results for forgery attacks are provided in the Appendix~\ref{sec:blackbox forgery}. We summarize four key observations: First, VideoSeal is significantly more vulnerable to removal attacks compared to StegaStamp and REVMark. This is primarily because VideoSeal is less robust to Gaussian noise, as shown in Figure~\ref{subfig:videoseal gaussian} in the Appendix, and the perturbations introduced by Square Attack exhibit noise-like patterns that mimic the effect of Gaussian noise, making them particularly effective against the less noise-robust VideoSeal. StegaStamp and REVMark require larger perturbations for successful watermark removal, as shown in Figure~\ref{fig:square attack larger} in the Appendix. Second, among aggregation strategies, detection-threshold aggregation is the most robust against watermark removal, and logit-level aggregation consistently outperforms BA-level aggregation, which aligns with our findings in the white-box setting. Third, across generative models, videos generated by SVD exhibit greater robustness to watermark removal attacks, whereas videos generated by Sora are more vulnerable. Fourth, videos in the cartoon style are more robust, while those in the sci-fi style are more vulnerable to watermark removal attacks.

\begin{figure}[t!]
    \centering
    \begin{subfigure}{.23\linewidth}
        \centering
        \includegraphics[width=\linewidth]{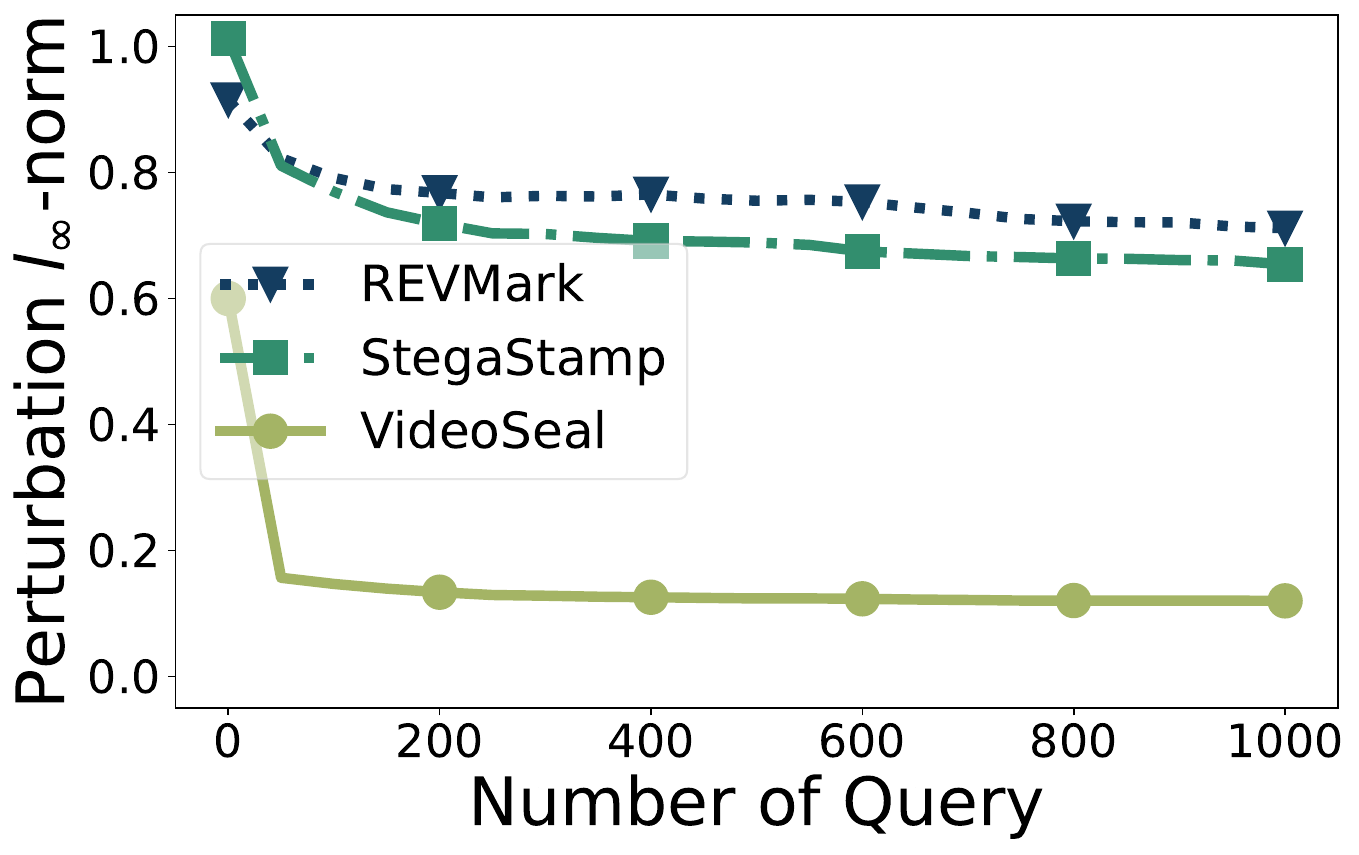}
        \caption{Watermarking}
    \end{subfigure}
    \begin{subfigure}{.23\linewidth}
        \centering
        \includegraphics[width=\linewidth]{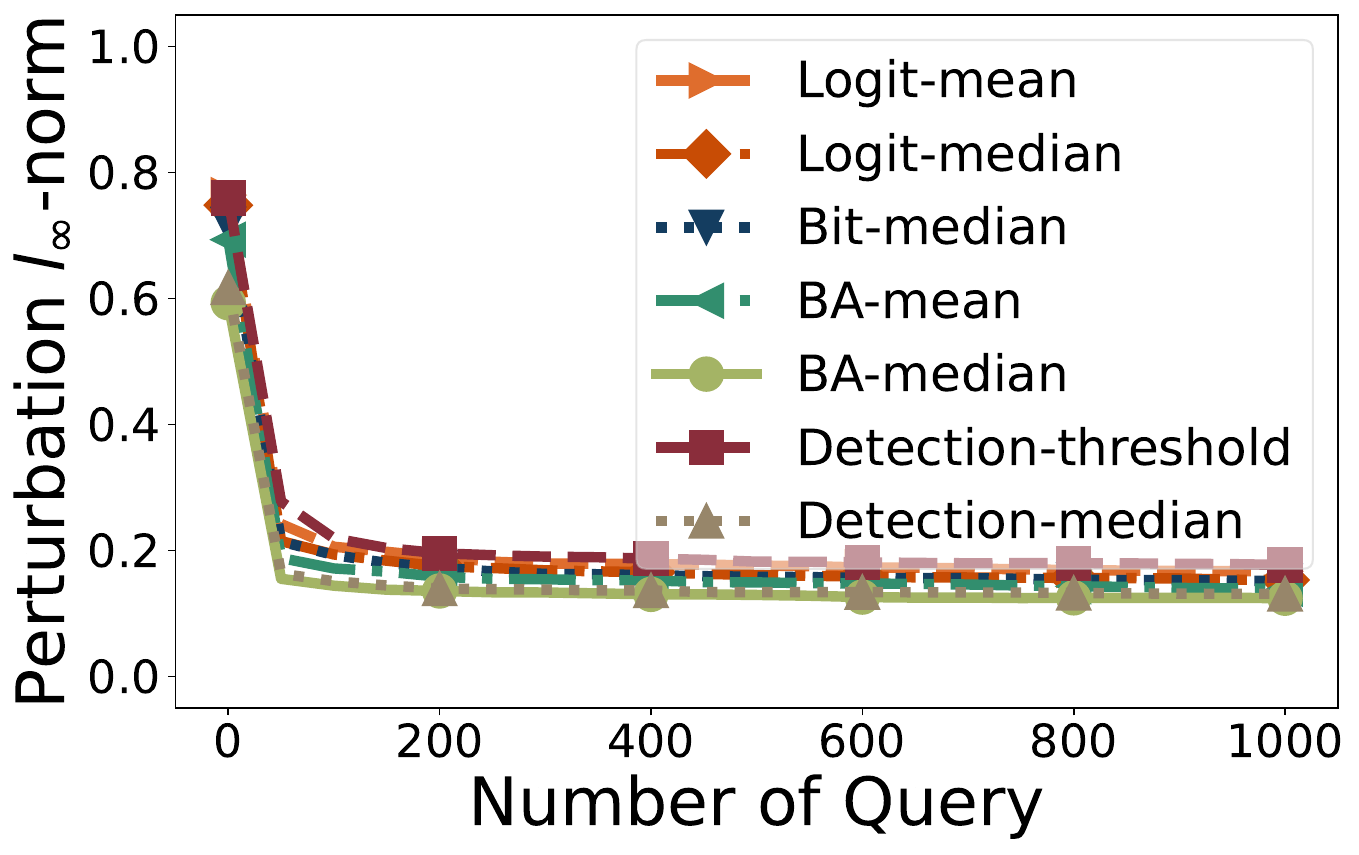}
        \caption{Aggregation}
    \end{subfigure}
    \begin{subfigure}{.23\linewidth}
        \centering
        \includegraphics[width=\linewidth]{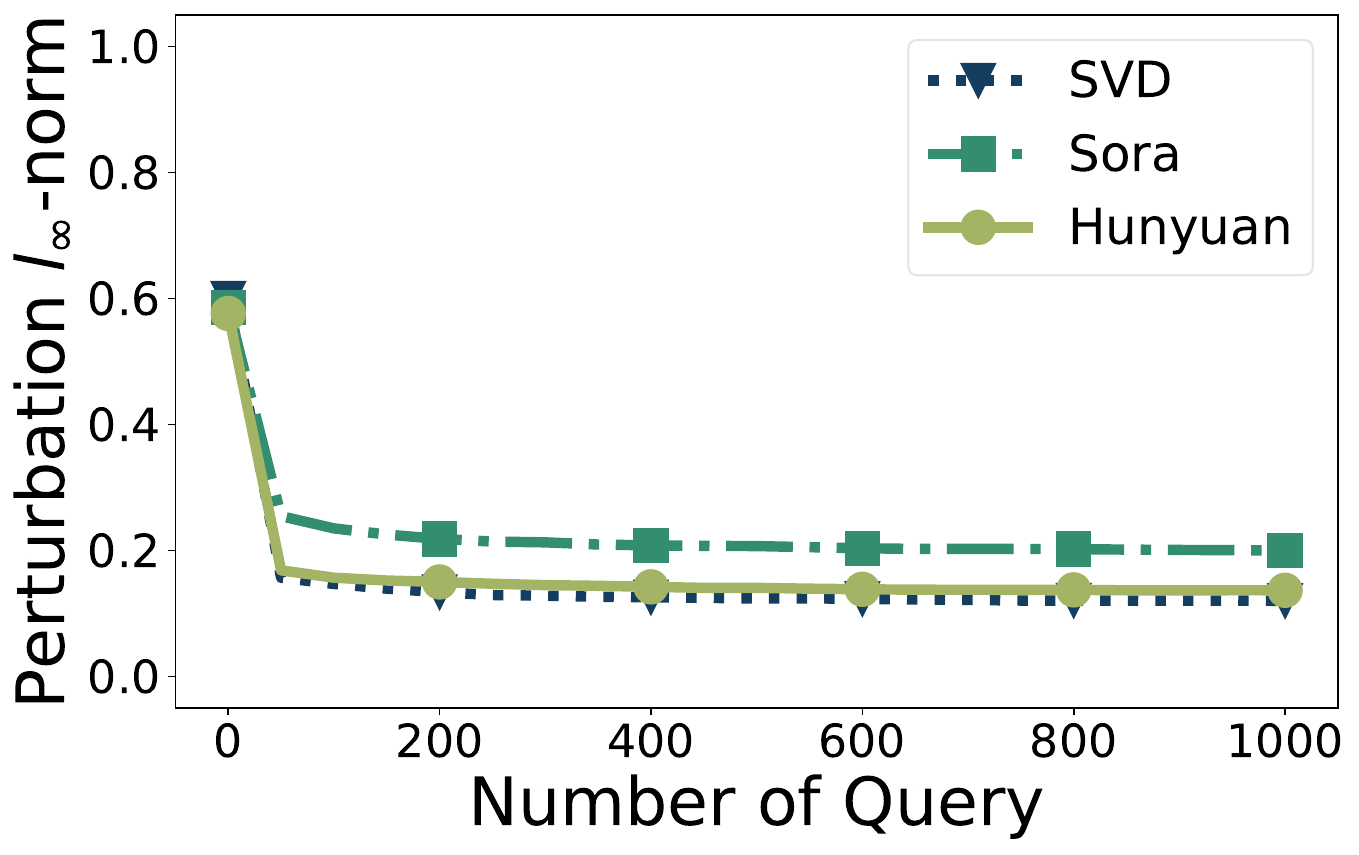}
        \caption{Model}
    \end{subfigure}
    \begin{subfigure}{.23\linewidth}
        \centering
        \includegraphics[width=\linewidth]{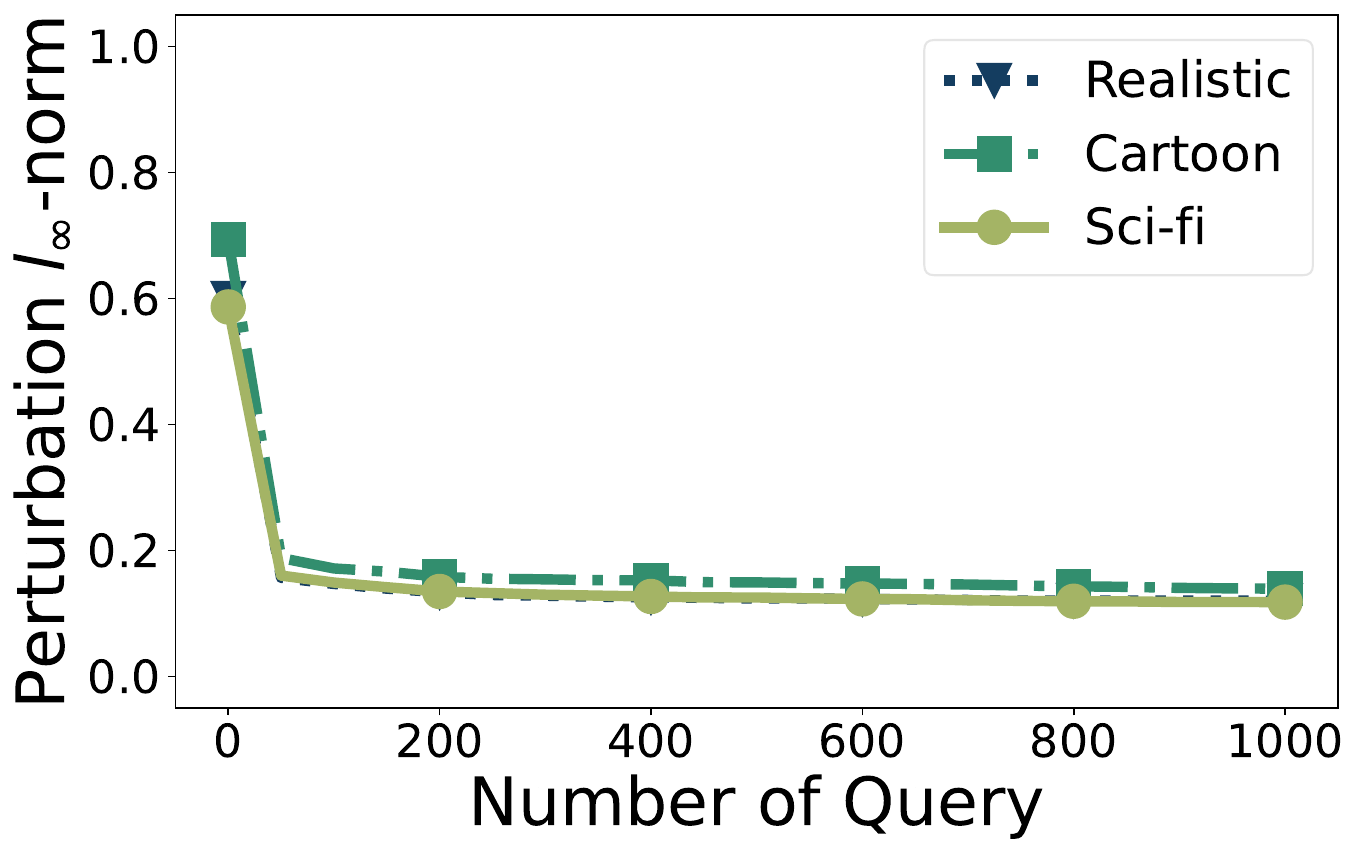}
        \caption{Style}
    \end{subfigure}
    \caption{\label{fig:triangle attack removal}Triangle Attack watermark removal results.}
    \vspace{-2mm}
\end{figure}

\myparatight{Triangle Attack (label-based)} In our experiments, we extend Triangle Attack~\cite{wang2022triangle} to the video setting and follow its default configuration. Figure~\ref{fig:triangle attack removal} shows the results for watermark removal; results for watermark forgery are provided in the Appendix~\ref{sec:blackbox forgery}. We summarize several key findings: First, VideoSeal requires much smaller perturbations to be successfully attacked, primarily due to the initialization process. We iteratively add Gaussian noise to the watermarked videos until an initial perturbed video is misclassified as unwatermarked. Since VideoSeal is not robust to Gaussian noise, the $l_{\infty}$ norm of the initial perturbation tends to be relatively small. Second, we observe similar trends across aggregation strategies as in previous experiments. Third, videos generated by Sora are more robust against watermark removal under Triangle Attack. Fourth, the perturbation size decreases most significantly within the first 100 queries, after which it drops slowly. We observe no significant difference in robustness across different video styles.

\subsection{Robustness against Common Video Perturbations}

\begin{figure}[t!]
    \centering
    \begin{subfigure}{.23\linewidth}
        \centering
        \includegraphics[width=\linewidth]{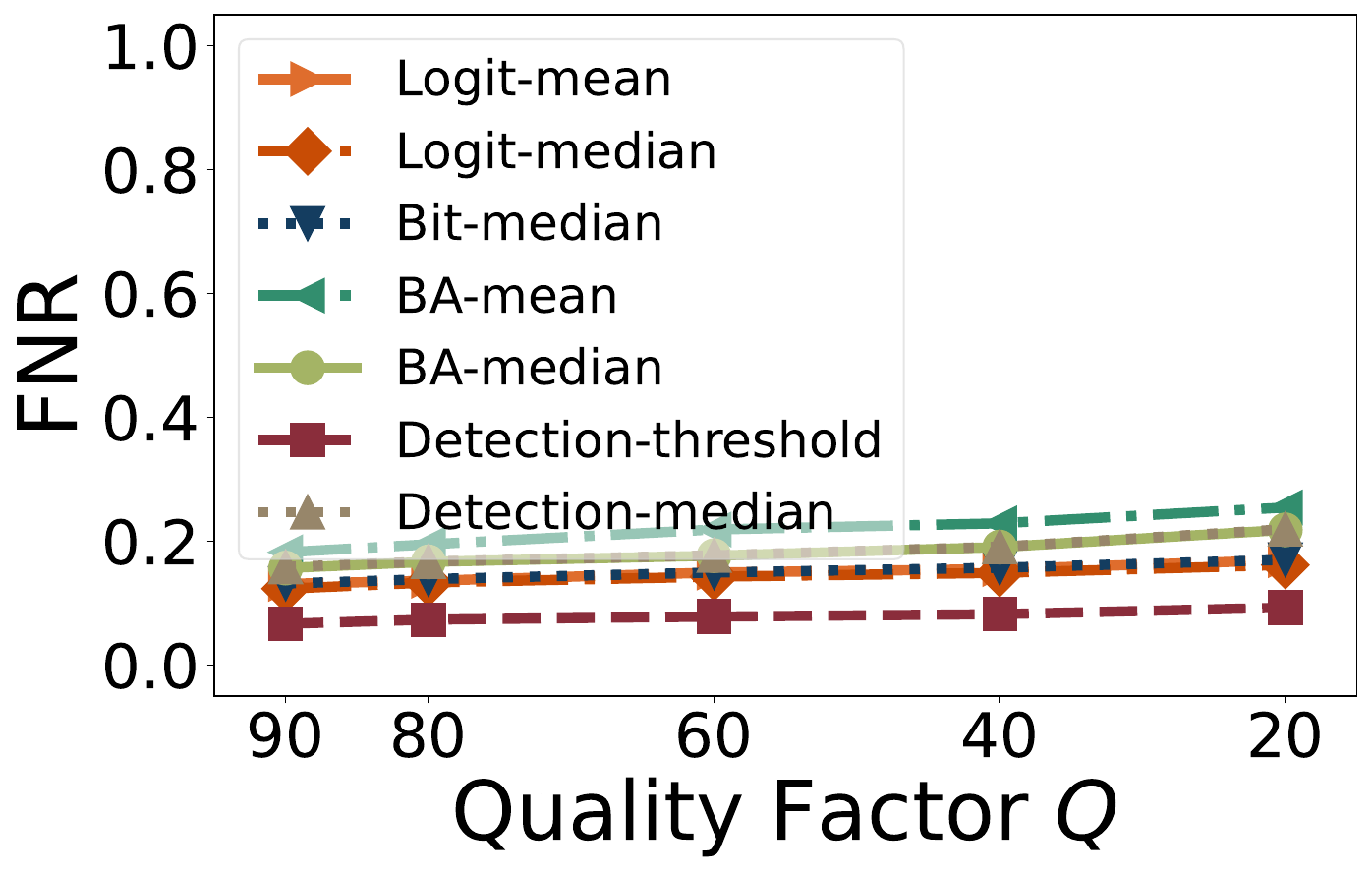}
        \caption{JPEG}
    \end{subfigure}
    \begin{subfigure}{.23\linewidth}
        \centering
        \includegraphics[width=\linewidth]{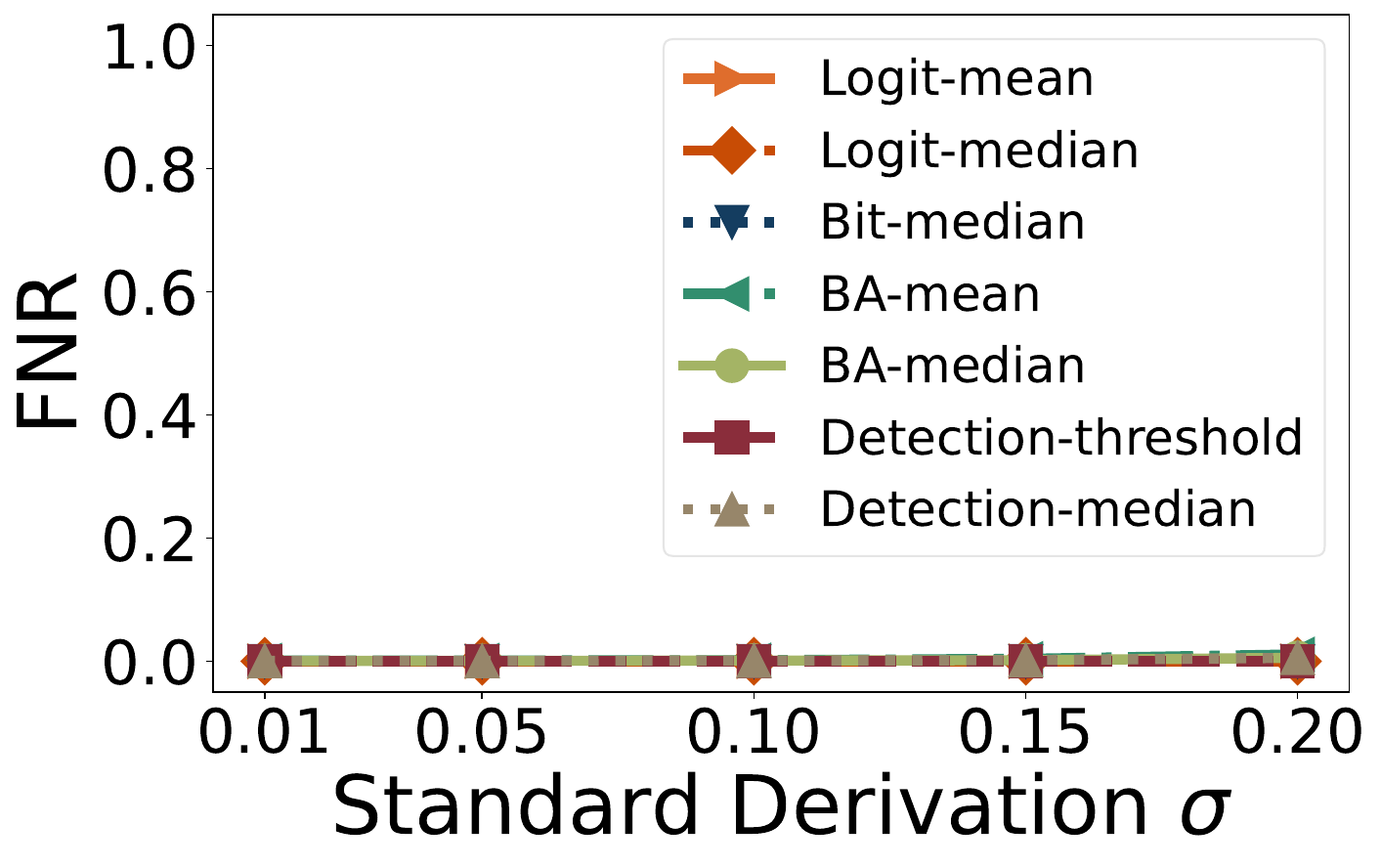}
        \caption{Gaussian Noise}
    \end{subfigure}
    \begin{subfigure}{.23\linewidth}
        \centering
        \includegraphics[width=\linewidth]{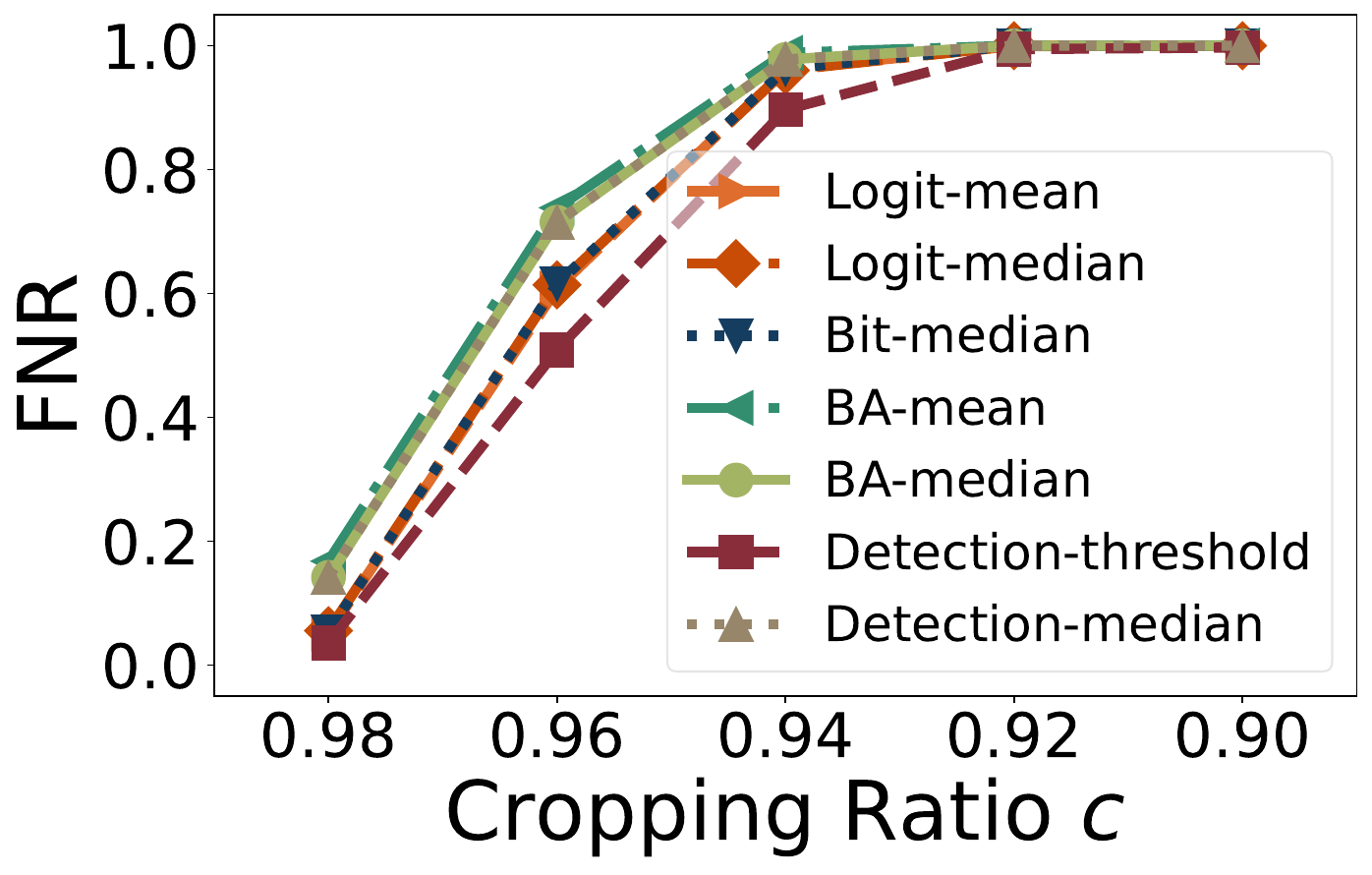}
        \caption{Cropping}
    \end{subfigure}
    \begin{subfigure}{.23\linewidth}
        \centering
        \includegraphics[width=\linewidth]{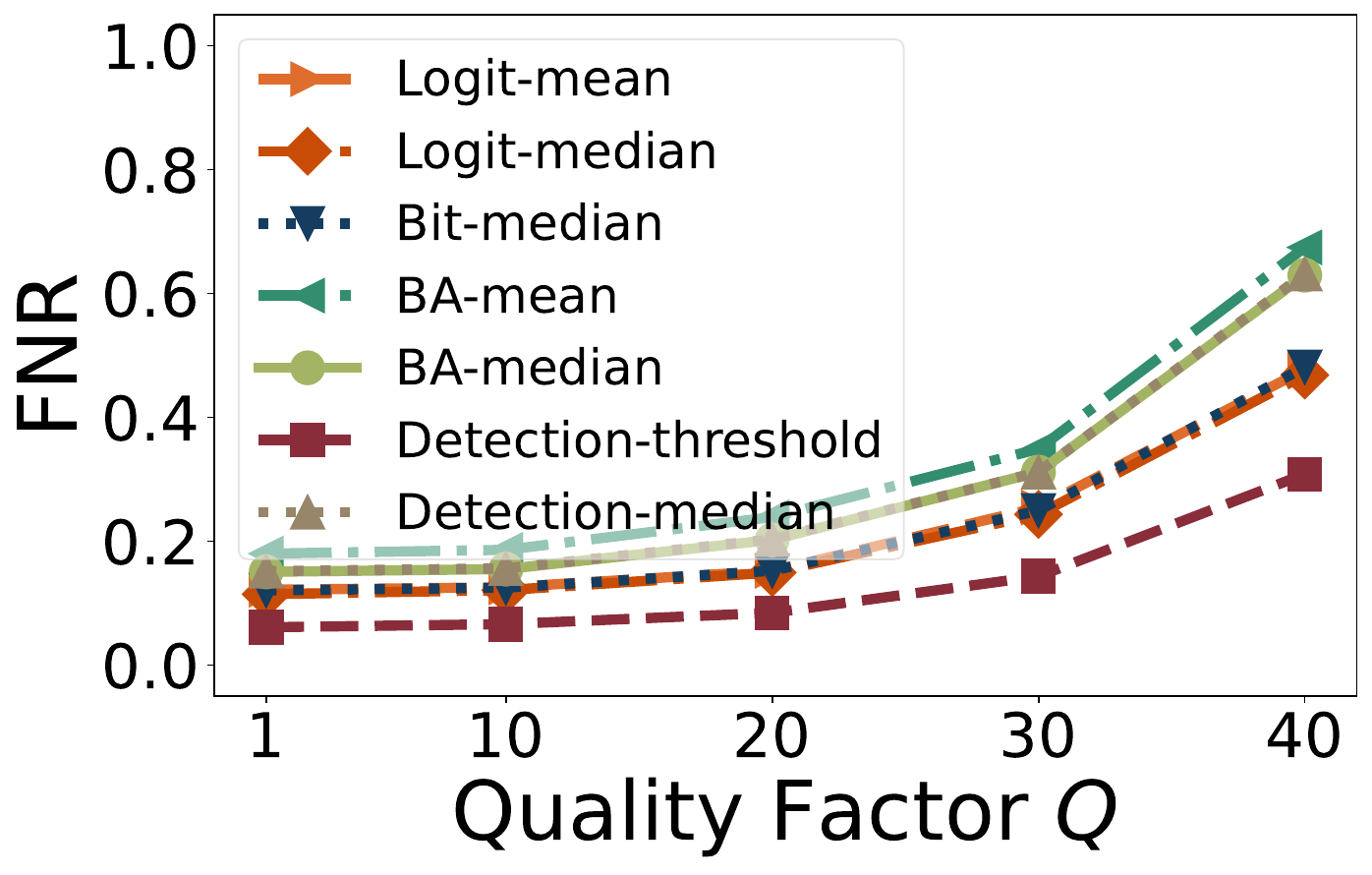}
        \caption{MPEG-4}
    \end{subfigure}
    \caption{\label{fig:aggregation_main}Common perturbation watermark removal results for StegaStamp with different aggregation.}
    \vspace{-2mm}
\end{figure}

Figures~\ref{fig:aggregation_main} and Figure~\ref{fig:watermark}–\ref{fig:utility} in the Appendix present results under common video perturbations. We summarize several key observations: First, existing video watermarking methods are generally robust to common video perturbations, particularly when video quality is preserved or the perturbation type is included in adversarial training~\cite{goodfellow2014explaining}. For example, all evaluated methods are robust to Gaussian blurring, as this perturbation maintains visual quality and is commonly used during adversarial training. Second, the robustness of watermarks varies across different types of perturbations. Specifically, all methods are robust to frame averaging, frame switching, and frame removal perturbations, as these operations minimally alter the video content and the watermark detection are not heavily dependent on temporal consistency. In contrast, watermarking methods are more vulnerable to both frame-level and video-level compression such as JPEG and MPEG-4. Third, when perturbations are large enough to noticeably degrade visual quality, video watermarks can be removed. This is because large perturbations can distort the watermark structure, making it difficult for the decoder to extract the correct watermark. For instance, when MPEG-4 compression is applied with a quality factor of $Q = 40$, the FNR begins to increase for all methods. Fourth, existing watermarking methods are robust to watermark forgery using common perturbations, as shown in Figure~\ref{fig:watermark fpr} in the Appendix. In particular, the FPRs remain near zero regardless of the applied perturbation. This robustness is likely because the added perturbations do not mimic the structural patterns of valid watermarks, making watermark forgery substantially more difficult than watermark removal in the no-box setting. A more detailed analysis can be found in Appendix~\ref{sec:nobox analysis}.
\section{Conclusion}
In this work, we introduce \benchmarkname, the first systematic benchmark for evaluating the robustness of video watermarking methods against both watermark removal and forgery perturbations. Our study includes a comprehensive AI-generated dataset called \datasetname, created using three video generative models. We evaluate four state-of-the-art video watermarking methods under 12 types of perturbations across white-box, black-box, and no-box threat scenarios. Experimental results show that existing video watermarks are not robust to a wide range of perturbations. In addition, we extend image watermarking methods to the video domain and propose seven aggregation strategies, among which logit-level aggregation consistently outperforms BA-level aggregation. This benchmark fosters further research toward developing more robust video watermarking.

\bibliographystyle{plain}
\bibliography{refs}

\begin{thebibliography}{10}

\bibitem{fakevideo}
Bobby Allyn.
\newblock Deepfake video of zelenskyy could be 'tip of the iceberg' in info war.
\newblock https://www.npr.org/2022/03/16/1087062648/deepfake-video-zelenskyy-experts-war-manipulation-ukraine-russia.
\newblock Online; accessed March 16, 2022.

\bibitem{an2024waves}
Bang An, Mucong Ding, Tahseen Rabbani, Aakriti Agrawal, Yuancheng Xu, Chenghao Deng, Sicheng Zhu, Abdirisak Mohamed, Yuxin Wen, Tom Goldstein, et~al.
\newblock Waves: Benchmarking the robustness of image watermarks.
\newblock In {\em International Conference on Machine Learning}, 2024.

\bibitem{andriushchenko2020square}
Maksym Andriushchenko, Francesco Croce, Nicolas Flammarion, and Matthias Hein.
\newblock Square attack: a query-efficient black-box adversarial attack via random search.
\newblock In {\em European Conference on Computer Vision}, 2020.

\bibitem{christodorescu2024securing}
Mihai Christodorescu, Ryan Craven, Soheil Feizi, Neil Gong, Mia Hoffmann, Somesh Jha, Zhengyuan Jiang, Mehrdad~Saberi Kamarposhti, John Mitchell, Jessica Newman, et~al.
\newblock Securing the future of genai: Policy and technology.
\newblock {\em arXiv}, 2024.

\bibitem{chu2020learning}
Mengyu Chu, You Xie, Jonas Mayer, Laura Leal-Taix{\'e}, and Nils Thuerey.
\newblock Learning temporal coherence via self-supervision for gan-based video generation.
\newblock {\em ACM Transactions on Graphics (TOG)}, 2020.

\bibitem{dhariwal2021diffusion}
Prafulla Dhariwal and Alexander Nichol.
\newblock Diffusion models beat gans on image synthesis.
\newblock {\em Conference on Neural Information Processing Systems}, 2021.

\bibitem{fernandez2024video}
Pierre Fernandez, Hady Elsahar, I~Zeki Yalniz, and Alexandre Mourachko.
\newblock Video seal: Open and efficient video watermarking.
\newblock {\em arXiv}, 2024.

\bibitem{goodfellow2014explaining}
Ian~J Goodfellow, Jonathon Shlens, and Christian Szegedy.
\newblock Explaining and harnessing adversarial examples.
\newblock In {\em International Conference on Learning Representations}, 2015.

\bibitem{hore2010image}
Alain Hore and Djemel Ziou.
\newblock Image quality metrics: Psnr vs. ssim.
\newblock In {\em International Conference on Pattern Recognition}, 2010.

\bibitem{hu2025videoshield}
Runyi Hu, Jie Zhang, Yiming Li, Jiwei Li, Qing Guo, Han Qiu, and Tianwei Zhang.
\newblock Videoshield: Regulating diffusion-based video generation models via watermarking.
\newblock In {\em International Conference on Learning Representations}, 2025.

\bibitem{hu2024stable}
Yuepeng Hu, Zhengyuan Jiang, Moyang Guo, and Neil Gong.
\newblock Stable signature is unstable: Removing image watermark from diffusion models.
\newblock {\em arXiv}, 2024.

\bibitem{hu2024transfer}
Yuepeng Hu, Zhengyuan Jiang, Moyang Guo, and Neil Gong.
\newblock A transfer attack to image watermarks.
\newblock In {\em International Conference on Learning Representations}, 2025.

\bibitem{jiang2024watermark}
Zhengyuan Jiang, Moyang Guo, Yuepeng Hu, and Neil Zhenqiang~Gong.
\newblock Watermark-based detection and attribution of ai-generated content.
\newblock {\em arXiv}, 2024.

\bibitem{jiang2023evading}
Zhengyuan Jiang, Jinghuai Zhang, and Neil~Zhenqiang Gong.
\newblock Evading watermark based detection of ai-generated content.
\newblock In {\em ACM SIGSAC Conference on Computer and Communications Security}, 2023.

\bibitem{kay2017kinetics}
Will Kay, Joao Carreira, Karen Simonyan, Brian Zhang, Chloe Hillier, Sudheendra Vijayanarasimhan, Fabio Viola, Tim Green, Trevor Back, Paul Natsev, Mustafa Suleyman, and Andrew Zisserman.
\newblock The kinetics human action video dataset.
\newblock {\em arXiv}, 2017.

\bibitem{kong2024hunyuanvideo}
Weijie Kong, Qi~Tian, Zijian Zhang, Rox Min, Zuozhuo Dai, Jin Zhou, Jiangfeng Xiong, Xin Li, Bo~Wu, Jianwei Zhang, et~al.
\newblock Hunyuanvideo: A systematic framework for large video generative models.
\newblock {\em arXiv}, 2024.

\bibitem{lukas2023leveraging}
Nils Lukas, Abdulrahman Diaa, Lucas Fenaux, and Florian Kerschbaum.
\newblock Leveraging optimization for adaptive attacks on image watermarks.
\newblock In {\em International Conference on Learning Representations}, 2024.

\bibitem{madry2018towards}
Aleksander Madry, Aleksandar Makelov, Ludwig Schmidt, Dimitris Tsipras, and Adrian Vladu.
\newblock Towards deep learning models resistant to adversarial attacks.
\newblock In {\em International Conference on Learning Representations}, 2018.

\bibitem{nie2022diffusion}
Weili Nie, Brandon Guo, Yujia Huang, Chaowei Xiao, Arash Vahdat, and Animashree Anandkumar.
\newblock Diffusion models for adversarial purification.
\newblock In {\em International Conference on Machine Learning}, 2022.

\bibitem{gpt}
OpenAI.
\newblock Gpt-4.
\newblock https://chatgpt.com/.
\newblock Online; accessed March 14, 2023.

\bibitem{sora}
OpenAI.
\newblock Sora.
\newblock https://sora.chatgpt.com/explore.
\newblock Online; accessed November 21, 2023.

\bibitem{powell1964efficient}
Michael~JD Powell.
\newblock An efficient method for finding the minimum of a function of several variables without calculating derivatives.
\newblock {\em The Computer Journal}, 1964.

\bibitem{saberi2023robustness}
Mehrdad Saberi, Vinu~Sankar Sadasivan, Keivan Rezaei, Aounon Kumar, Atoosa Chegini, Wenxiao Wang, and Soheil Feizi.
\newblock Robustness of ai-image detectors: Fundamental limits and practical attacks.
\newblock In {\em International Conference on Learning Representations}, 2024.

\bibitem{svd}
Stability-AI.
\newblock Stable video diffusion.
\newblock https://github.com/Stability-AI/generative-models.
\newblock GitHub; accessed November 21, 2023.

\bibitem{szegedy2013intriguing}
Christian Szegedy, Wojciech Zaremba, Ilya Sutskever, Joan Bruna, Dumitru Erhan, Ian Goodfellow, and Rob Fergus.
\newblock Intriguing properties of neural networks.
\newblock In {\em International Conference on Learning Representations}, 2014.

\bibitem{tancik2020stegastamp}
Matthew Tancik, Ben Mildenhall, and Ren Ng.
\newblock Stegastamp: Invisible hyperlinks in physical photographs.
\newblock In {\em IEEE/CVF Conference on Computer Vision and Pattern Recognition}, 2020.

\bibitem{wang2022triangle}
Xiaosen Wang, Zeliang Zhang, Kangheng Tong, Dihong Gong, Kun He, Zhifeng Li, and Wei Liu.
\newblock Triangle attack: A query-efficient decision-based adversarial attack.
\newblock In {\em European Conference on Computer Vision}, 2022.

\bibitem{wang2004image}
Zhou Wang, Alan~C Bovik, Hamid~R Sheikh, and Eero~P Simoncelli.
\newblock Image quality assessment: from error visibility to structural similarity.
\newblock {\em IEEE Transactions on Image Processing}, 2004.

\bibitem{wen2023tree}
Yuxin Wen, John Kirchenbauer, Jonas Geiping, and Tom Goldstein.
\newblock Tree-ring watermarks: Fingerprints for diffusion images that are invisible and robust.
\newblock In {\em Conference on Neural Information Processing Systems}, 2023.

\bibitem{zhang2023novel}
Yulin Zhang, Jiangqun Ni, Wenkang Su, and Xin Liao.
\newblock A novel deep video watermarking framework with enhanced robustness to h. 264/avc compression.
\newblock In {\em ACM International Conference on Multimedia}, 2023.

\bibitem{zhao2023invisible}
Xuandong Zhao, Kexun Zhang, Zihao Su, Saastha~Vasan Vasan, Ilya Grishchenko, Christopher Kruegel, Giovanni Vigna, Yu-Xiang Wang, and Lei Li.
\newblock Invisible image watermarks are provably removable using generative ai.
\newblock In {\em Conference on Neural Information Processing Systems}, 2024.

\bibitem{zhu2018hidden}
Jiren Zhu, Russell Kaplan, Justin Johnson, and Li~Fei-Fei.
\newblock Hidden: Hiding data with deep networks.
\newblock In {\em European Conference on Computer Vision}, 2018.

\end{thebibliography}

\newpage
\appendix
\section{Appendix}
\subsection{\label{sec:compute}Experiments Compute Resources}
We conduct our experiments on 18 NVIDIA-RTX-6000 GPUs, each with 24 GB memory. The complete set of experiments requires about 300 GPU-hours to execute.

\subsection{\label{sec:aggregation strategies}A Detailed Explanation for Aggregation Strategies}
In image watermark detection, given an image $I$, the watermark decoder $Dec$ extracts a vector of logits $y$ from the image $I$, i.e., $y = Dec(I)$. These logits are then rounded to obtain the decoded watermark bitstring $w$:
\begin{align}
\mathbf{w} = \mathbb{I}\left( \mathbf{y} \geq 0.5 \right), \quad \mathbf{w} \in \{0,1\}^n \label{eq:round}
\end{align}
where $\mathbb{I}(\cdot)$ denotes the element-wise indicator function, and both $w$ and $y$ have length $n$. The bitwise accuracy ($BA$) between the decoded watermark $w$ and the ground-truth watermark $w_g$ is compared against a predefined detection threshold $\tau$: the image $I$ is detected as watermarked if $BA(w, w_g) \geq \tau$, and as unwatermarked otherwise. 

In frame-level video watermark detection, given a video $x$ with $F$ frames, each frame is treated as an individual image. The watermark decoder $Dec$ is applied to each frame to decode logits $y_i$, where $y_i$ denotes the logits decoded from the $i$-th frame, for $i \in \{1, 2, \dots, F\}$. To obtain a final video-level detection result, we propose seven aggregation strategies based on different ways of aggregating these frame-level decoded logits.

\myparatight{Logit-mean} The watermark decoder $Dec$ extracts decoded logits $y_i$ from the $i$-th frame of the video $x$, and we compute the average of these logits to obtain the aggregated logits:
\begin{align}
\mathbf{y} = \frac{1}{F} \sum_{i=1}^{F} \mathbf{y_i}, \nonumber
\end{align}
then, the decoded watermark $w$ is obtained using Equation~\ref{eq:round}. The video $x$ is detected as watermarked if $BA(w, w_g) \geq \tau$; otherwise, it is considered unwatermarked.

\myparatight{Logit-median} The watermark decoder $Dec$ extracts decoded logits $y_i$ from the $i$-th frame of the video $x$, and we compute the geometric median of these logits to obtain the aggregated logits:
\begin{align}
\mathbf{y} = \arg\min_{\mathbf{z} \in \mathbb{R}^d} \sum_{i=1}^{F} \left| \mathbf{z} - \mathbf{y_i} \right|_2, \nonumber
\end{align}
we then apply the same procedure as in logit-mean to obtain the decoded watermark $w$ and make the final detection decision.

\myparatight{Bit-median} The watermark decoder $Dec$ extracts decoded logits $y_i$ from the $i$-th frame of the video $x$, and each set of logits is rounded to obtain a decoded watermark bitstring for that frame:
\begin{align}
    \mathbf{w_i} = \mathbb{I}\left( \mathbf{y_i} \geq 0.5 \right), \quad \mathbf{w_i} \in \{0,1\}^n, \label{eq:i round}
\end{align}
we then take a majority vote across frames at each bit position to produce the final decoded watermark:
\begin{align}
    \mathbf{w}[j] = 
    \begin{cases}
    1, & \text{if } \sum_{i=1}^{F} \mathbf{w_i}[j] \geq \frac{F}{2} \\
    0, & \text{otherwise}
    \end{cases}, \quad \forall j \in \{1, \dots, n\}, \nonumber
\end{align}
the video $x$ is detected as watermarked if $BA(w, w_g) \geq \tau$; otherwise, it is considered unwatermarked. Note that majority voting yields the same result as taking the median for binary values.

\myparatight{BA-mean} The watermark decoder $Dec$ extracts decoded logits $y_i$ from the $i$-th frame of the video $x$. These logits $y_i$ are rounded to obtain the decoded watermark $w_i$, as defined in Equation~\ref{eq:i round}. We then compute the bitwise accuracy $BA(w_i, w_g)$ between $w_i$ and the ground-truth watermark $w_g$ for each frame, and take the average of these bitwise accuracy scores:
\begin{align}
    BA = \frac{1}{F} \sum_{i=1}^{F} BA(\mathbf{w_i}, \mathbf{w_g}), \nonumber
\end{align}
then the video $x$ is detected as watermarked if $BA \geq \tau$ or unwatermarked otherwise.

\myparatight{BA-median} Following the same procedure as in BA-mean aggregation, we calculate the bitwise accuracy $BA(w_i, w_g)$ between $w_i$ and the ground-truth watermark $w_g$ for the $i$-th frame, and then take the median of these bitwise accuracy values:
\begin{align}
    BA = \operatorname{median}\{ BA(\mathbf{w_1}, \mathbf{w_g}), BA(\mathbf{w_2}, \mathbf{w_g}), \dots, BA(\mathbf{w_F}, \mathbf{w_g}) \}, \nonumber
\end{align}
where $\operatorname{median}$ denotes the statistical median over the $F$ per-frame accuracy values. The video $x$ is detected as watermarked if $BA \geq \tau$; otherwise, it is considered unwatermarked.

\myparatight{Detection-median} Following the same procedure as in BA-mean, we calculate the bitwise accuracy $BA(w_i, w_g)$ between $w_i$ and the ground-truth watermark $w_g$ for the $i$-th frame. We then compare each $BA(w_i, w_g)$ with the detection threshold $\tau$ to obtain the detection result $d_i$ for the $i$-th frame:
\begin{align}
    d_i = 
    \begin{cases}
    1, & \text{if } BA(\mathbf{w_i}, \mathbf{w_g}) \geq \tau \\
    0, & \text{otherwise}
    \end{cases}, \label{eq:detection}
\end{align}
we then take a majority vote among the frame-level detection results $d_i$, for $i \in \{1, 2, \dots, F\}$, to obtain the aggregated video-level detection result. That is, the video $x$ is classified as watermarked if $\sum_{i=1}^{F} d_i \geq \frac{F}{2}$.

\myparatight{Detection-threshold} In this aggregation strategy, we set a detection-level threshold $k$. Specifically, a video $x$ is detected as watermarked if at least $k$ frames are detected as watermarked. Following the same procedure as in detection-median, we obtain the frame-level detection results $d_i$ using Equation~\ref{eq:detection}, and classify the image $x$ as watermarked if $\sum_{i=1}^{F} d_i \geq k$.

The value of $k$ is selected to ensure a low theoretical false positive rate (FPR), which is kept below 0.01\% in this work. We assume that the probability of a non-watermarked frame being falsely detected as watermarked is $P$ (details on how to compute $P$ given $\tau$ are provided in Appendix~\ref{sec:select tau}). Based on this assumption, the value of $k$ is determined as follows:
\begin{align}
    k = \arg\min_{m \in \{0, 1, \dots, F\}} \left\{ \Pr(B \geq m) \leq 10^{-4} \right\}, \nonumber
\end{align}
where $B$ follows binomial distribution with parameter $F$ and $P$, i.e.,  $B \sim \mathrm{Binomial}(F, P)$.

\subsection{\label{sec:blackbox aggregation}Implementation Details for Aggregation Strategies in Black-box Perturbations}
Square Attack~\cite{andriushchenko2020square} and Triangle Attack~\cite{wang2022triangle} were originally developed for image classification tasks. To adapt them to the video watermark removal and forgery setting, we introduce two key modifications for each method.

\myparatight{Square Attack} First, Square Attack's official implementation takes a batch of images and an image classifier as input, perturbs each image individually, and aims to mislead the classification results. In our adaptation, the attack takes a video and a video watermark detector as input. The video is treated as a batch of frames, and a video-level perturbation is crafted to either remove or forge a watermark.

Second, Square Attack is a score-based attack that iteratively crafts perturbations based on score feedback. In its original form, the scores correspond to class probabilities output by an image classifier. In our experiments, video watermark detection is a binary classification task, and we redefine the scoring function according to the aggregation strategy used. For logit-level, bit-level, and BA-level aggregation strategies, we define the score as the bitwise accuracy BA after aggregation. For detection-level aggregation strategies, the score is defined as the number of frames detected as watermarked. We then optimize the perturbation to minimize the score for watermark removal, or maximize it for watermark forgery.

\myparatight{Triangle Attack} First, the original Triangle Attack takes an image of shape $[1, C, H, W]$ and an image classifier as input for each attack iteration, where $C$ is the number of channels (typically $C=3$ for RGB images), and $H$ and $W$ denote the height and width of the image, respectively. In the video setting, a video has shape $[F, C, H, W]$, where $F$ is the number of frames. To adapt to this format, we reshape the video into a tensor of shape $[1, F \times C, H, W]$, effectively treating the video as an image with an extended channel dimension. We then search for a video-level perturbation to remove or forge the video watermark.

Second, as a label-based attack, Triangle Attack crafts perturbations by checking whether the perturbed input retains or flips a desired target label. In our setting, watermark detection is a binary classification problem with labels "watermarked" and "unwatermarked". The detection label is produced by the watermark detector via different aggregation strategies. For watermark removal, we start with an initial video that is classified as unwatermarked and iteratively search for a smaller perturbation that preserves this label. For watermark forgery, we perform the reverse: we begin with a video that is classified as watermarked and aim to iteratively reduce the perturbation magnitude while ensuring the perturbed video remains classified as watermarked.

\subsection{\label{sec:select tau}Selecting Detection Threshold $\tau$} 
In image (or frame-level) detection, given an image $x$, the watermark decoder $Dec$ extracts a decoded watermark $w$ from it. The image is classified as watermarked if the bitwise accuracy between the decoded watermark and the ground-truth watermark $w_g$ satisfies $BA(w, w_g) \geq \tau$, where $\tau$ is a predefined detection threshold. A key consideration is how to select the threshold $\tau$ such that the \emph{false positive rate (FPR)}—the probability that an unwatermarked image is incorrectly classified as watermarked—is bounded by a small target value $\eta$ (e.g., $\eta = 10^{-4}$).

To introduce randomness, we assume that the watermarking service provider randomly selects the ground-truth watermark $w_g$. As a result, for an unwatermarked image, the decoded watermark $w$ is independent of $w_g$, and each bit matches with probability $0.5$. Consequently, the bitwise accuracy $BA(w, w_g)$ follows a scaled binomial distribution: $BA(w, w_g) \sim \frac{1}{n} \cdot \mathrm{Binomial}(n, 0.5),$ where $n$ is the watermark length. Given a detection threshold $\tau$, the theoretical FPR can be computed as:
\begin{align}
    FPR(\tau) = \Pr(BA(\mathbf{w},\mathbf{w_g})>\tau) = \sum_{k=\lceil n\tau \rceil}^n {n \choose k} \frac{1}{2^n}, \nonumber
\end{align}
where $n$ is the watermark length. To ensure that the FPR is less than a desired threshold $\eta$, the detection threshold $\tau$ can be selected as follows:
\begin{align}
    \tau=\arg\min_{c} \sum_{k=\lceil nc \rceil}^n {n \choose k} \frac{1}{2^n} < \eta. \nonumber
\end{align}
For instance, given $\eta = 10^{-4}$, the detection threshold $\tau$ should be set to $\frac{67}{96}$ when $n = 96$, and to $\frac{27}{32}$ when $n = 32$.

\subsection{\label{sec:blackbox forgery}Forgery Results for Black-box Perturbations}
For forgery attacks, we evaluate on the real-world Kinetics-400 dataset. To maintain consistency with the removal attack setting described in the main text, we conduct experiments on 40 videos. Each video is trimmed to 14 frames—the same number used for videos generated by SVD—and BA-mean aggregation is used by default. We find that existing video watermarking methods are robust against watermark forgery perturbations in the black-box setting.

\begin{figure}
    \centering
    \begin{subfigure}{.4\linewidth}
        \centering
        \includegraphics[width=\linewidth]{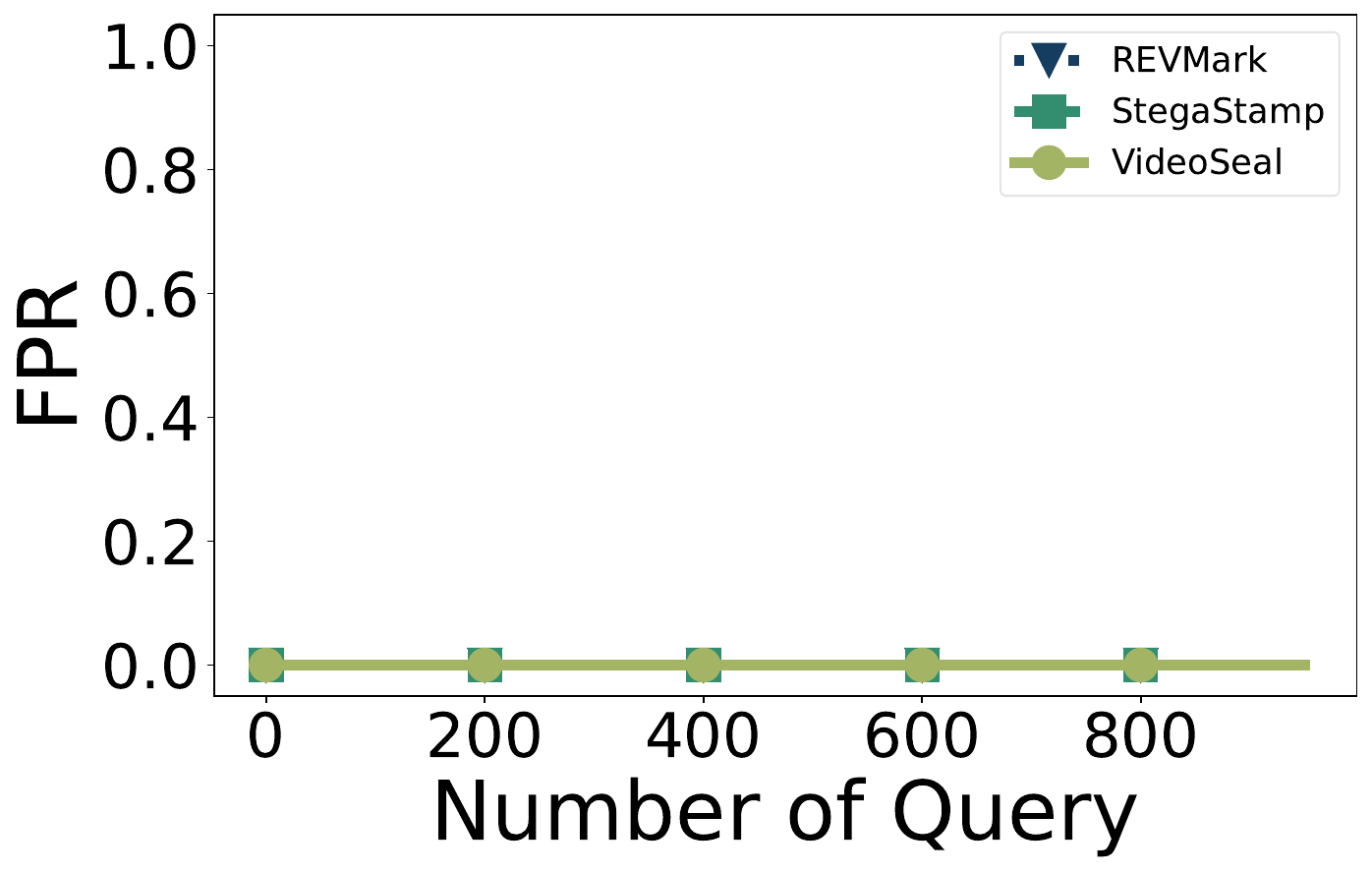}
        \caption{Watermarking}
    \end{subfigure}
    \begin{subfigure}{.4\linewidth}
        \centering
        \includegraphics[width=\linewidth]{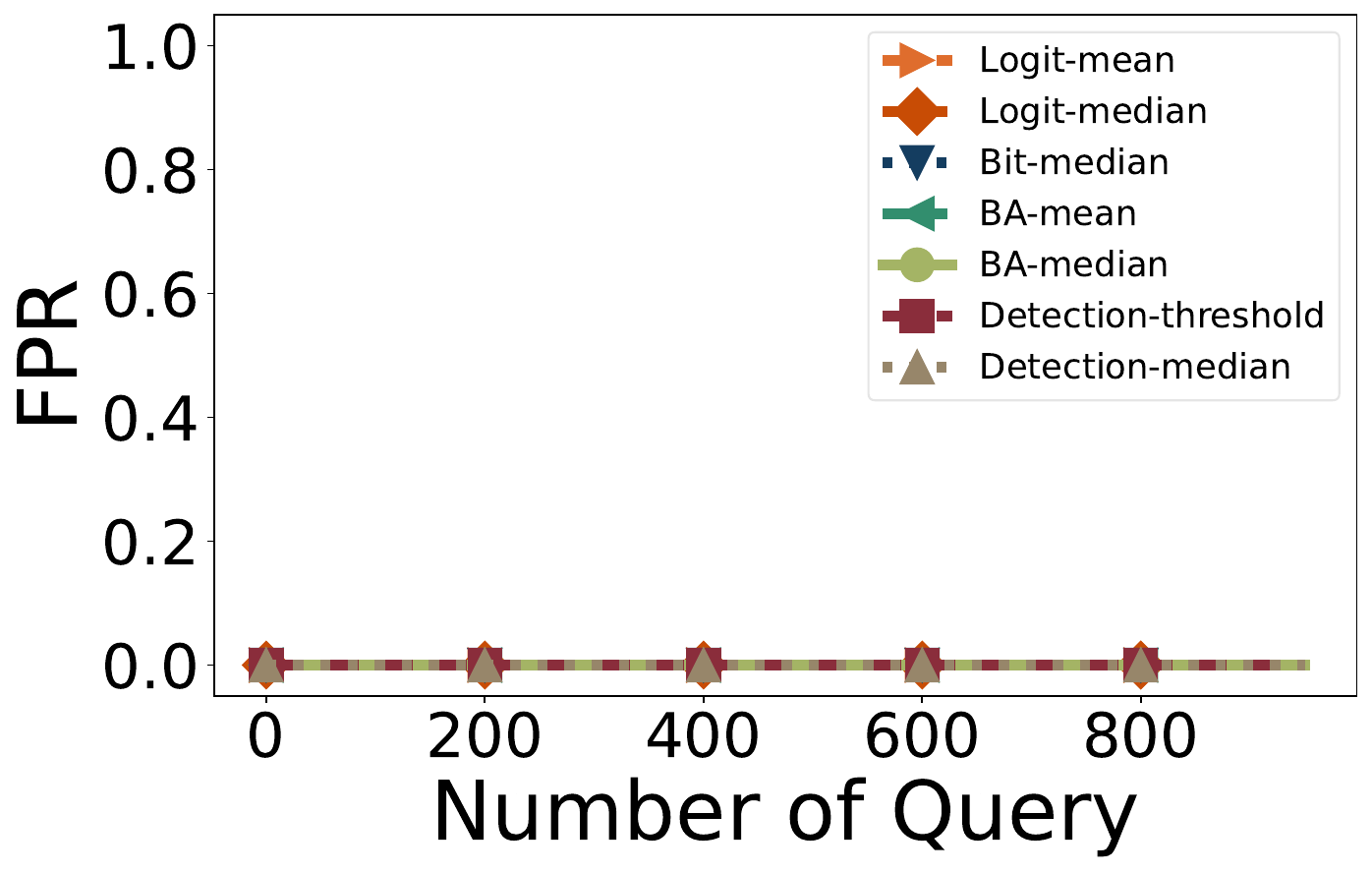}
        \caption{Aggregation}
    \end{subfigure}
    \caption{\label{fig:square forgery}Square Attack watermark forgery results. Perturbations are $l_{\infty}$ bounded by 0.05.}
\end{figure}

\myparatight{Square Attack} Figure~\ref{fig:square forgery} presents the results of Square Attack for watermark forgery, where the perturbation size is bounded by an $l_{\infty}$ norm of 0.05. We observe that all current video watermarking methods—including VideoSeal with different aggregation strategies—maintain FPRs close to zero, even after 1,000 queries. An intuitive explanation is as follows: If watermark detection is viewed as a binary classification task with "watermarked" and "non-watermarked" classes, the decision space corresponding to the "non-watermarked" class is likely much larger than that of the "watermarked" class. This makes it relatively easier to remove a watermark by crafting a sufficiently large perturbation. In contrast, forging a watermark becomes substantially more difficult, as it requires the attacker to precisely locate the decision boundary between the two classes.

\begin{figure}
    \centering
    \begin{subfigure}{.4\linewidth}
        \centering
        \includegraphics[width=\linewidth]{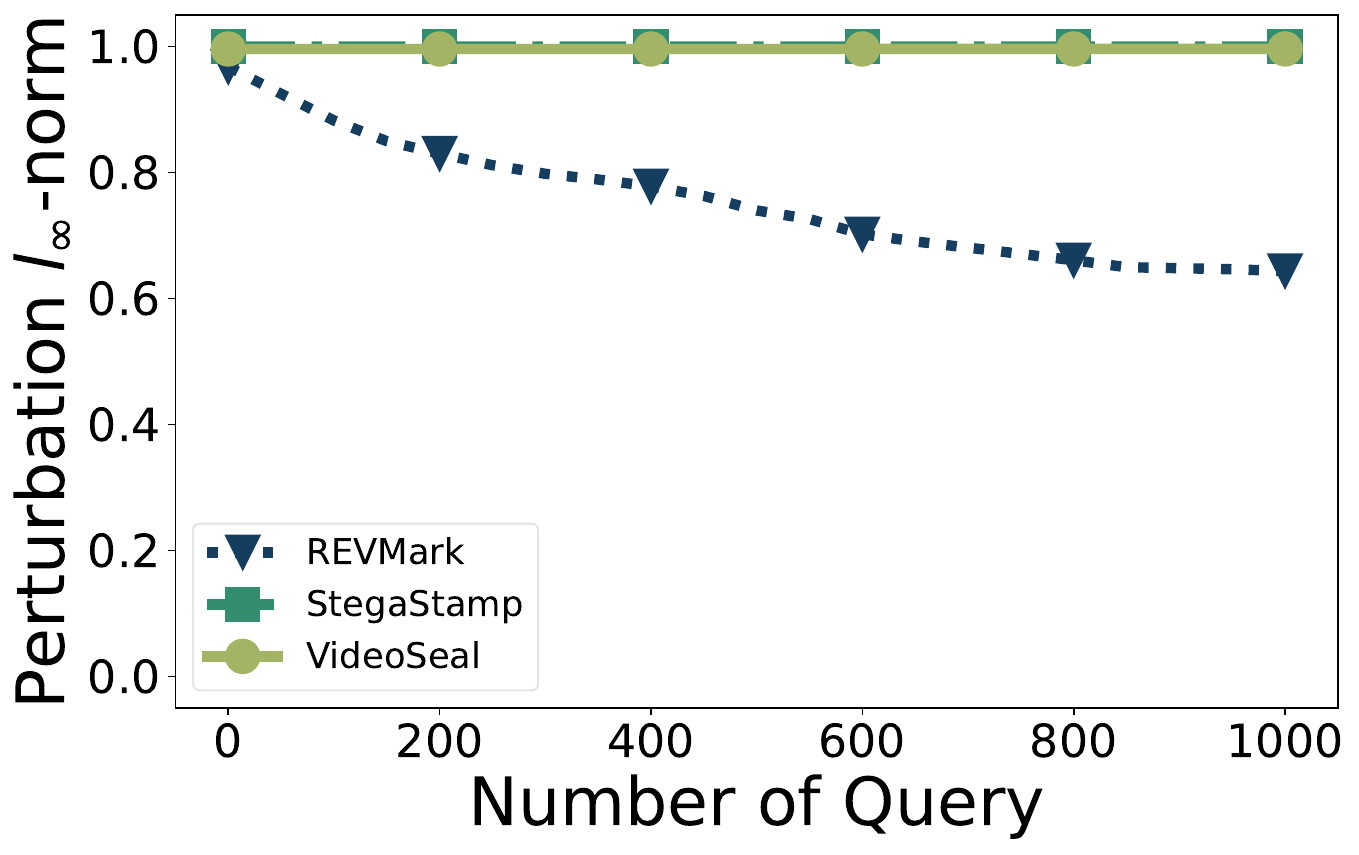}
        \caption{Watermarking}
    \end{subfigure}
    \begin{subfigure}{.4\linewidth}
        \centering
        \includegraphics[width=\linewidth]{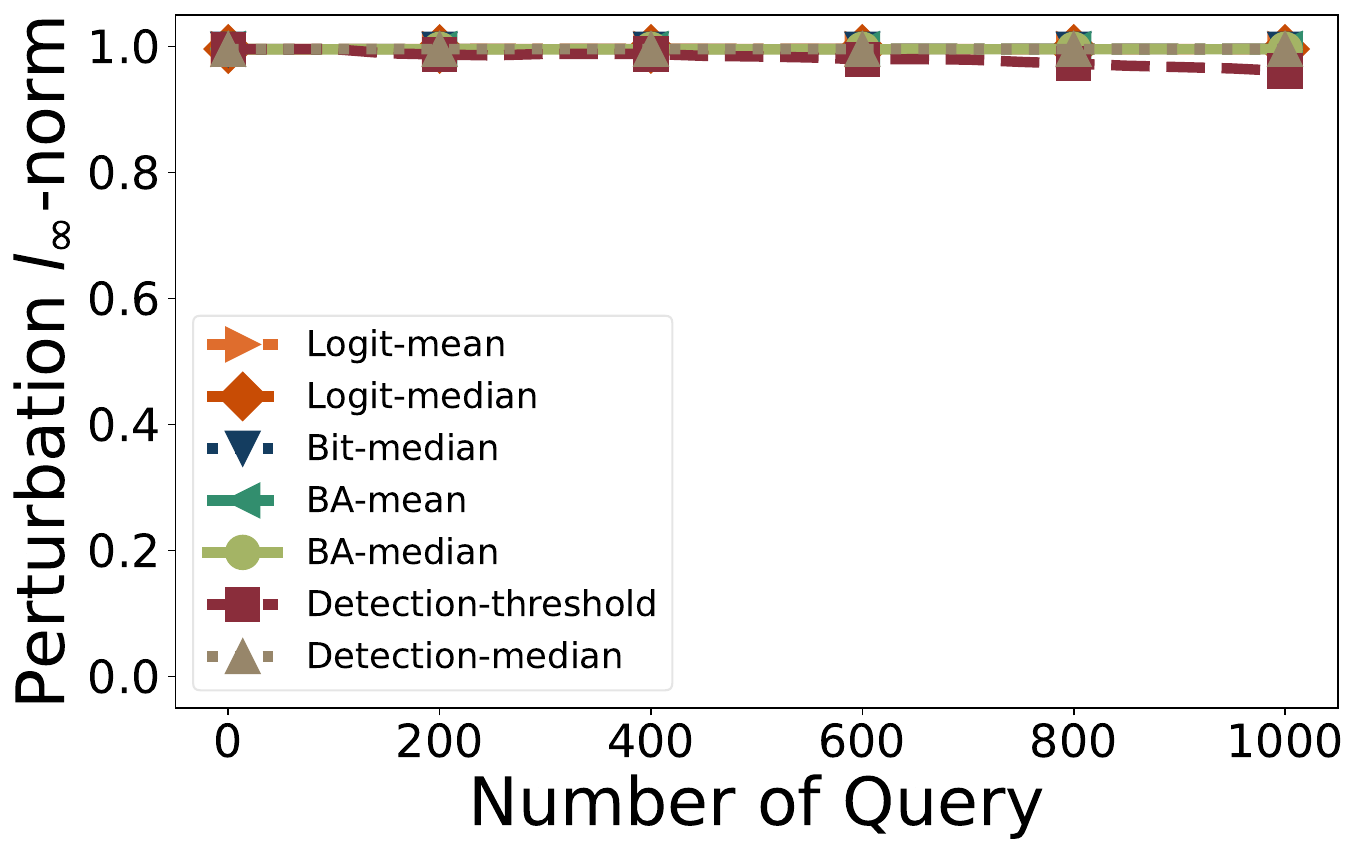}
        \caption{Aggregation}
    \end{subfigure}
    \caption{\label{fig:triangle forgery}Triangle Attack watermark forgery results.}
\end{figure}

\myparatight{Triangle Attack} Figure~\ref{fig:triangle forgery} presents the results of Triangle Attack for watermark forgery. Since Triangle Attack requires a watermarked video as the starting point to perturb a target unwatermarked video, we assume that the attacker does not have access to the watermark encoder but may use unrelated watermarked videos for initialization. Specifically, we generate a random video and embed a watermark into it using the watermark encoder to serve as the initialization.

\begin{figure}
    \centering
    \includegraphics[width=0.5\linewidth]{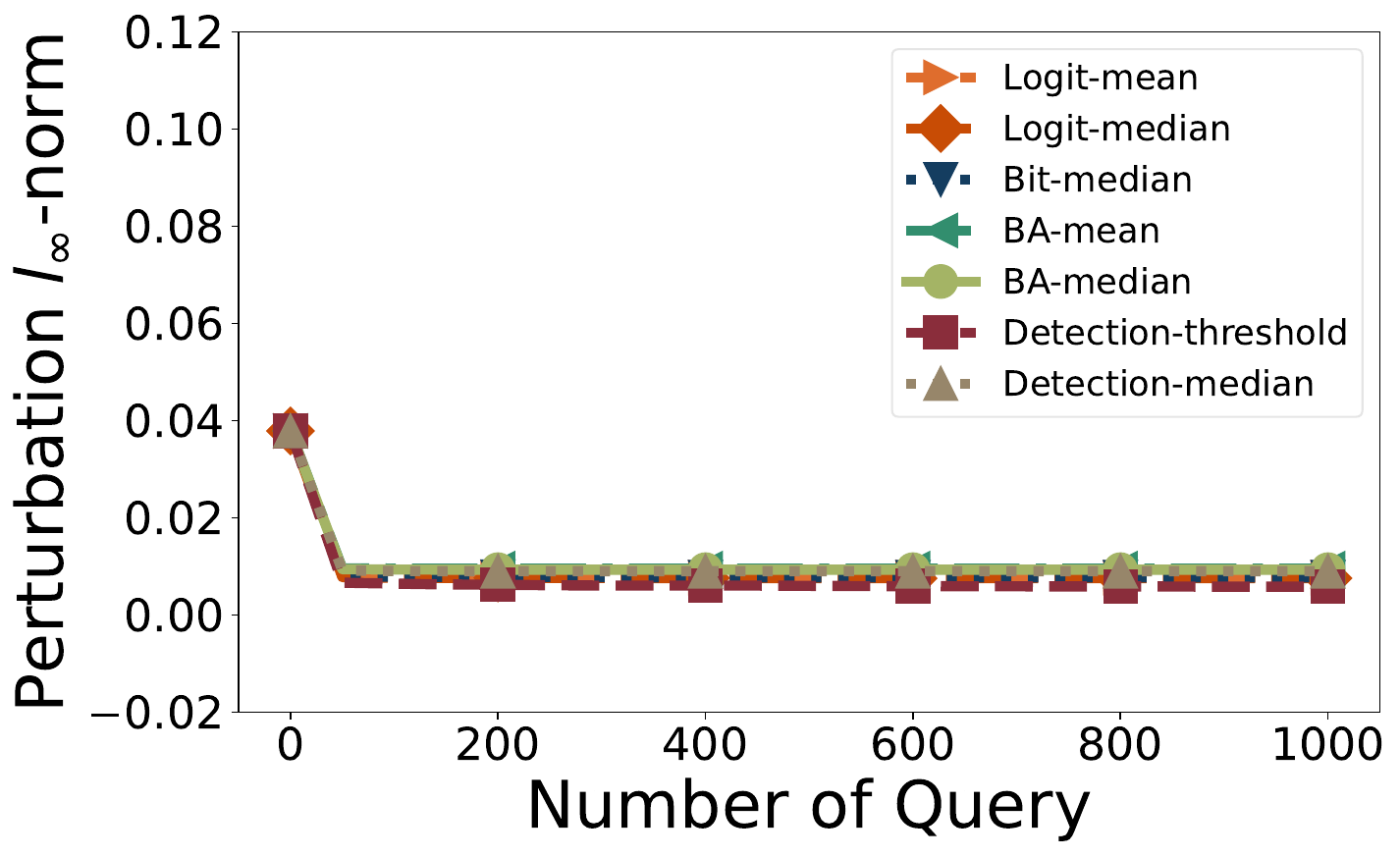}
    \caption{\label{fig:triangle forgery w}Triangle Attack watermark forgery results when watermarked versions are used as initialization.}
\end{figure}

Across the three evaluated video watermarking methods, all demonstrate robustness against forgery perturbations. The average $l_{\infty}$-norm of perturbations for StegaStamp and VideoSeal remains consistently at 1, indicating that Triangle Attack completely fails to forge watermarks for these methods. For REVMark, the average $l_{\infty}$-norm of perturbations decreases as the number of queries increases; however, a value of 0.6 still reflects a large perturbation that significantly degrades the video's visual quality. Among VideoSeal with different aggregation strategies, only the detection-threshold strategy shows a slight decrease in perturbation norm, as it is the least robust to forgery attacks (as previously discussed). Nonetheless, all aggregation strategies for VideoSeal remain robust overall against Triangle Attack in the forgery setting. Figure~\ref{fig:triangle forgery w} presents the results of Triangle Attack when watermarked versions of the target videos are used as initialization. While this setting is rarely practical—since an attacker with access to the watermark encoder could directly generate watermarked videos—it serves to highlight the importance of initialization in the attack process. The results demonstrate that Triangle Attack is highly sensitive to initialization and that finding a suitable starting point is significantly more challenging in watermark forgery than in watermark removal.

\subsection{\label{sec:nobox analysis}Detailed Analysis for No-box Perturbations}
\myparatight{Comparison across watermarking methods}
Figure~\ref{fig:watermark} in the Appendix shows FNR results under various common perturbations for different video watermarking methods. The FNR values are computed by averaging over different aggregation strategies, generative models, and video styles. We highlight several key observations: Overall, VideoShield appears to be more robust against various video perturbations. However, in some cases—particularly under cropping and Gaussian noise perturbations—its FNR is higher than that of REVMark. VideoSeal performs well when video quality is preserved, but its FNR increases dramatically under strong Gaussian noise perturbations. For instance, the FNR approaches 1 when Gaussian noise with standard deviation $\sigma = 0.15$ is applied. REVMark and StegaStamp are generally robust against common perturbations such as blurring and frame manipulation but show vulnerability to JPEG and MPEG-4 compression, even when the visual quality of the video is preserved. Cropping is found to be a particularly effective perturbation for watermark removal. Among all methods, only VideoSeal demonstrates robustness against cropping-based attacks.

\myparatight{Comparison across aggregation strategies}
Figure~\ref{fig:aggregation_main} in the main text, along with Figures~\ref{fig:aggregation stegastamp other} and~\ref{fig:aggregation videoseal} in the Appendix, presents FNR results under various video perturbations using different watermark aggregation strategies. The FNR values are averaged across generative models, and video styles for StegaStamp and VideoSeal. Surprisingly, although BA-level aggregation strategies are commonly used in image watermarking~\cite{tancik2020stegastamp,fernandez2024video}, they exhibit the lowest robustness in the context of video watermarking, as indicated by their higher FNRs. In contrast, detection-threshold aggregation achieves the lowest FNR among all strategies. Given that the false positive rate (FPR) remains close to zero across all strategies, detection-threshold aggregation may be considered the most robust approach—despite its known vulnerability to forgery attacks in adversarial settings. Beyond detection-threshold, logit-level aggregation strategies also yield lower FNRs compared to BA-level strategies, further highlighting their relative robustness in video watermarking applications.

\myparatight{Comparison across generative models}
Figure~\ref{fig:models} in the Appendix presents FNR results under various video perturbations across different generative models. The FNR values are averaged over different watermarking methods, aggregation strategies, and video styles. Overall, we do not observe significant differences in FNR among AI-generated videos from different generative models. More specifically, videos generated by Hunyuan Video tend to be more robust against cropping and MPEG-4 compression, but are more vulnerable to JPEG and Gaussian noise perturbations. In contrast, videos generated by Stable Video Diffusion show greater robustness to Gaussian noise but are more susceptible to cropping and MPEG-4 compression. Despite these differences, there is no consistent or significant gap in robustness across the generative models.

\myparatight{Comparison across styles}
Figure~\ref{fig:styles} in the Appendix presents FNR results under various perturbations for different video styles. The FNR values are averaged across different watermarking methods, aggregation strategies, and generative models. We observe that videos in the realistic and sci-fi styles exhibit nearly identical FNRs, which is consistent with the design goal of watermarking methods to be content-independent. However, videos in the cartoon style show noticeably higher FNRs under JPEG and MPEG-4 compression. This can be attributed to the fact that cartoon frames are typically simpler, with less texture and lower pixel variability, making the subtle pixel-level changes introduced by watermarks more susceptible to removal during compression.

\subsection{\label{sec:limitation}Discussion and Limitations}
\myparatight{Adversarial robustness of frame-based detection} In this work, we extend an existing image watermarking method (StegaStamp) to the video domain by applying it at the frame level, and we similarly treat VideoSeal as a frame-based method. We evaluate the robustness of these approaches against adversarial perturbations in both white-box and black-box settings. Our findings show that frame-based video watermarking methods inherit the (non-)robustness of their underlying image watermarking counterparts. For example, image watermarking methods such as StegaStamp and HiDDeN~\cite{zhu2018hidden} (which forms the foundation of VideoSeal) are vulnerable in the white-box setting and fail to withstand black-box removal attacks when the attacker is allowed multiple queries. These vulnerabilities are consistent with our observations in this video watermarking benchmark. To mitigate these weaknesses, future video watermarking methods may need to incorporate temporal information across frames, rather than relying solely on frame-level detection, to achieve better robustness.

\myparatight{Adversarial perturbations} Our results show that adversarial perturbations are significantly more effective at removing or forging watermarks compared to common video perturbations. However, these attacks typically require more knowledge about the watermarking system or computational resources. For example, white-box attacks assume access to the internal parameters of the watermark detector, which may only be feasible if the detection model is publicly released by the service provider or if the attacker is an insider. Despite these constraints, evaluating robustness in the white-box setting provides valuable insight into the worst-case vulnerability of a watermarking method. In contrast, black-box attacks require only query access to the watermark detector’s API. While such attacks are query-expensive and time-consuming, they remain practical and highly effective—especially in scenarios where an attacker aims to target a specific video rather than performing large-scale attacks.

\myparatight{More robust video watermarks} Our experimental results show that while existing video watermarking methods are generally robust when there are no perturbations, they remain vulnerable to adversarial perturbations and certain common video perturbations such as MPEG-4 compression and cropping. These findings highlight the need for designing more robust video watermarking techniques that can withstand both common and adversarial perturbations in real-world scenarios.


\begin{table}[]
\centering
\caption{Example base prompts from \datasetname. To generate videos in different styles, we prepend the base prompts with style-specific prefixes: "In the realistic style, ", "In the cartoon style, ", or "In the sci-fi style, ".}
\label{tab:exmaple prompt}
\renewcommand{\arraystretch}{1.2}
\resizebox{\textwidth}{!}{
\begin{tabular}{|c|c|c|}
\hline
 & Index & Prompts
 \\ \hline
\multirow{5}{*}{Fast Motion} & 1     & Generate a dynamic video with rapid frame changes featuring a massive volcanic eruption with lava flows and ash clouds.       \\ \cline{2-3} 
& 2     & Generate a dynamic video with rapid frame changes featuring a high-speed car crash with flying debris and shattered glass.    \\ \cline{2-3} 
& 3     & Generate a dynamic video with rapid frame changes featuring a dazzling fireworks display with vibrant explosions.             \\ \cline{2-3} 
& 4     & Generate a dynamic video with rapid frame changes featuring stormy ocean waves crashing against cliffs in a chaotic sequence. \\ \cline{2-3} 
& 5     & Generate a dynamic video with rapid frame changes featuring an urban chase scene with vehicles weaving through traffic.       \\ \hline

\multirow{5}{*}{Slow Motion} & 1     & Generate a slow, evolving video with subtle frame changes featuring a pond with fish making subtle ripples.                   \\ \cline{2-3} 
& 2     & Generate a slow, evolving video with subtle frame changes featuring a timelapse of fog rolling into a valley.                 \\ \cline{2-3} 
& 3     & Generate a slow, evolving video with subtle frame changes featuring a slow timelapse of a bustling market square.             \\ \cline{2-3}  
& 4     & Generate a slow, evolving video with subtle frame changes featuring grasses moving softly in a light breeze.                  \\ \cline{2-3} 
& 5     & Generate a slow, evolving video with subtle frame changes featuring the gradual formation of frost on a window.
\\ \hline
\end{tabular}}
\end{table}

\begin{table}[]
\centering
\caption{Watermark removal results (measured by FNR) for different video watermarking methods using various aggregation strategies under no perturbations. REVMark and VideoShield do not perform frame-level watermark extraction, so aggregation strategies are not applicable to them. Note that VideoShield relies on access to DDIM inversion of the video generative model; thus, it is only evaluated on videos generated by SVD.}
\renewcommand{\arraystretch}{1.2}
\resizebox{\textwidth}{!}{
\begin{tabular}{|c|c|c|c|c|c|c|c|c|c|c|}
\hline
\multicolumn{2}{|c|}{\multirow{2}{*}{\textbf{Methods}}} &
\multicolumn{3}{c|}{\textbf{SVD}} &
\multicolumn{3}{c|}{\textbf{Sora}} &
\multicolumn{3}{c|}{\textbf{HunyuanVideo}} \\
\cline{3-11}
\multicolumn{2}{|c|}{} & Realistic & Cartoon & Sci-fi & Realistic & Cartoon & Sci-fi & Realistic & Cartoon & Sci-fi \\
\hline

\multicolumn{2}{|c|}{\textbf{REVMark}} & 0.000 & 0.000 & 0.000 & 0.000 & 0.000 & 0.000 & 0.000 & 0.000 & 0.000 \\
\hline

\multirow{7}{*}{\textbf{StegaStamp}} & logit-mean & 0.000 & 0.000 & 0.000 & 0.000 & 0.000 & 0.000 & 0.000 & 0.000 & 0.000 \\ \cline{2-11}
& logit-median & 0.000 & 0.000 & 0.000 & 0.000 & 0.000 & 0.000 & 0.000 & 0.000 & 0.000 \\ \cline{2-11}
& bit-median & 0.000 & 0.000 & 0.000 & 0.000 & 0.000 & 0.000 & 0.000 & 0.000 & 0.000 \\ \cline{2-11}
& BA-mean & 0.005 & 0.005 & 0.000 & 0.000 & 0.000 & 0.000 & 0.000 & 0.000 & 0.000 \\ \cline{2-11}
& BA-median & 0.005 & 0.000 & 0.000 & 0.000 & 0.000 & 0.000 & 0.000 & 0.000 & 0.000 \\ \cline{2-11}
& detection-threshold & 0.000 & 0.000 & 0.000 & 0.000 & 0.000 & 0.000 & 0.000 & 0.000 & 0.000 \\ \cline{2-11}
& detection-median & 0.005 & 0.000 & 0.000 & 0.000 & 0.000 & 0.000 & 0.000 & 0.000 & 0.000 \\
\hline

\multirow{7}{*}{\textbf{VideoSeal}} & logit-mean & 0.000 & 0.000 & 0.000 & 0.000 & 0.000 & 0.000 & 0.000 & 0.000 & 0.000 \\ \cline{2-11}
& logit-median & 0.000 & 0.000 & 0.000 & 0.000 & 0.000 & 0.000 & 0.000 & 0.000 & 0.000 \\ \cline{2-11}
& bit-median & 0.000 & 0.000 & 0.000 & 0.000 & 0.000 & 0.000 & 0.000 & 0.000 & 0.000 \\ \cline{2-11} \cline{2-11}
& BA-mean & 0.000 & 0.005 & 0.000 & 0.000 & 0.000 & 0.000 & 0.000 & 0.000 & 0.000 \\ \cline{2-11}
& BA-median & 0.000 & 0.005 & 0.000 & 0.000 & 0.000 & 0.000 & 0.000 & 0.000 & 0.000 \\ \cline{2-11}
& detection-threshold & 0.000 & 0.000 & 0.000 & 0.000 & 0.000 & 0.000 & 0.000 & 0.000 & 0.000 \\ \cline{2-11}
& detection-median & 0.000 & 0.005 & 0.000 & 0.000 & 0.000 & 0.000 & 0.000 & 0.000 & 0.000 \\
\hline

\multicolumn{2}{|c|}{\textbf{VideoShield}} & 0.000 & 0.000 & 0.000 & - & - & - & - & - & - \\
\hline
\end{tabular}}
\label{tab:fnr-no-perturbation}
\end{table}

\begin{table}[]
\centering
\caption{Watermark forgery results (measured by FPR) for different video watermarking methods using various aggregation strategies under no perturbations. FPRs are computed on 1,000 real videos from the Kinetics-400 dataset. The term "default" refers to the aggregation strategy originally used in each method. StegaStamp does not have a default strategy, as it is designed for image watermarking. VideoSeal uses BA-mean as its default aggregation strategy.}
\resizebox{14cm}{!}{
\begin{tabular}{lccccccccc}
    \toprule
    Method & default & logit-mean & logit-median & bit-median & BA-mean & BA-median & detection-threshold & detection-median \\
    \midrule
    REVMark & 0.000 & - & - & - & - & - & - & - \\
    \midrule
    StegaStamp & - & 0.000 & 0.000 & 0.000 & 0.000 & 0.000 & 0.000 & 0.000 \\
    \midrule
    VideoSeal & 0.000 & 0.000 & 0.000 & 0.000 & 0.000 & 0.000 & 0.000 & 0.000 \\
    \midrule
    VideoShield & 0.000 & - & - & - & - & - & - & - \\
    \bottomrule
\end{tabular}}
\label{tab:fpr-no-perturbation}
\end{table}

\begin{figure}[]
    \centering
    \begin{subfigure}{.4\linewidth}
        \centering
        \includegraphics[width=\linewidth]{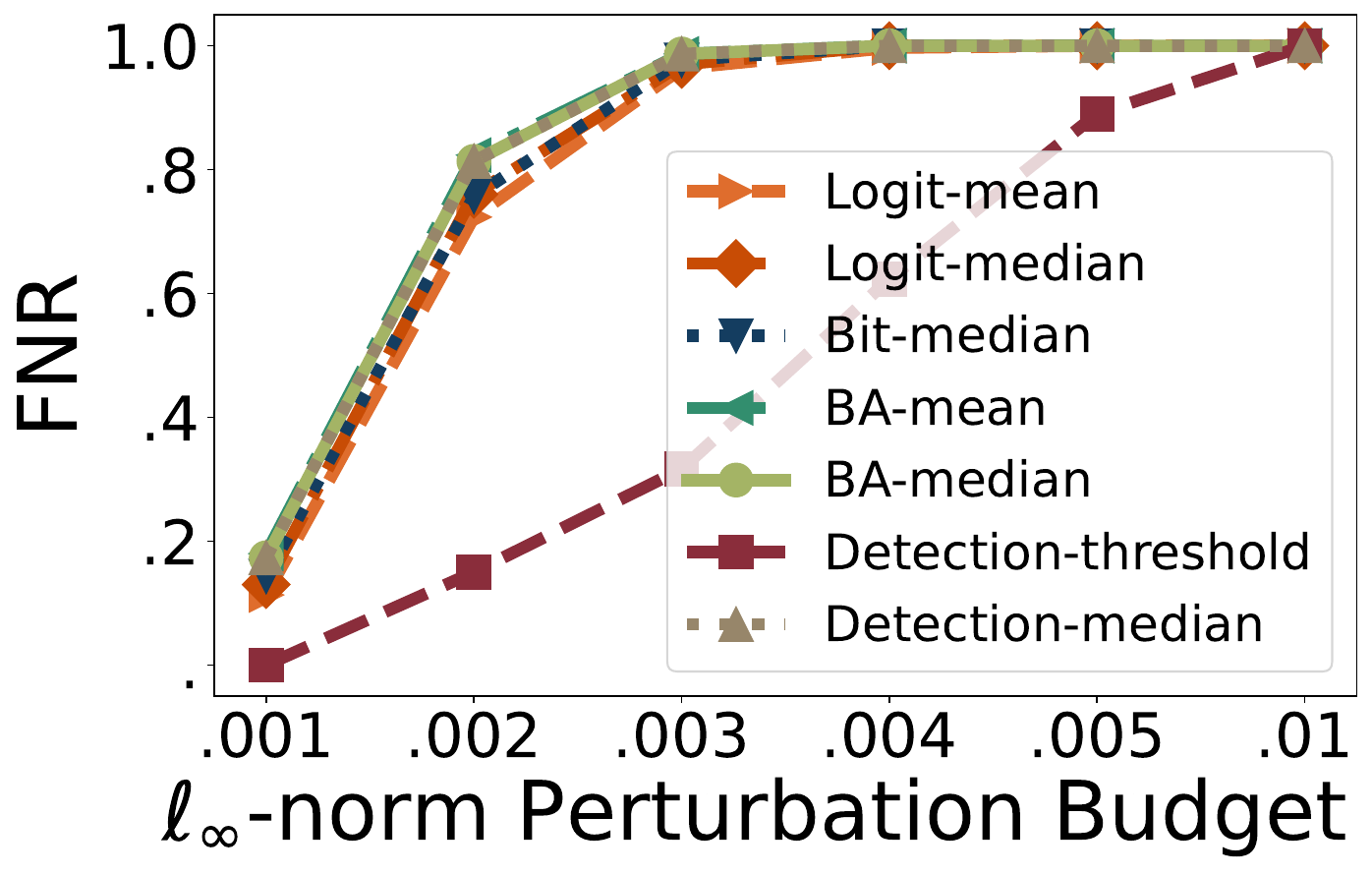}
        \caption{Removal Attack}
    \end{subfigure}
    \begin{subfigure}{.4\linewidth}
        \centering
        \includegraphics[width=\linewidth]{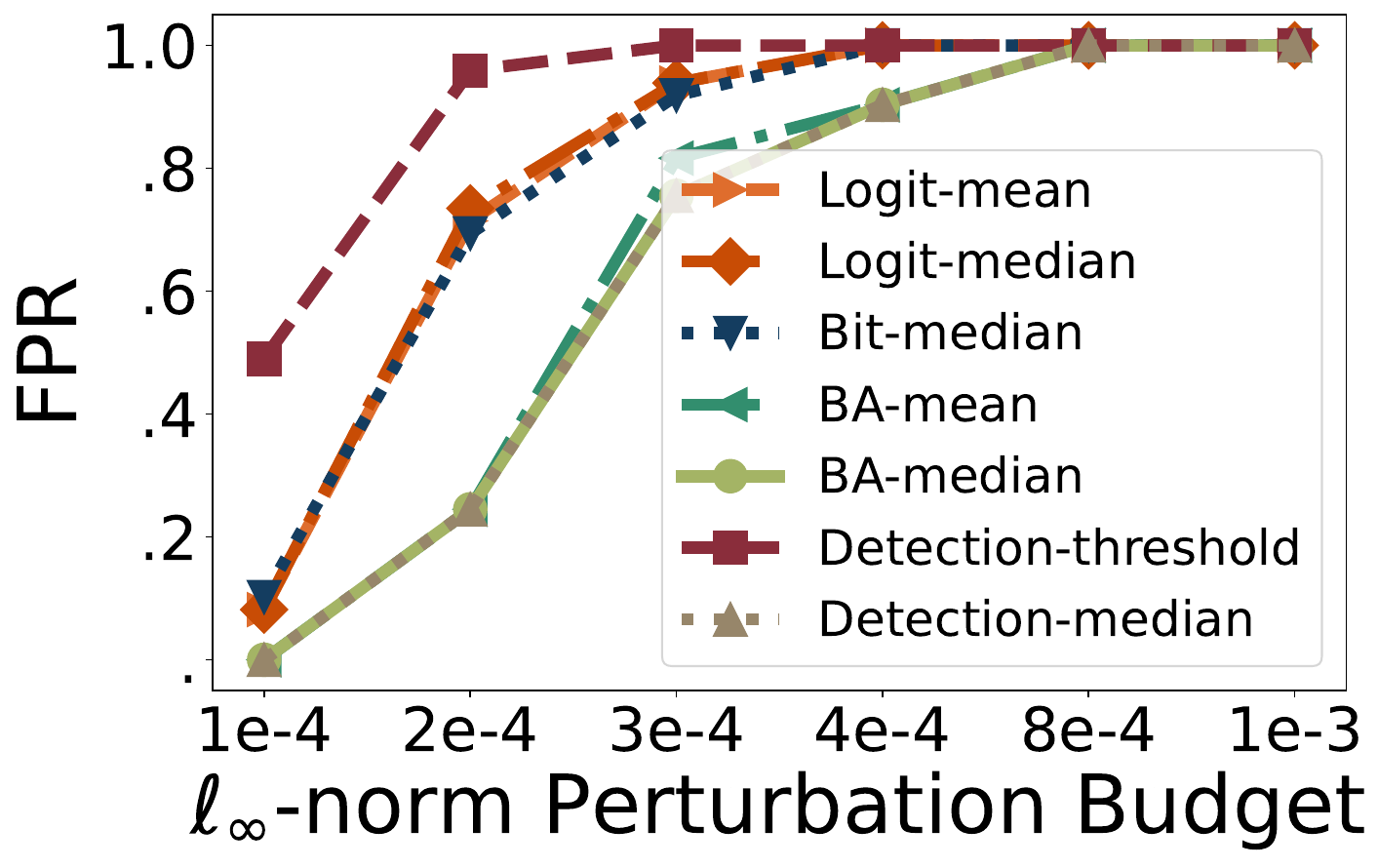}
        \caption{Forgery Attack}
    \end{subfigure}
    \caption{\label{fig:scenario1 aggregation VideoSeal}White-box attack results for VideoSeal using different aggregation strategies in the first scenario.}
\end{figure}

\begin{figure}[]
    \centering
    \begin{subfigure}{.4\linewidth}
        \centering
        \includegraphics[width=\linewidth]{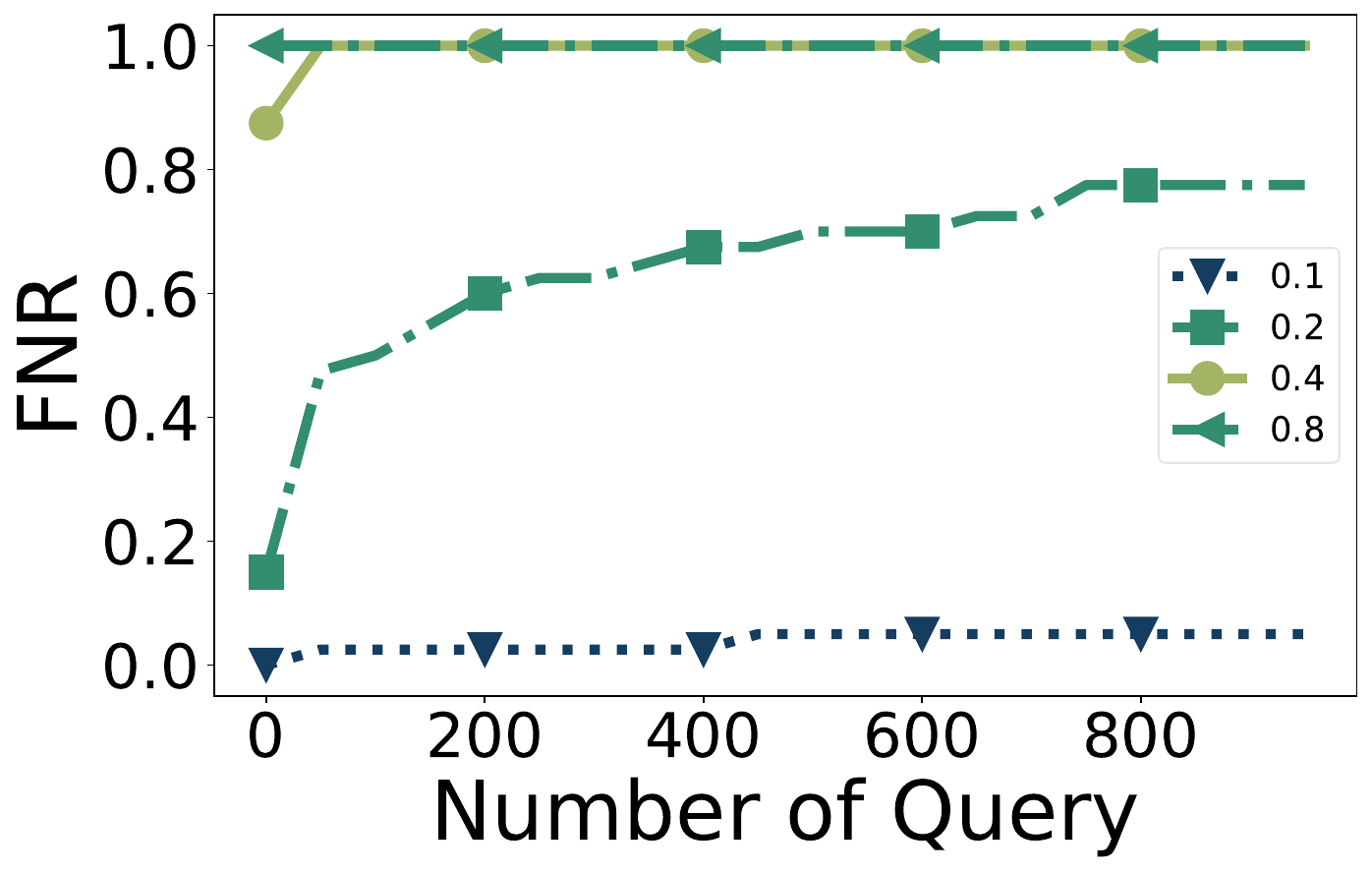}
        \caption{REVMark}
    \end{subfigure}
    \begin{subfigure}{.4\linewidth}
        \centering
        \includegraphics[width=\linewidth]{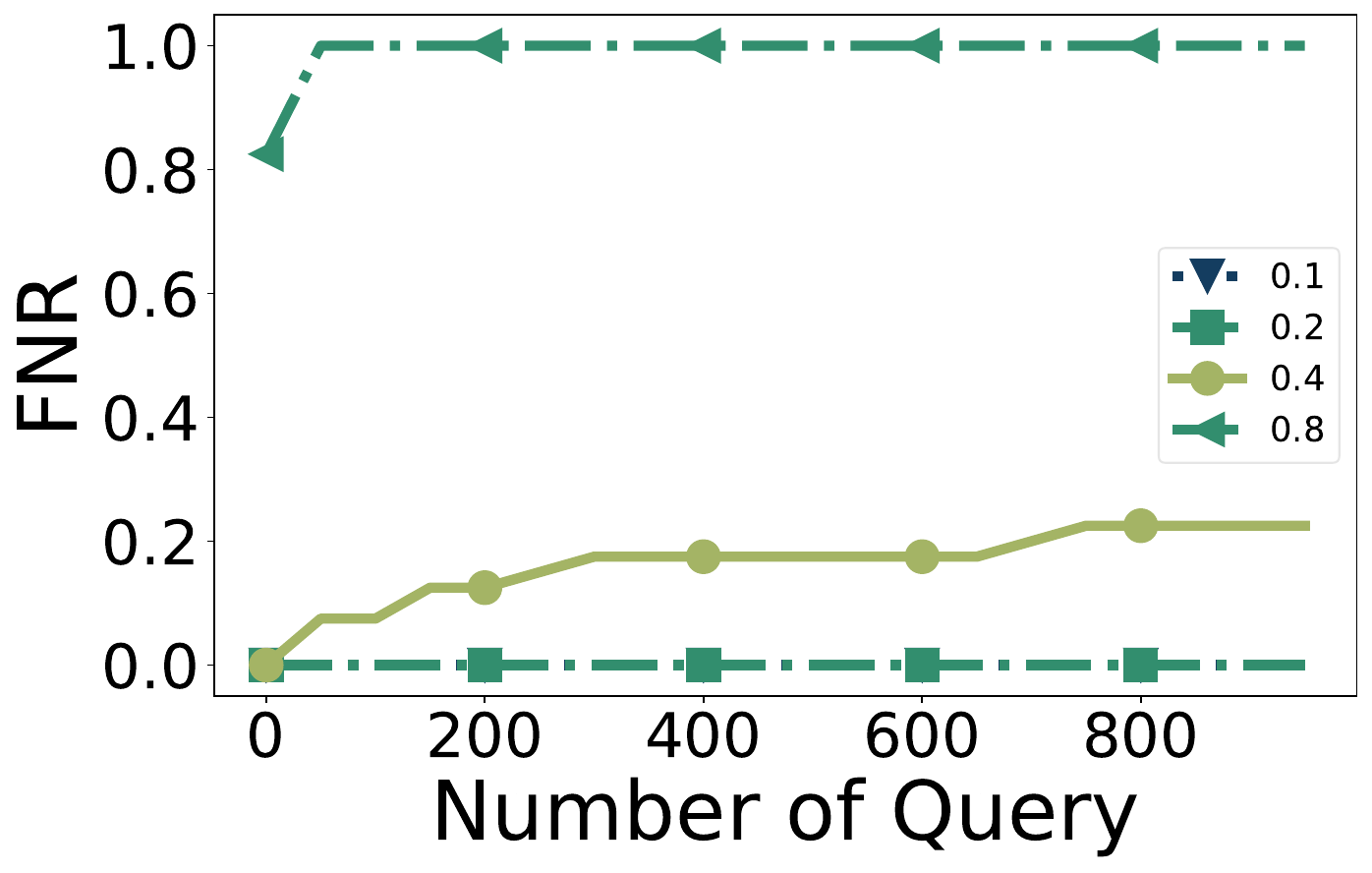}
        \caption{StegaStamp}
    \end{subfigure}
    \caption{\label{fig:square attack larger}Square Attack watermark removal results with larger perturbation bounds. Legend indicates the $l_{\infty}$ bound of perturbations.}
\end{figure}

\begin{figure}[]
    \centering
    \begin{subfigure}{.23\linewidth}
        \centering
        \includegraphics[width=\linewidth]{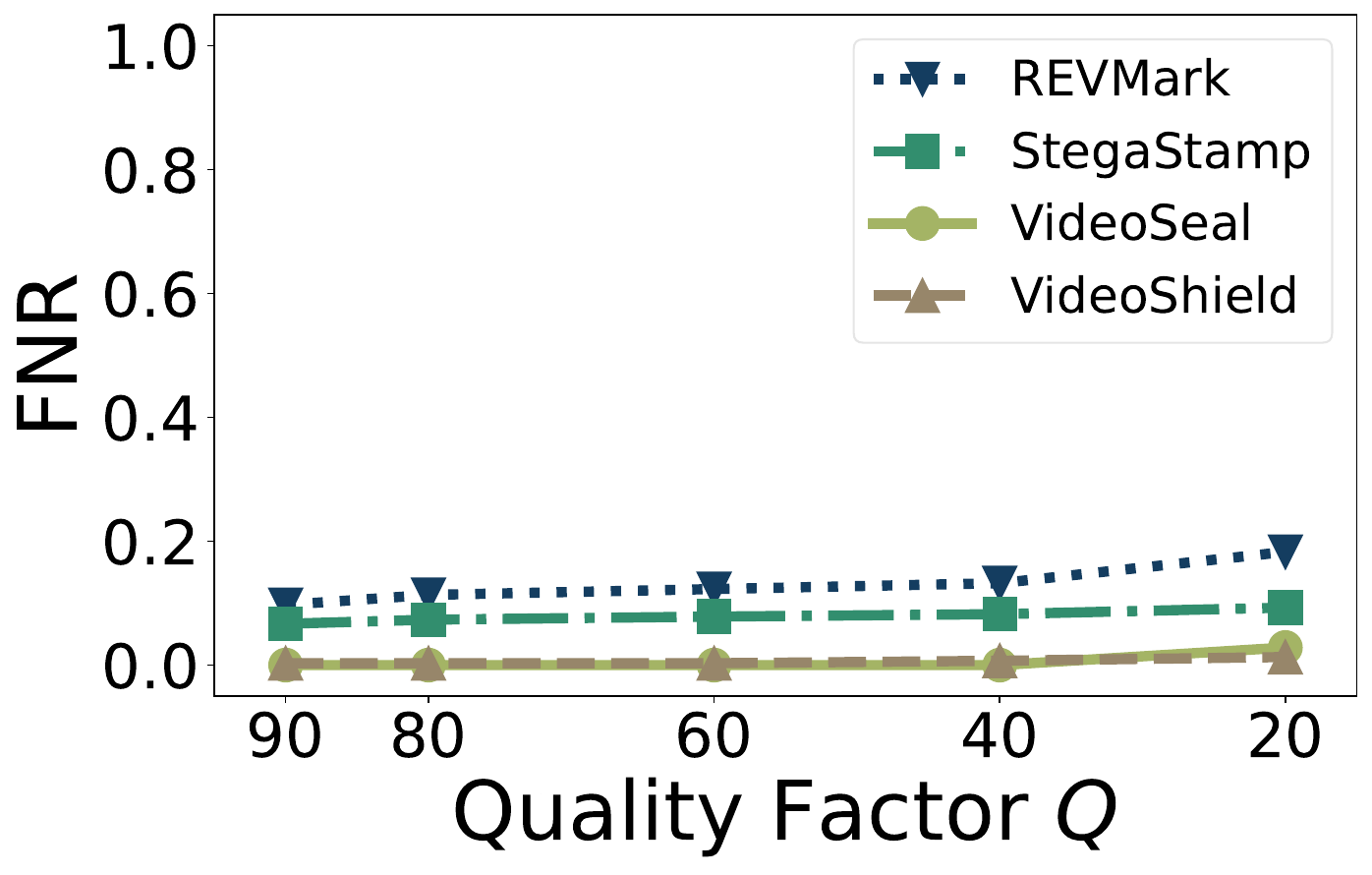}
        \caption{JPEG}
    \end{subfigure}
    \begin{subfigure}{.23\linewidth}
        \centering
        \includegraphics[width=\linewidth]{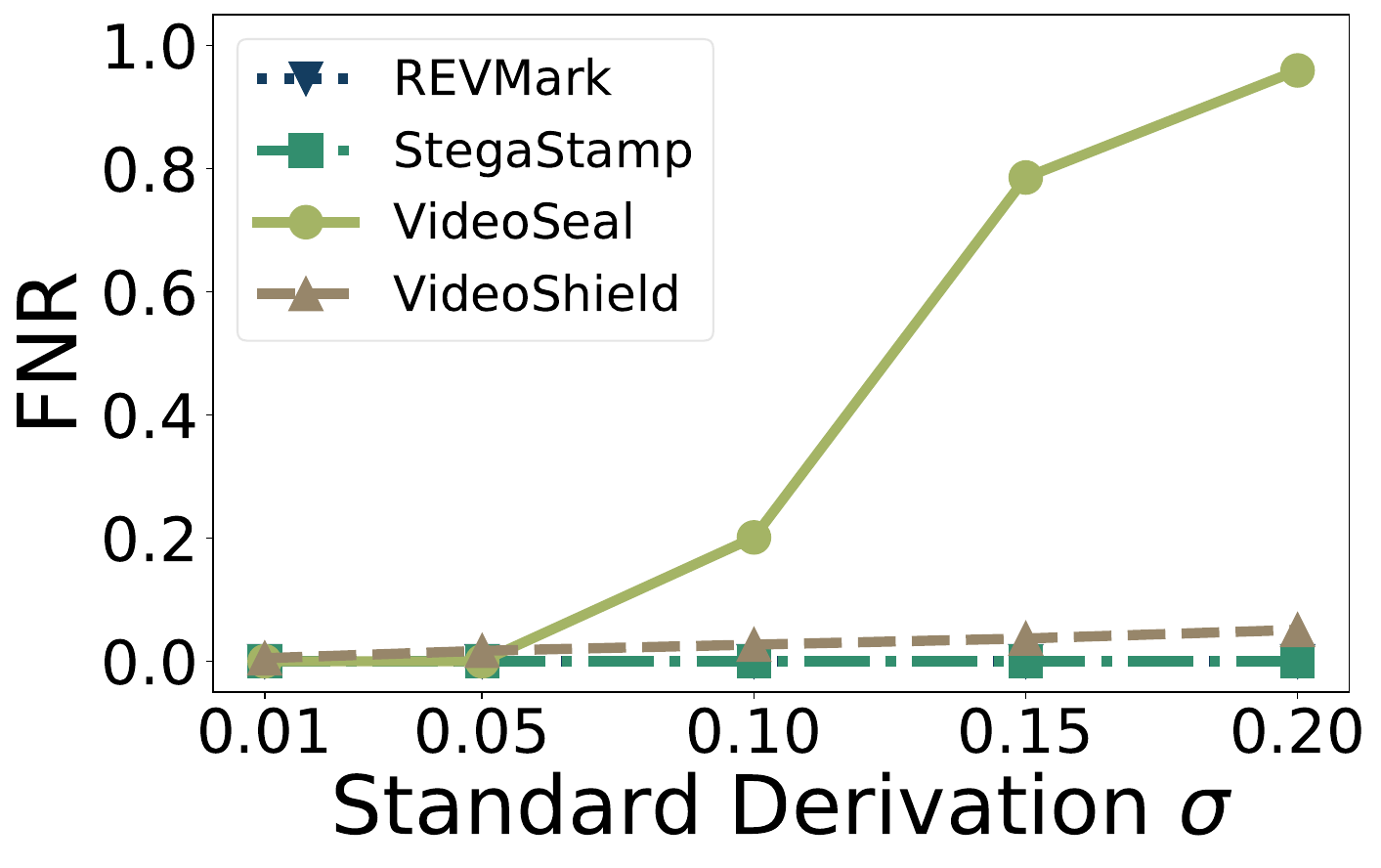}
        \caption{Gaussian Noise}
        \label{subfig:videoseal gaussian}
    \end{subfigure}
    \begin{subfigure}{.23\linewidth}
        \centering
        \includegraphics[width=\linewidth]{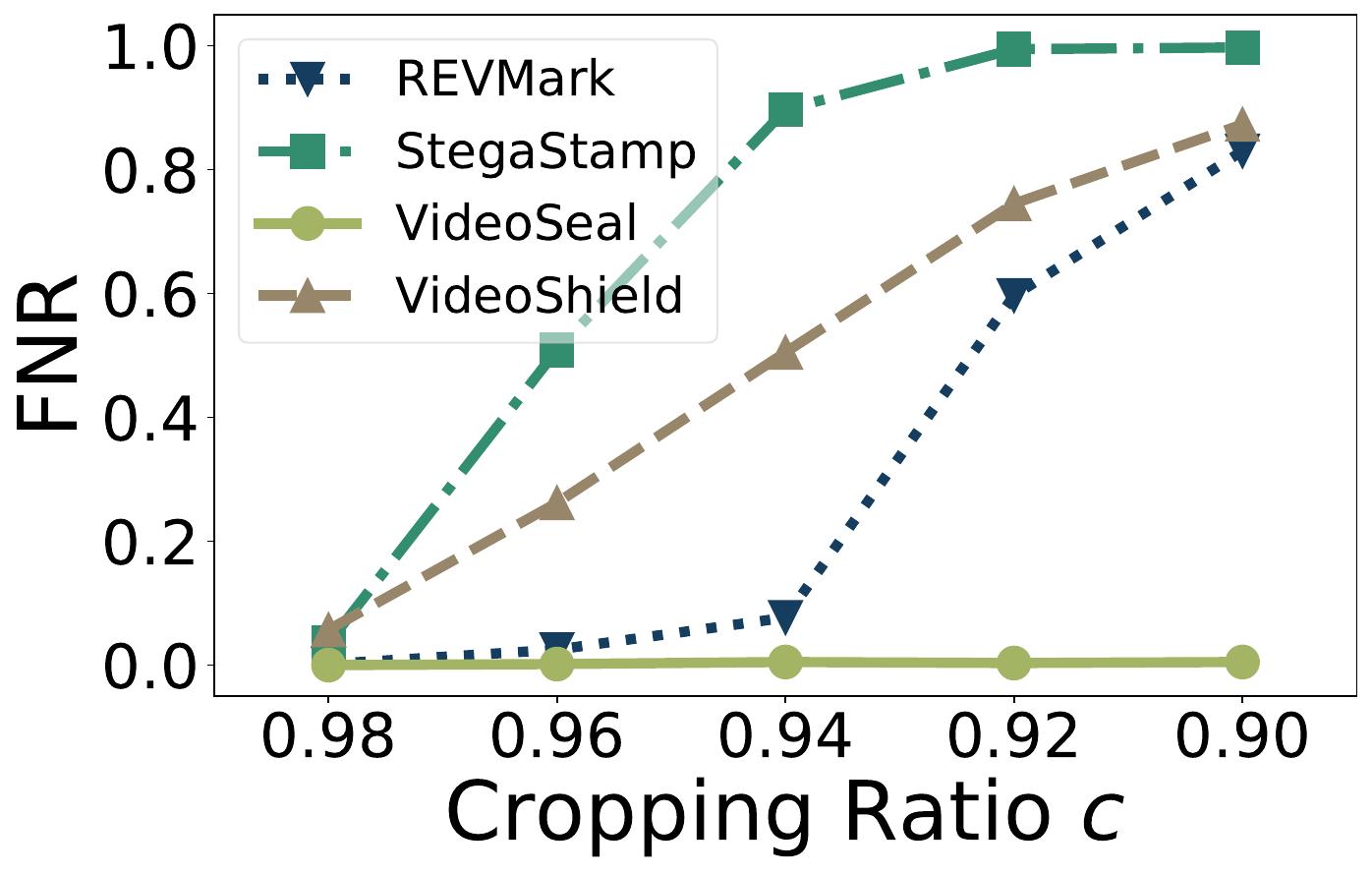}
        \caption{Cropping}
    \end{subfigure}
    \begin{subfigure}{.23\linewidth}
        \centering
        \includegraphics[width=\linewidth]{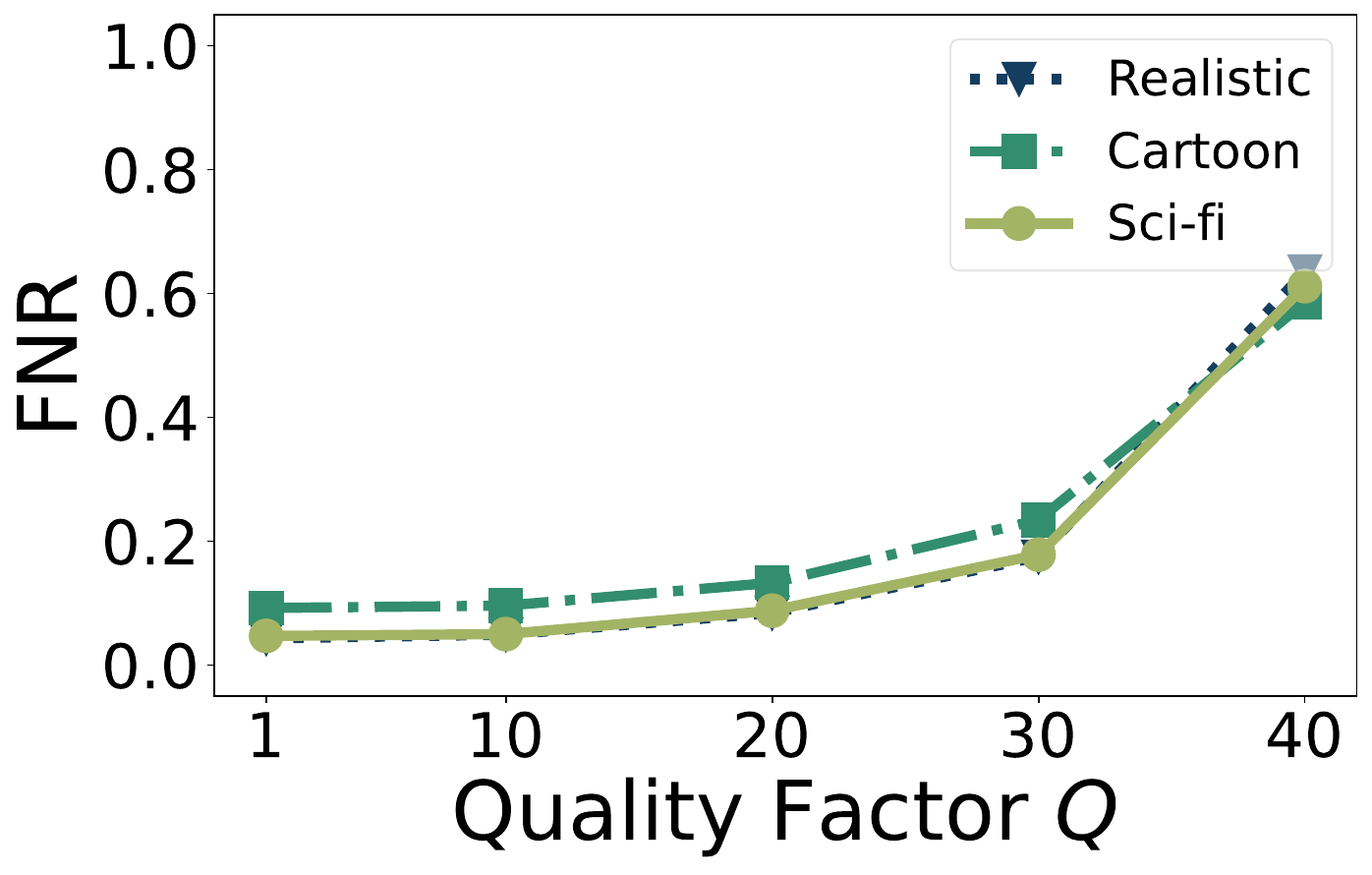}
        \caption{MPEG-4}
    \end{subfigure} \\
    
    \begin{subfigure}{.23\linewidth}
        \centering
        \includegraphics[width=\linewidth]{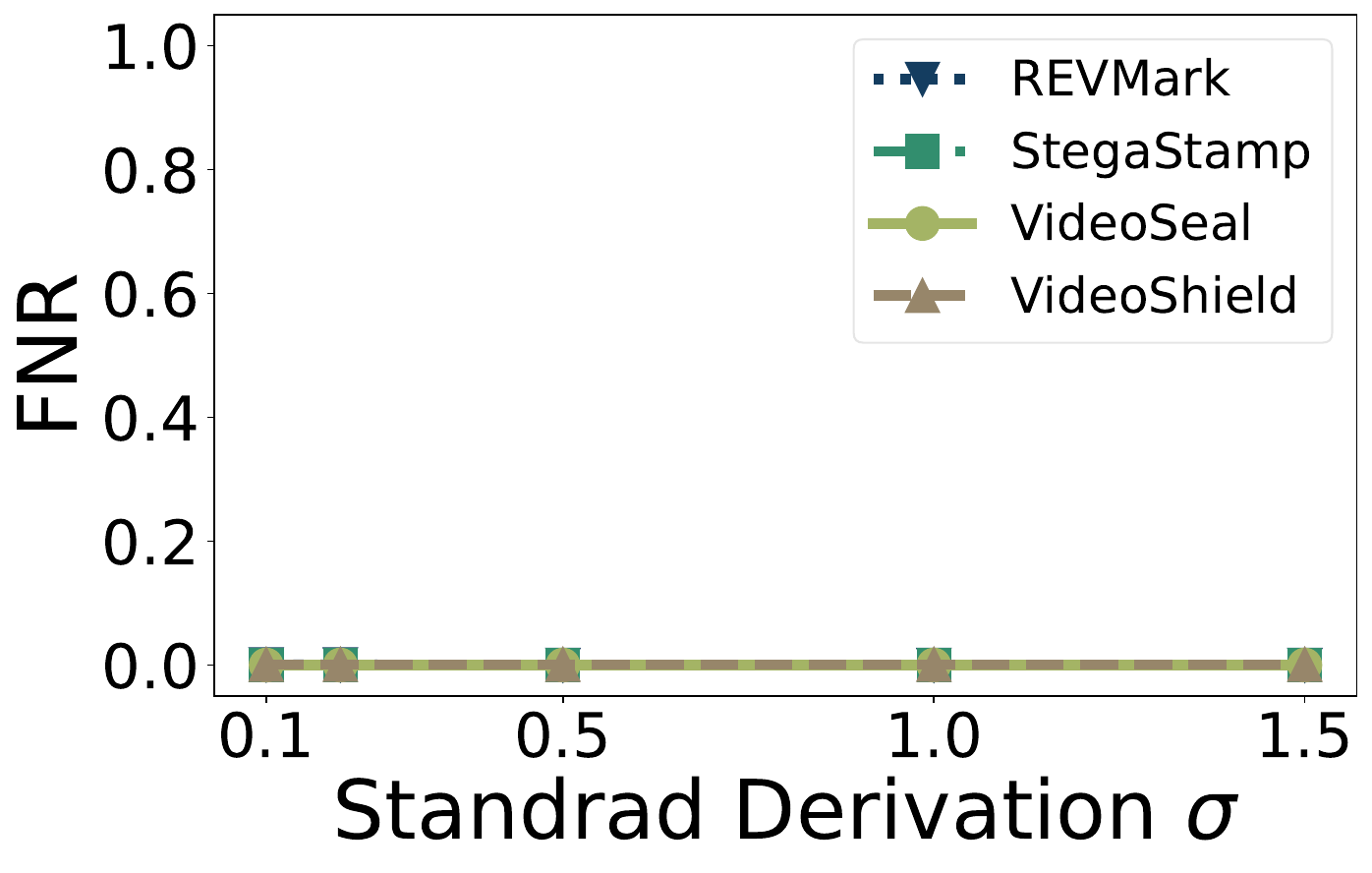}
        \caption{Gaussian Blur}
    \end{subfigure}
    \begin{subfigure}{.23\linewidth}
        \centering
        \includegraphics[width=\linewidth]{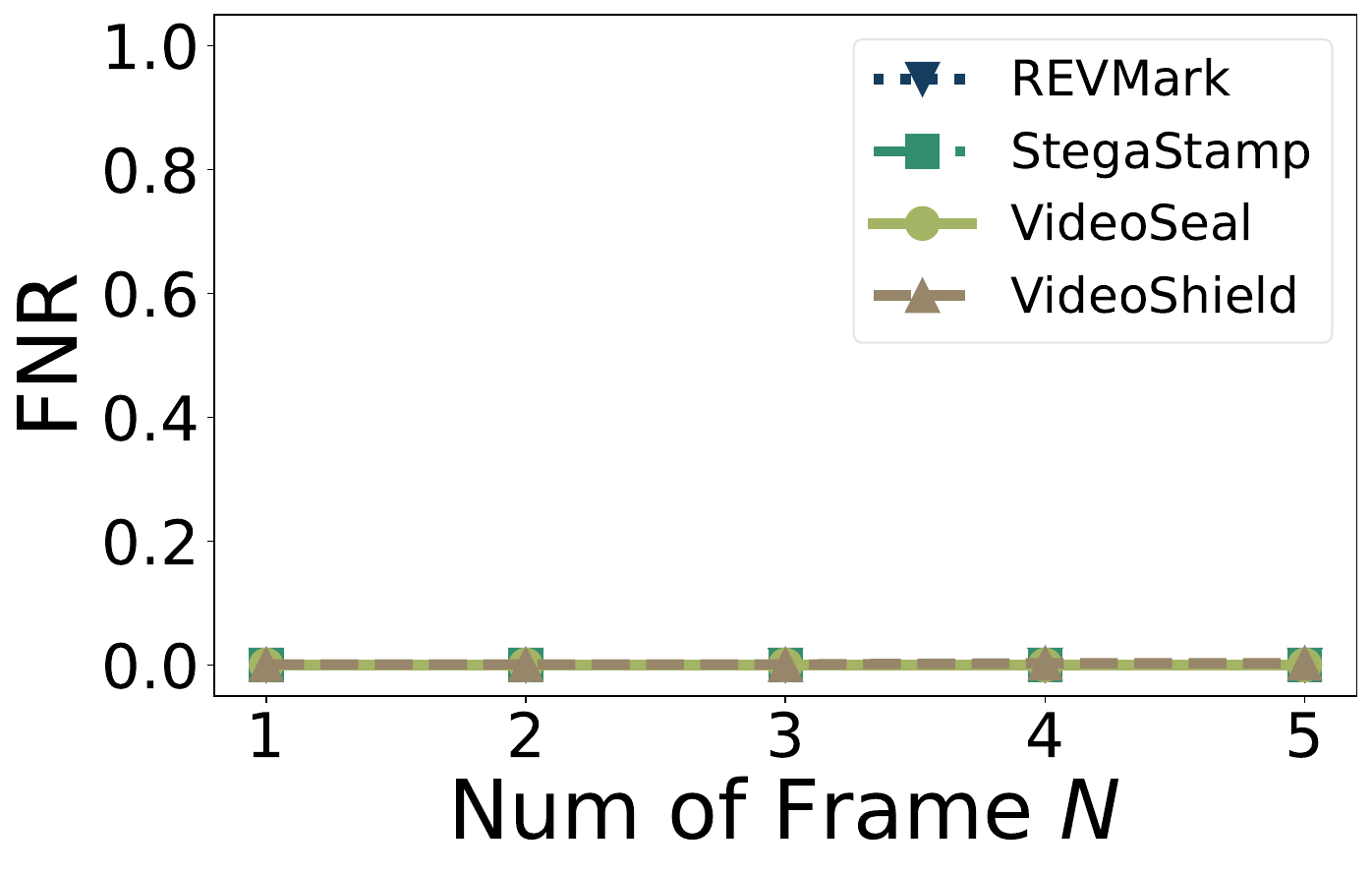}
        \caption{Frame Average}
    \end{subfigure}
    \begin{subfigure}{.23\linewidth}
        \centering
        \includegraphics[width=\linewidth]{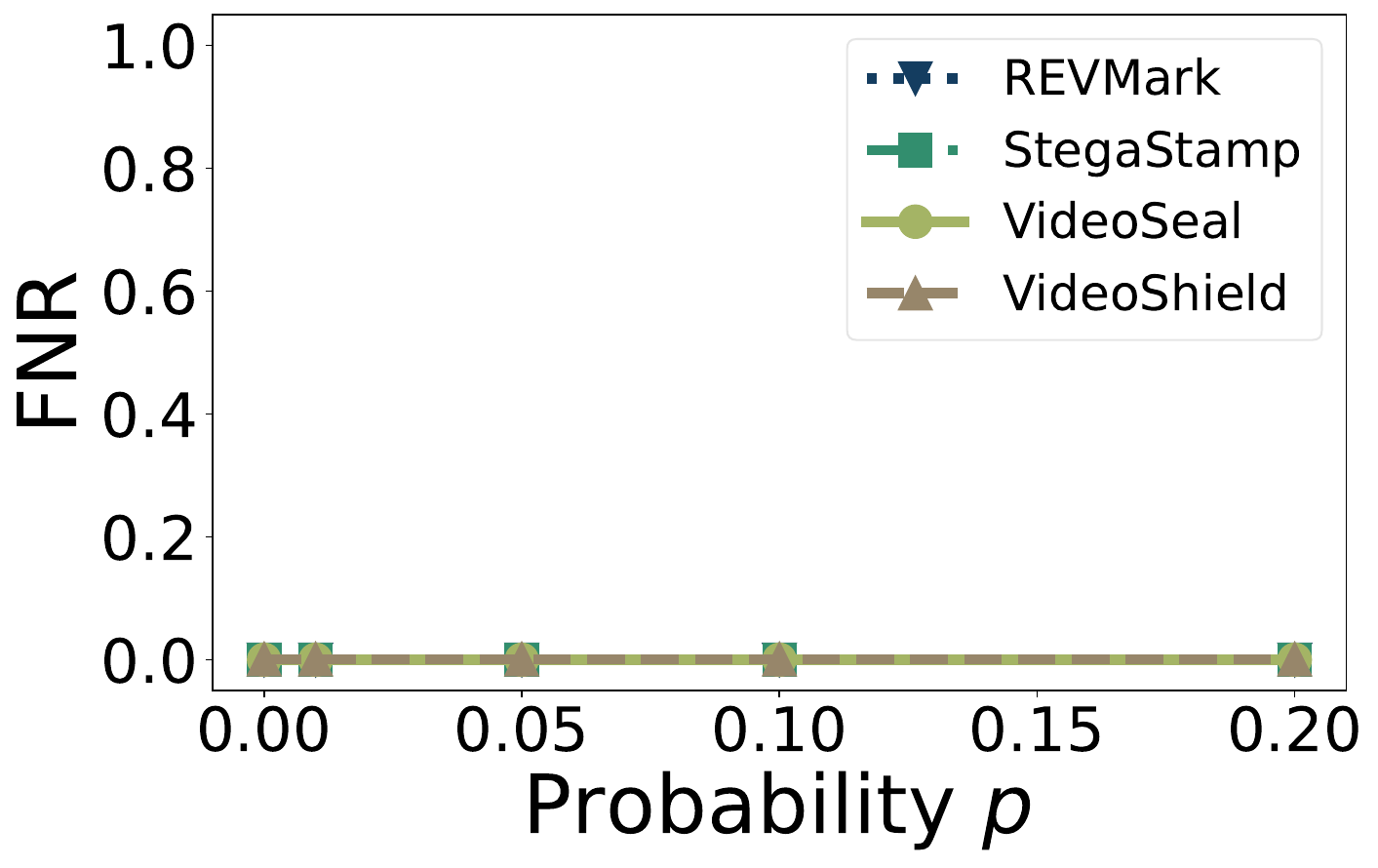}
        \caption{Frame Switch}
    \end{subfigure}
    \begin{subfigure}{.23\linewidth}
        \centering
        \includegraphics[width=\linewidth]{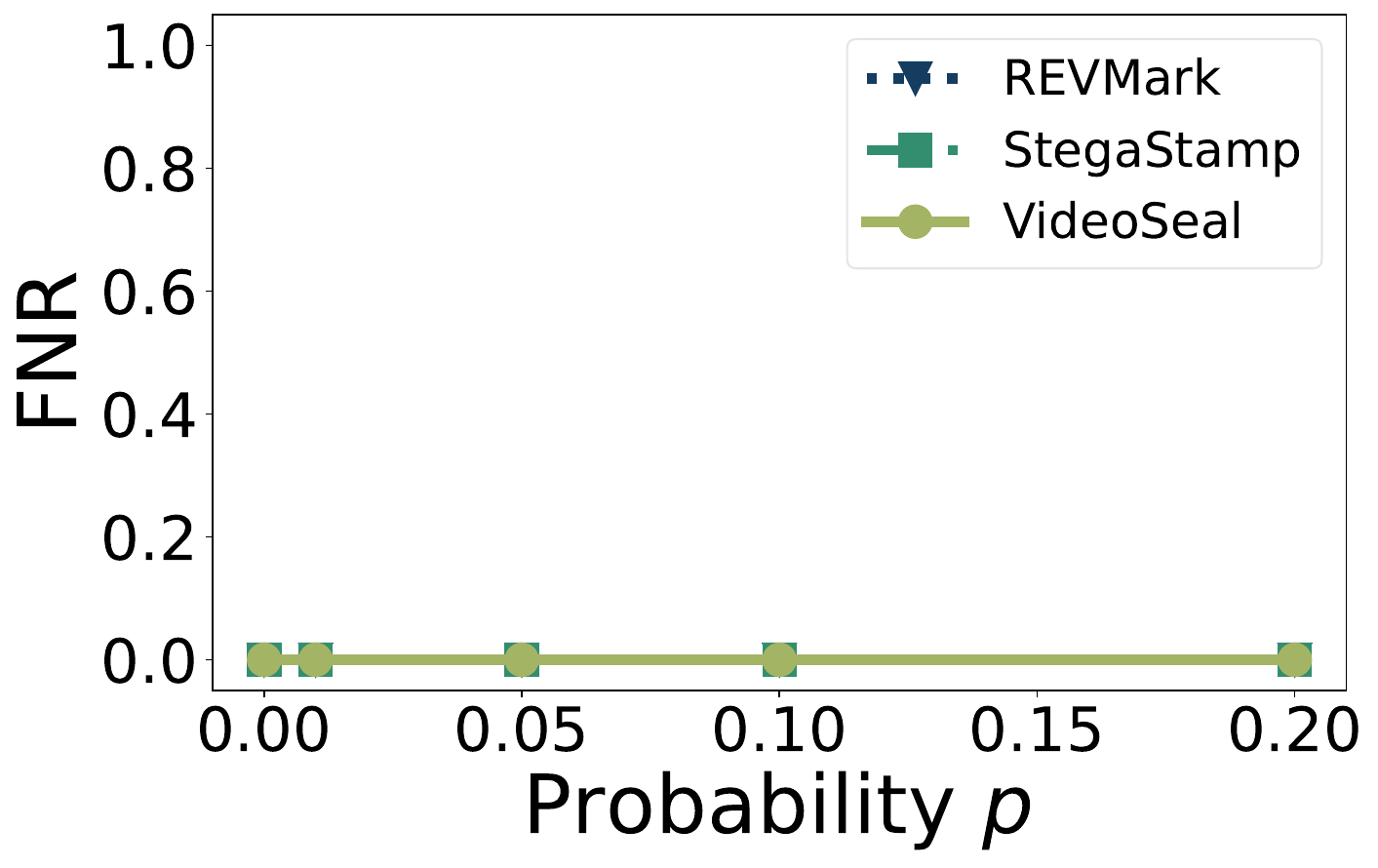}
        \caption{Frame Removal}
    \end{subfigure}
    \caption{\label{fig:watermark}Common perturation watermark removal results for different video watermarking methods. For StegaStamp and VideoSeal, we report results using their best-performing aggregation strategies. FPRs are averaged over videos generated by three generative models and across different video styles. Note that VideoShield does not report results for Frame Removal, as this perturbation changes the video's shape, rendering the perturbed video invalid as input for VideoShield’s detection.}
\end{figure}

\begin{figure}[]
    \centering
    \begin{subfigure}{.23\linewidth}
        \centering
        \includegraphics[width=\linewidth]{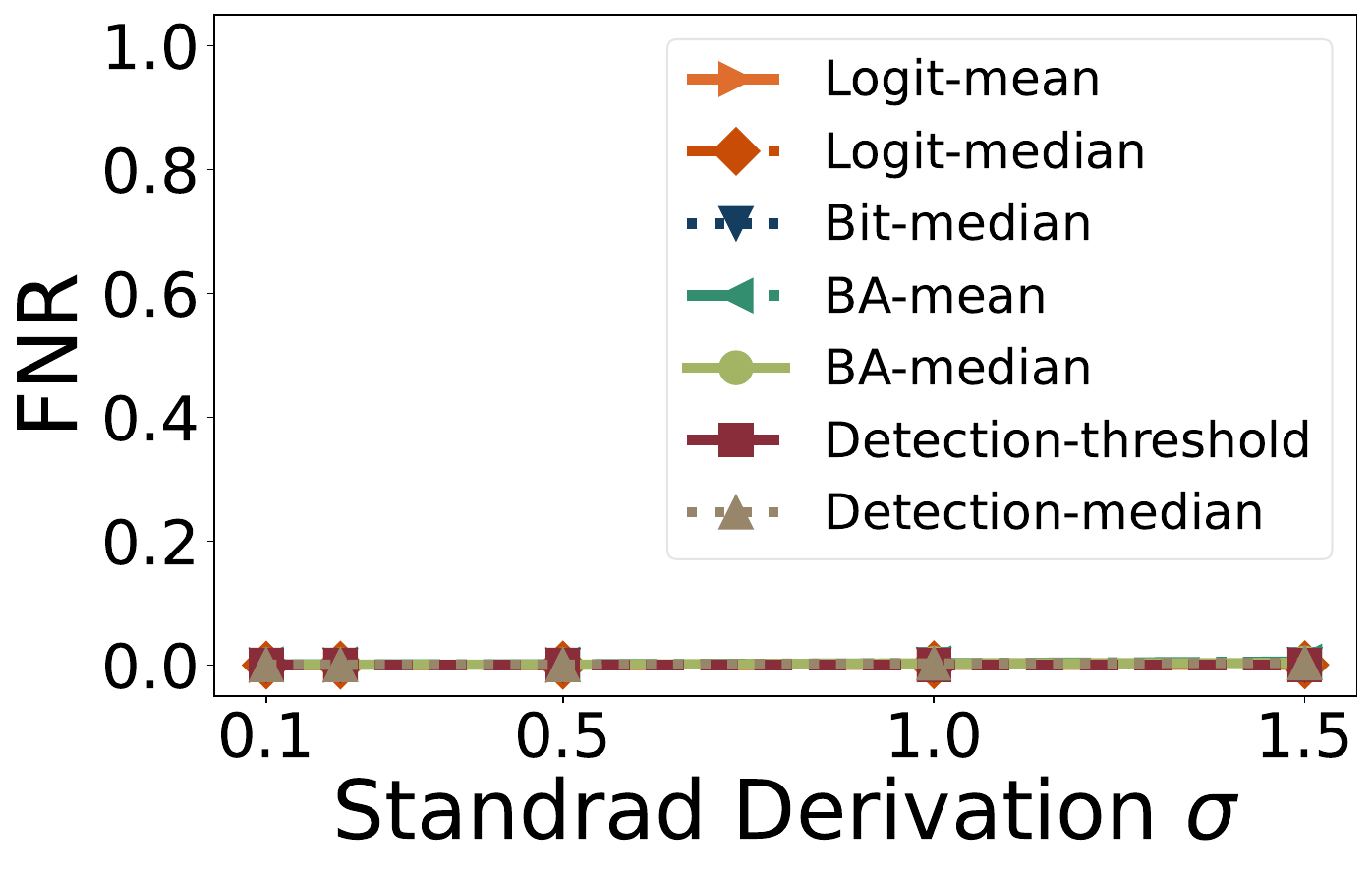}
        \caption{Gaussian Blur}
    \end{subfigure}
    \begin{subfigure}{.23\linewidth}
        \centering
        \includegraphics[width=\linewidth]{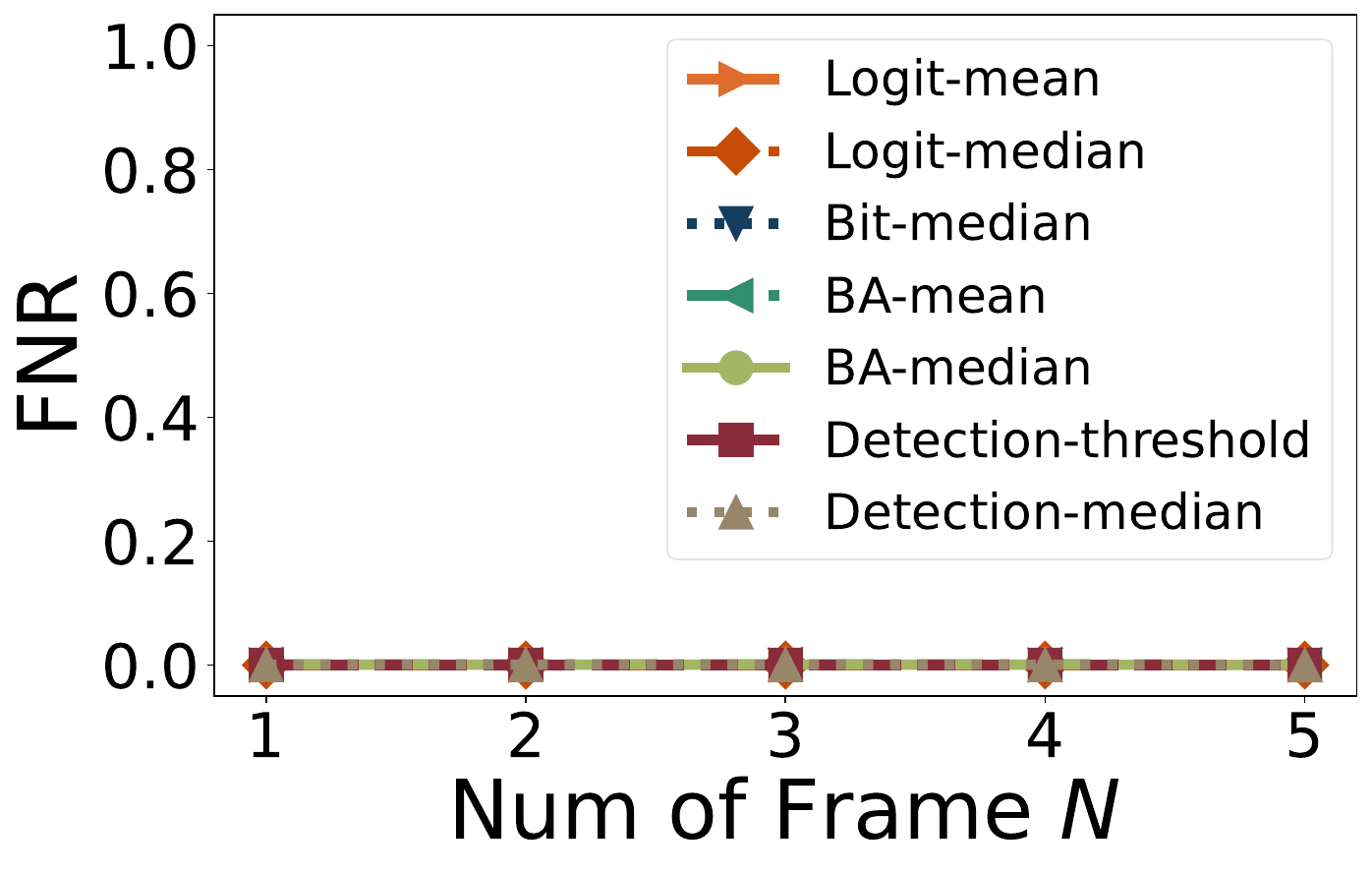}
        \caption{Frame Average}
    \end{subfigure}
    \begin{subfigure}{.23\linewidth}
        \centering
        \includegraphics[width=\linewidth]{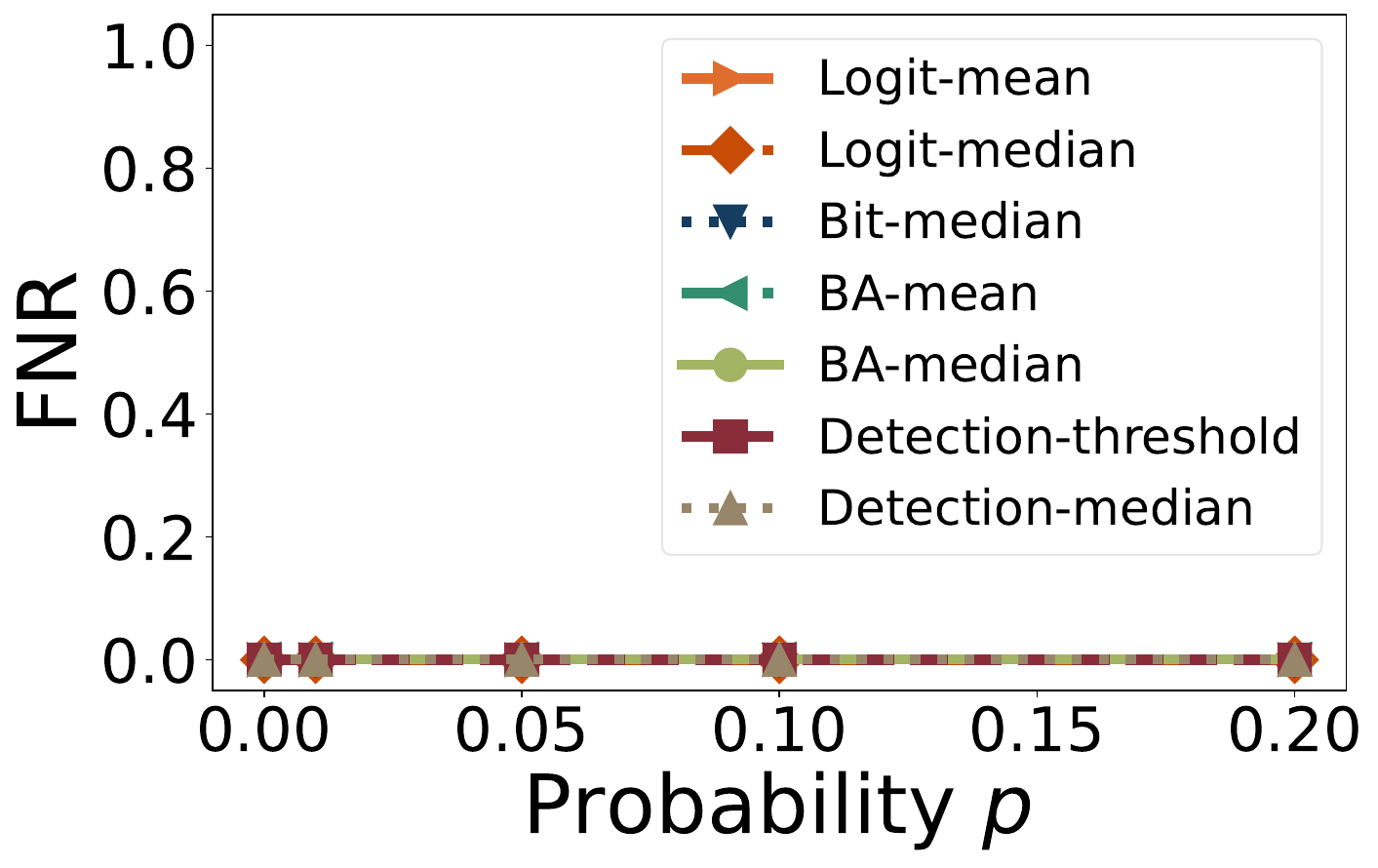}
        \caption{Frame Switch}
    \end{subfigure}
    \begin{subfigure}{.23\linewidth}
        \centering
        \includegraphics[width=\linewidth]{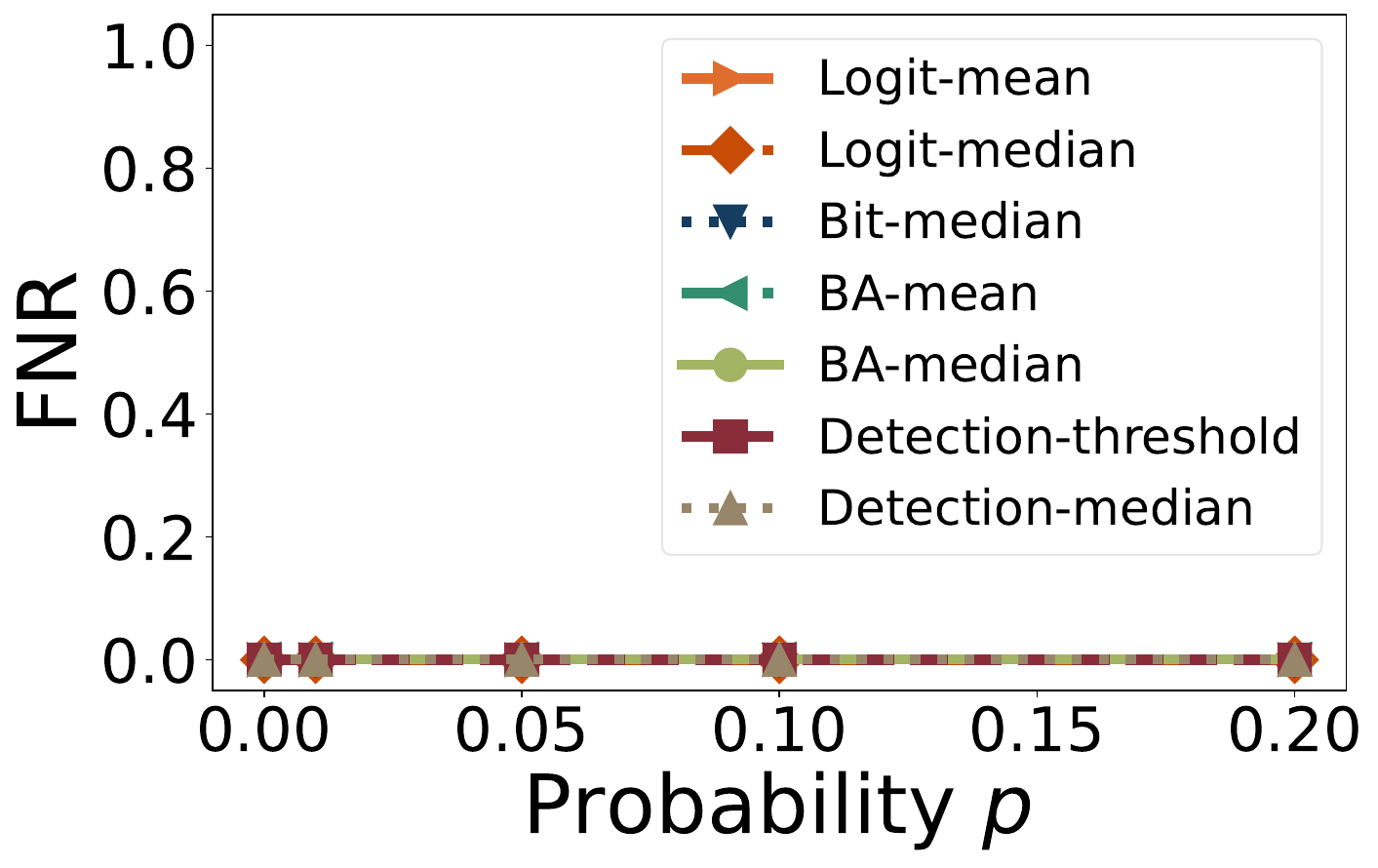}
        \caption{Frame Removal}
    \end{subfigure}
    \caption{\label{fig:aggregation stegastamp other}Other common perturbation watermark removal results for StegaStamp with different aggregation strategies.}
\end{figure}

\begin{figure}[]
    \centering
    \begin{subfigure}{.23\linewidth}
        \centering
        \includegraphics[width=\linewidth]{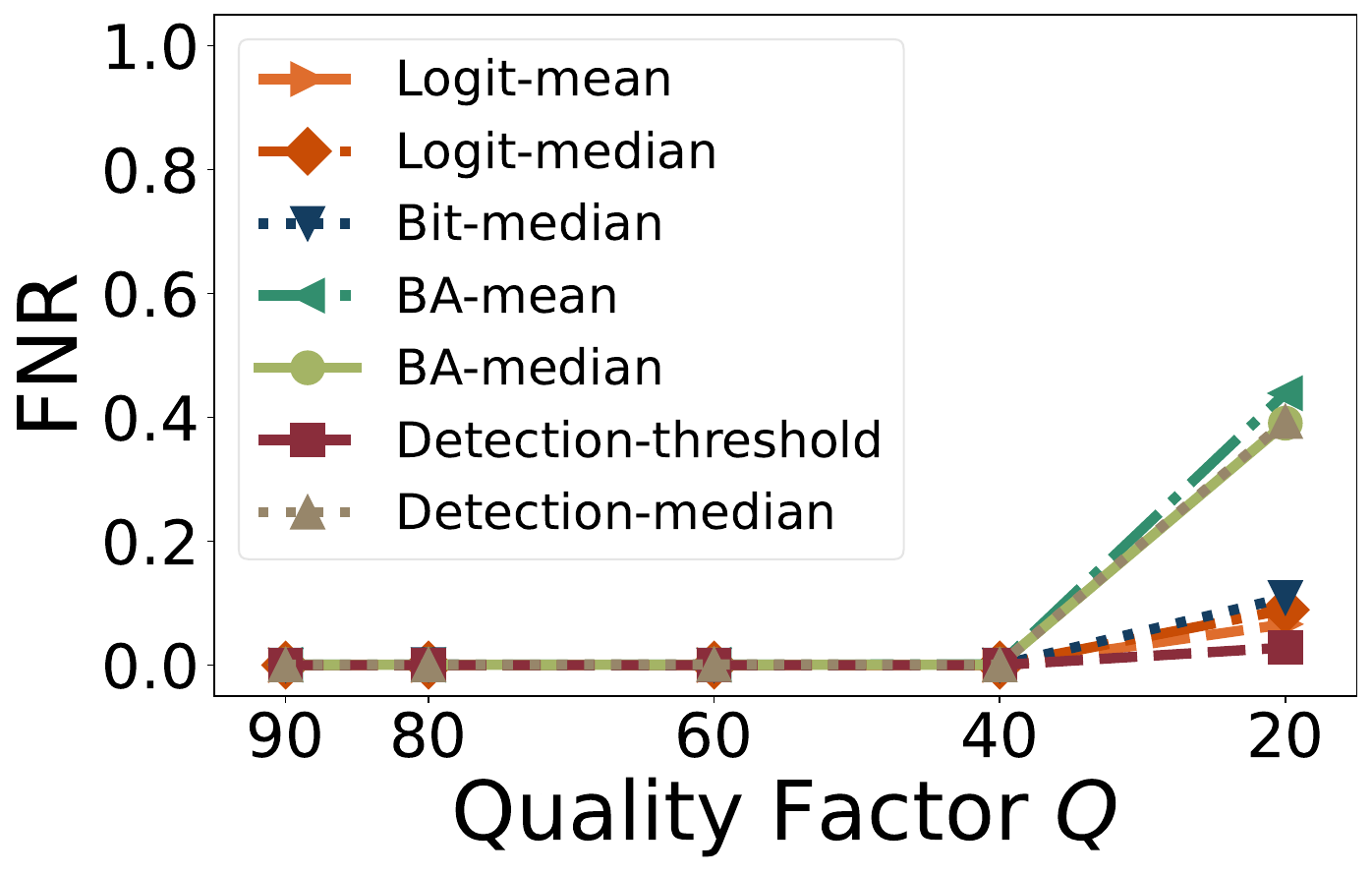}
        \caption{JPEG}
    \end{subfigure}
    \begin{subfigure}{.23\linewidth}
        \centering
        \includegraphics[width=\linewidth]{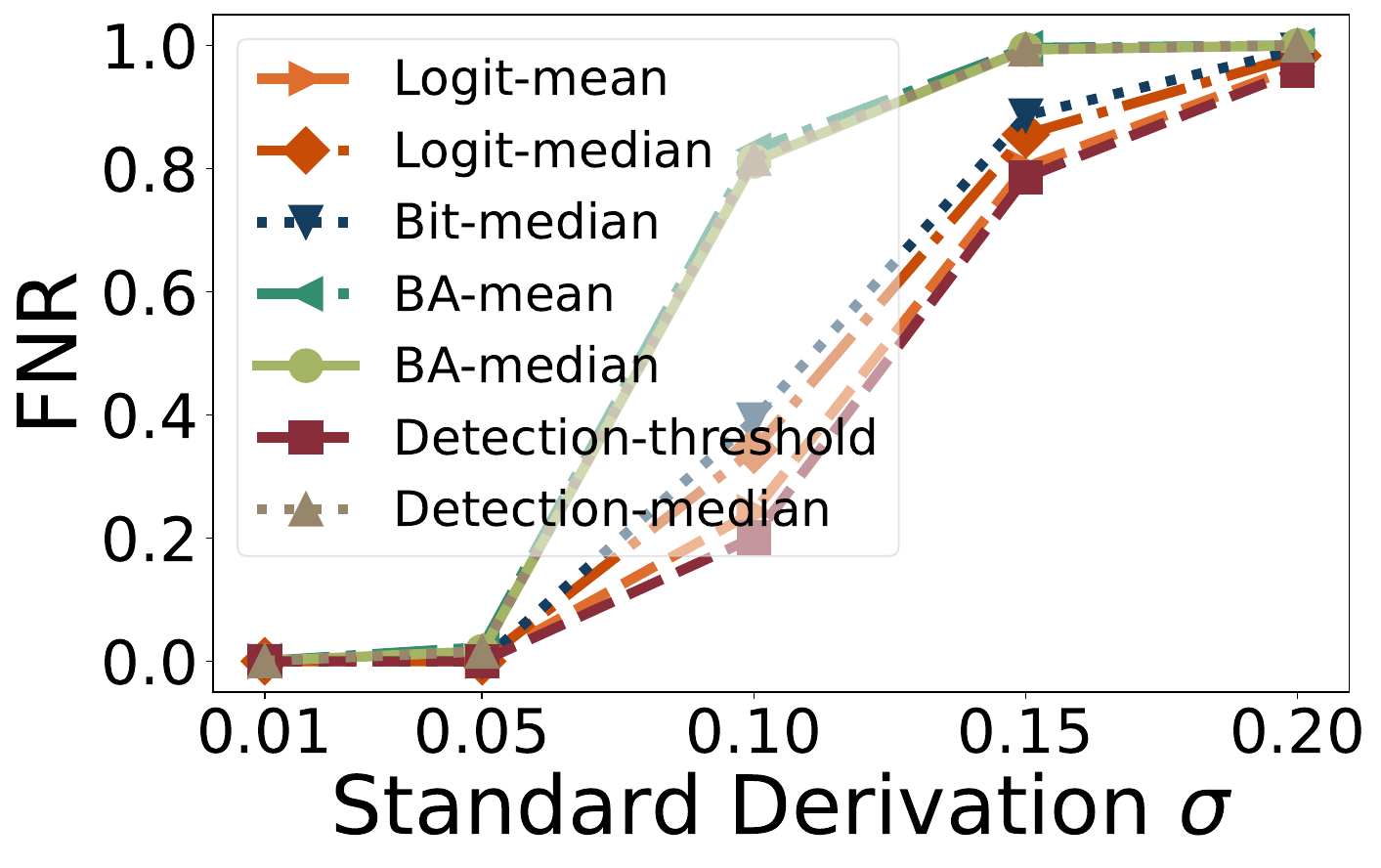}
        \caption{Gaussian Noise}
    \end{subfigure}
    \begin{subfigure}{.23\linewidth}
        \centering
        \includegraphics[width=\linewidth]{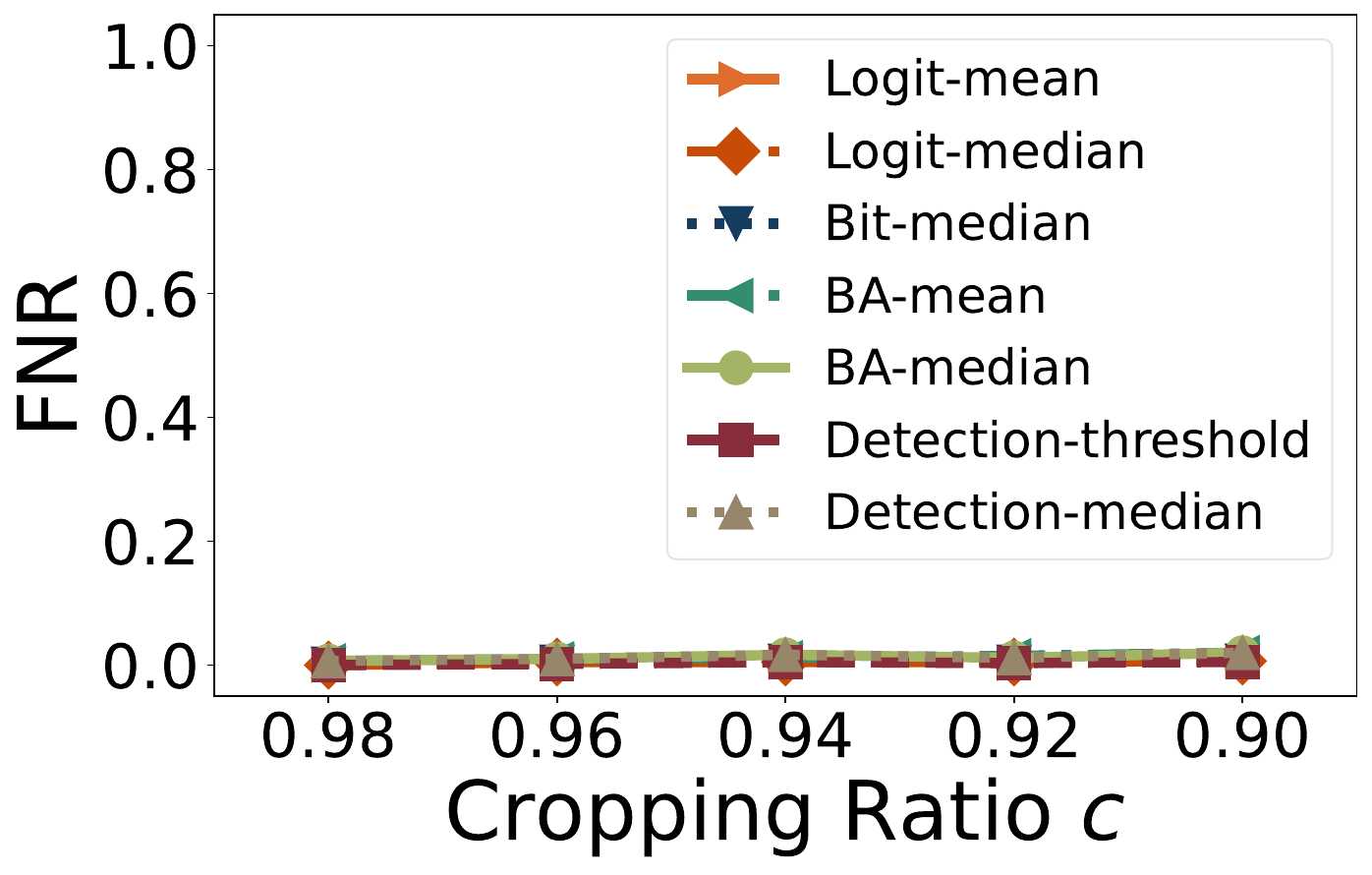}
        \caption{Cropping}
    \end{subfigure}
    \begin{subfigure}{.23\linewidth}
        \centering
        \includegraphics[width=\linewidth]{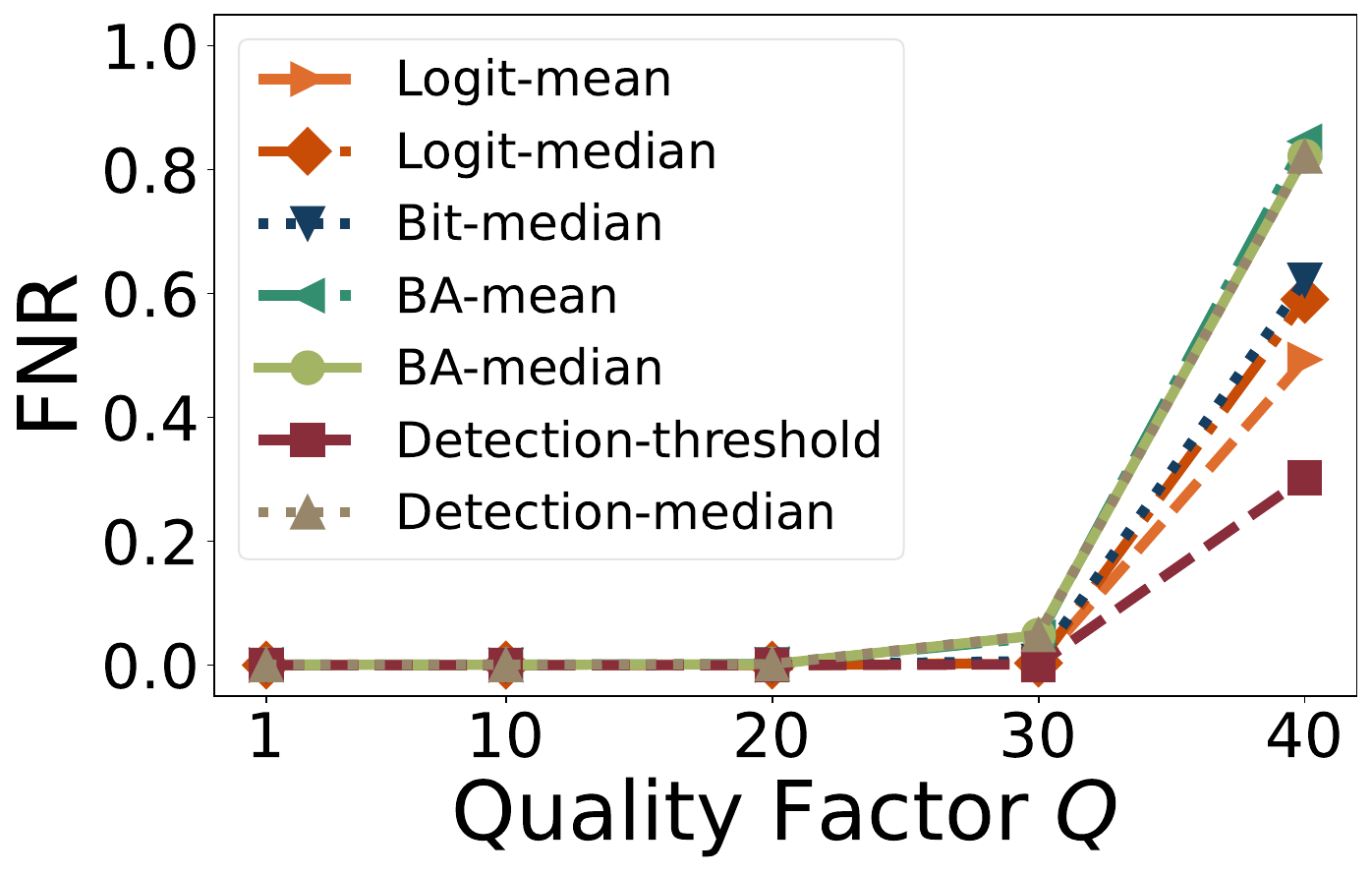}
        \caption{MPEG-4}
    \end{subfigure} \\

    \begin{subfigure}{.23\linewidth}
        \centering
        \includegraphics[width=\linewidth]{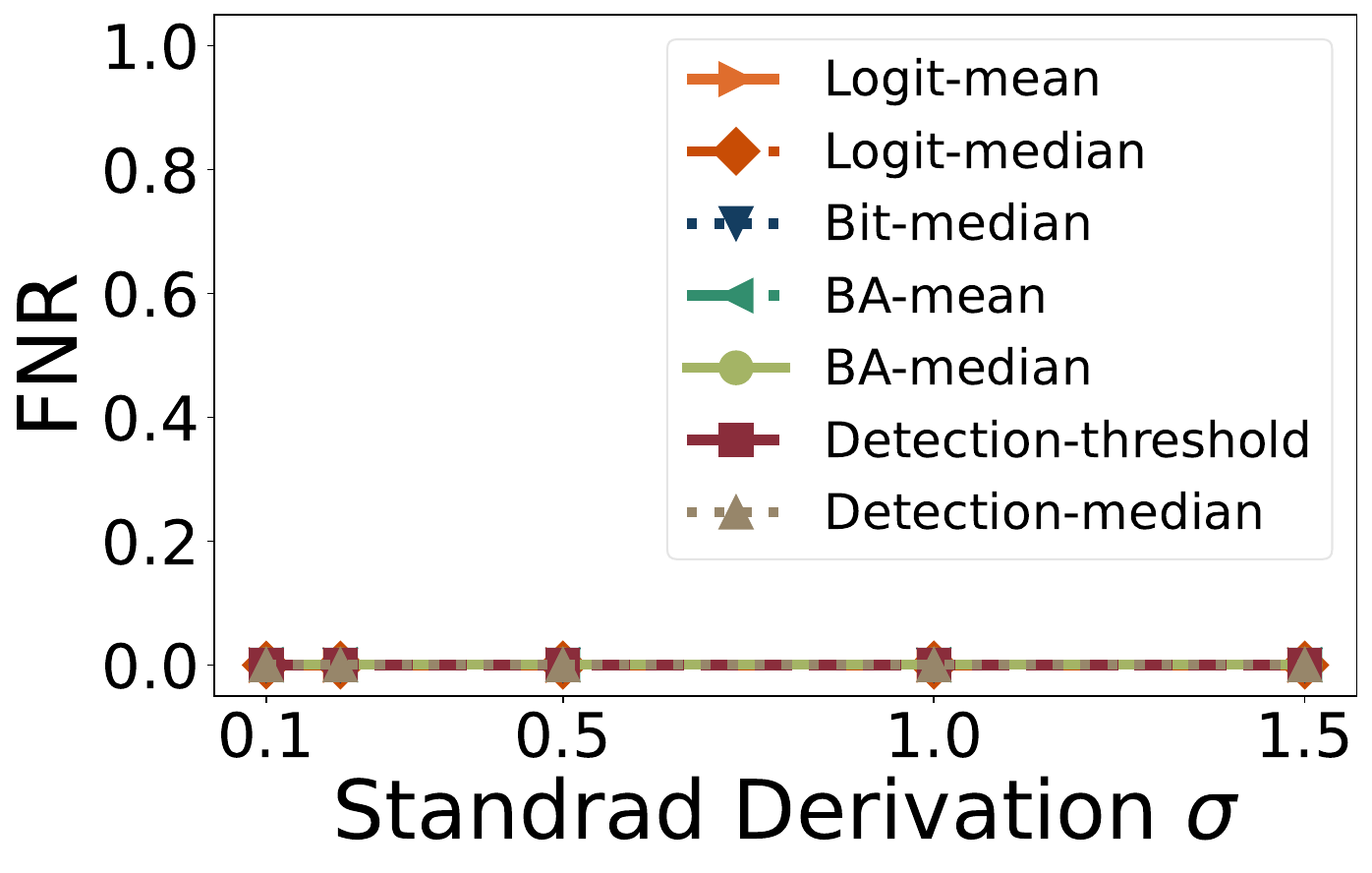}
        \caption{Gaussian Blur}
    \end{subfigure}
    \begin{subfigure}{.23\linewidth}
        \centering
        \includegraphics[width=\linewidth]{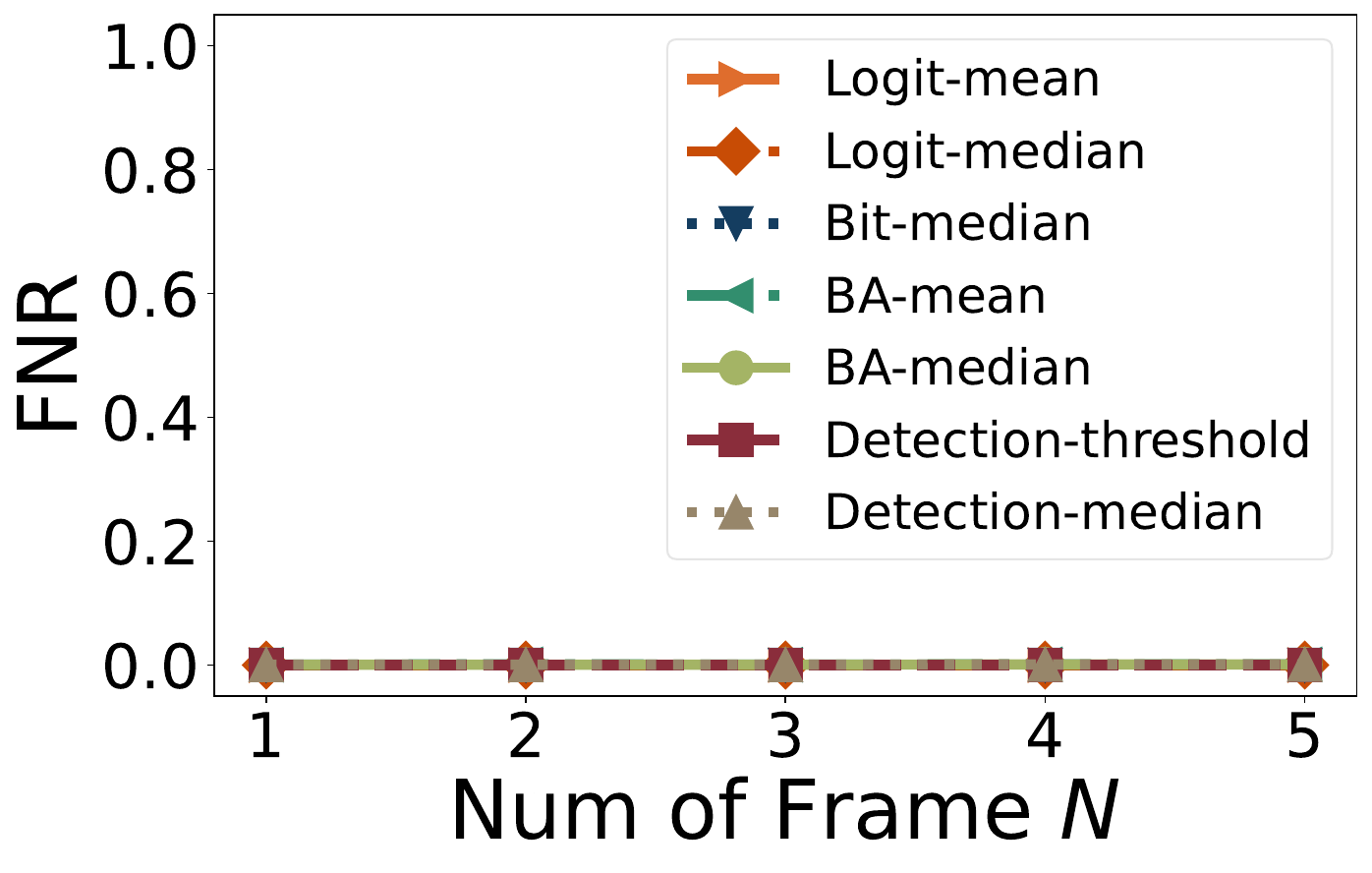}
        \caption{Frame Average}
    \end{subfigure}
    \begin{subfigure}{.23\linewidth}
        \centering
        \includegraphics[width=\linewidth]{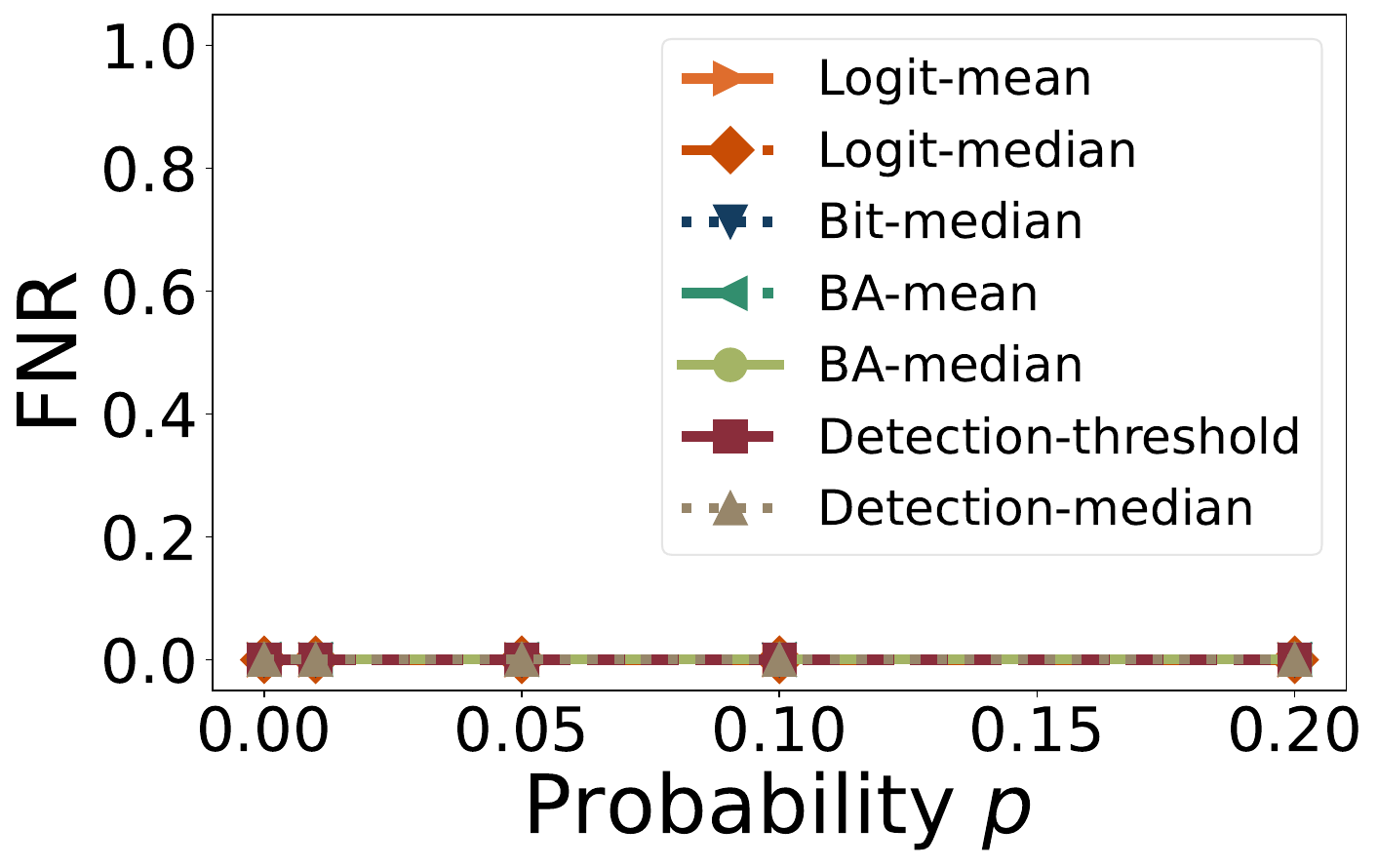}
        \caption{Frame Switch}
    \end{subfigure}
    \begin{subfigure}{.23\linewidth}
        \centering
        \includegraphics[width=\linewidth]{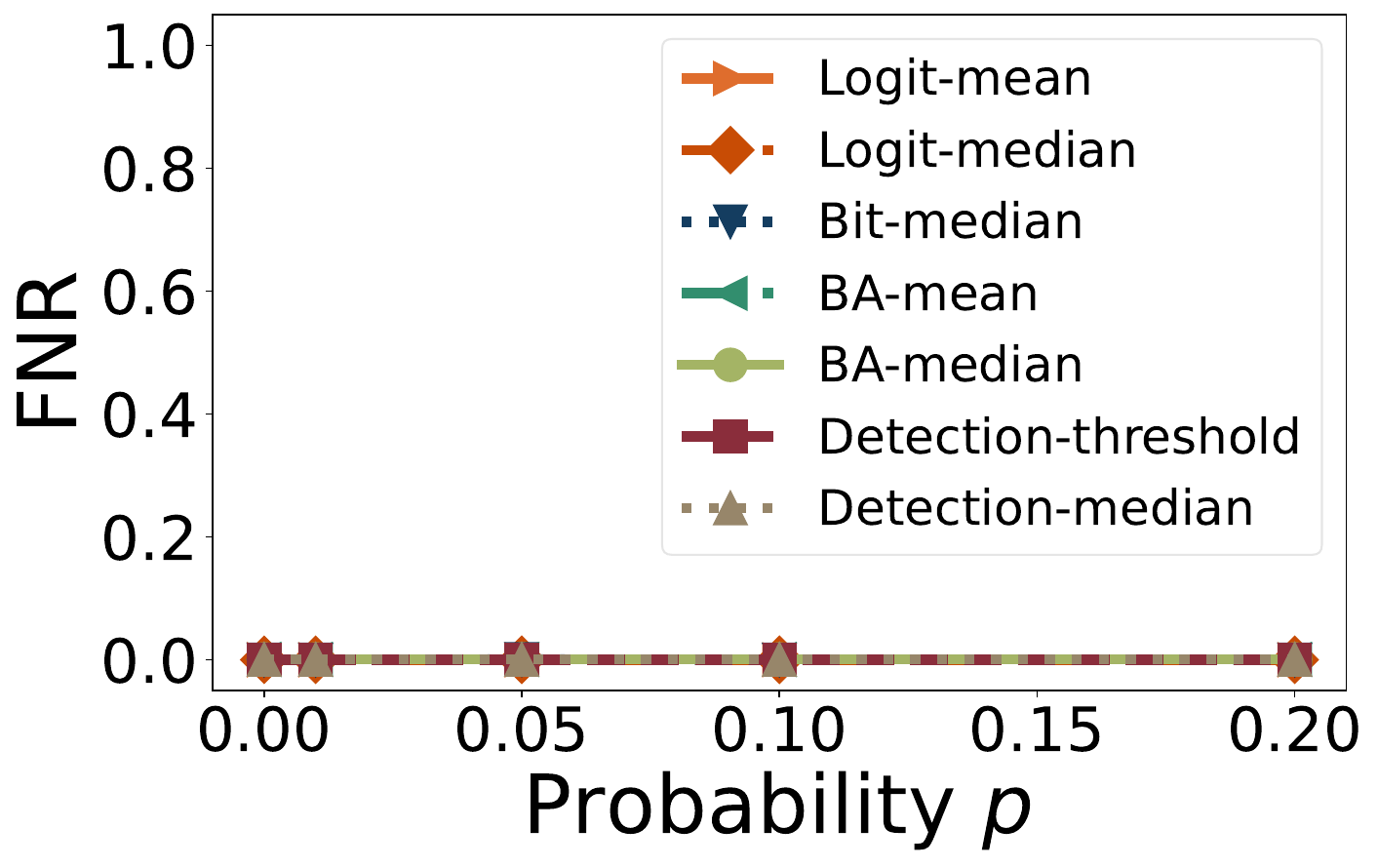}
        \caption{Frame Removal}
    \end{subfigure}
    \caption{\label{fig:aggregation videoseal}Common perturbation watermark removal results for VideoSeal with different aggregation strategies.}
\end{figure}

\begin{figure}[]
    \centering
    \begin{subfigure}{.23\linewidth}
        \centering
        \includegraphics[width=\linewidth]{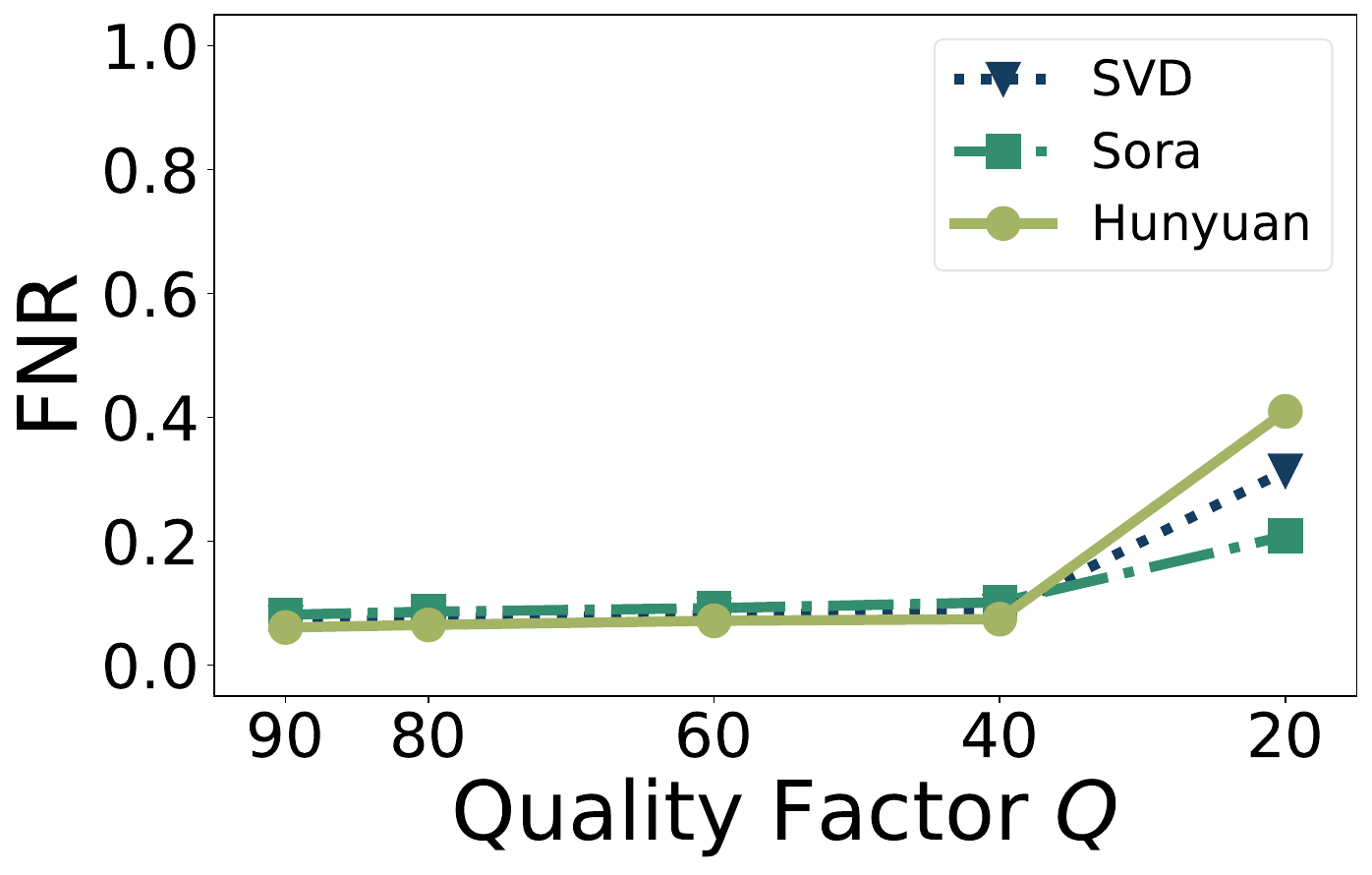}
        \caption{JPEG}
    \end{subfigure}
    \begin{subfigure}{.23\linewidth}
        \centering
        \includegraphics[width=\linewidth]{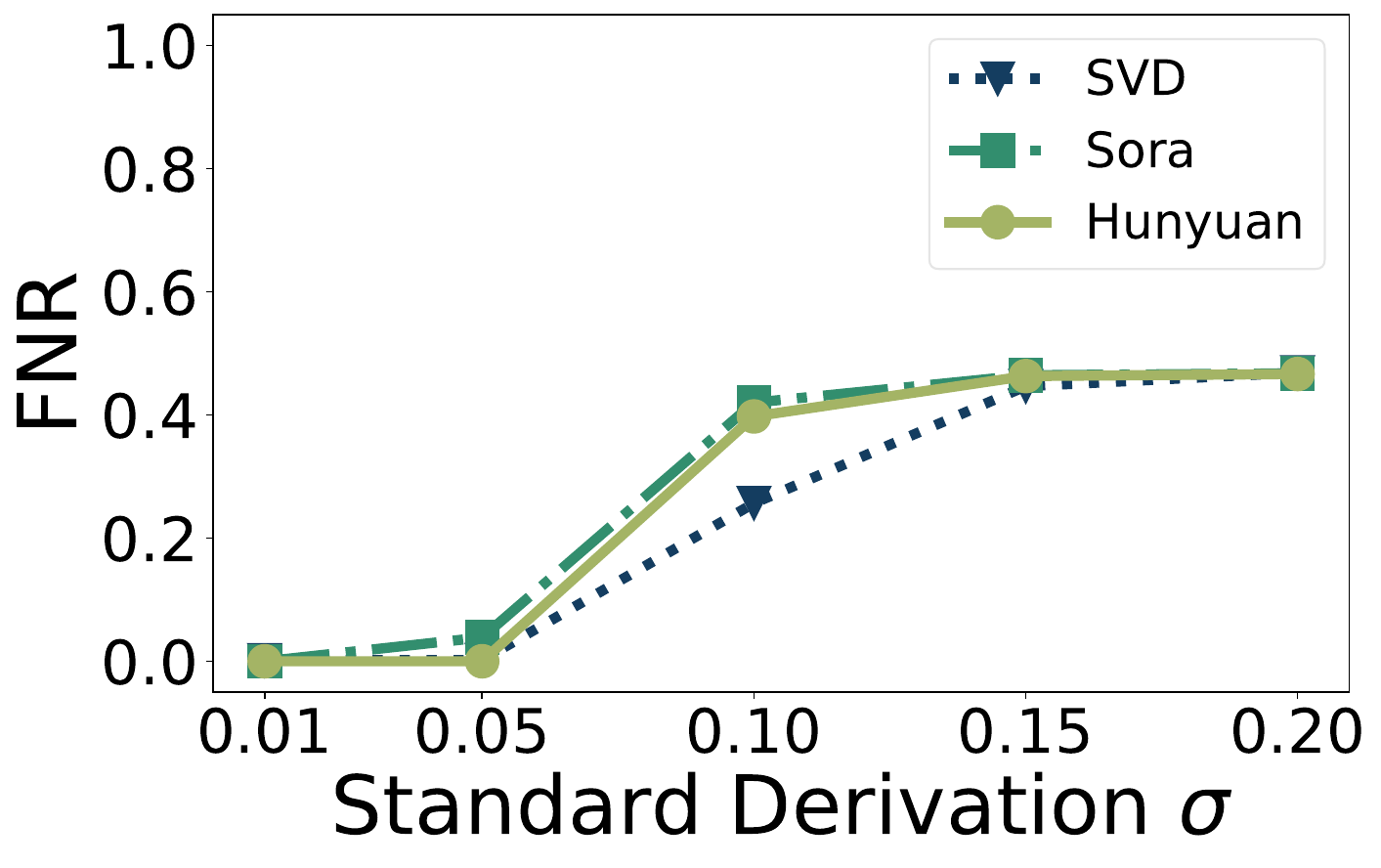}
        \caption{Gaussian Noise}
    \end{subfigure}
    \begin{subfigure}{.23\linewidth}
        \centering
        \includegraphics[width=\linewidth]{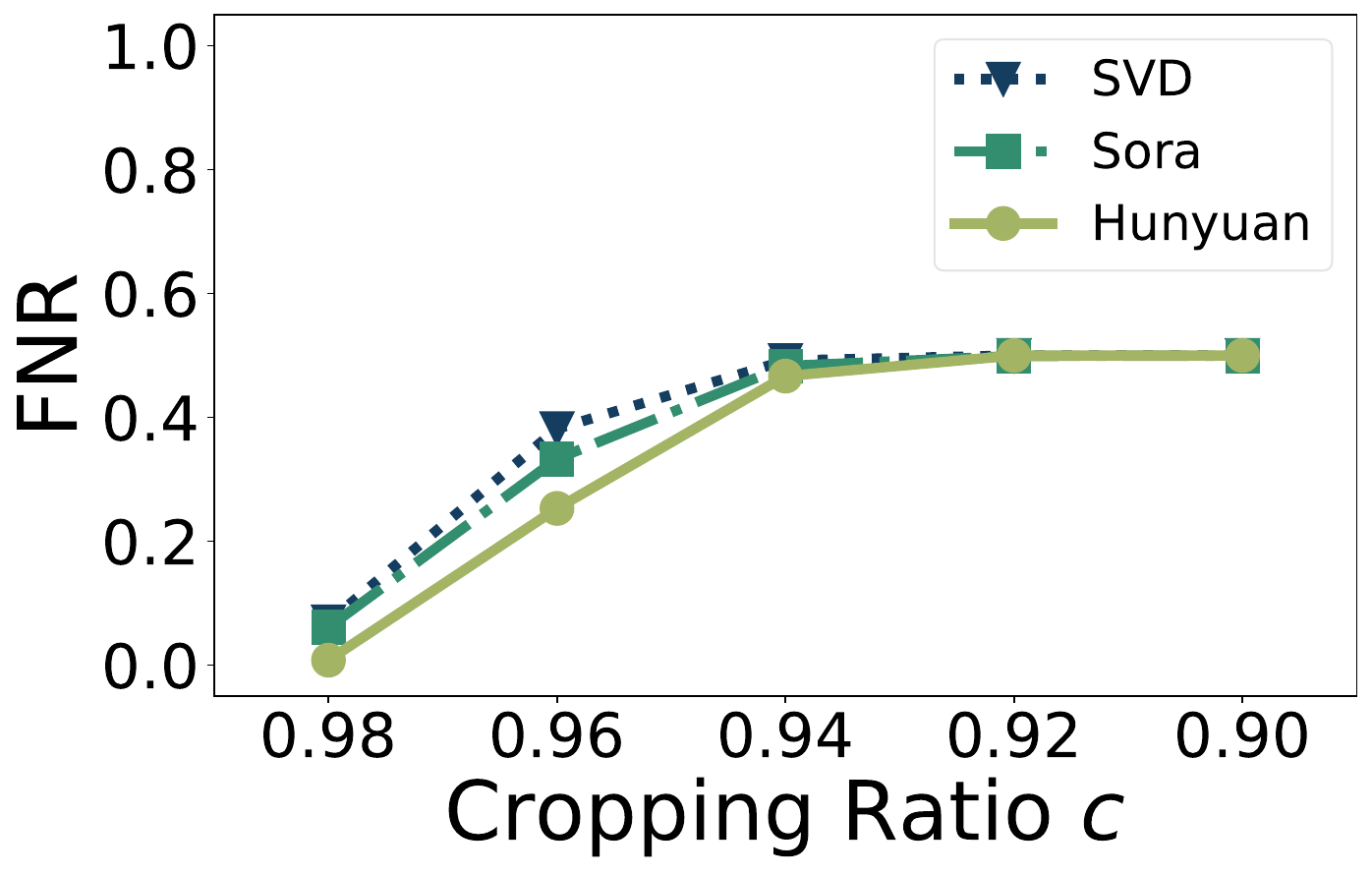}
        \caption{Cropping}
    \end{subfigure}
    \begin{subfigure}{.23\linewidth}
        \centering
        \includegraphics[width=\linewidth]{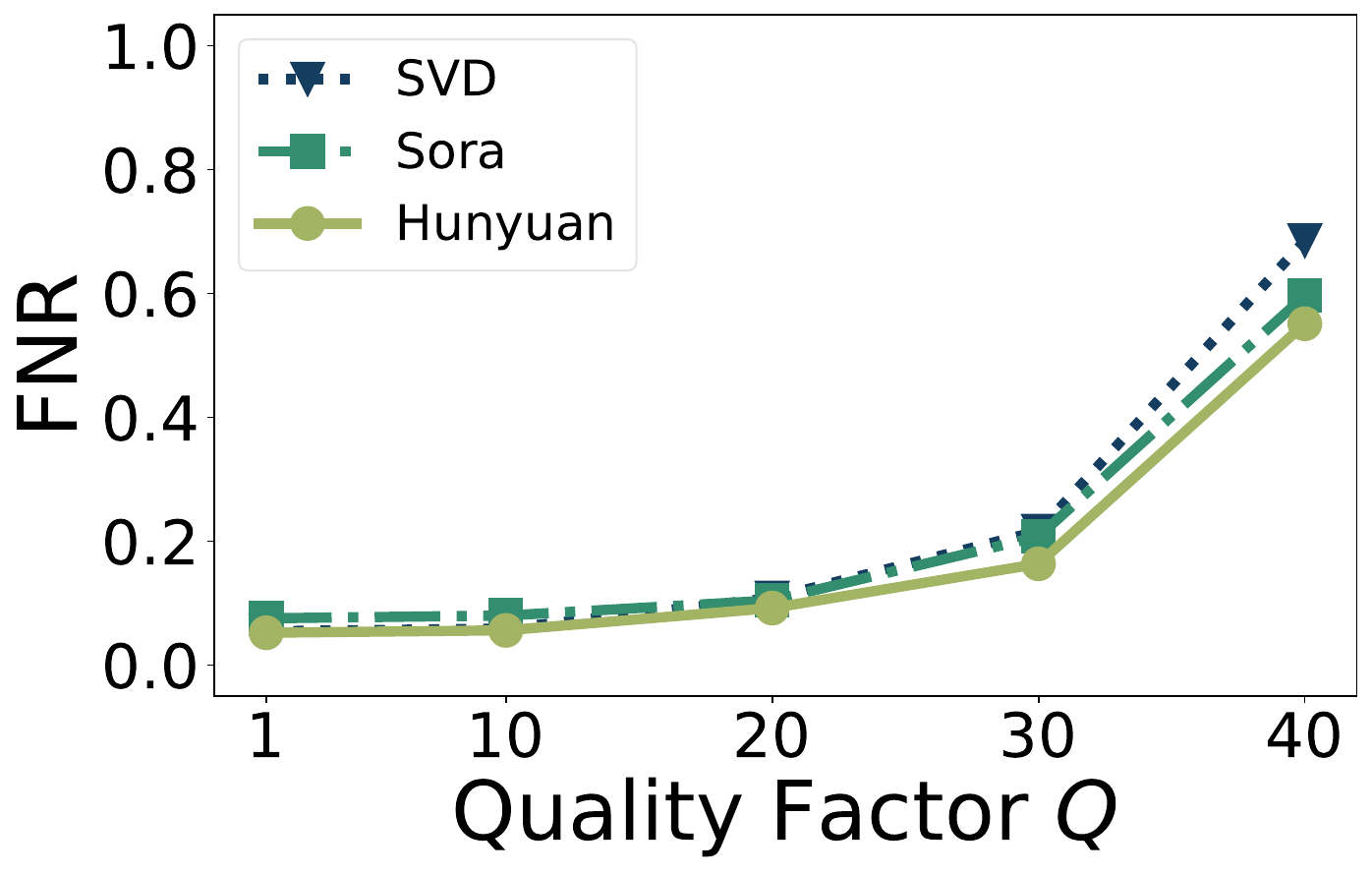}
        \caption{MPEG-4}
    \end{subfigure} \\

    \begin{subfigure}{.23\linewidth}
        \centering
        \includegraphics[width=\linewidth]{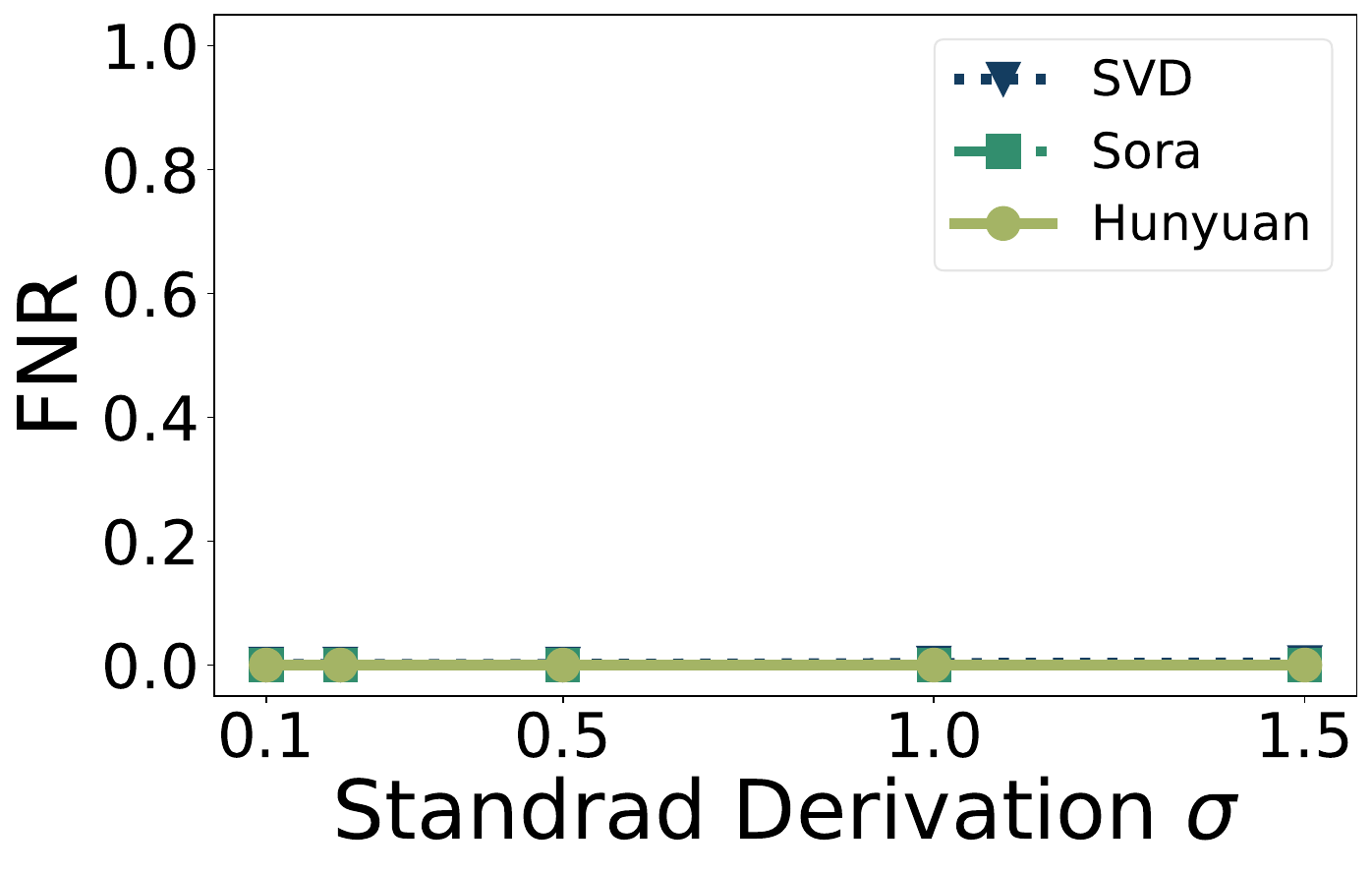}
        \caption{Gaussian Blur}
    \end{subfigure}
    \begin{subfigure}{.23\linewidth}
        \centering
        \includegraphics[width=\linewidth]{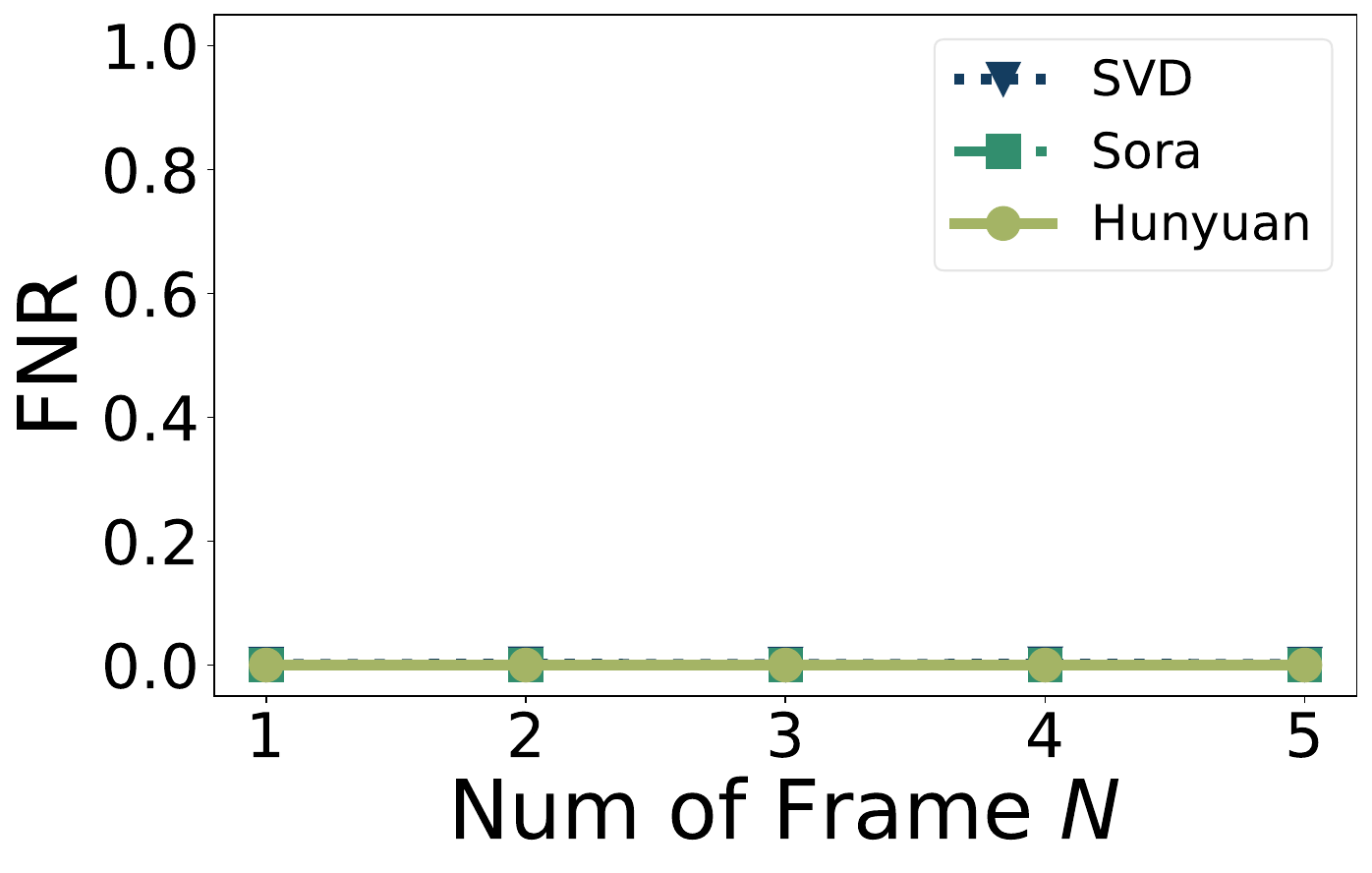}
        \caption{Frame Average}
    \end{subfigure}
    \begin{subfigure}{.23\linewidth}
        \centering
        \includegraphics[width=\linewidth]{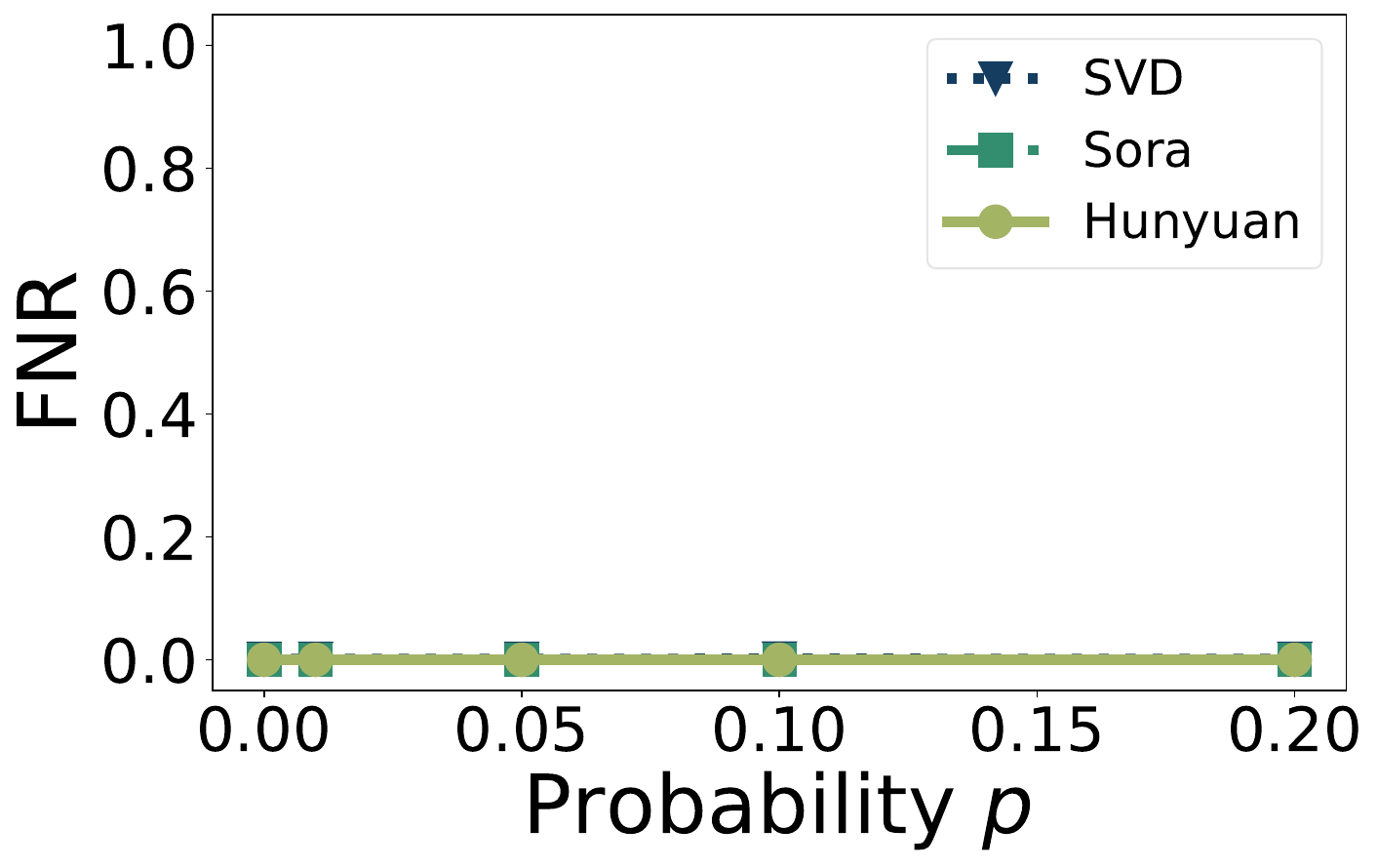}
        \caption{Frame Switch}
    \end{subfigure}
    \begin{subfigure}{.23\linewidth}
        \centering
        \includegraphics[width=\linewidth]{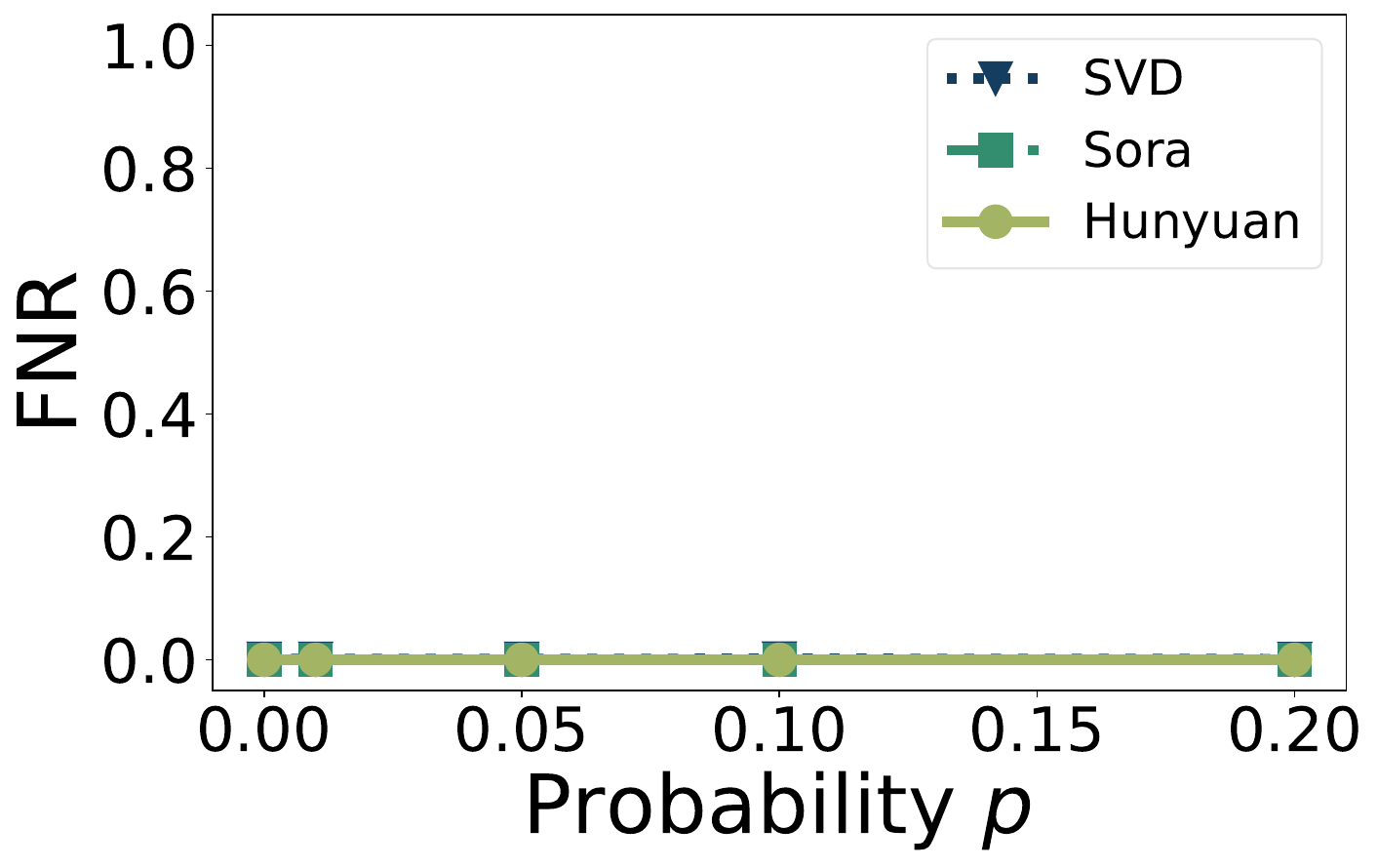}
        \caption{Frame Removal}
    \end{subfigure}
    \caption{\label{fig:models}Common perturbation watermark removal results across videos generated by different generative models. FNRs are averaged on all watermarking methods with various aggregation strategies and styles.}
\end{figure}

\begin{figure}[]
    \centering
    \begin{subfigure}{.23\linewidth}
        \centering
        \includegraphics[width=\linewidth]{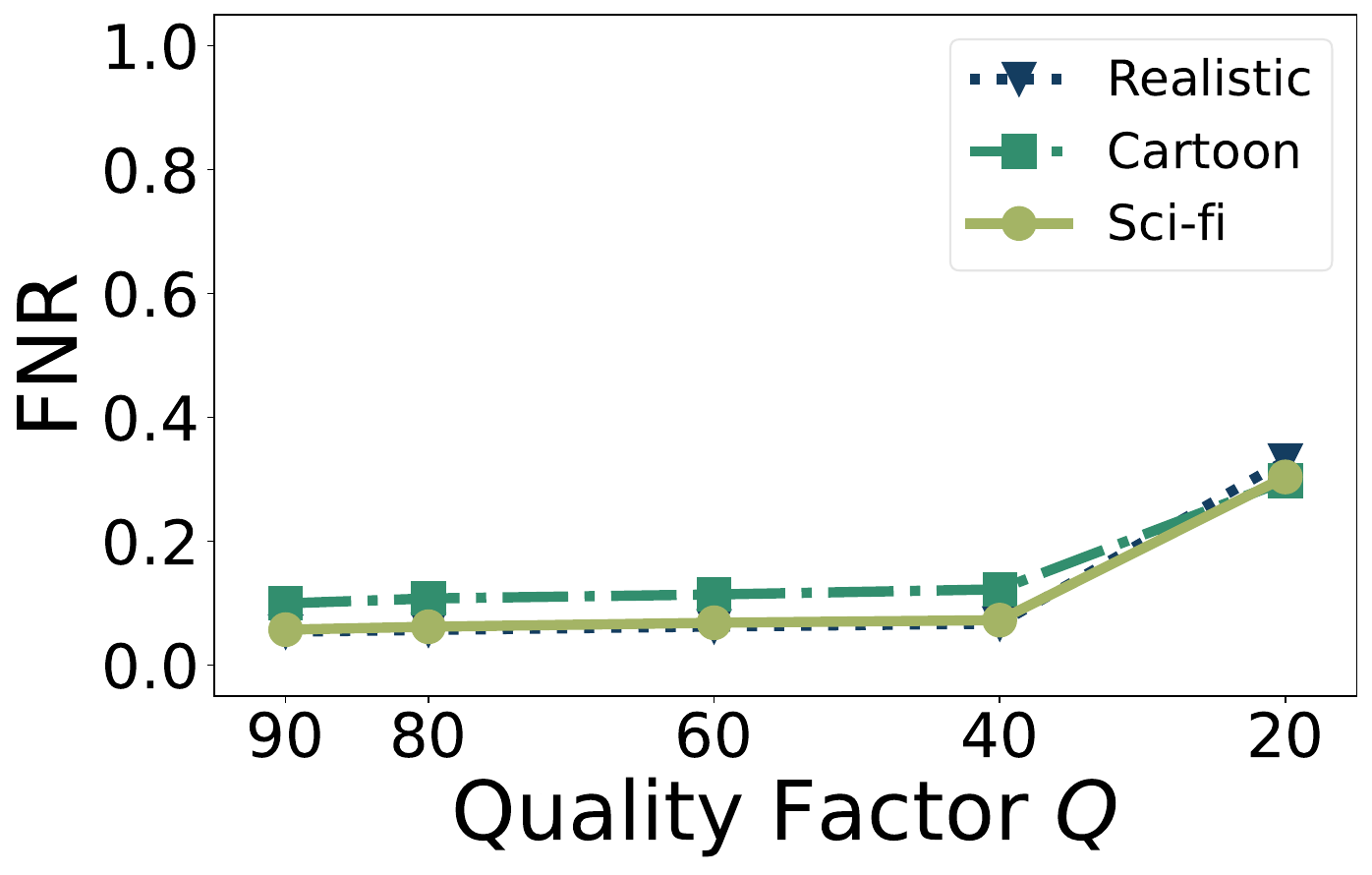}
        \caption{JPEG}
    \end{subfigure}
    \begin{subfigure}{.23\linewidth}
        \centering
        \includegraphics[width=\linewidth]{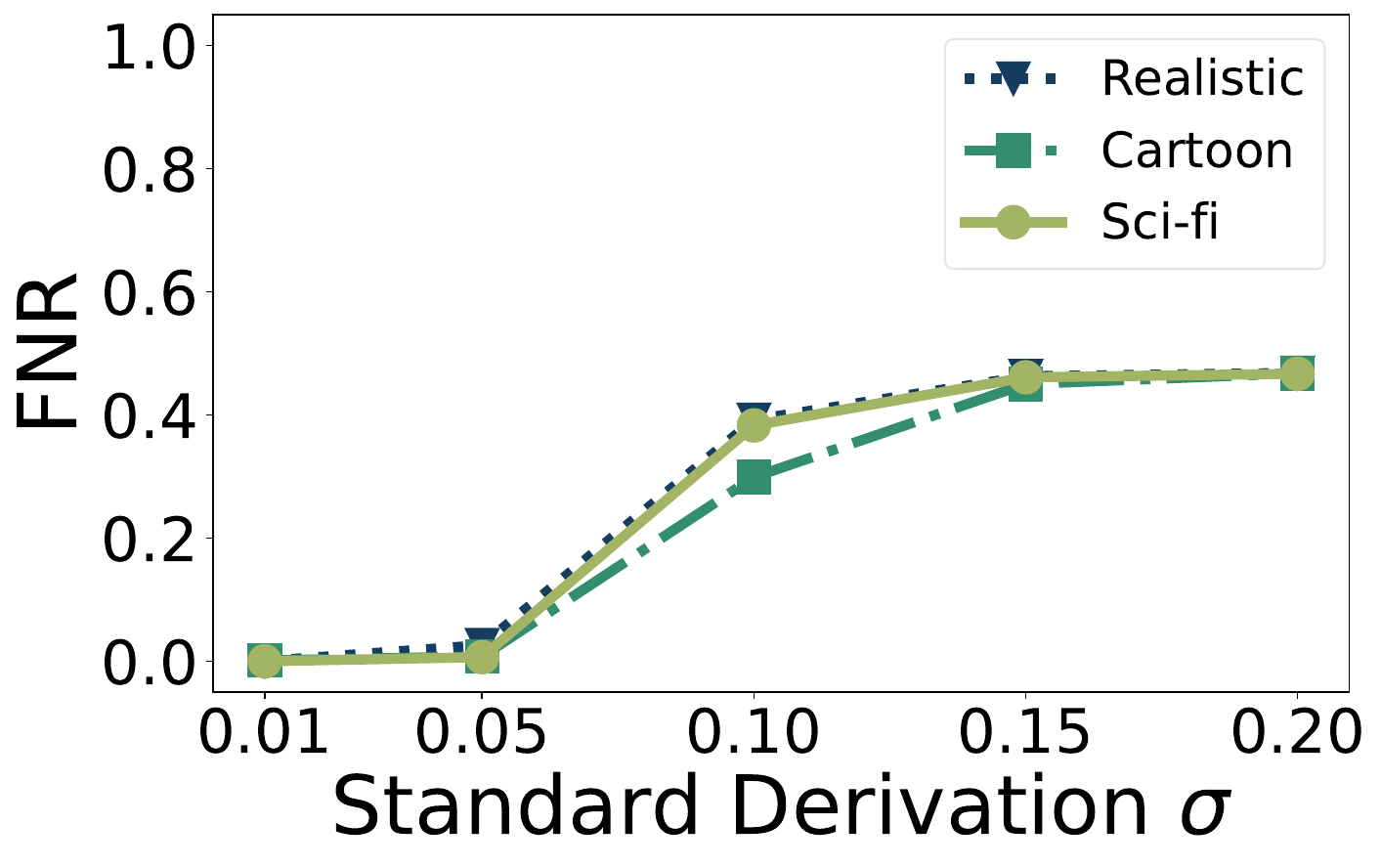}
        \caption{Gaussian Noise}
    \end{subfigure}
    \begin{subfigure}{.23\linewidth}
        \centering
        \includegraphics[width=\linewidth]{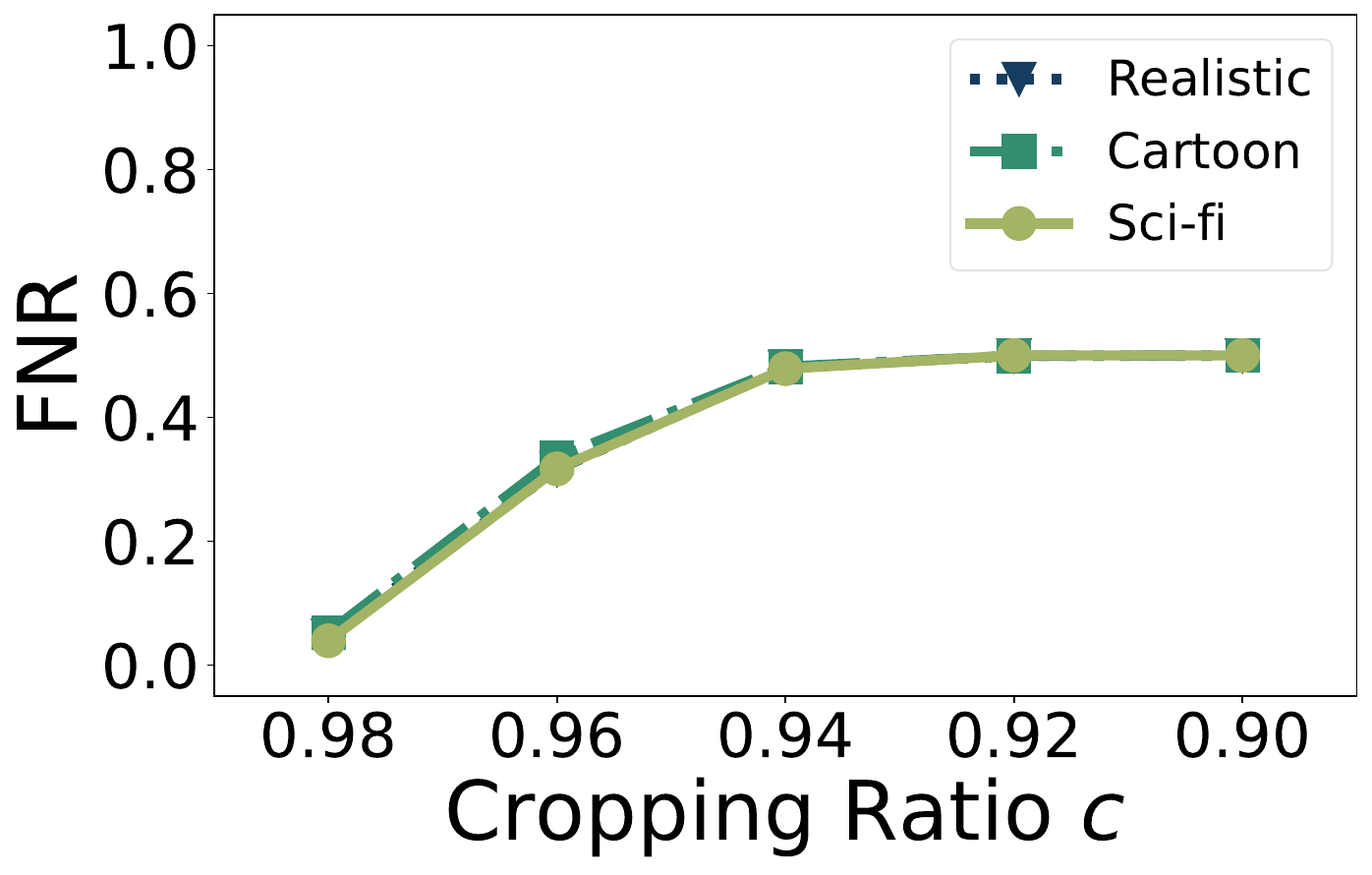}
        \caption{Cropping}
    \end{subfigure}
    \begin{subfigure}{.23\linewidth}
        \centering
        \includegraphics[width=\linewidth]{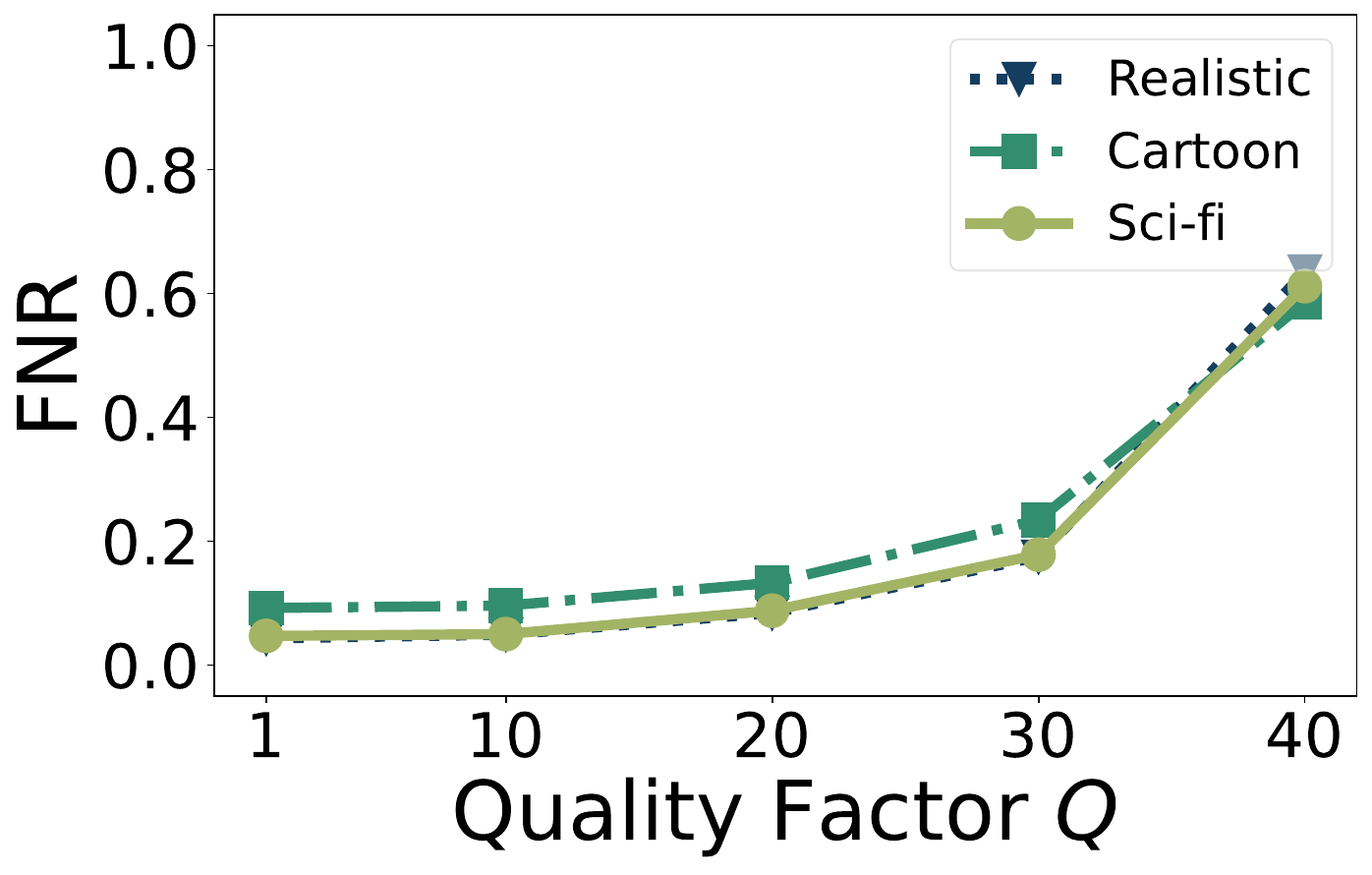}
        \caption{MPEG-4}
    \end{subfigure} \\

    \begin{subfigure}{.23\linewidth}
        \centering
        \includegraphics[width=\linewidth]{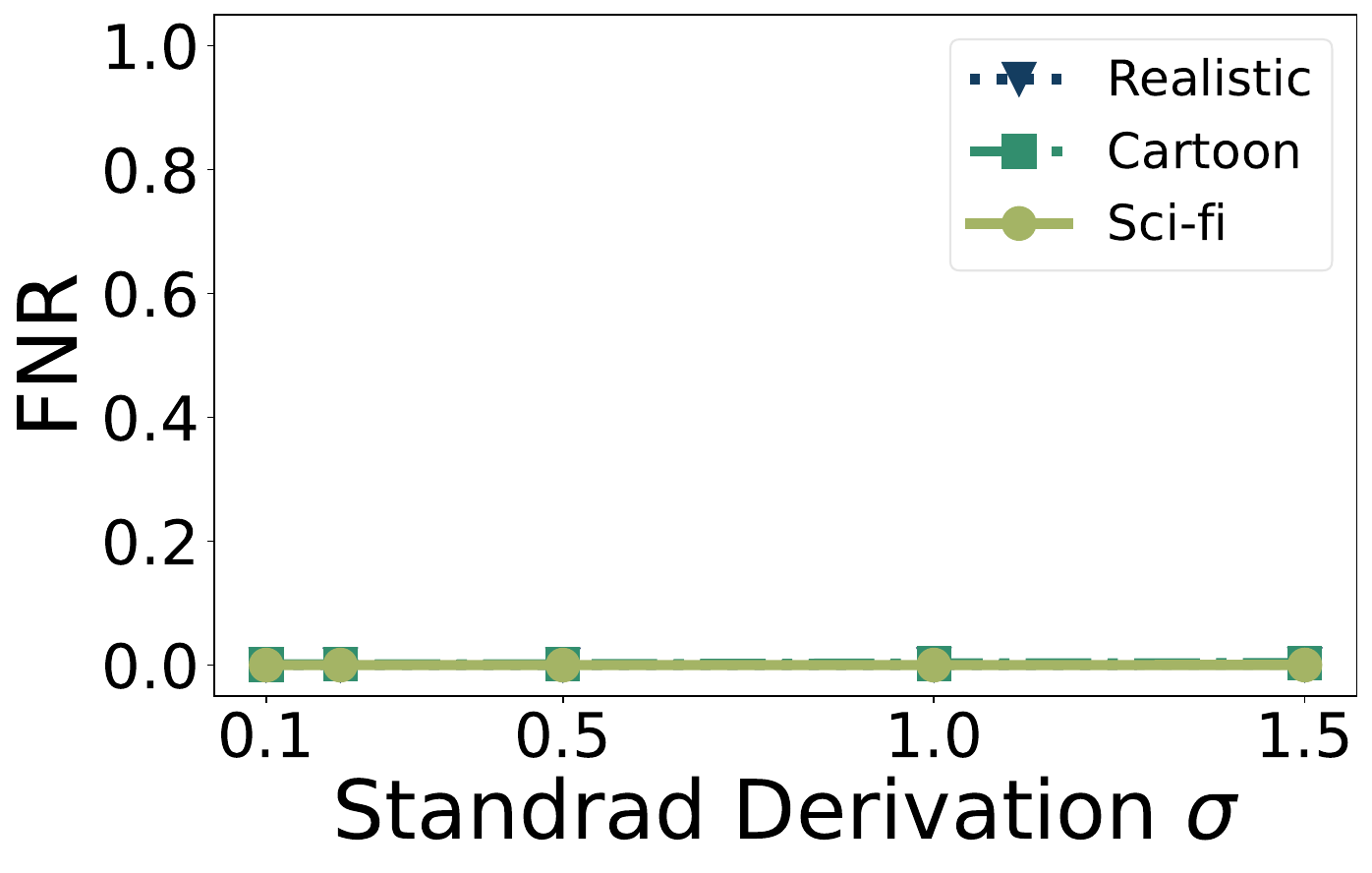}
        \caption{Gaussian Blur}
    \end{subfigure}
    \begin{subfigure}{.23\linewidth}
        \centering
        \includegraphics[width=\linewidth]{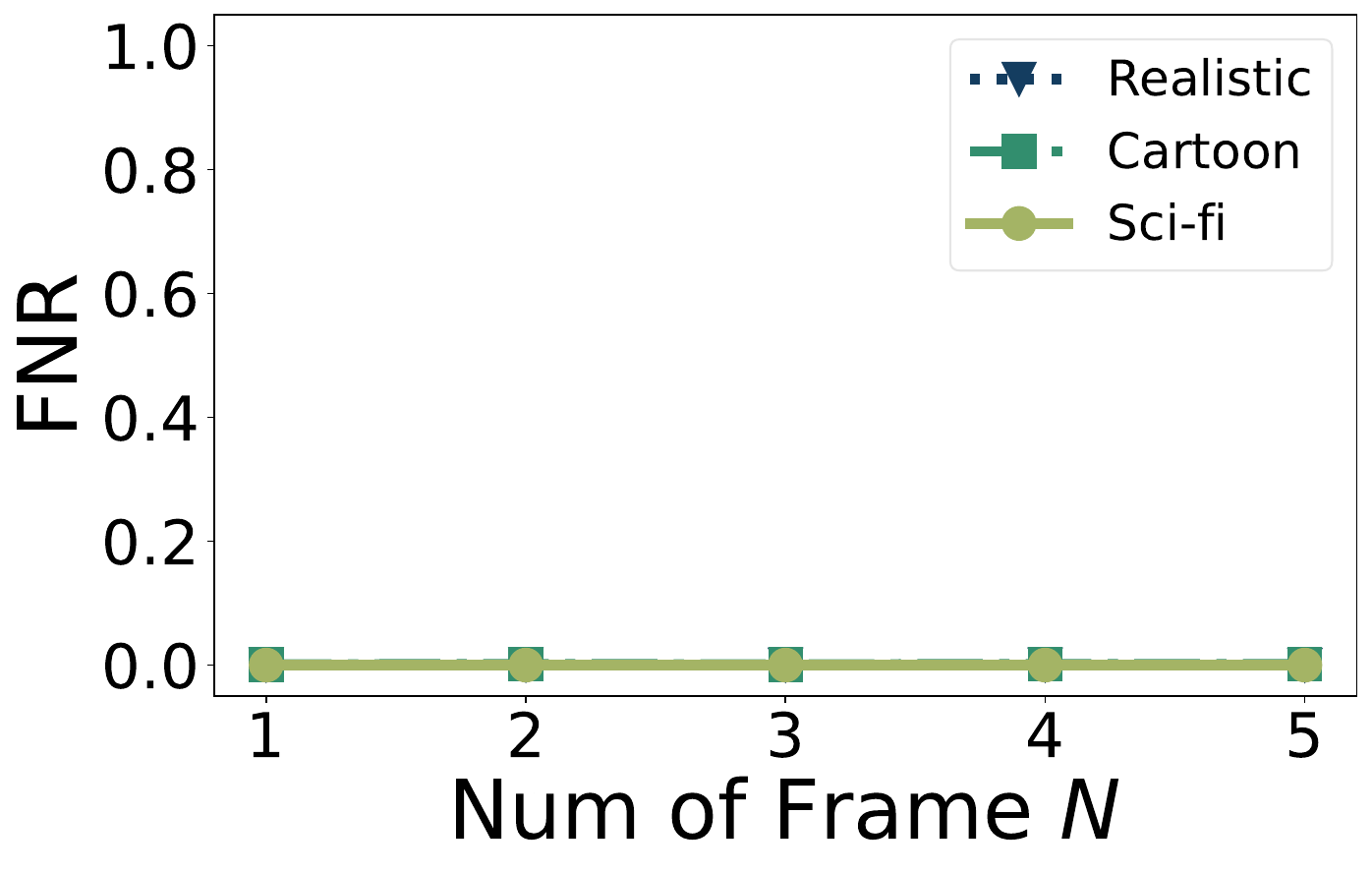}
        \caption{Frame Average}
    \end{subfigure}
    \begin{subfigure}{.23\linewidth}
        \centering
        \includegraphics[width=\linewidth]{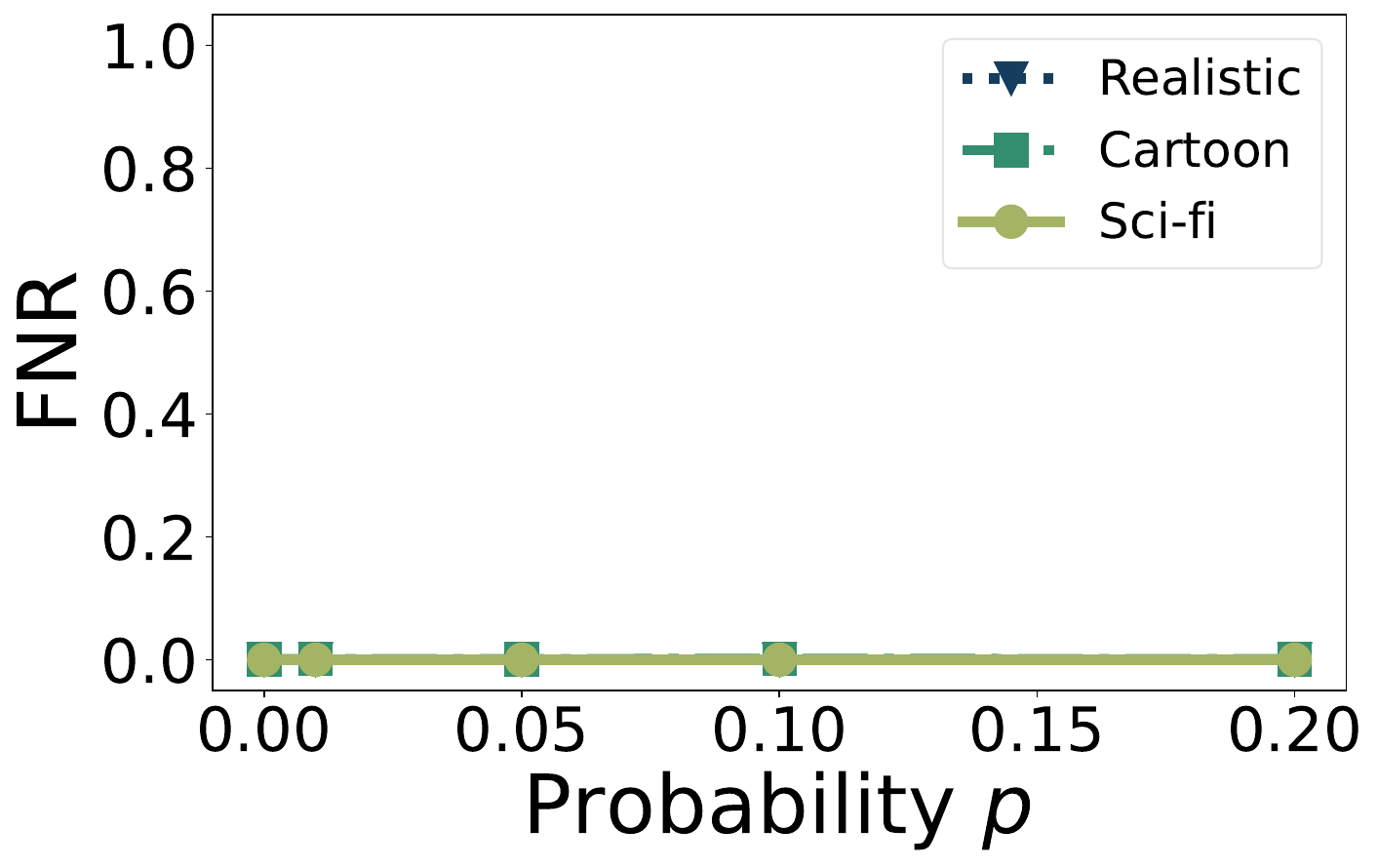}
        \caption{Frame Switch}
    \end{subfigure}
    \begin{subfigure}{.23\linewidth}
        \centering
        \includegraphics[width=\linewidth]{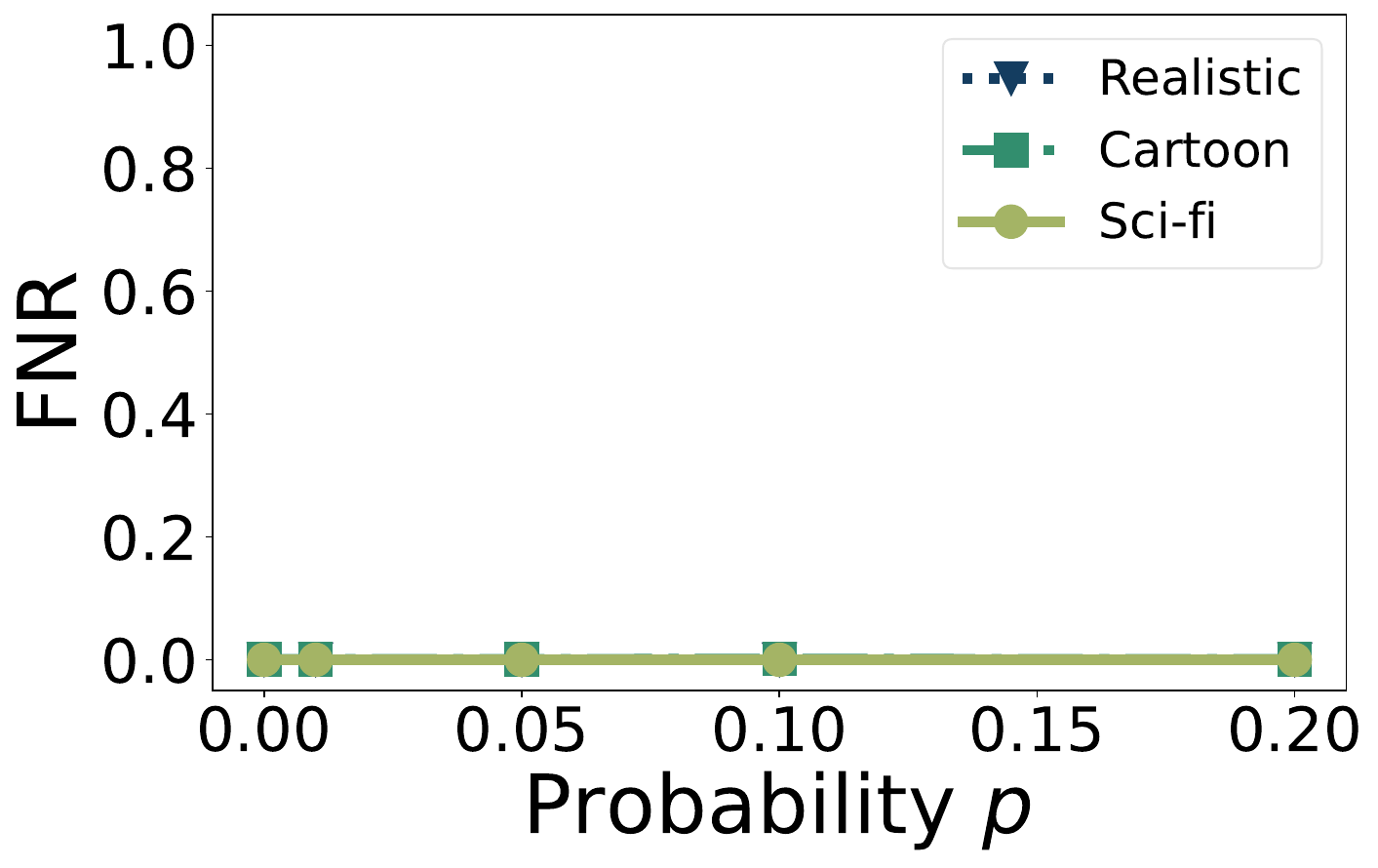}
        \caption{Frame Removal}
    \end{subfigure}
    
    \caption{\label{fig:styles}Common perturbation watermark removal results across video styles. FNRs are averaged on all watermarking methods with various aggregation strategies and generative models.}
\end{figure}

\begin{figure}[]
    \centering
    \begin{subfigure}{.23\linewidth}
        \centering
        \includegraphics[width=\linewidth]{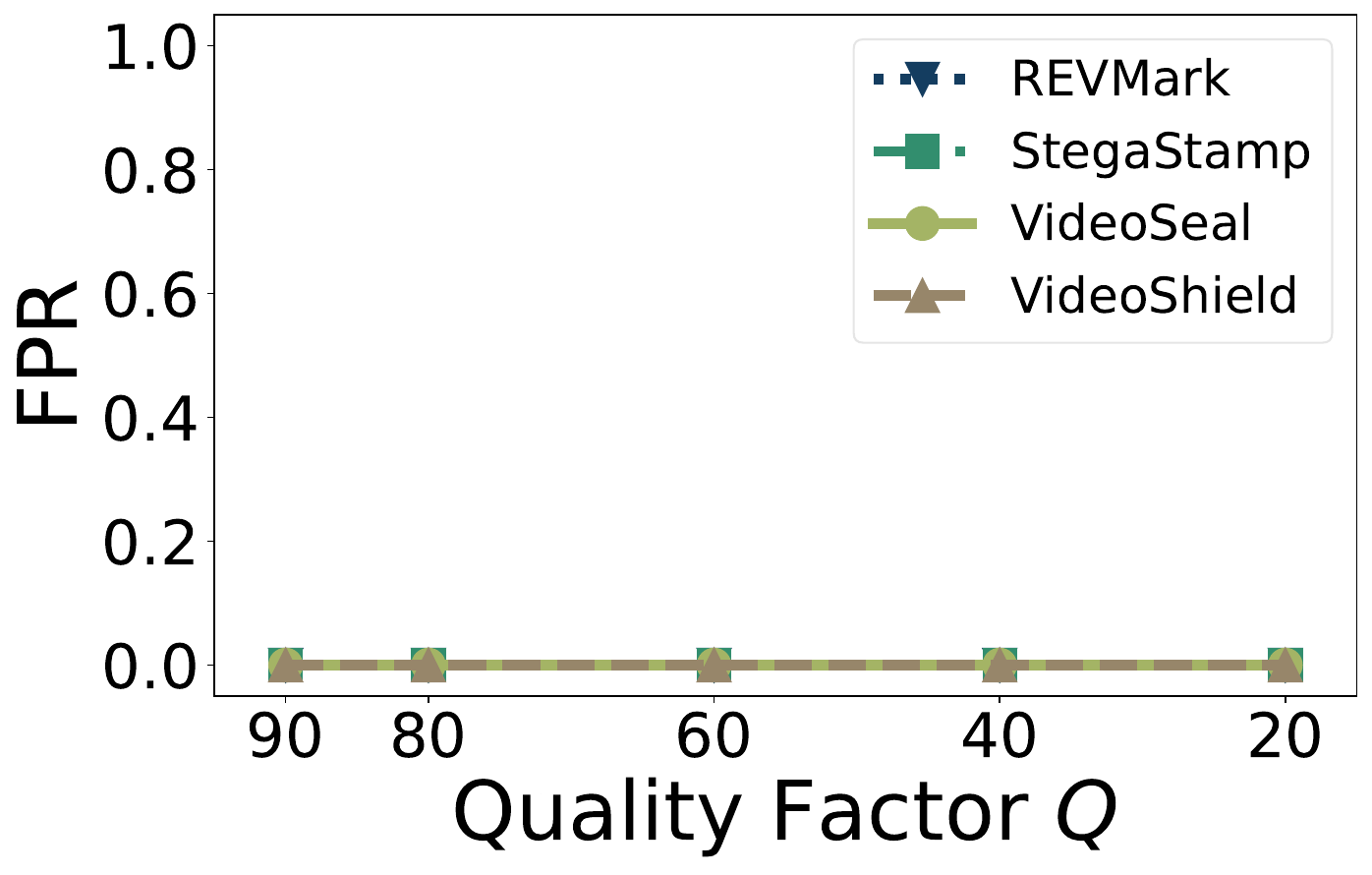}
        \caption{JPEG}
    \end{subfigure}
    \begin{subfigure}{.23\linewidth}
        \centering
        \includegraphics[width=\linewidth]{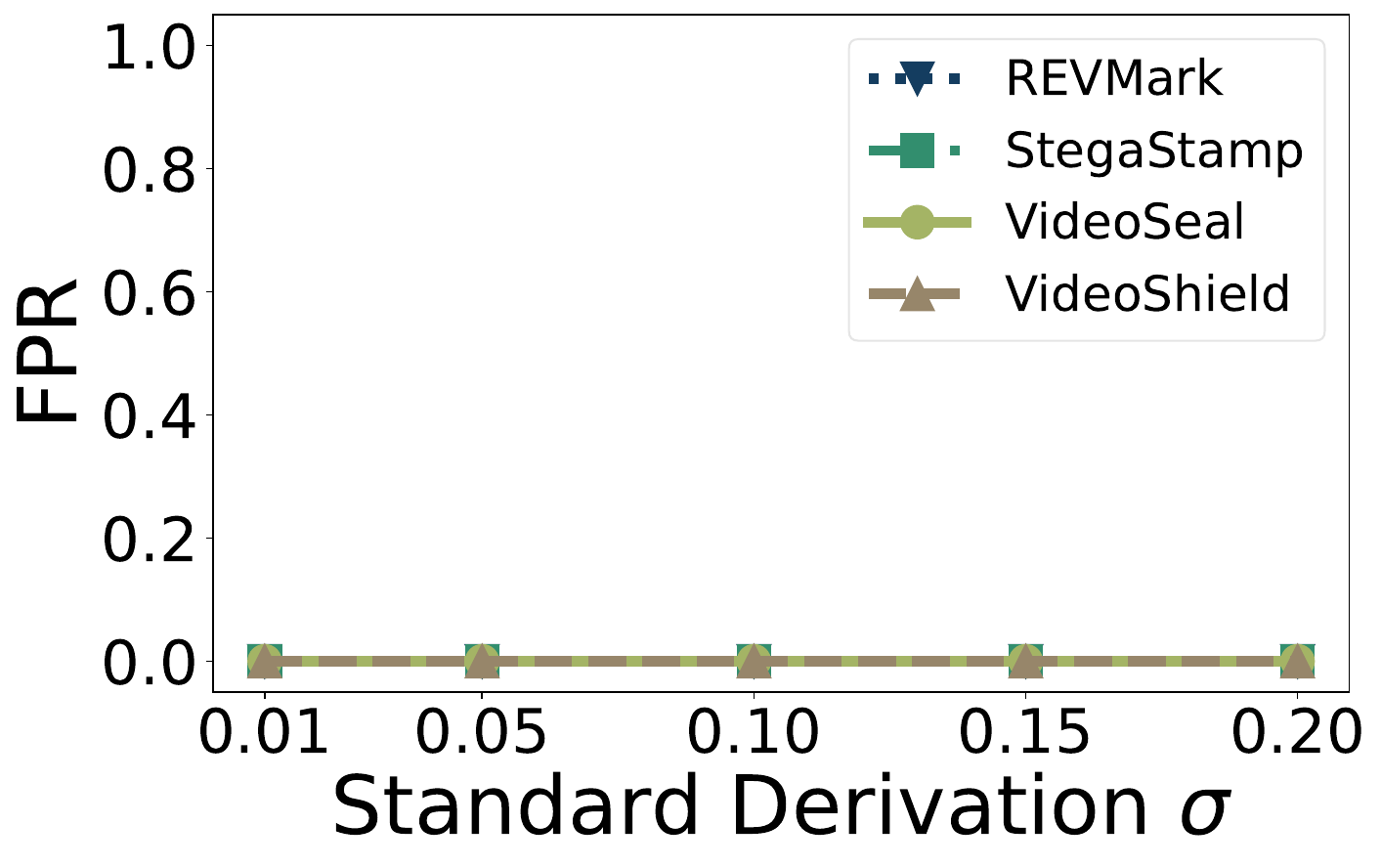}
        \caption{Gaussian Noise}
    \end{subfigure}
    \begin{subfigure}{.23\linewidth}
        \centering
        \includegraphics[width=\linewidth]{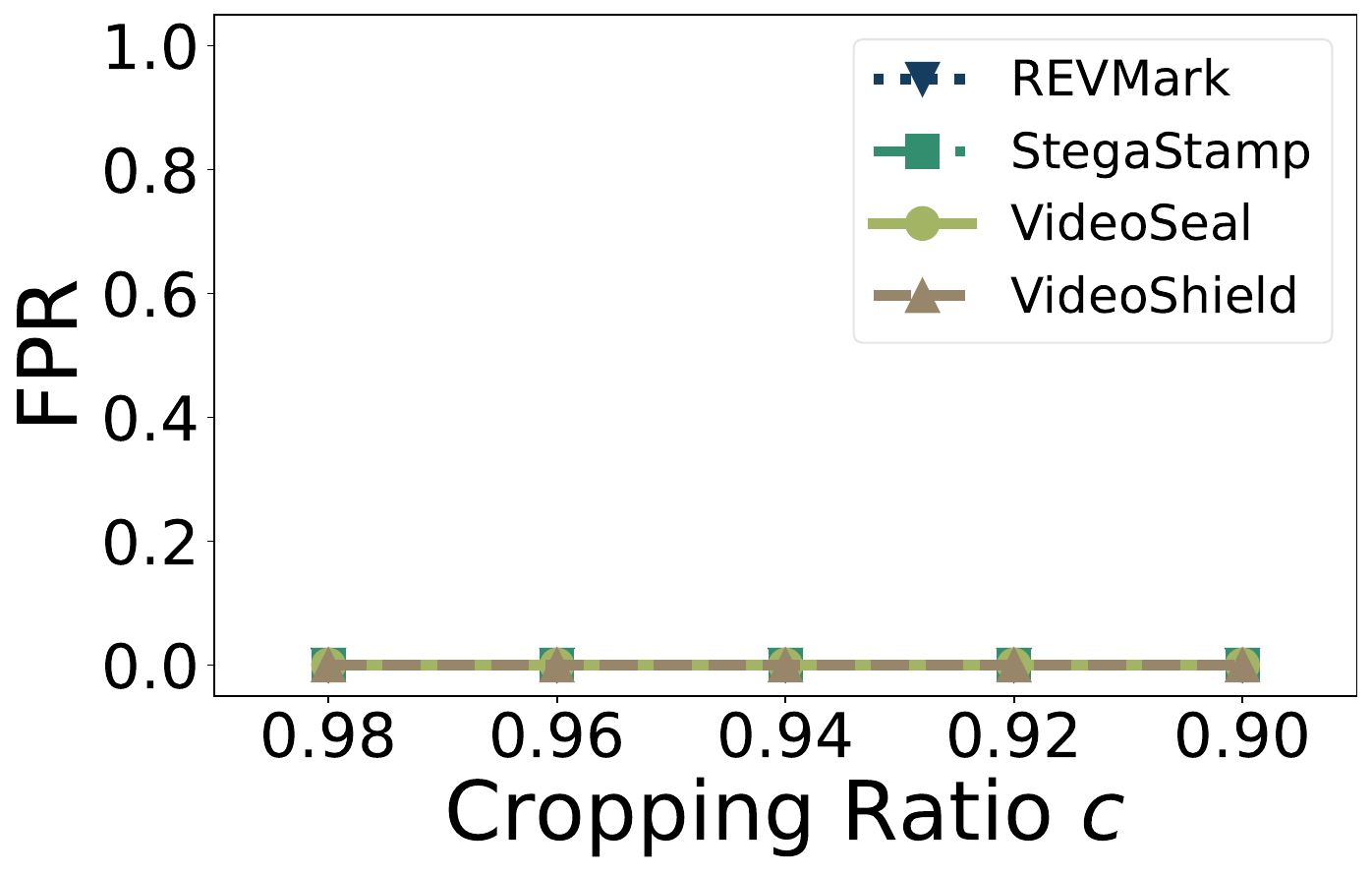}
        \caption{Cropping}
    \end{subfigure}
    \begin{subfigure}{.23\linewidth}
        \centering
        \includegraphics[width=\linewidth]{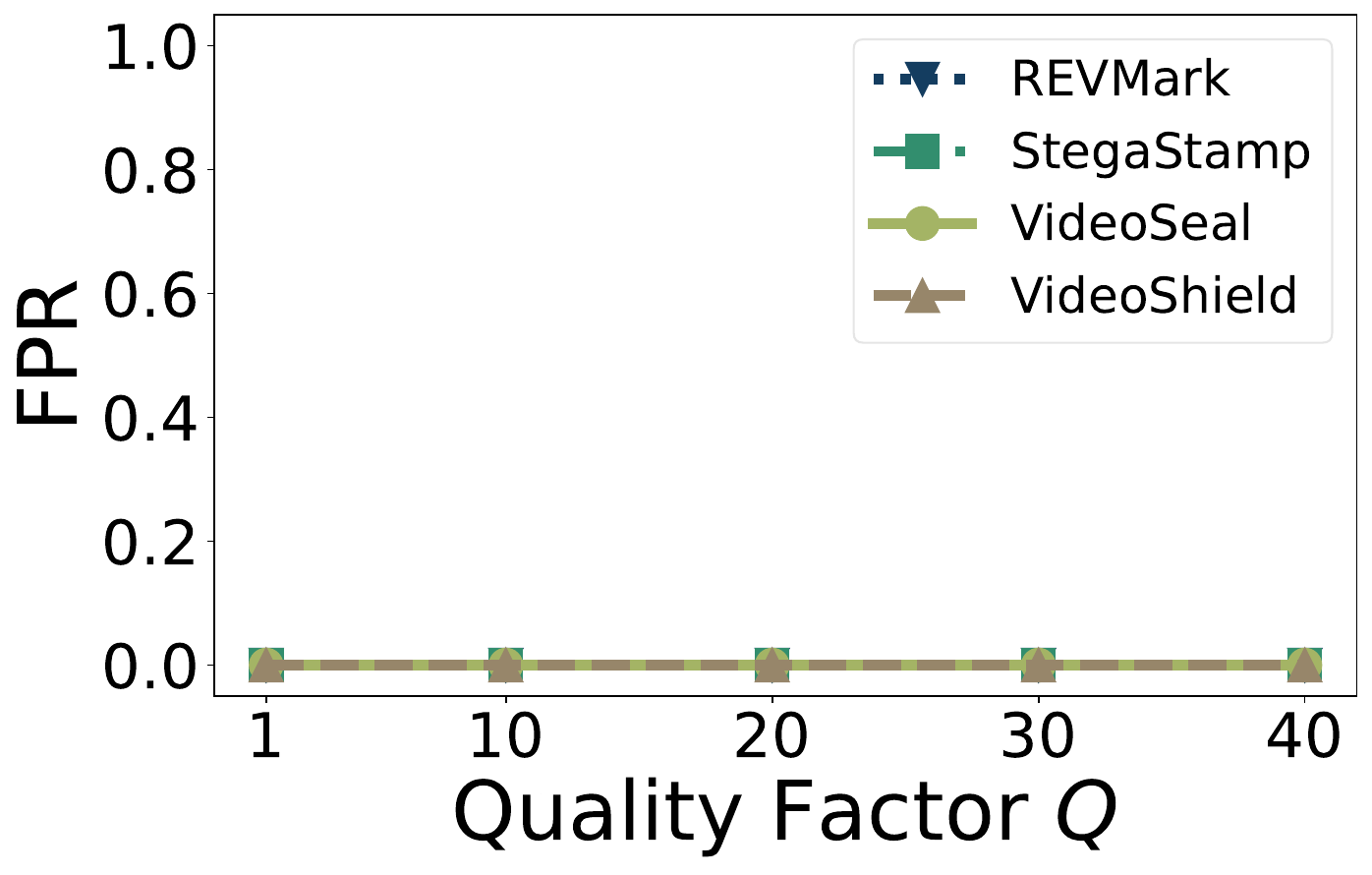}
        \caption{MPEG-4}
    \end{subfigure} \\

    \begin{subfigure}{.23\linewidth}
        \centering
        \includegraphics[width=\linewidth]{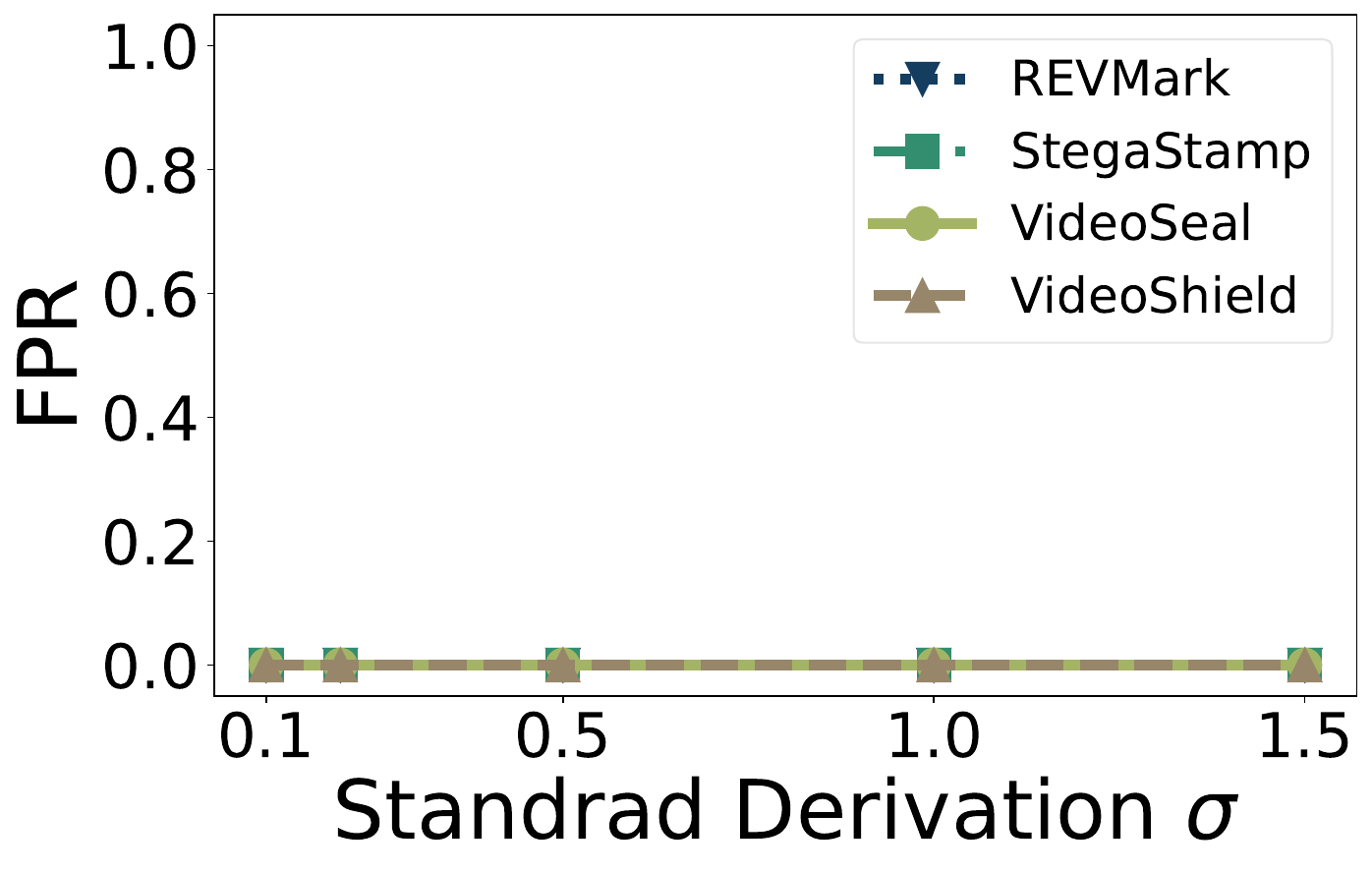}
        \caption{Gaussian Blur}
    \end{subfigure}
    \begin{subfigure}{.23\linewidth}
        \centering
        \includegraphics[width=\linewidth]{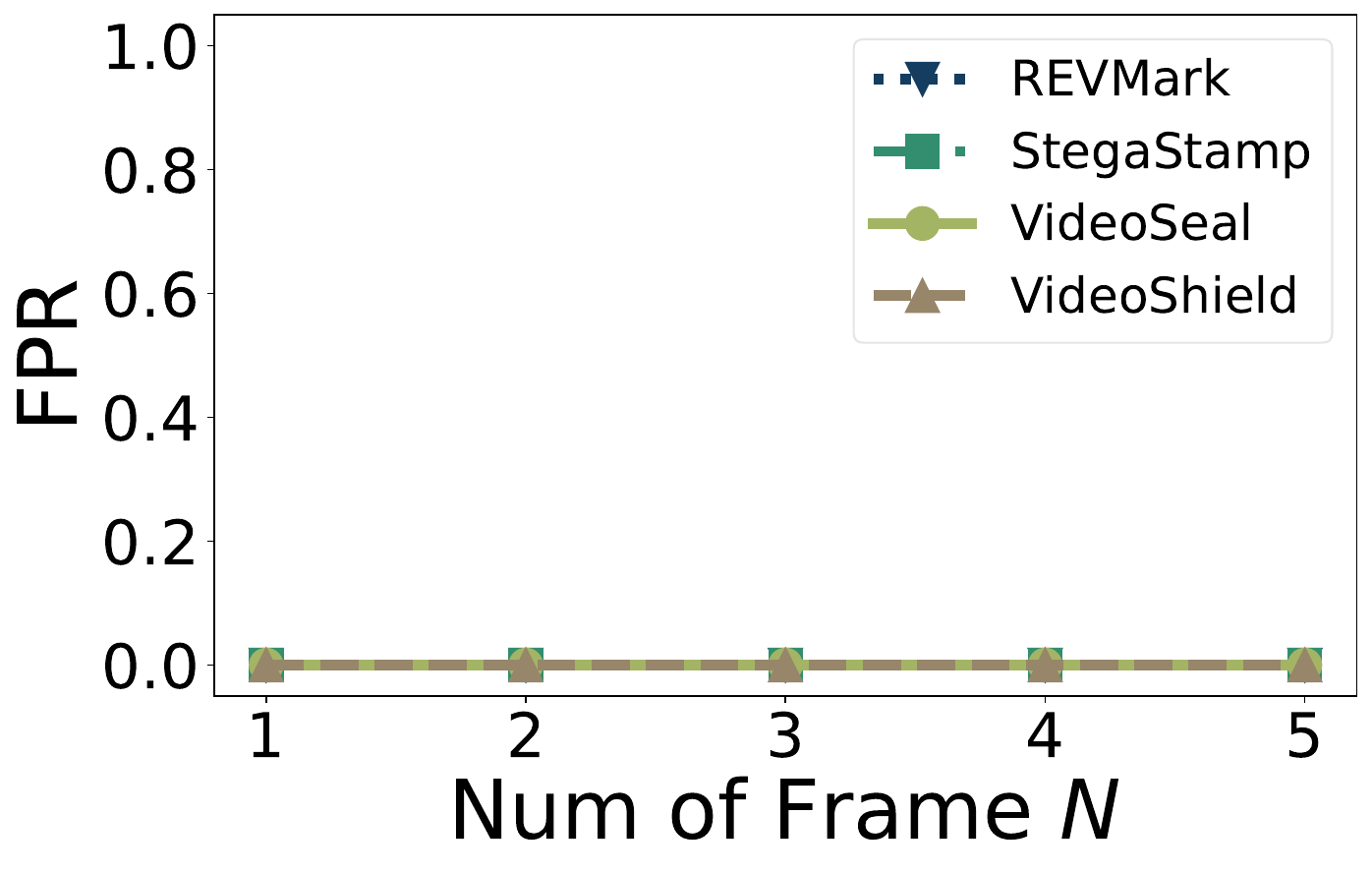}
        \caption{Frame Average}
    \end{subfigure}
    \begin{subfigure}{.23\linewidth}
        \centering
        \includegraphics[width=\linewidth]{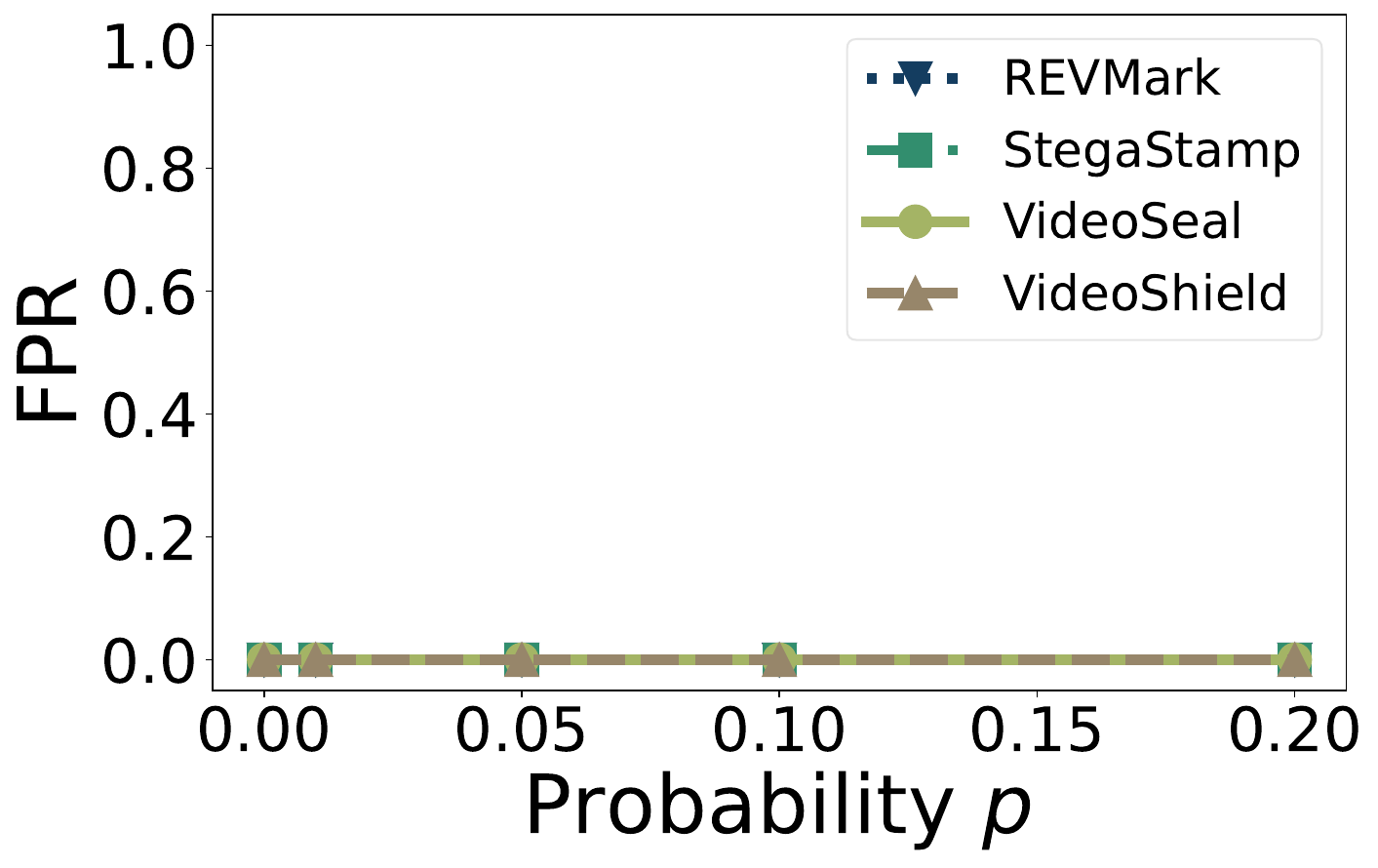}
        \caption{Frame Switch}
    \end{subfigure}
    \begin{subfigure}{.23\linewidth}
        \centering
        \includegraphics[width=\linewidth]{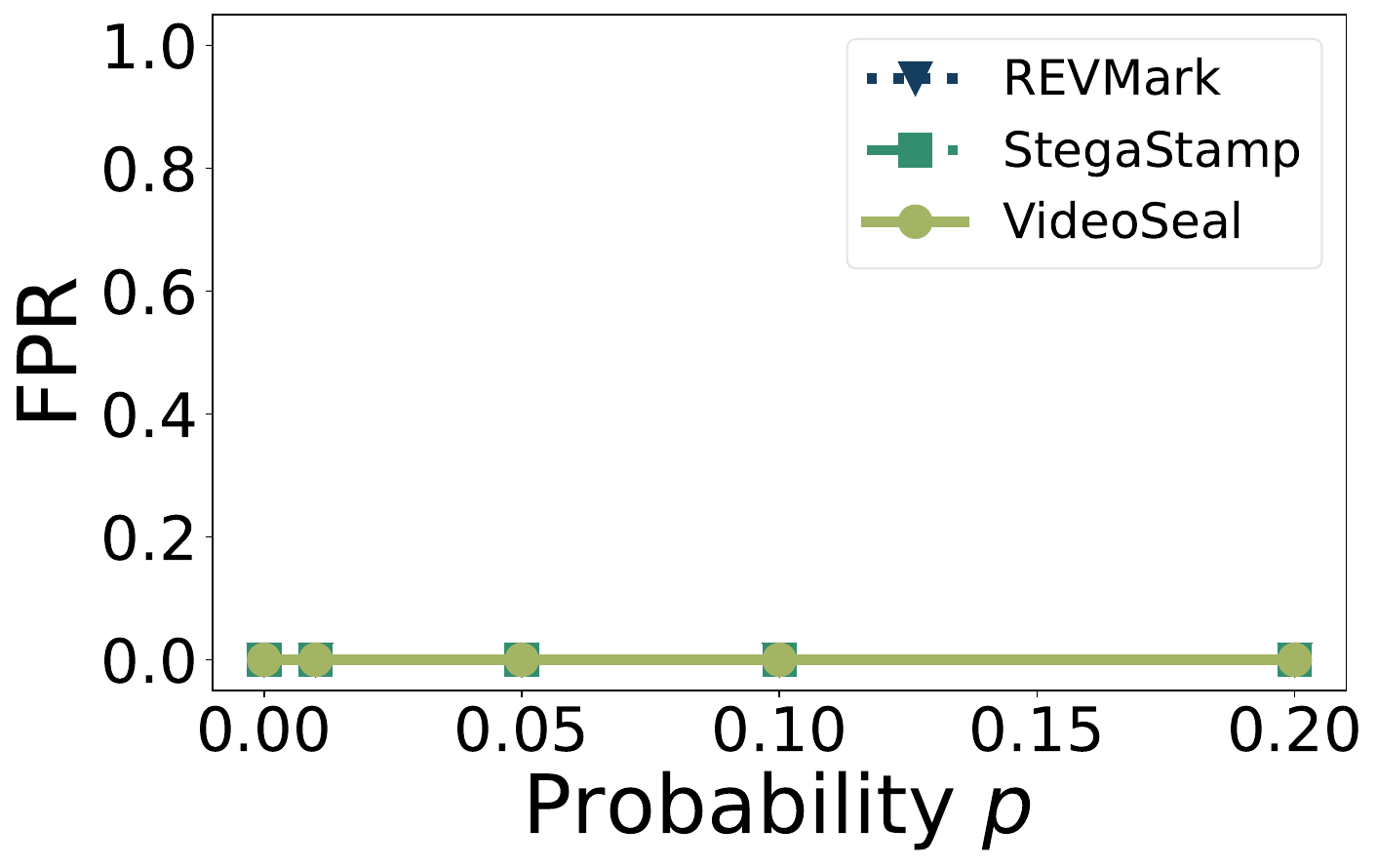}
        \caption{Frame Removal}
    \end{subfigure}
    \caption{\label{fig:watermark fpr}Common perturbation watermark forgery results for different video watermarking methods. For StegaStamp and VideoSeal, we report results using their best-performing aggregation strategies. FPRs are averaged over 1000 real videos from Kinetics-400 dataset.}
\end{figure}

\begin{figure}[]
    \centering
    \begin{subfigure}{.3\linewidth}
        \centering
        \includegraphics[width=\linewidth]{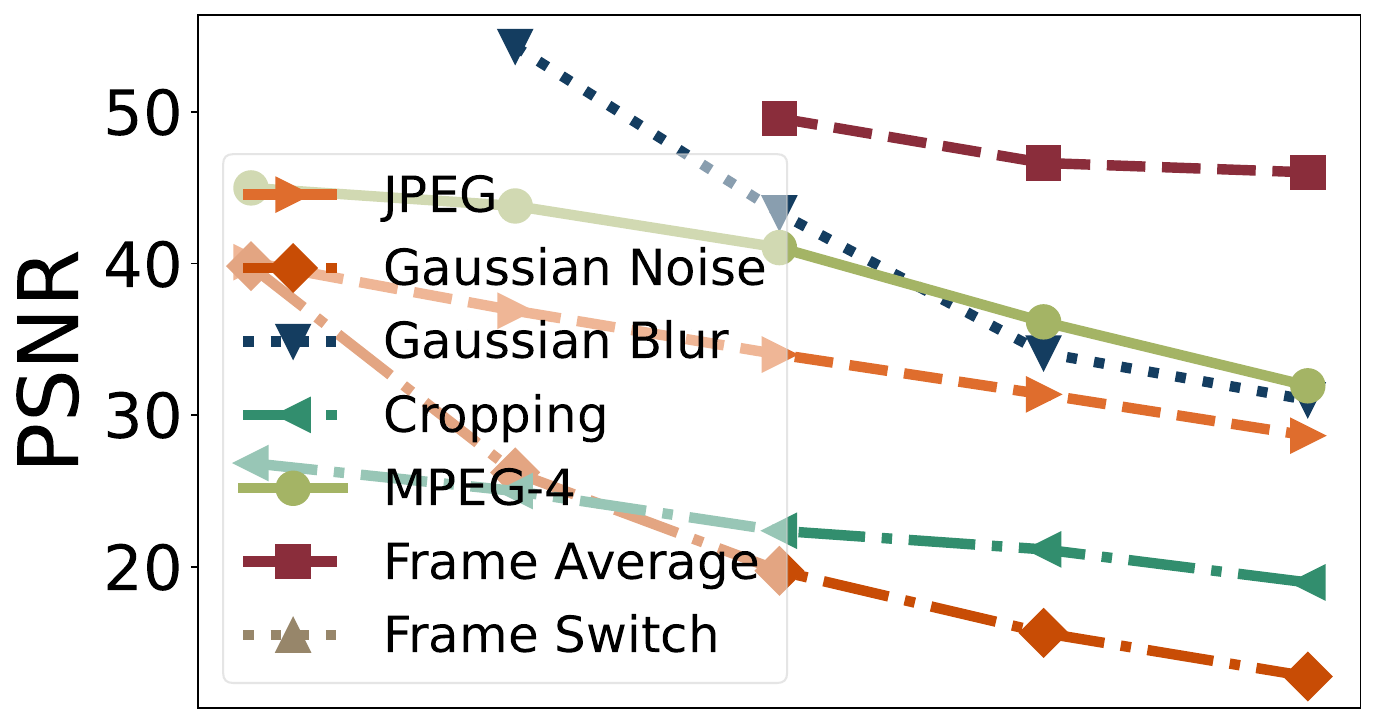}
        \caption{PSNR}
    \end{subfigure}
    \begin{subfigure}{.3\linewidth}
        \centering
        \includegraphics[width=\linewidth]{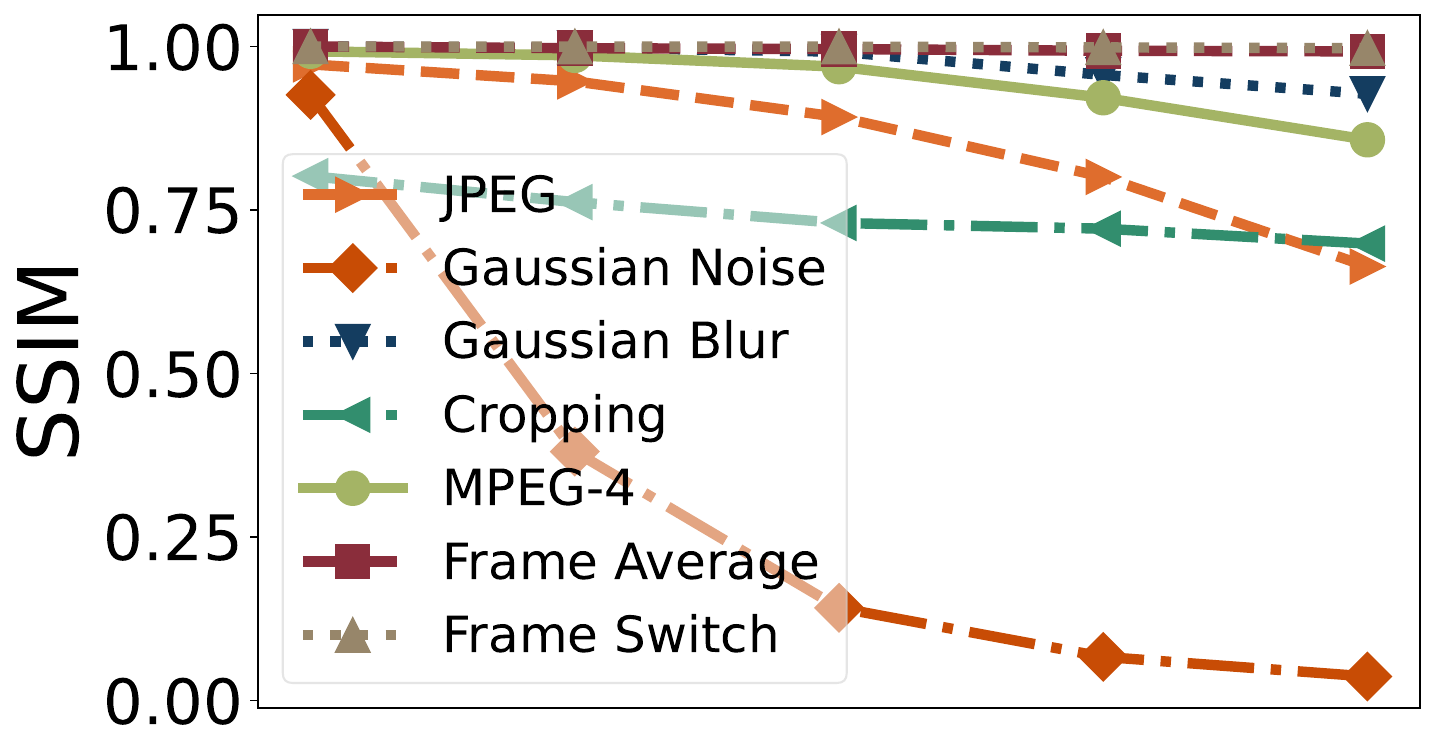}
        \caption{SSIM}
    \end{subfigure}
    \begin{subfigure}{.3\linewidth}
        \centering
        \includegraphics[width=\linewidth]{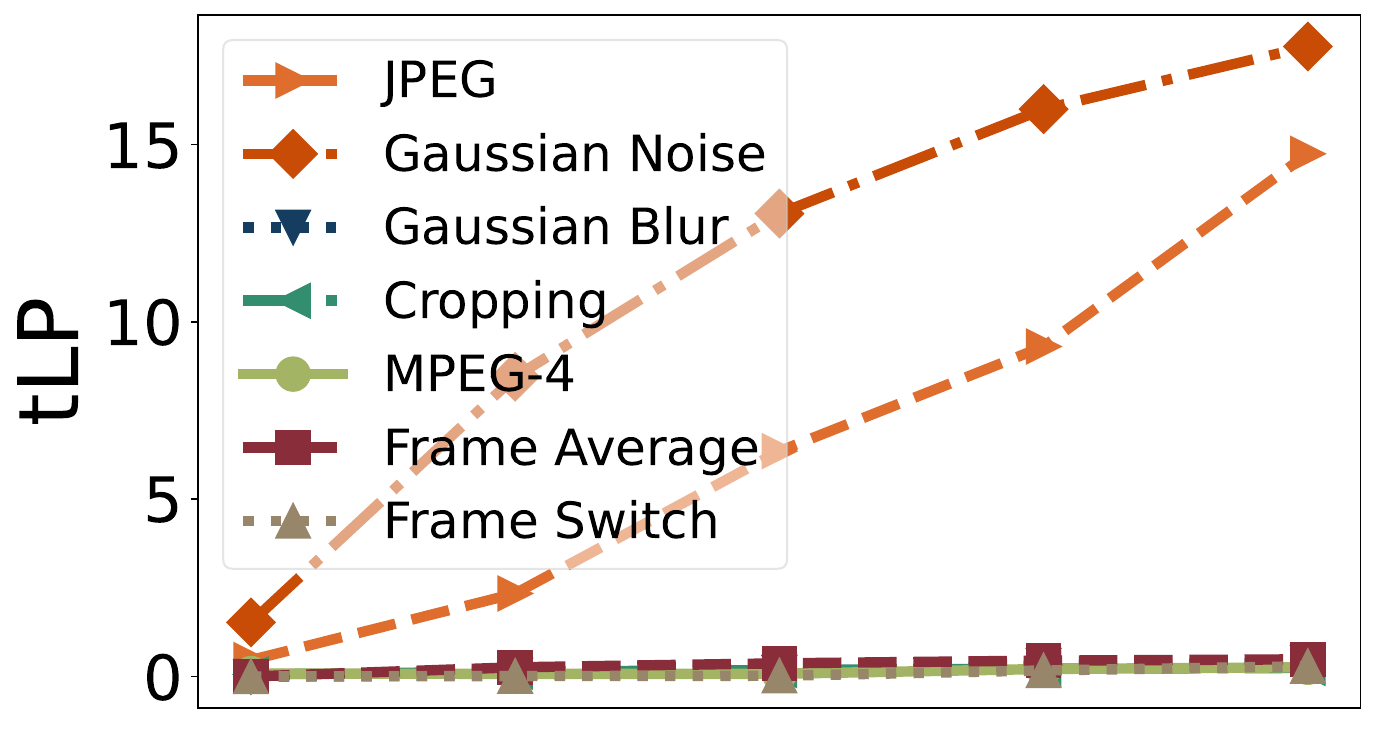}
        \caption{tLP}
    \end{subfigure}
    \caption{\label{fig:utility}Common perturbation utility results. A missing point in the PSNR subfigure indicates a PSNR value of $\infty$. We observe that Gaussian Noise, Cropping, and JPEG are the top-3 most impactful perturbations in the no-box setting, as they degrade the video's visual quality the most. In contrast, Frame Switch, Frame Average, and Gaussian Blur preserve video quality best. Note that results for Frame Removal are not reported, as this perturbation alters the video's shape, making it incompatible with direct computation of utility metrics.}
\end{figure}

\begin{figure}[]
    \centering

    \begin{subfigure}{.23\linewidth}
        \centering
        \includegraphics[width=\linewidth]{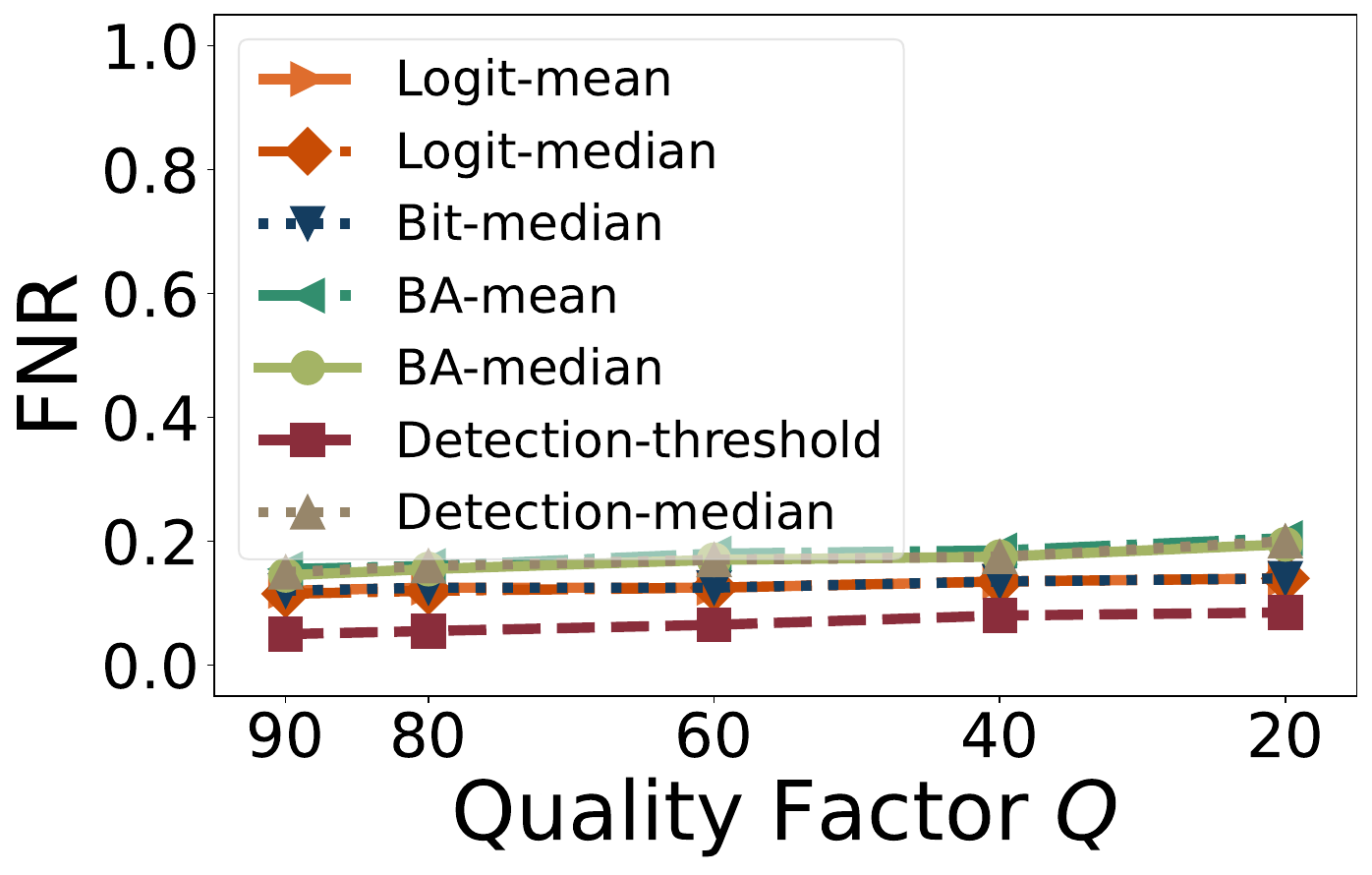}
        \caption{JPEG}
    \end{subfigure}
    \begin{subfigure}{.23\linewidth}
        \centering
        \includegraphics[width=\linewidth]{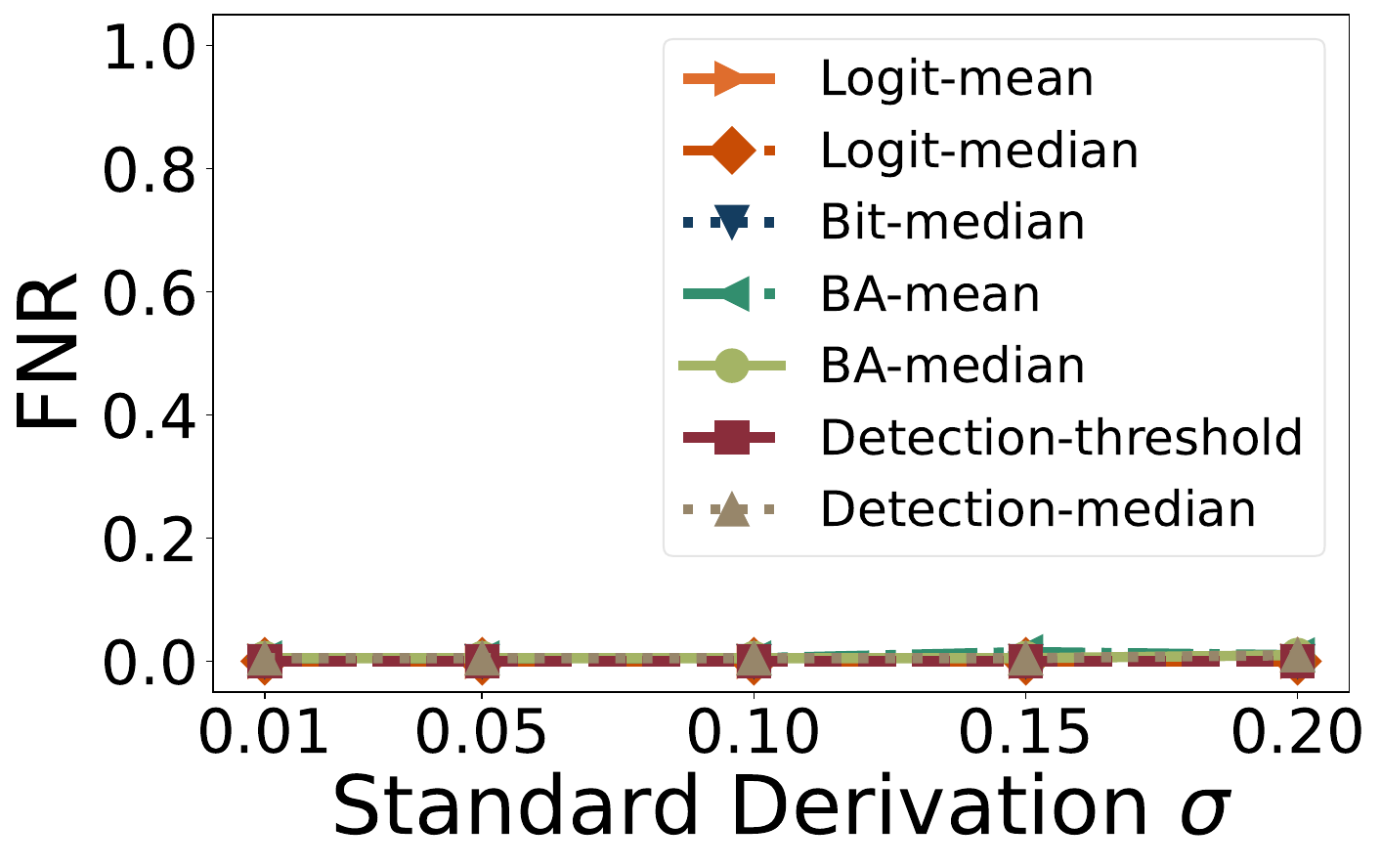}
        \caption{Gaussian Noise}
    \end{subfigure}
    \begin{subfigure}{.23\linewidth}
        \centering
        \includegraphics[width=\linewidth]{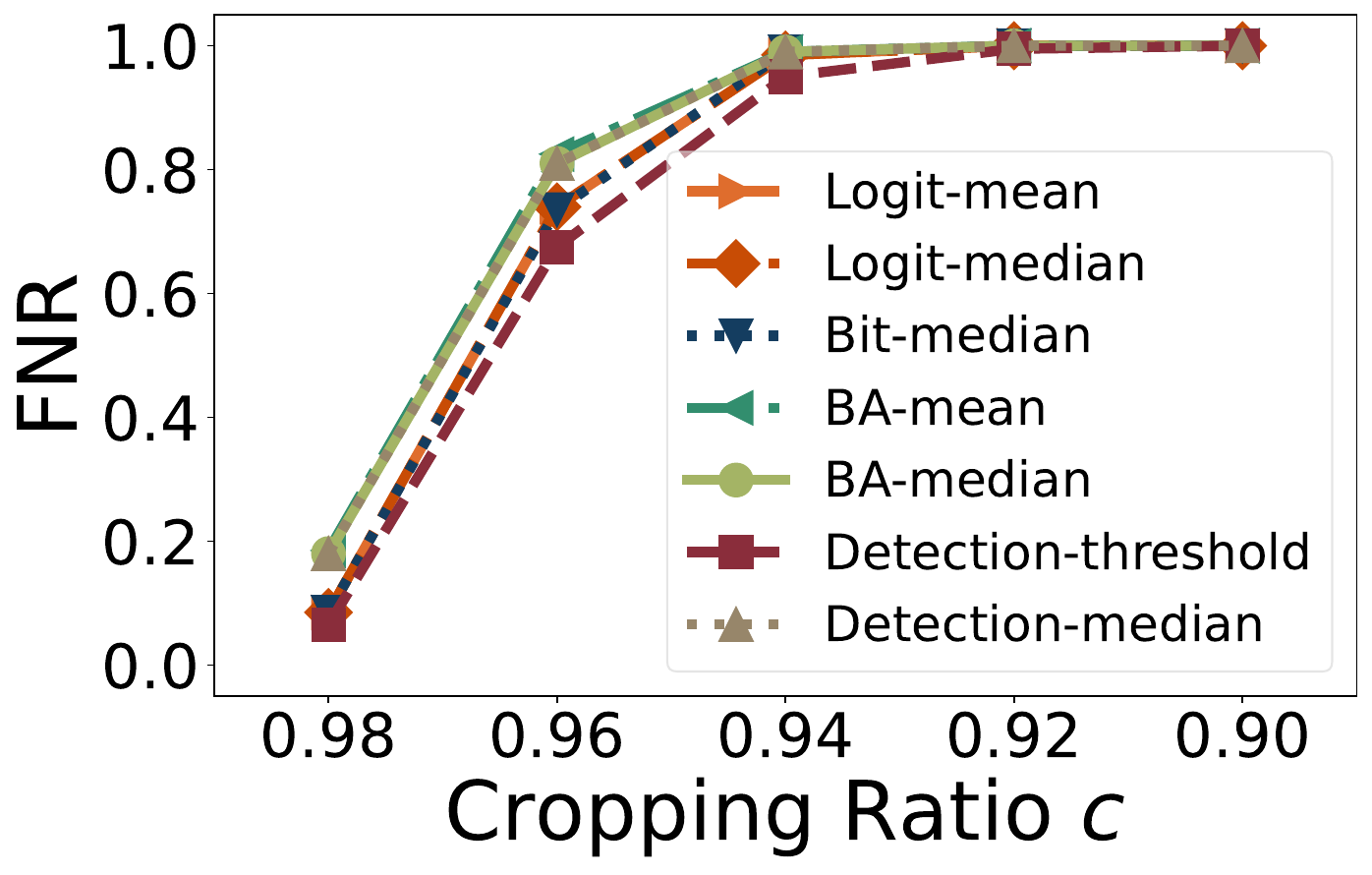}
        \caption{Cropping}
    \end{subfigure}
    \begin{subfigure}{.23\linewidth}
        \centering
        \includegraphics[width=\linewidth]{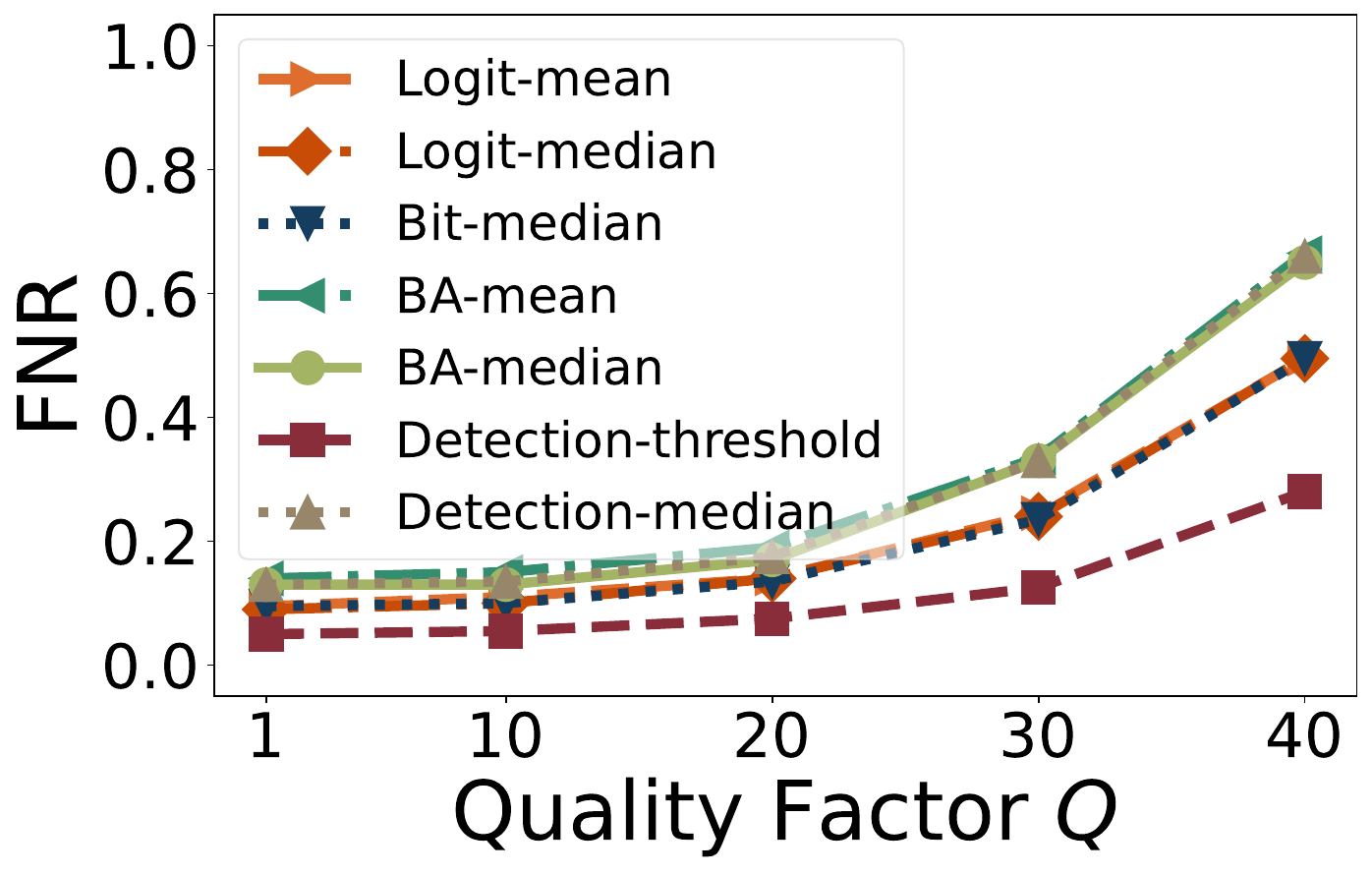}
        \caption{MPEG-4}
    \end{subfigure} \\
    
    \begin{subfigure}{.23\linewidth}
        \centering
        \includegraphics[width=\linewidth]{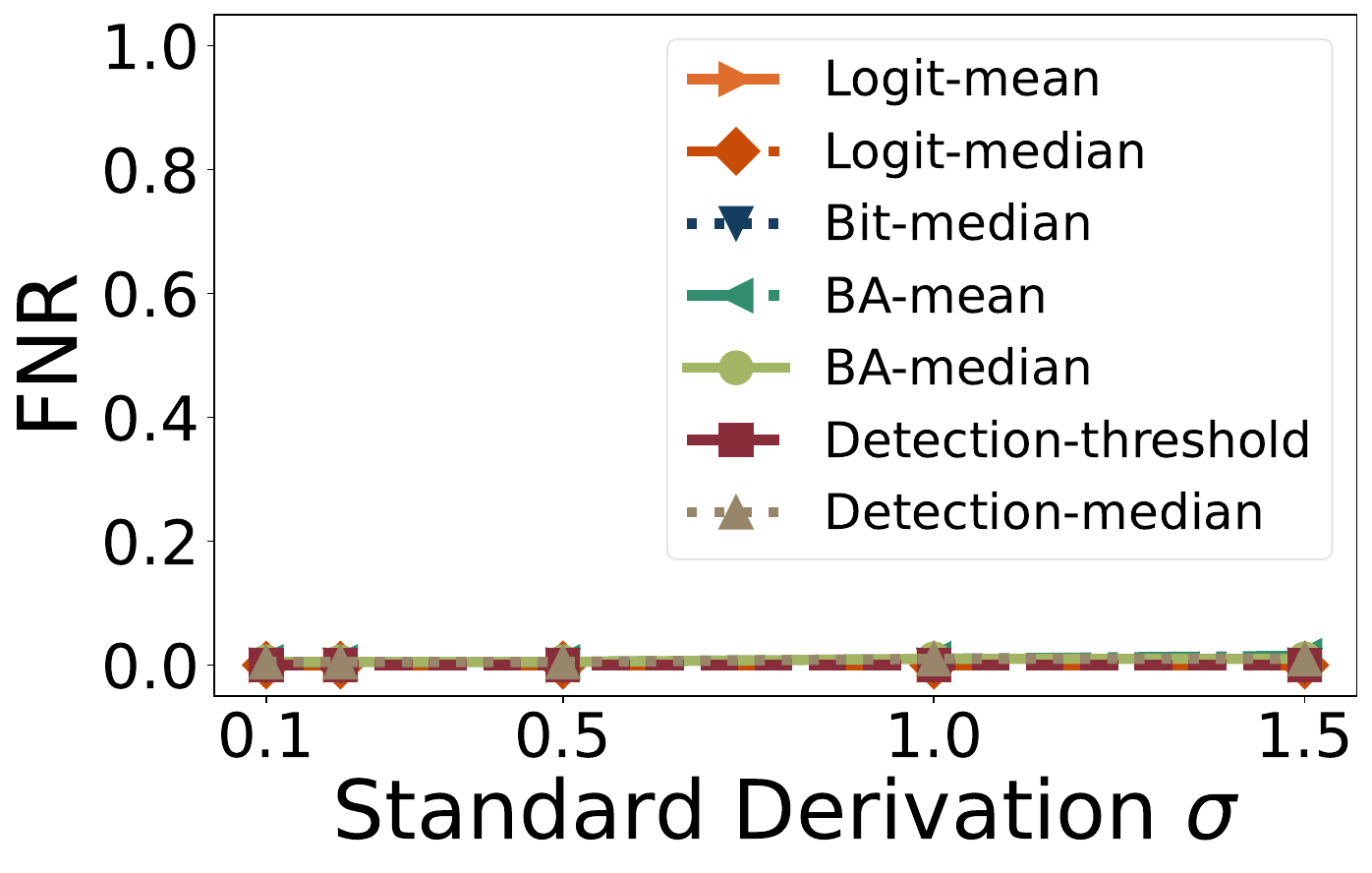}
        \caption{Gaussian Blur}
    \end{subfigure}
    \begin{subfigure}{.23\linewidth}
        \centering
        \includegraphics[width=\linewidth]{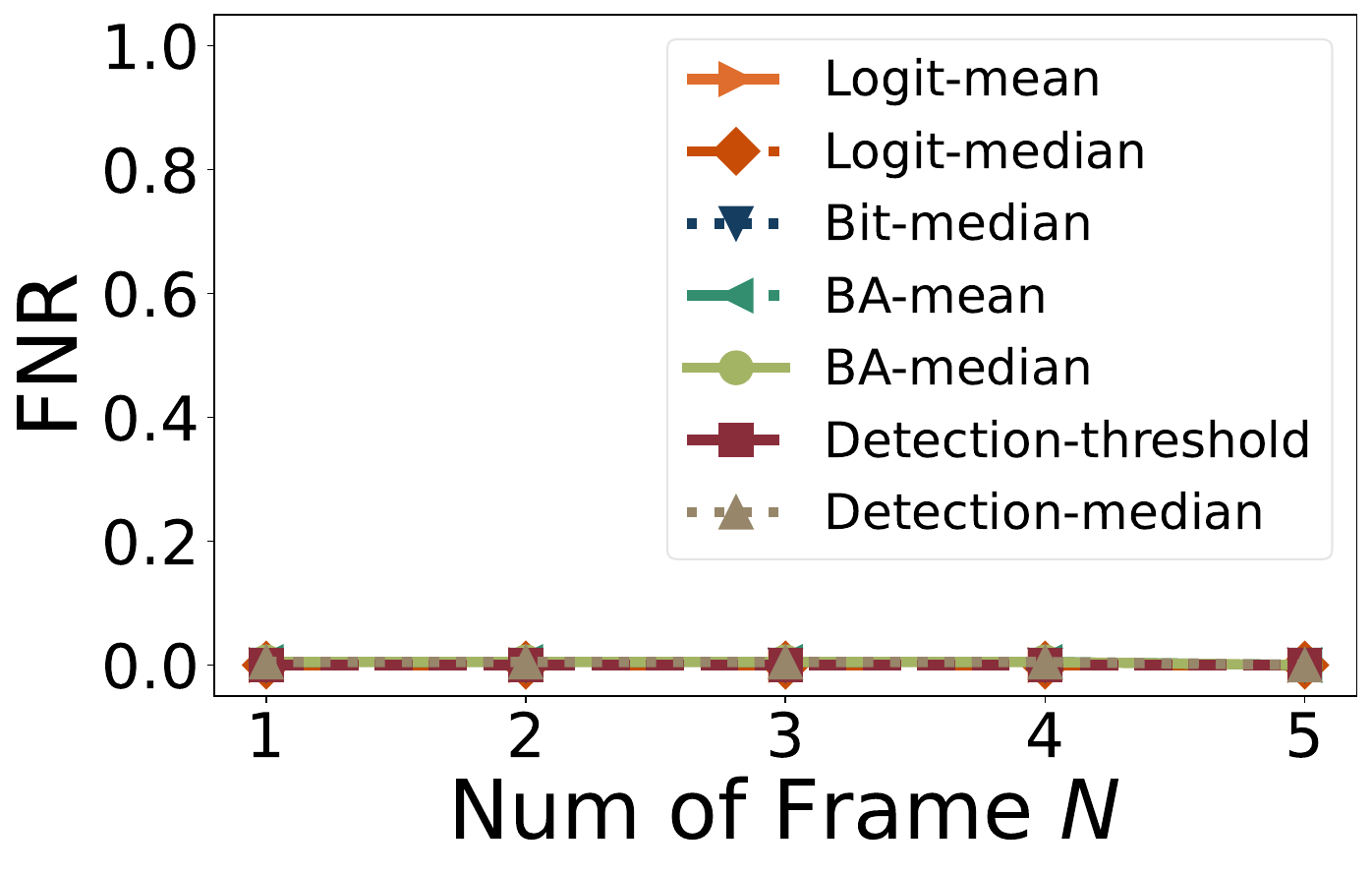}
        \caption{Frame Average}
    \end{subfigure}
    \begin{subfigure}{.23\linewidth}
        \centering
        \includegraphics[width=\linewidth]{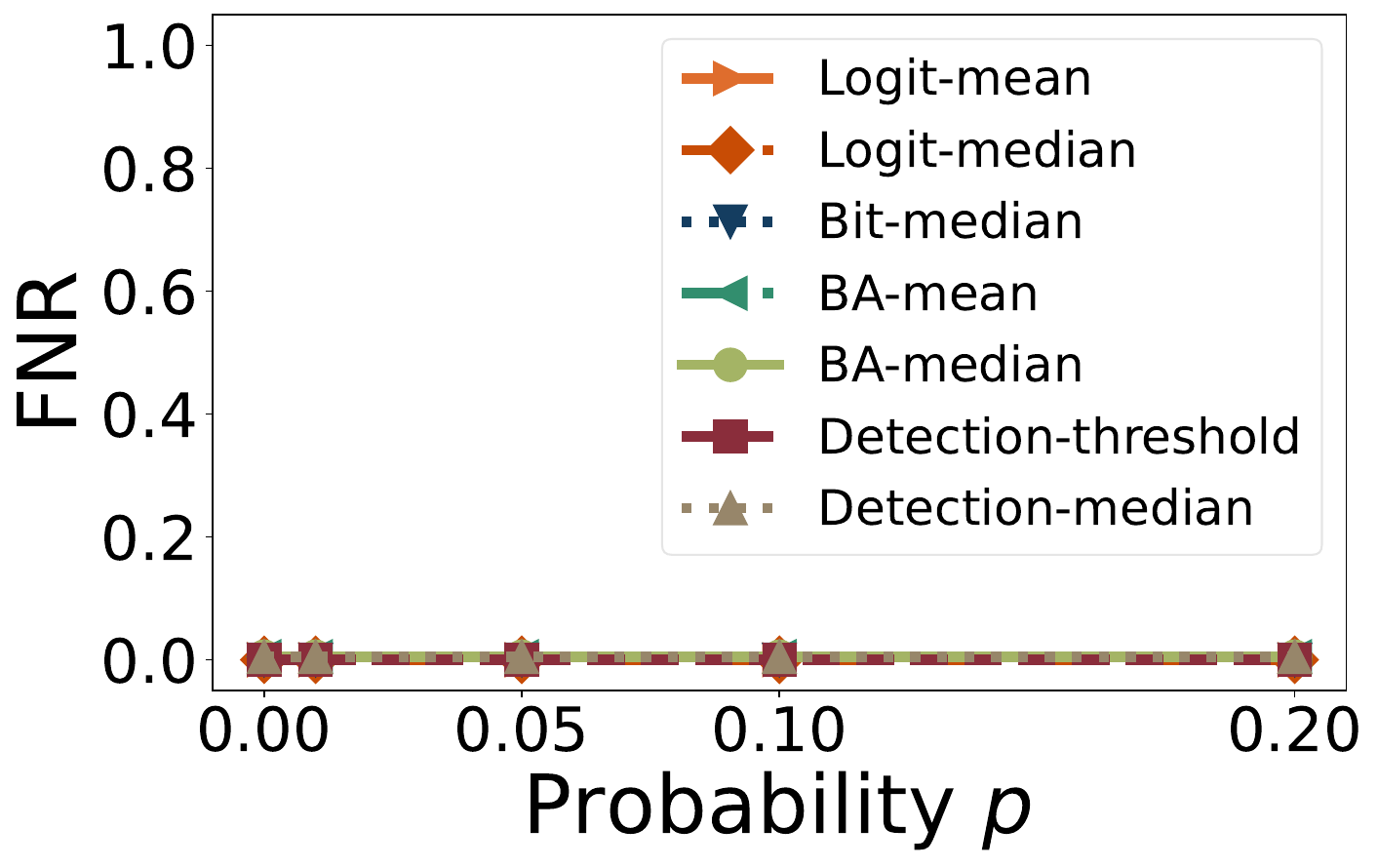}
        \caption{Frame Switch}
    \end{subfigure}
    \begin{subfigure}{.23\linewidth}
        \centering
        \includegraphics[width=\linewidth]{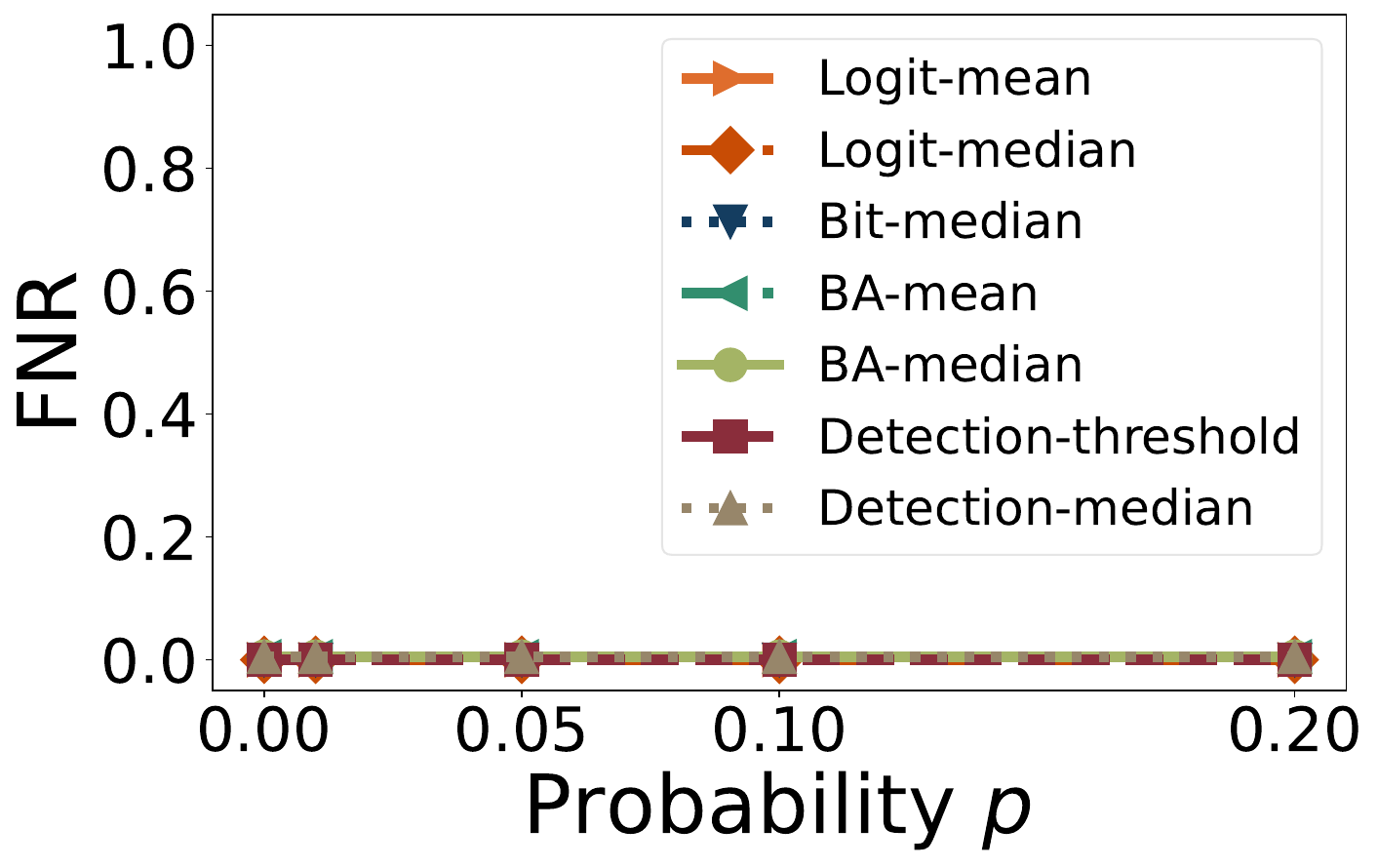}
        \caption{Frame Removal}
    \end{subfigure} \\

    \begin{subfigure}{.9\linewidth}
    \centering
    \caption*{Realistic video style}
    \end{subfigure}

    \begin{subfigure}{.23\linewidth}
        \centering
        \includegraphics[width=\linewidth]{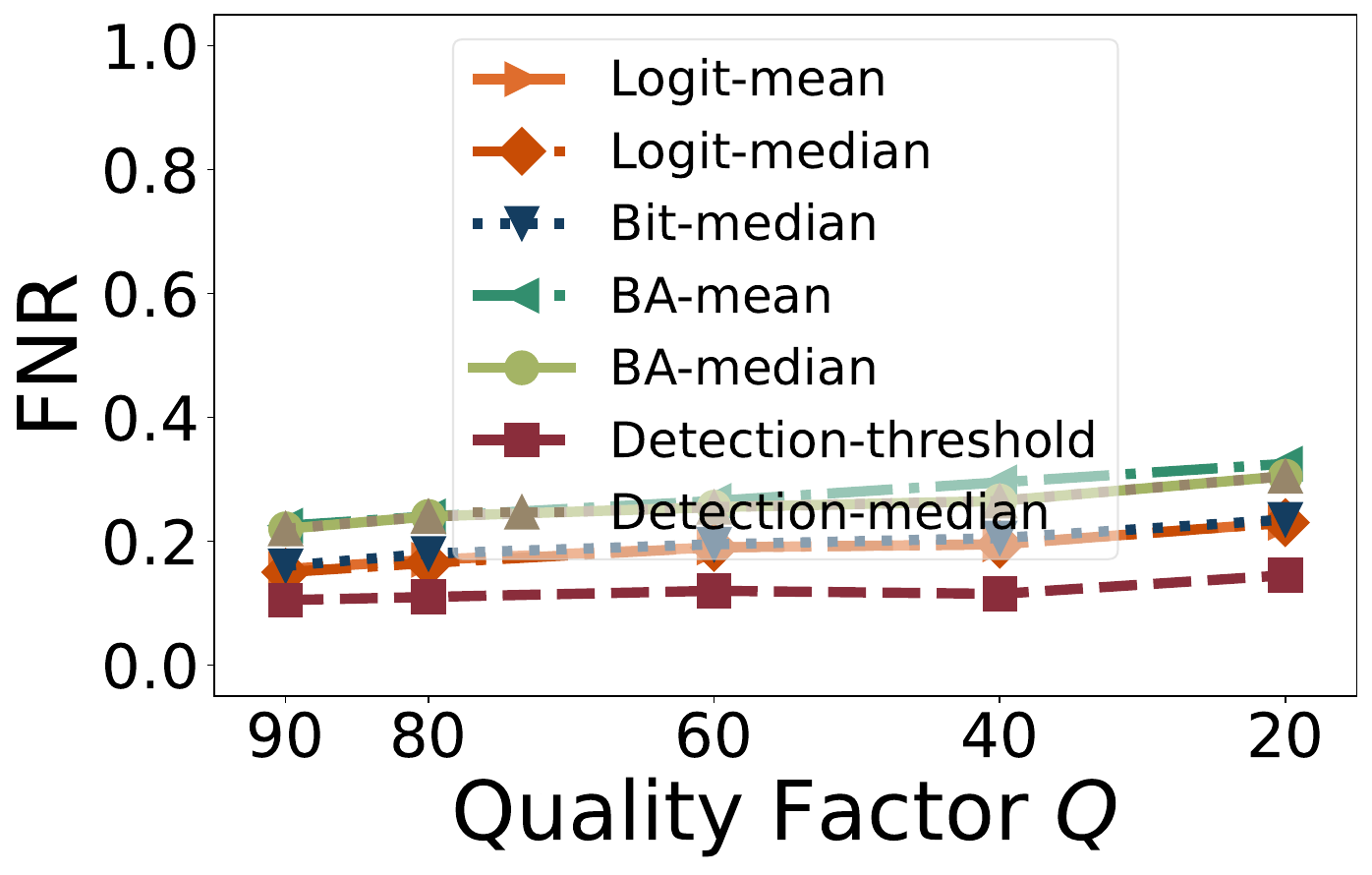}
        \caption{JPEG}
    \end{subfigure}
    \begin{subfigure}{.23\linewidth}
        \centering
        \includegraphics[width=\linewidth]{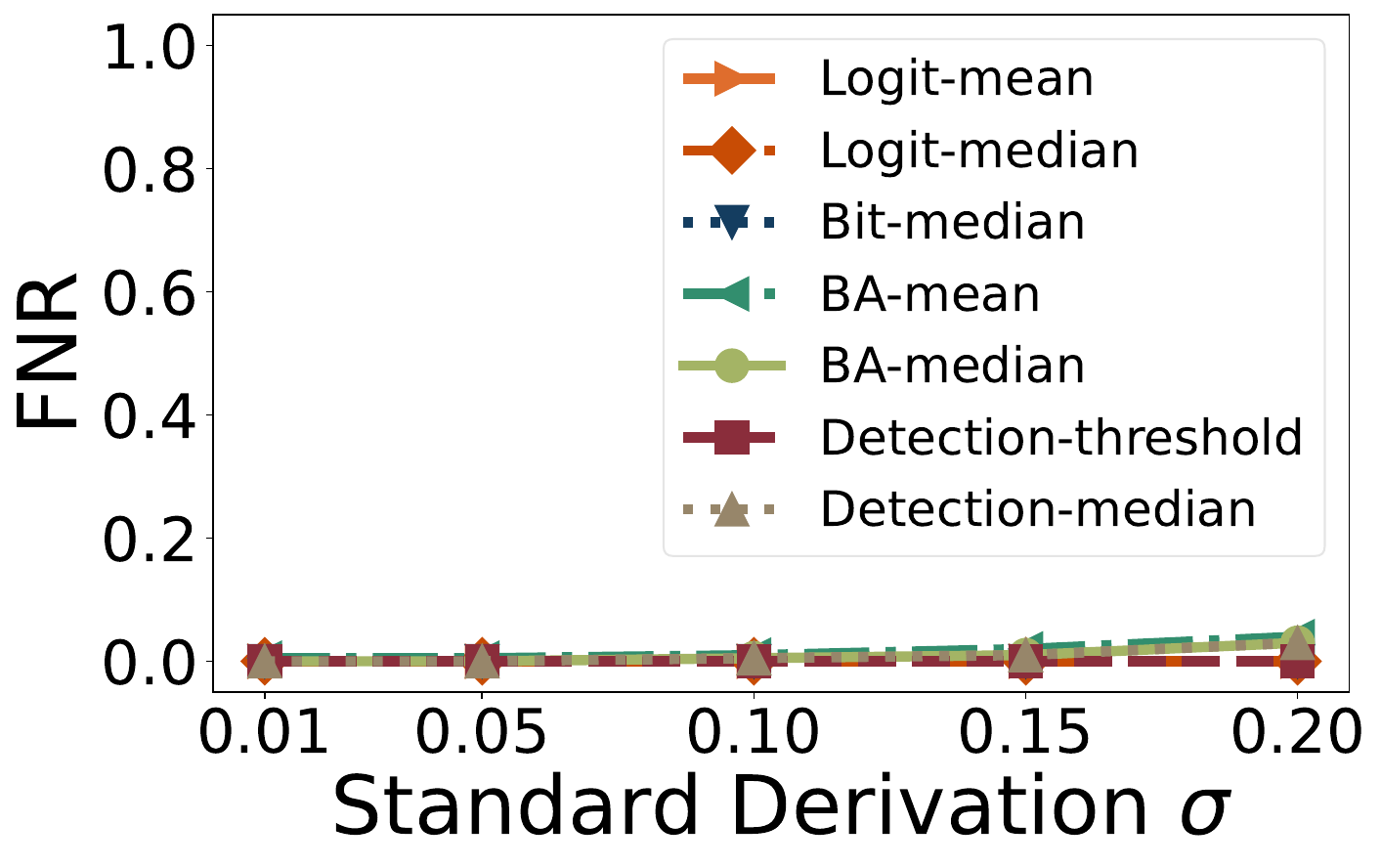}
        \caption{Gaussian Noise}
    \end{subfigure}
    \begin{subfigure}{.23\linewidth}
        \centering
        \includegraphics[width=\linewidth]{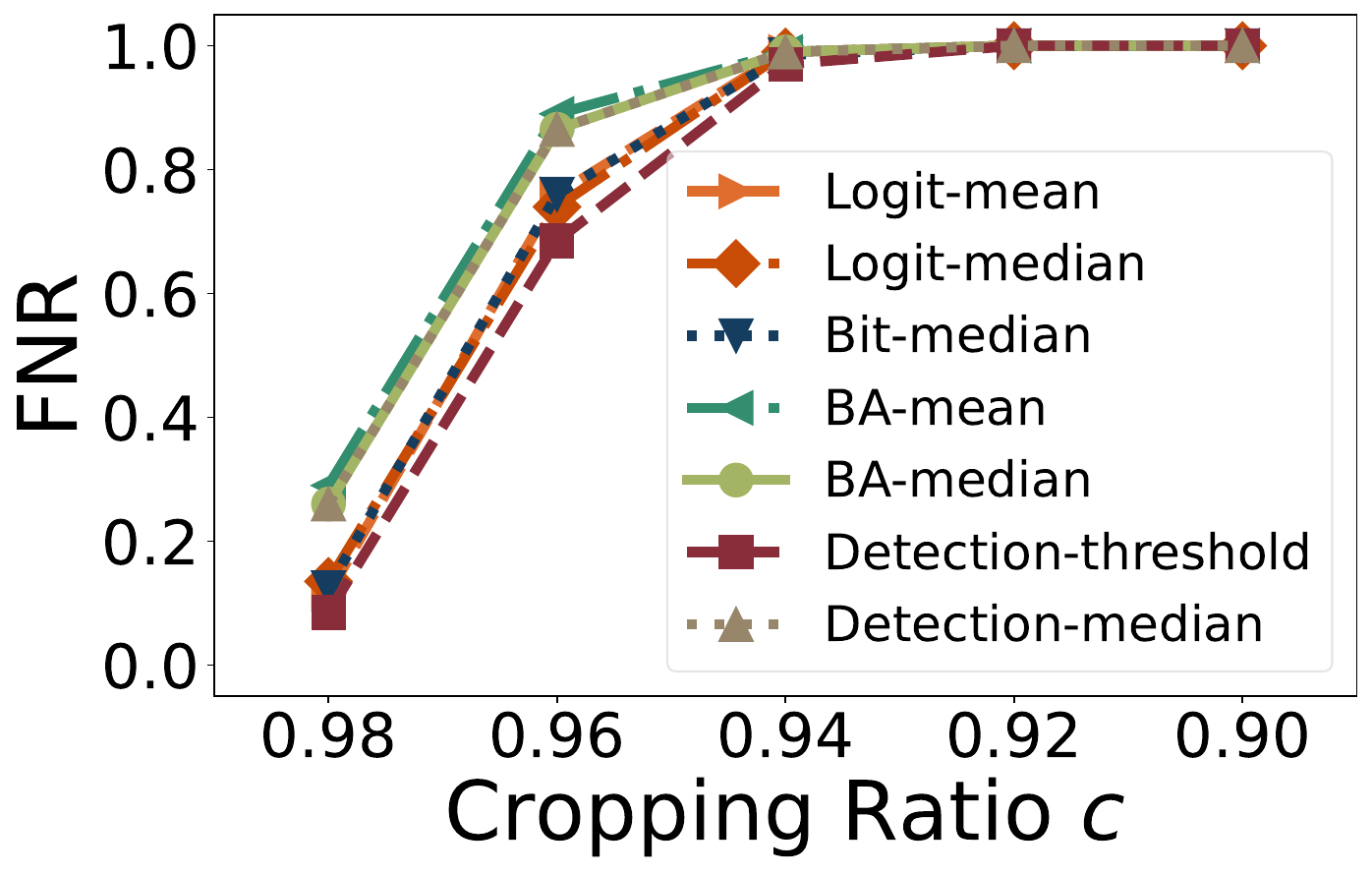}
        \caption{Cropping}
    \end{subfigure}
    \begin{subfigure}{.23\linewidth}
        \centering
        \includegraphics[width=\linewidth]{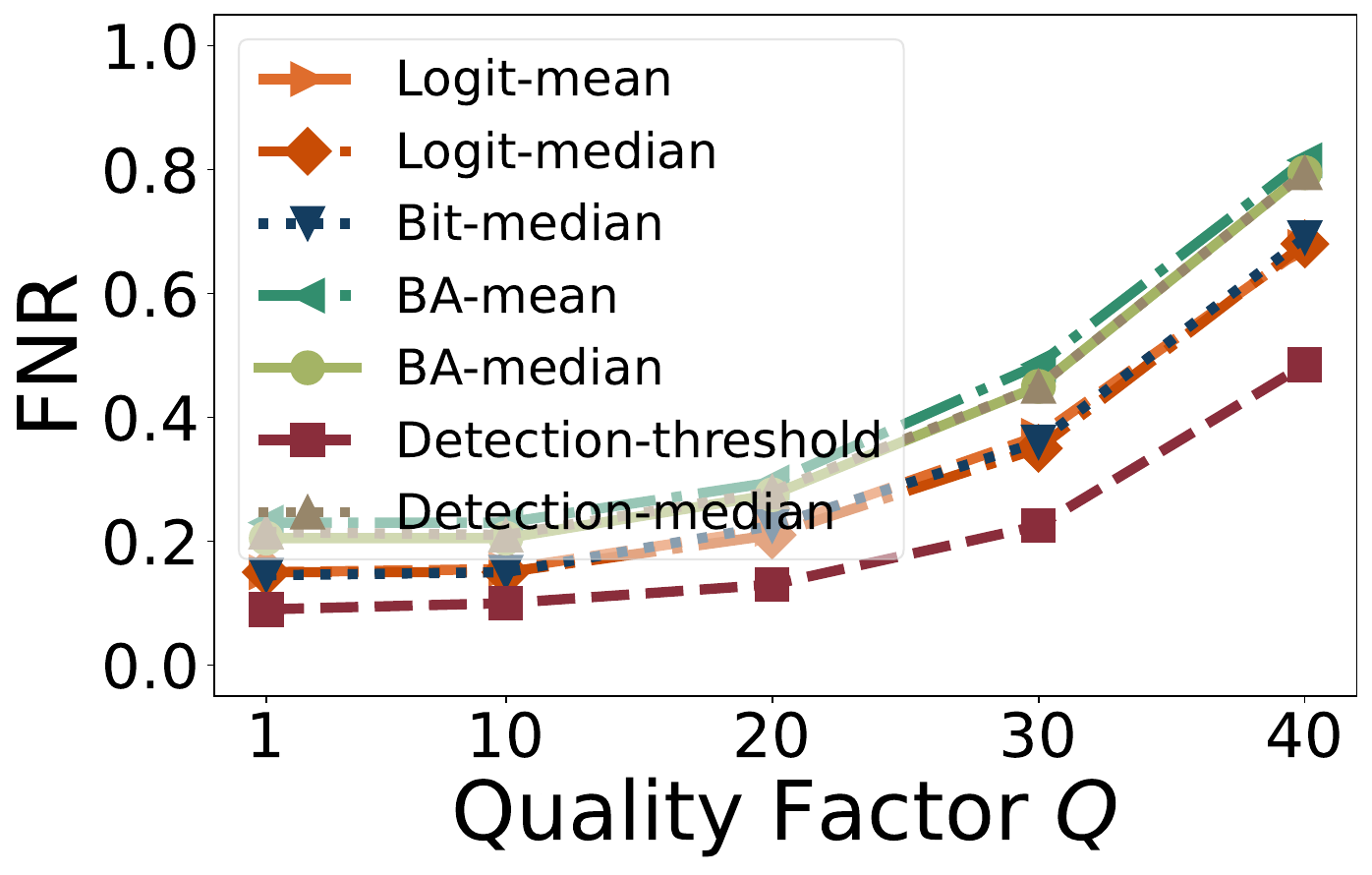}
        \caption{MPEG-4}
    \end{subfigure} \\
    
    \begin{subfigure}{.23\linewidth}
        \centering
        \includegraphics[width=\linewidth]{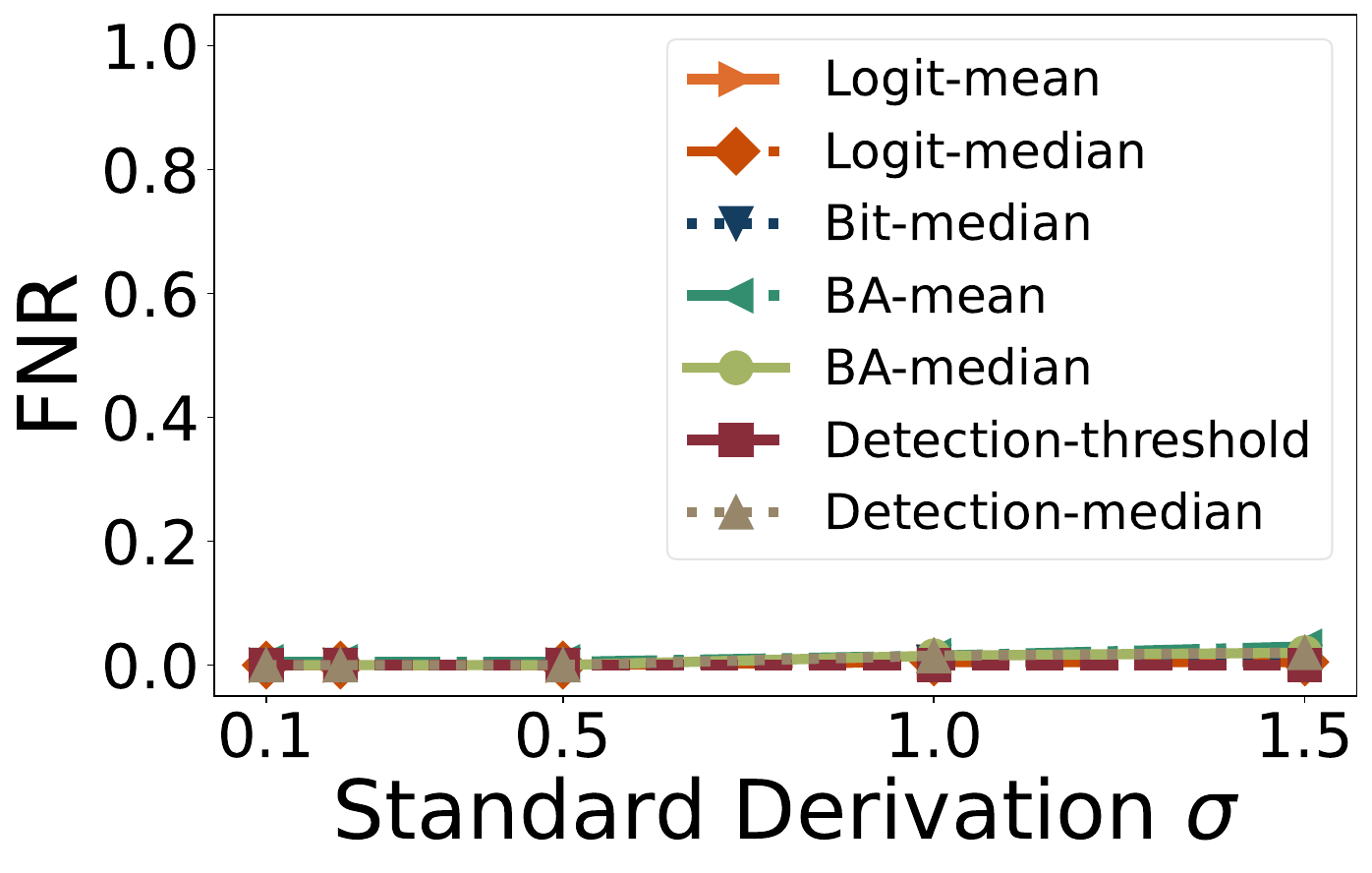}
        \caption{Gaussian Blur}
    \end{subfigure}
    \begin{subfigure}{.23\linewidth}
        \centering
        \includegraphics[width=\linewidth]{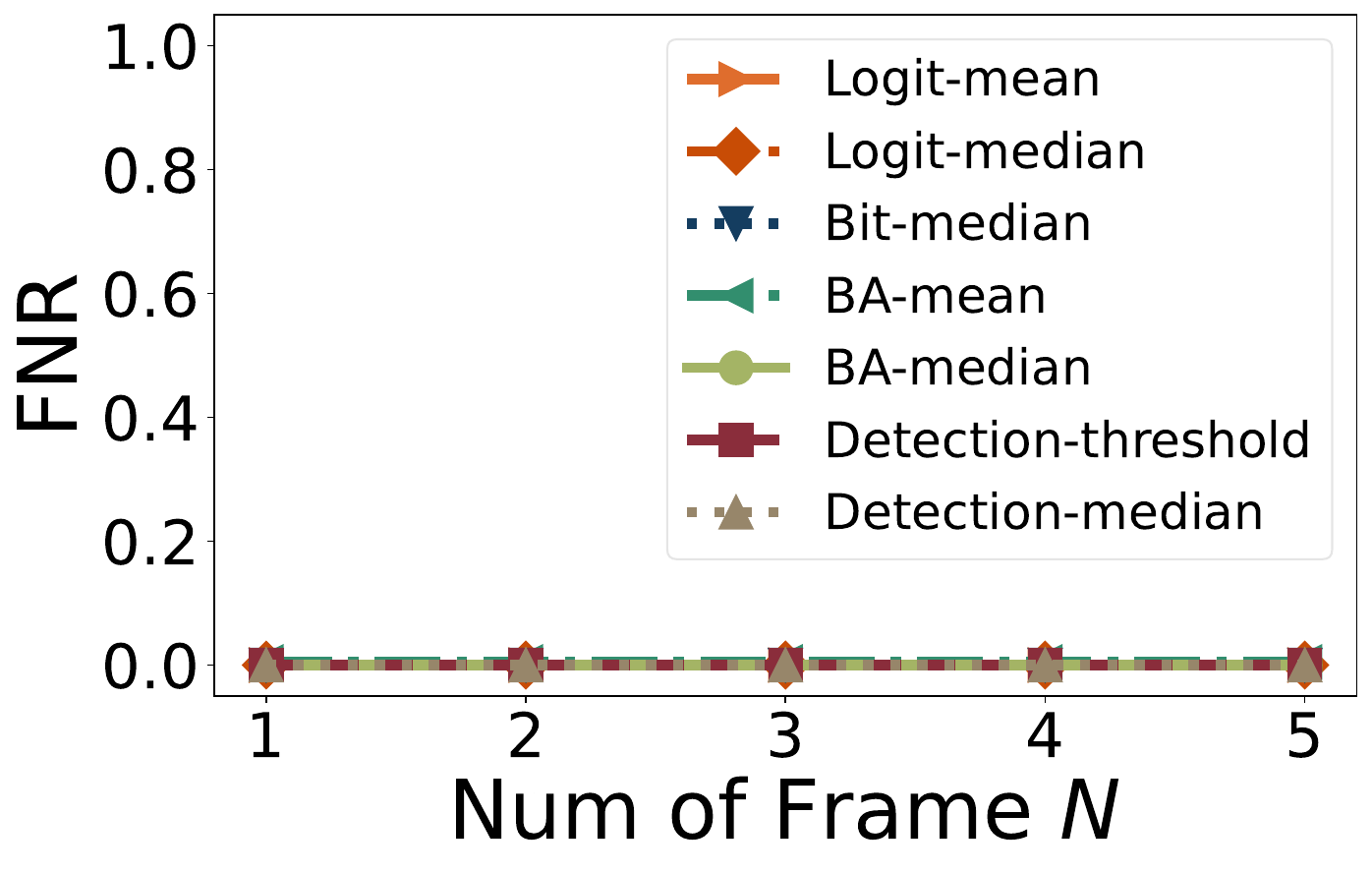}
        \caption{Frame Average}
    \end{subfigure}
    \begin{subfigure}{.23\linewidth}
        \centering
        \includegraphics[width=\linewidth]{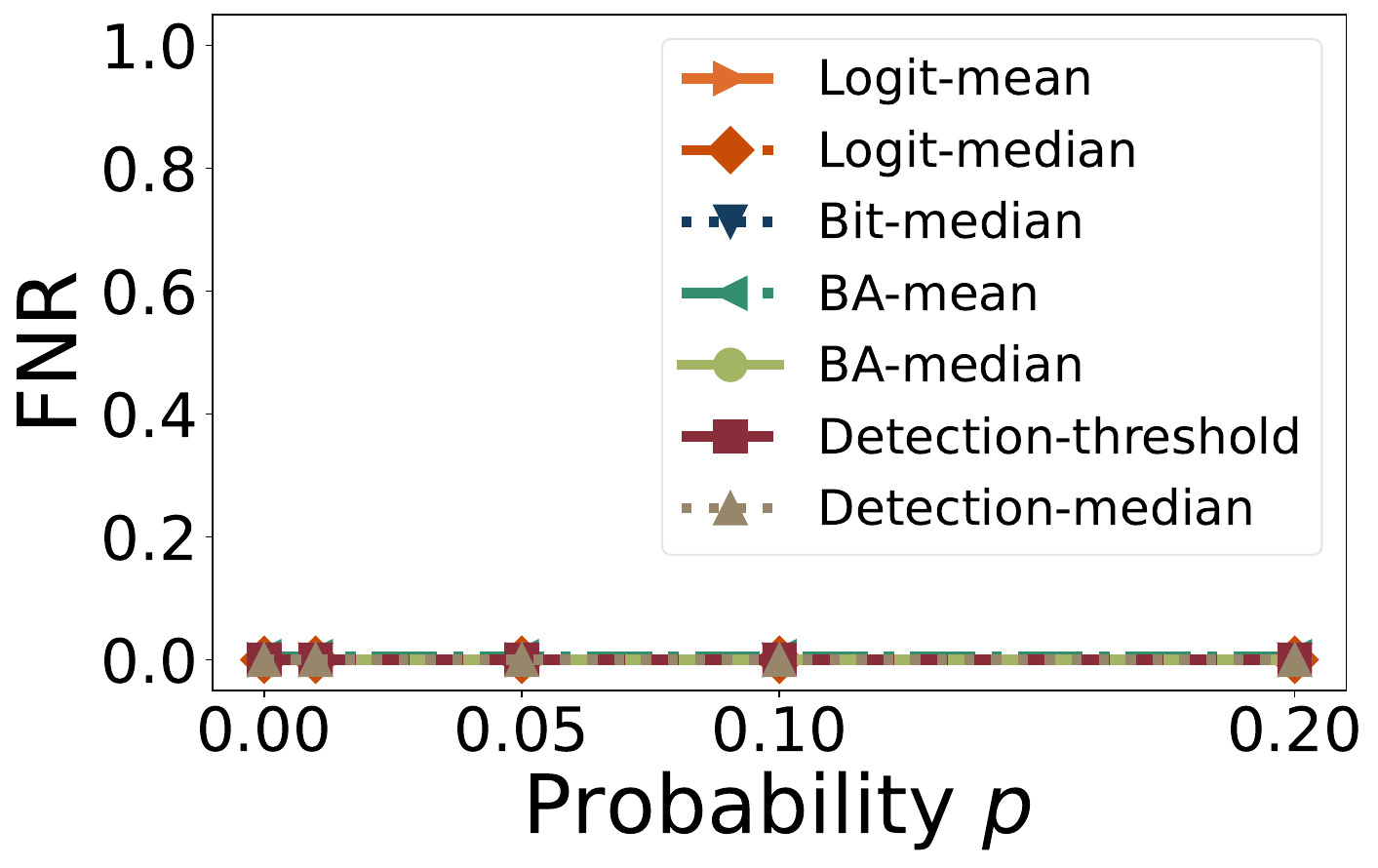}
        \caption{Frame Switch}
    \end{subfigure}
    \begin{subfigure}{.23\linewidth}
        \centering
        \includegraphics[width=\linewidth]{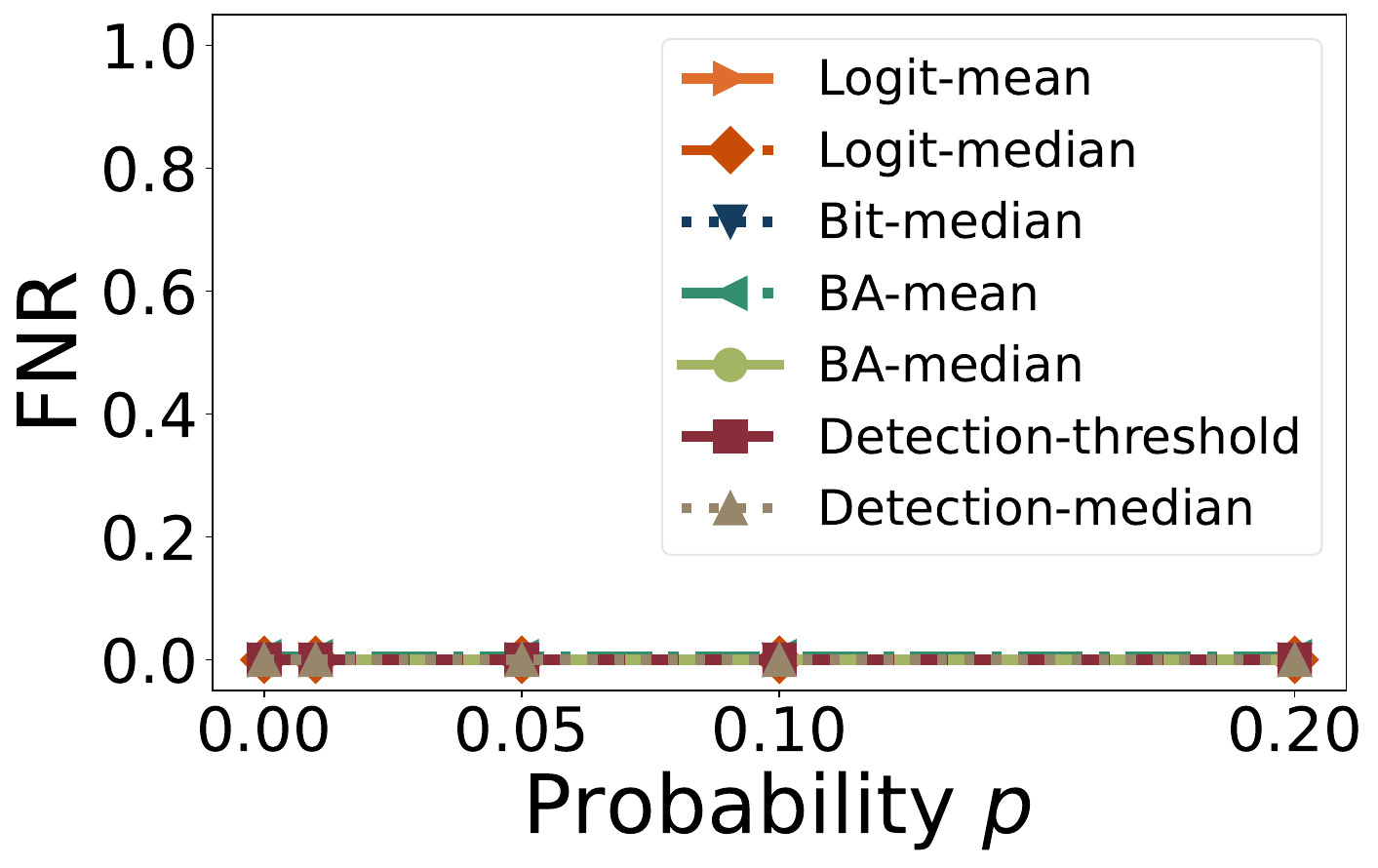}
        \caption{Frame Removal}
    \end{subfigure} \\

    \begin{subfigure}{.9\linewidth}
    \centering
    \caption*{Cartoon video style}
    \end{subfigure}

    \begin{subfigure}{.23\linewidth}
        \centering
        \includegraphics[width=\linewidth]{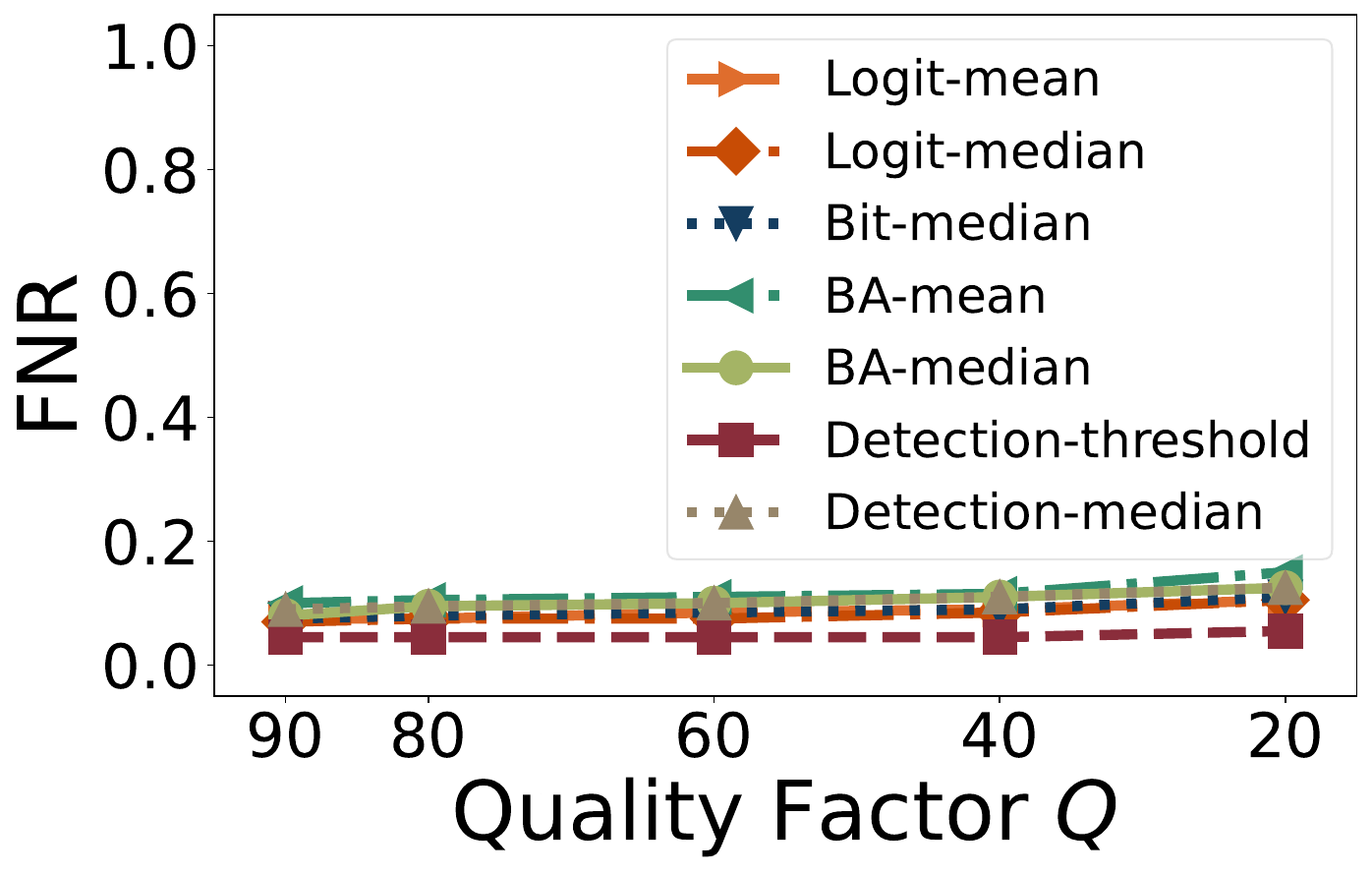}
        \caption{JPEG}
    \end{subfigure}
    \begin{subfigure}{.23\linewidth}
        \centering
        \includegraphics[width=\linewidth]{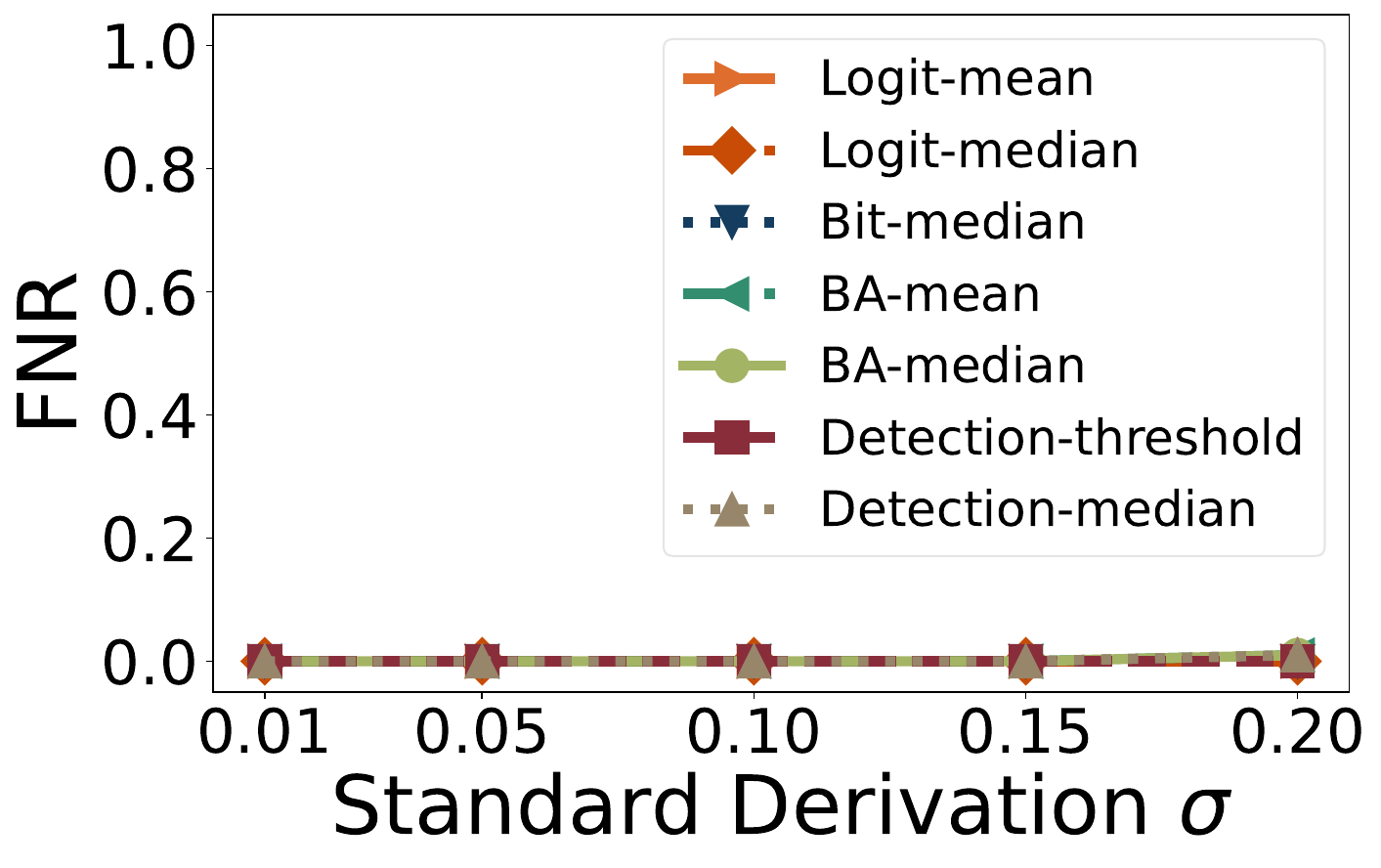}
        \caption{Gaussian Noise}
    \end{subfigure}
    \begin{subfigure}{.23\linewidth}
        \centering
        \includegraphics[width=\linewidth]{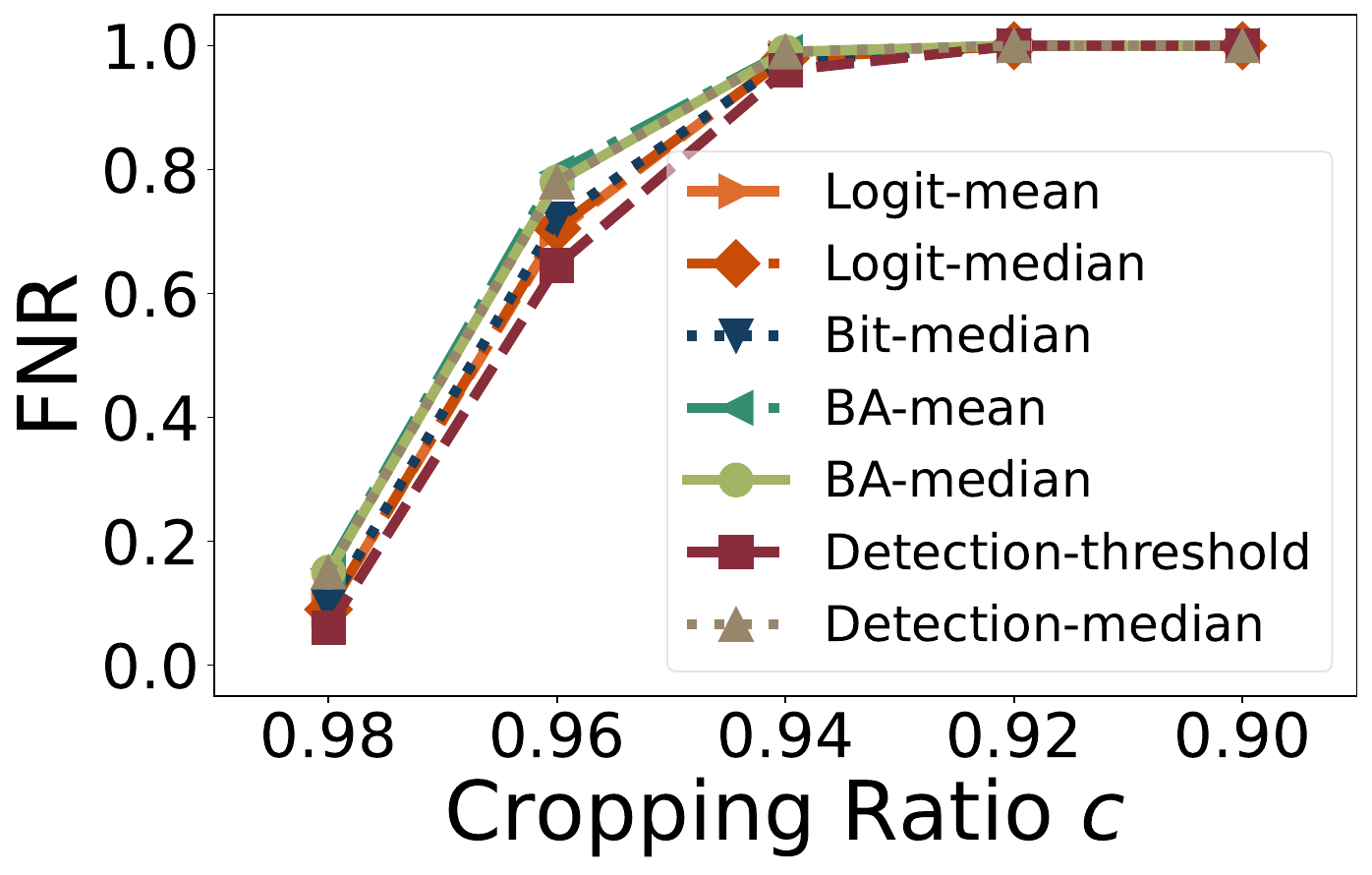}
        \caption{Cropping}
    \end{subfigure}
    \begin{subfigure}{.23\linewidth}
        \centering
        \includegraphics[width=\linewidth]{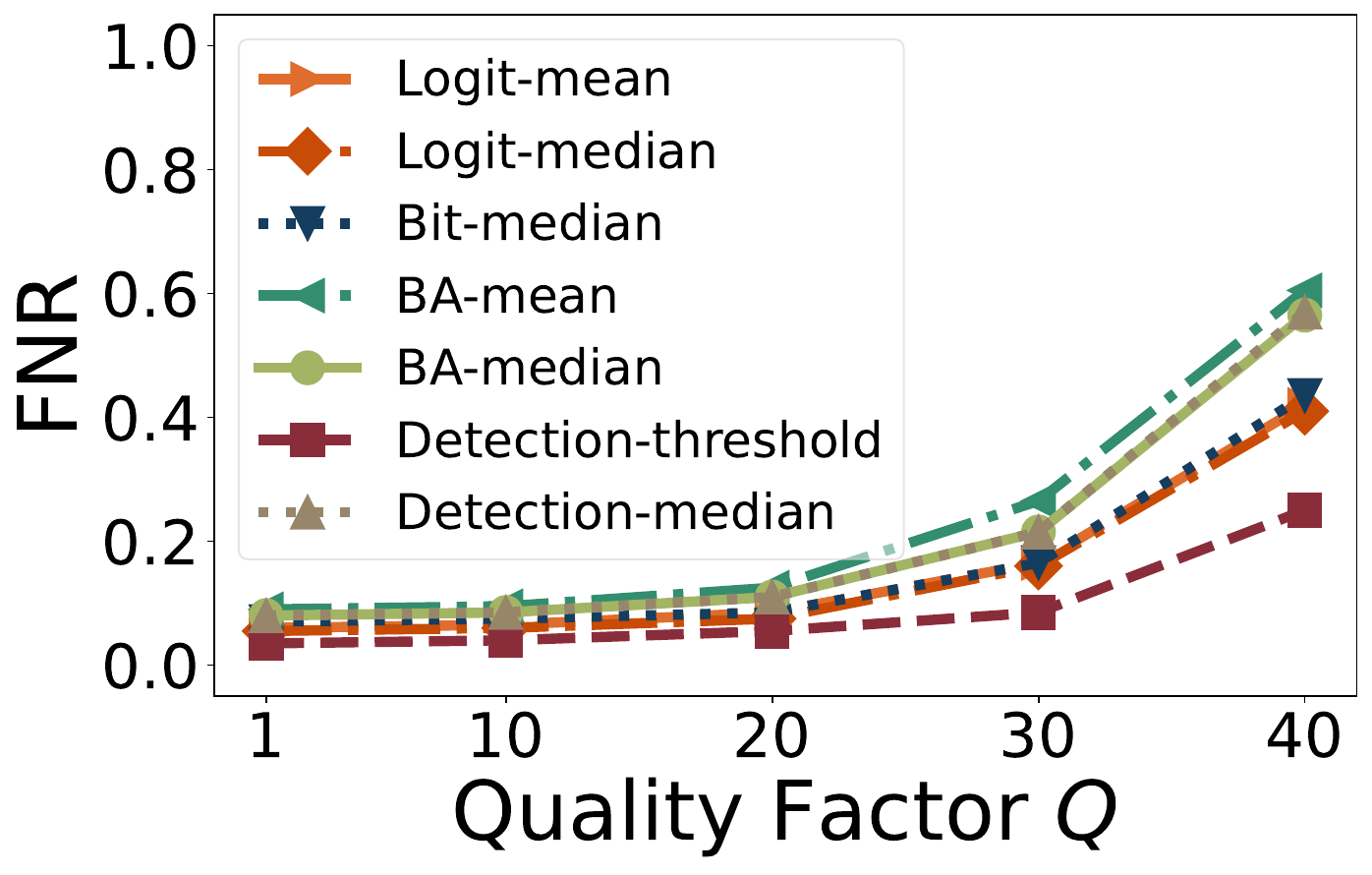}
        \caption{MPEG-4}
    \end{subfigure} \\
    
    \begin{subfigure}{.23\linewidth}
        \centering
        \includegraphics[width=\linewidth]{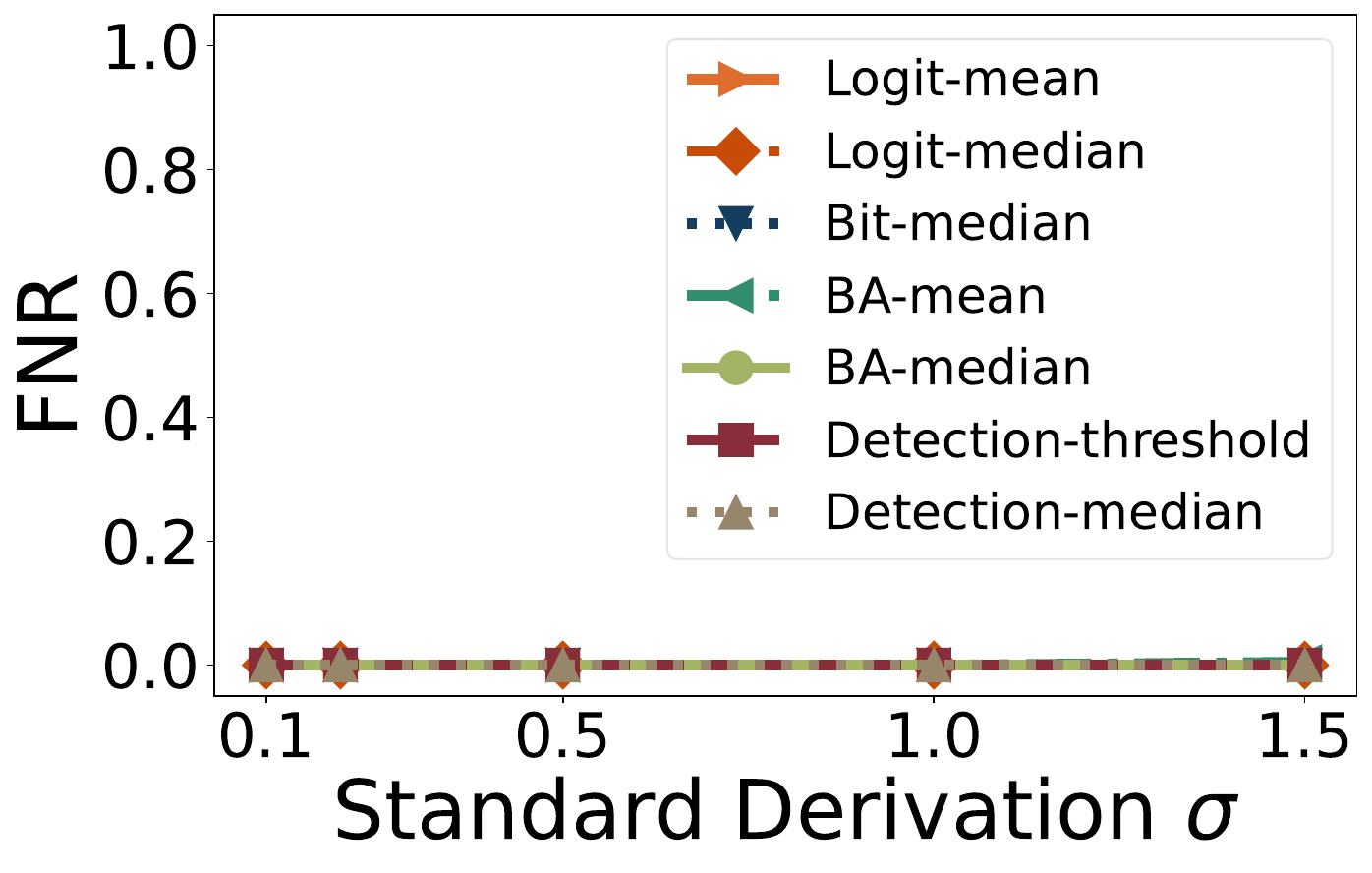}
        \caption{Gaussian Blur}
    \end{subfigure}
    \begin{subfigure}{.23\linewidth}
        \centering
        \includegraphics[width=\linewidth]{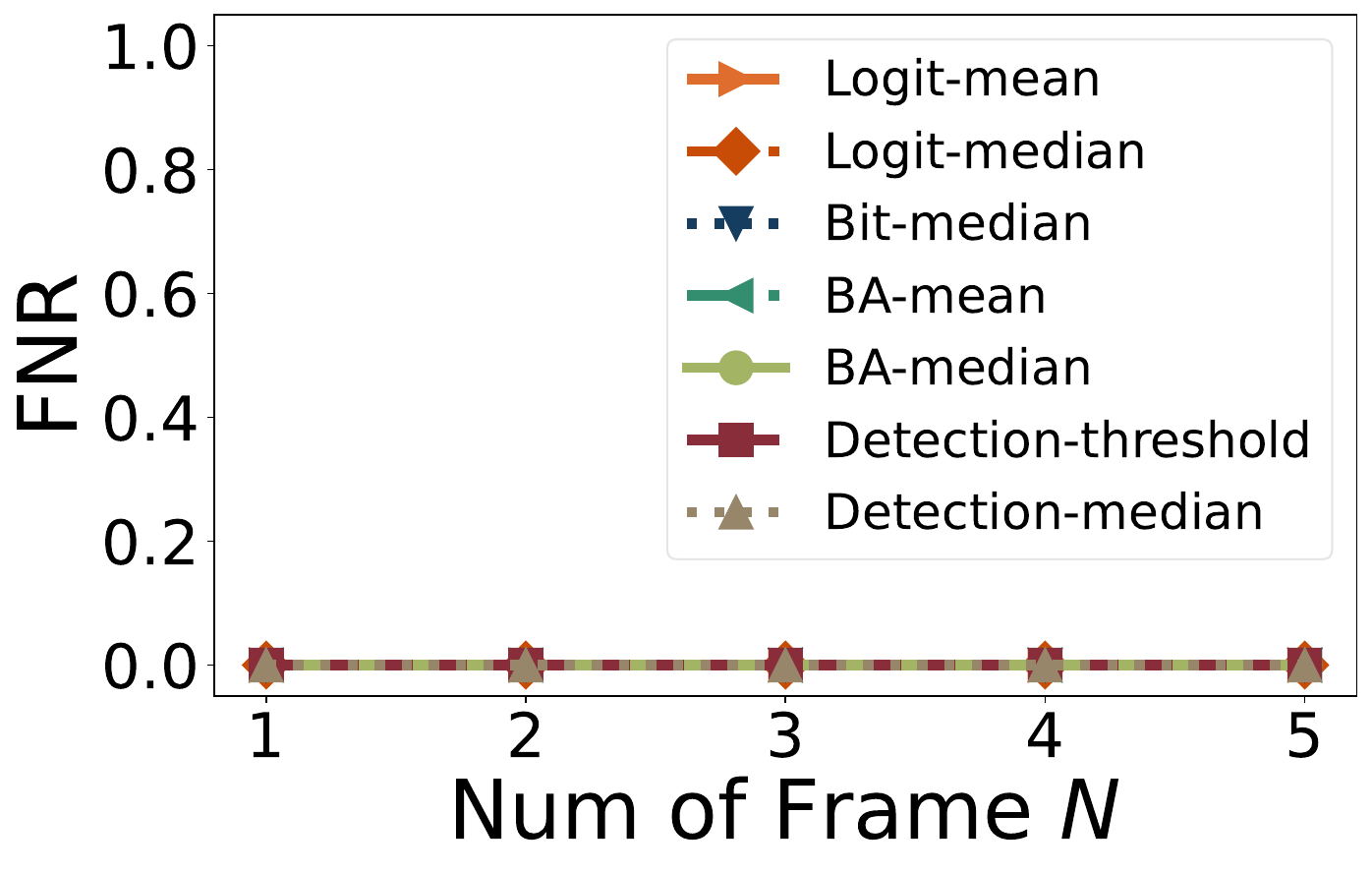}
        \caption{Frame Average}
    \end{subfigure}
    \begin{subfigure}{.23\linewidth}
        \centering
        \includegraphics[width=\linewidth]{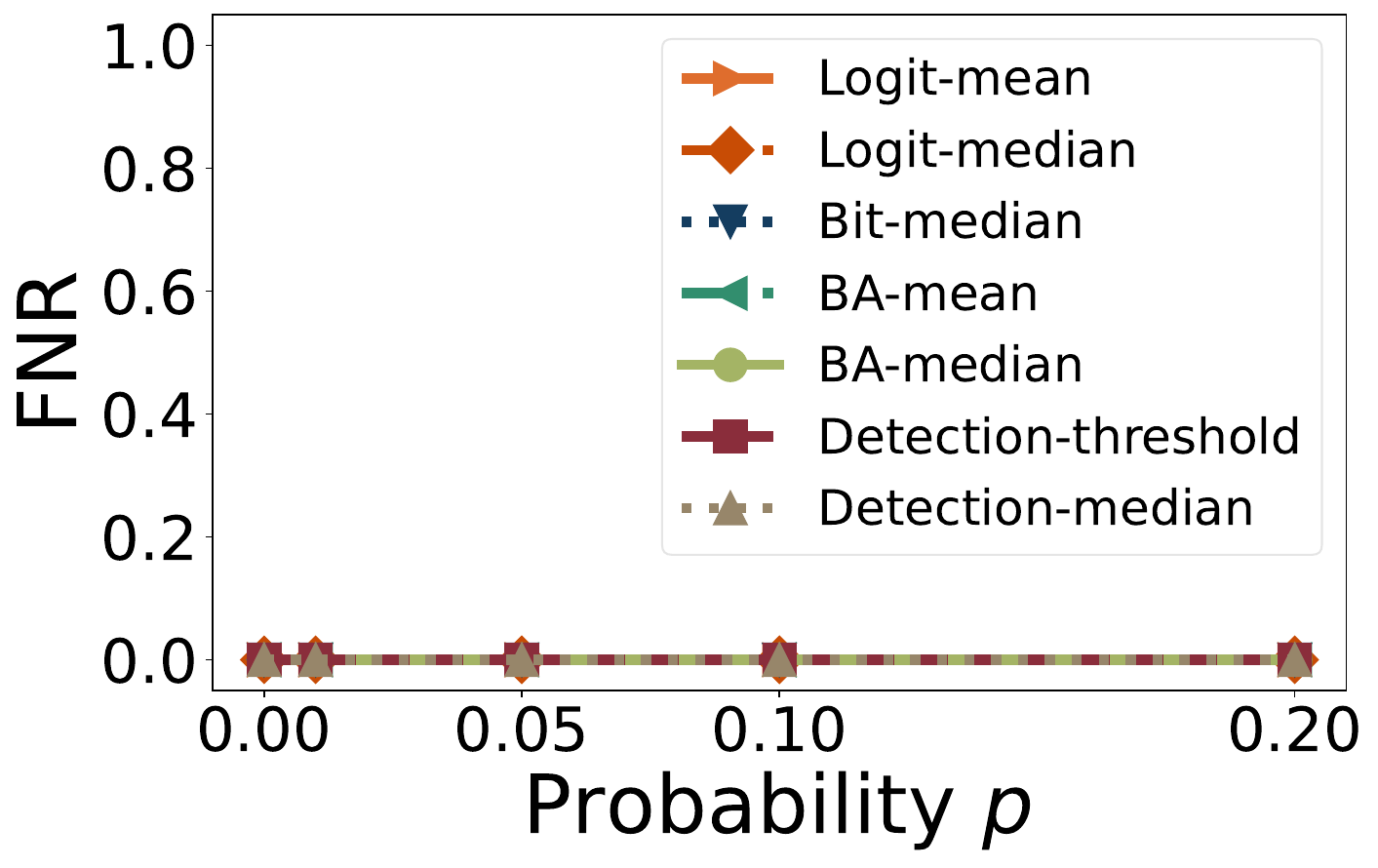}
        \caption{Frame Switch}
    \end{subfigure}
    \begin{subfigure}{.23\linewidth}
        \centering
        \includegraphics[width=\linewidth]{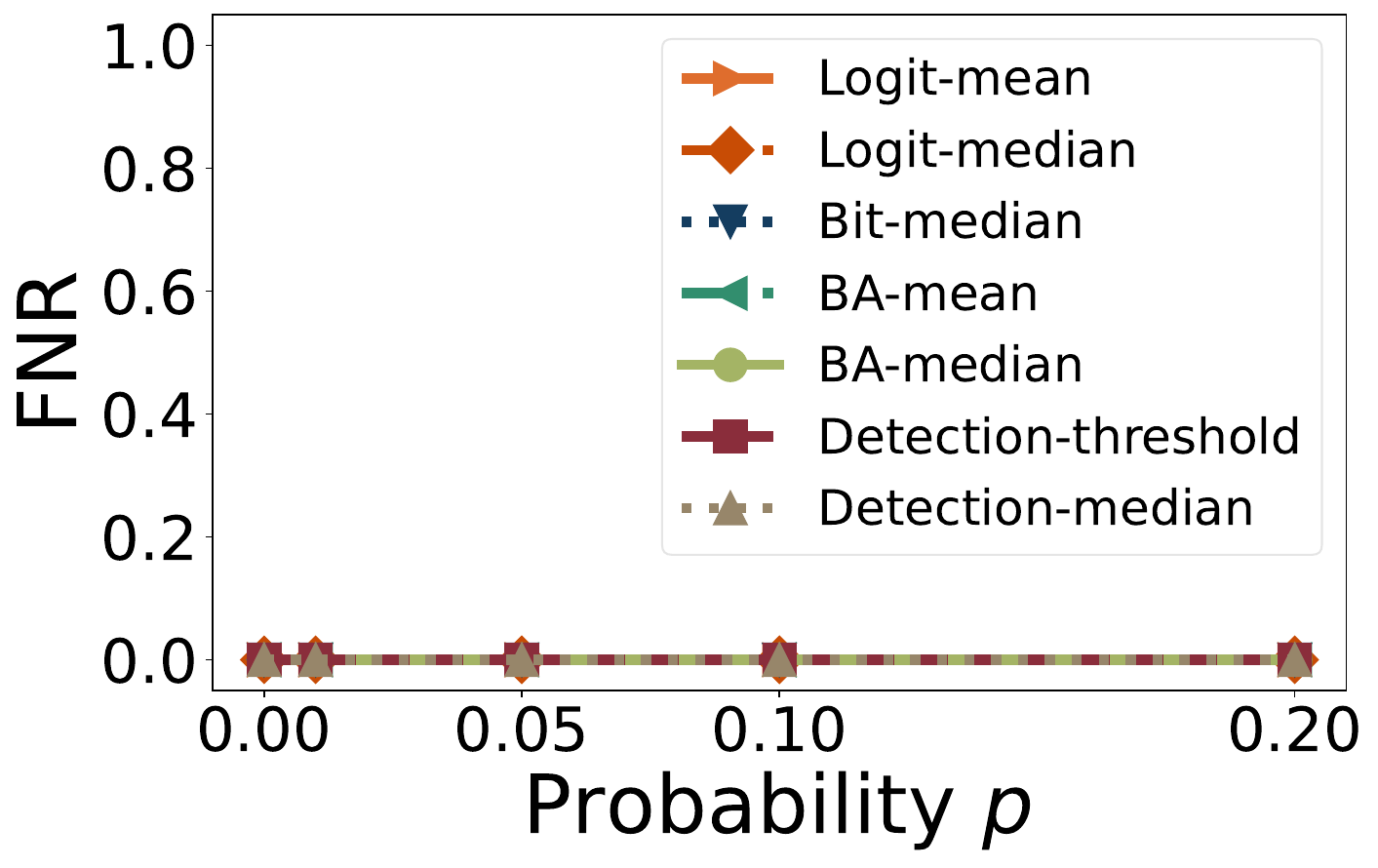}
        \caption{Frame Removal}
    \end{subfigure} \\

    \begin{subfigure}{.9\linewidth}
    \centering
    \caption*{Sci-fi video style}
    \end{subfigure}
    
    \caption{More fine-grained watermark removal results for StegaStamp on videos generated by Stable Video Diffusion.}
\end{figure}

\begin{figure}[]
    \centering

    \begin{subfigure}{.23\linewidth}
        \centering
        \includegraphics[width=\linewidth]{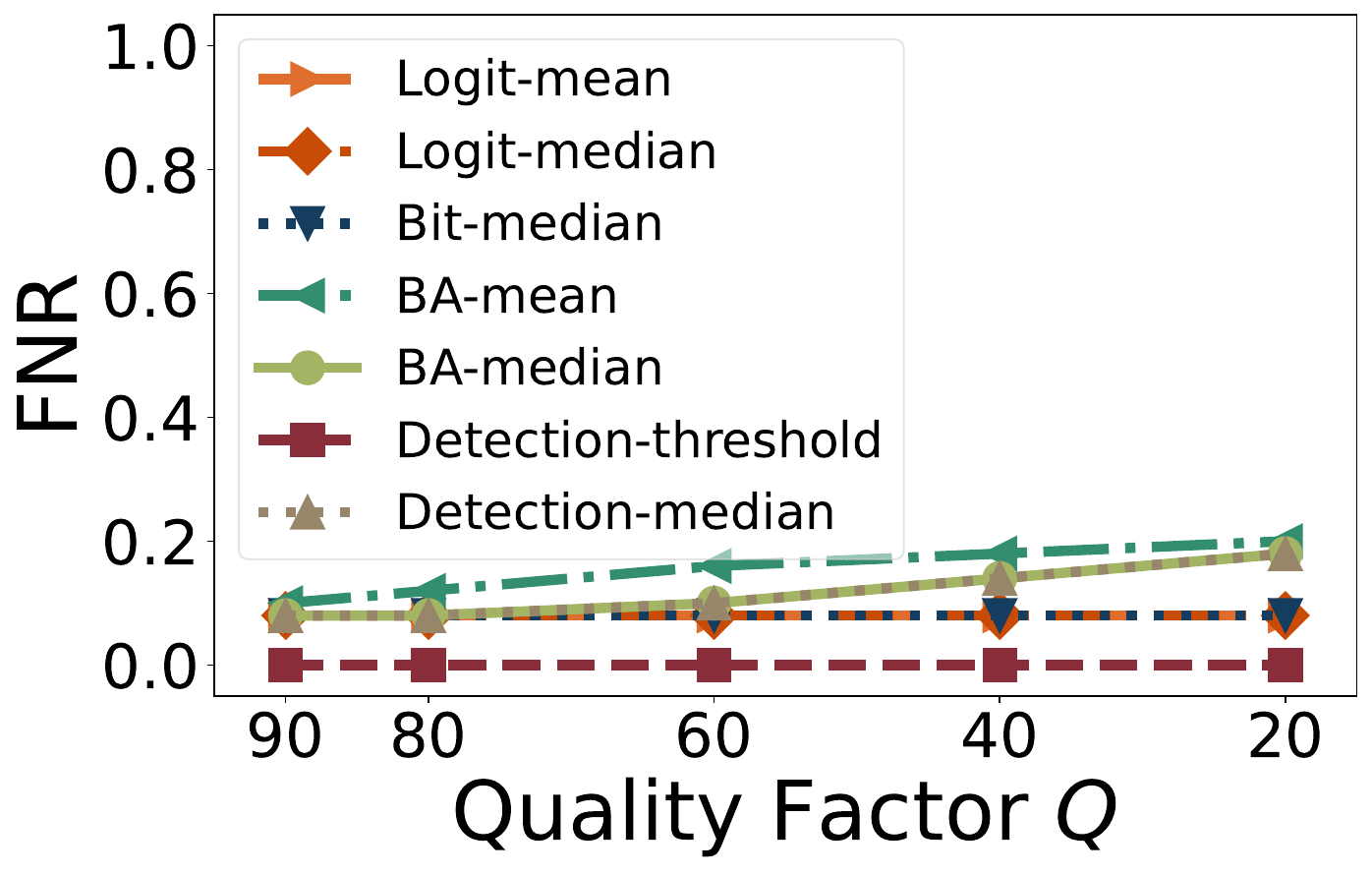}
        \caption{JPEG}
    \end{subfigure}
    \begin{subfigure}{.23\linewidth}
        \centering
        \includegraphics[width=\linewidth]{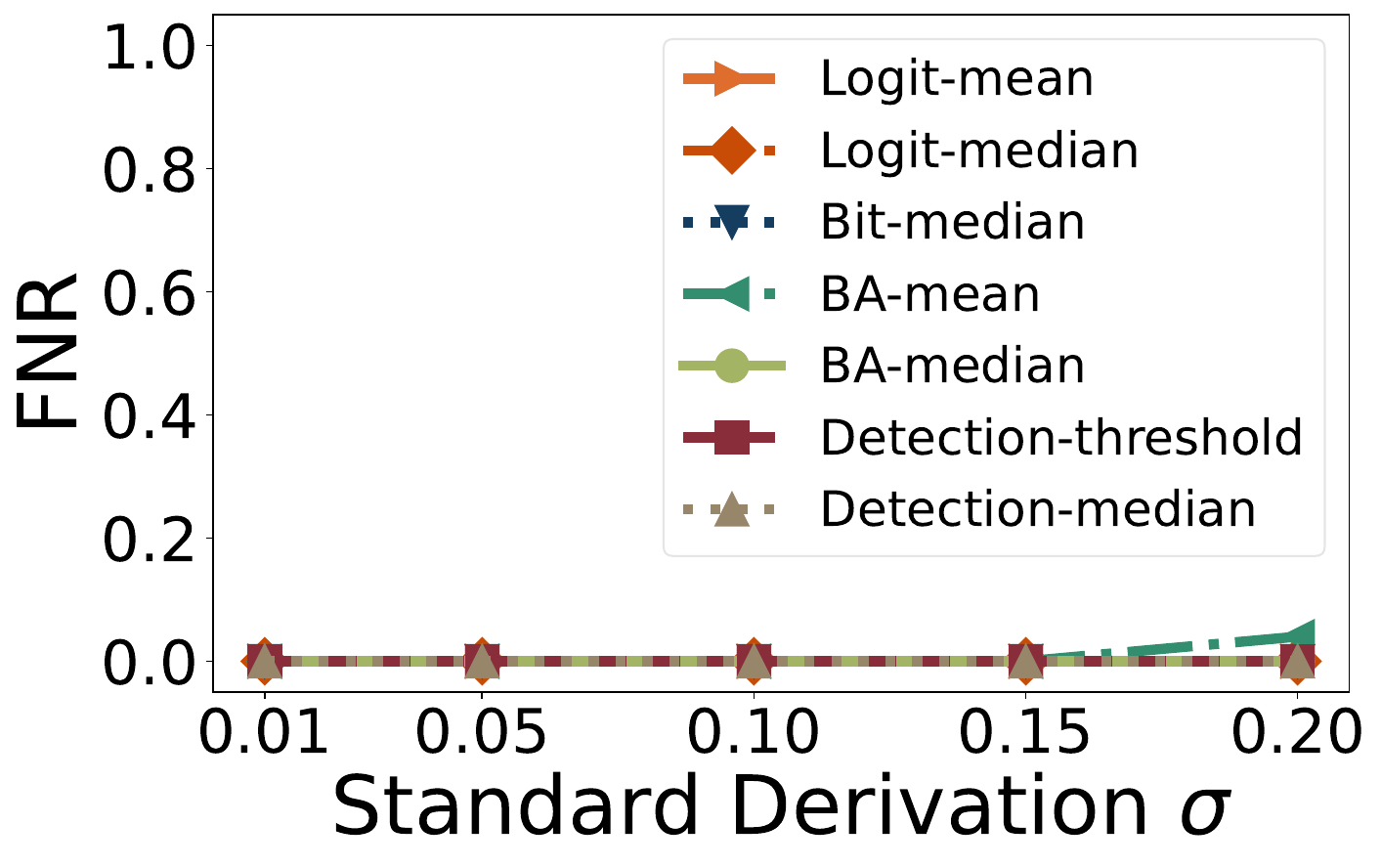}
        \caption{Gaussian Noise}
    \end{subfigure}
    \begin{subfigure}{.23\linewidth}
        \centering
        \includegraphics[width=\linewidth]{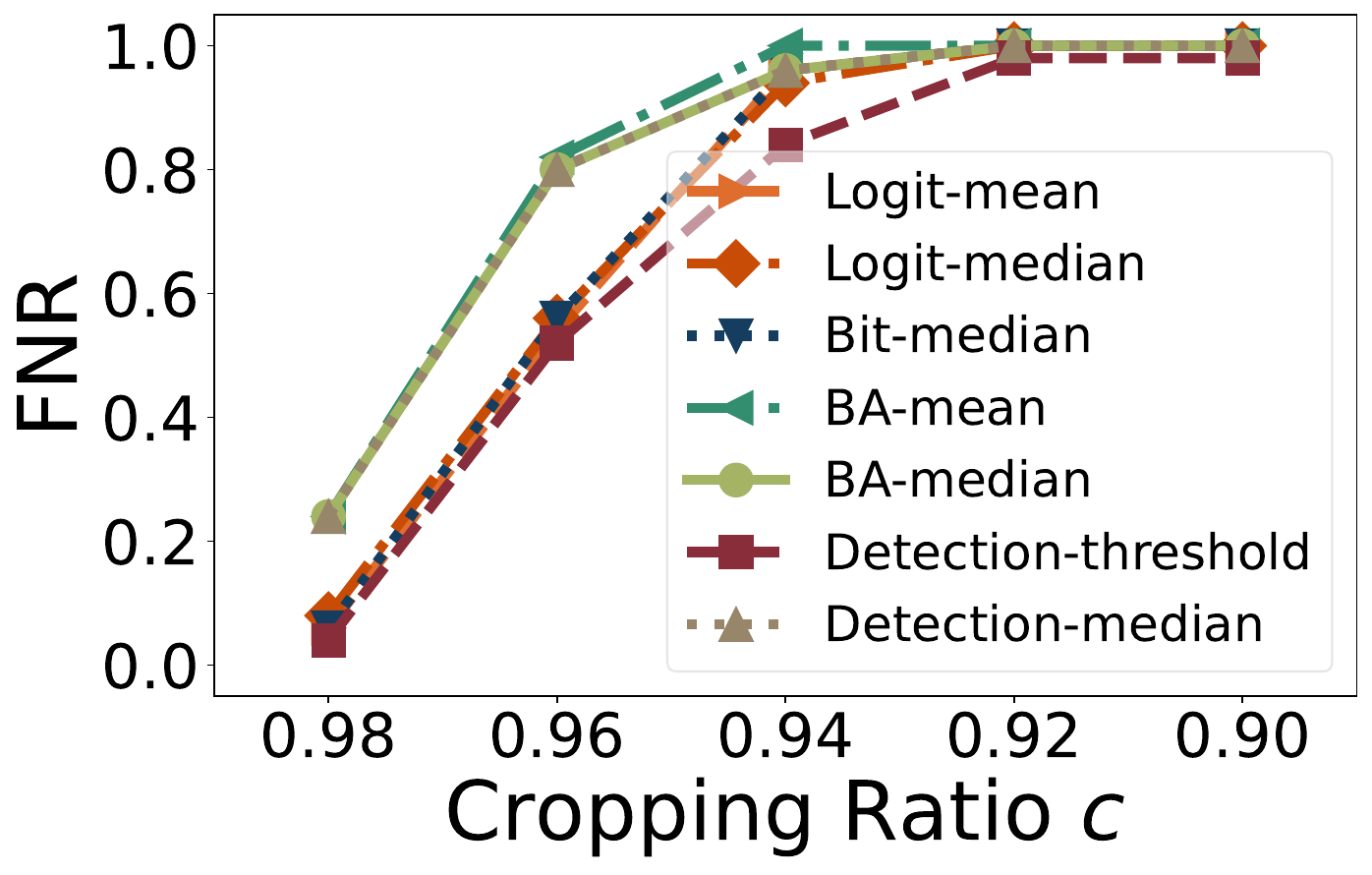}
        \caption{Cropping}
    \end{subfigure}
    \begin{subfigure}{.23\linewidth}
        \centering
        \includegraphics[width=\linewidth]{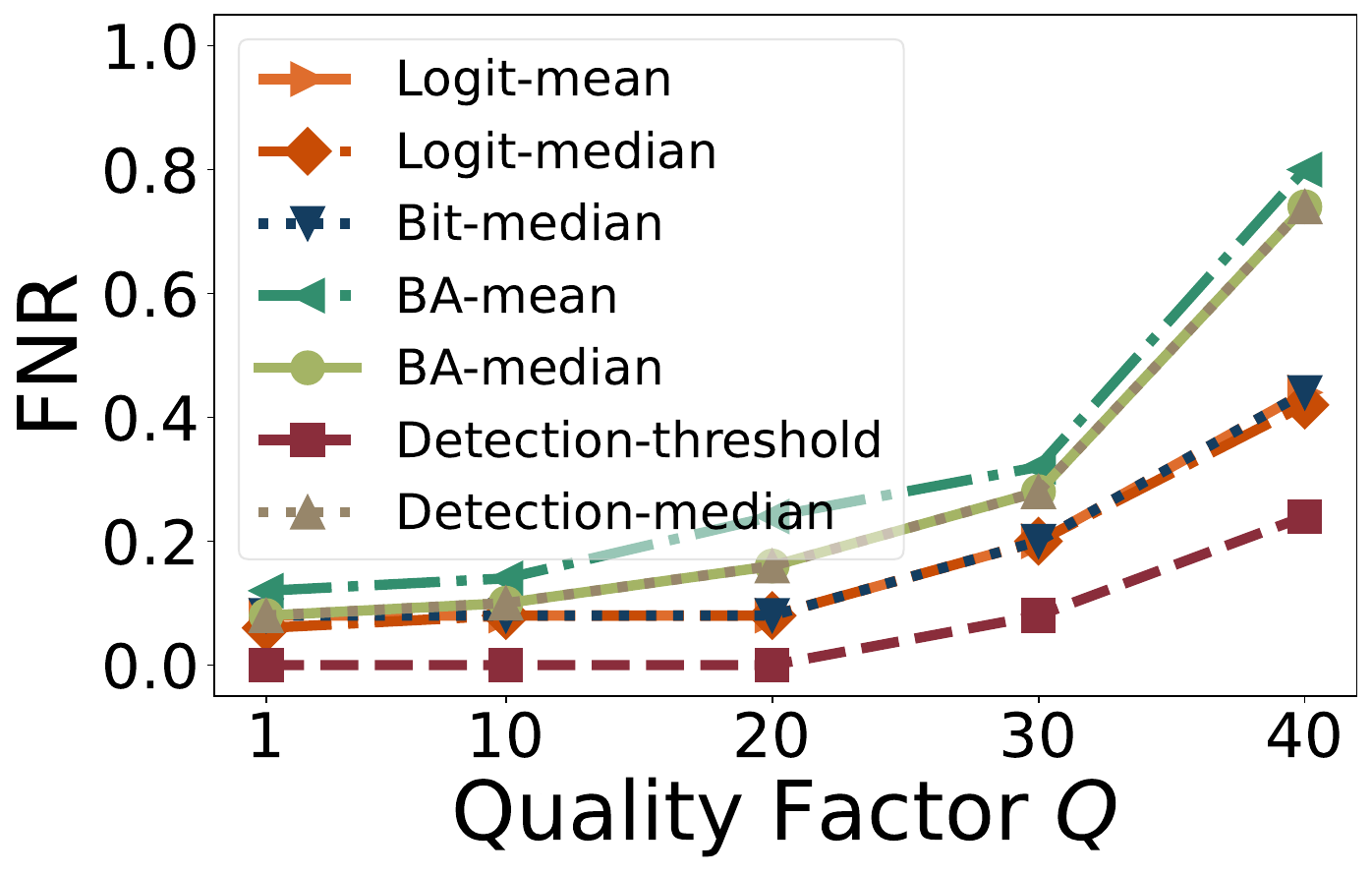}
        \caption{MPEG-4}
    \end{subfigure} \\
    
    \begin{subfigure}{.23\linewidth}
        \centering
        \includegraphics[width=\linewidth]{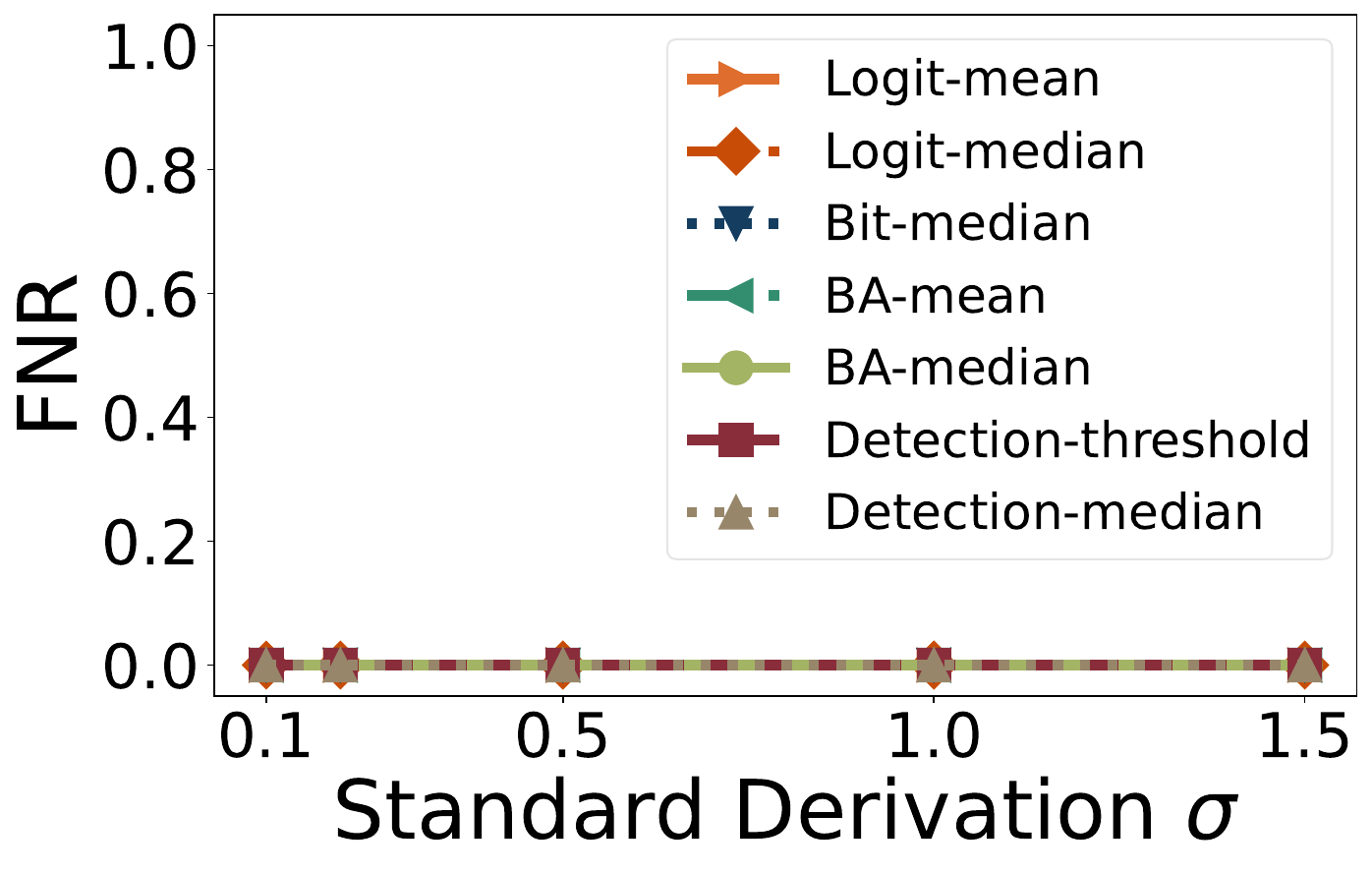}
        \caption{Gaussian Blur}
    \end{subfigure}
    \begin{subfigure}{.23\linewidth}
        \centering
        \includegraphics[width=\linewidth]{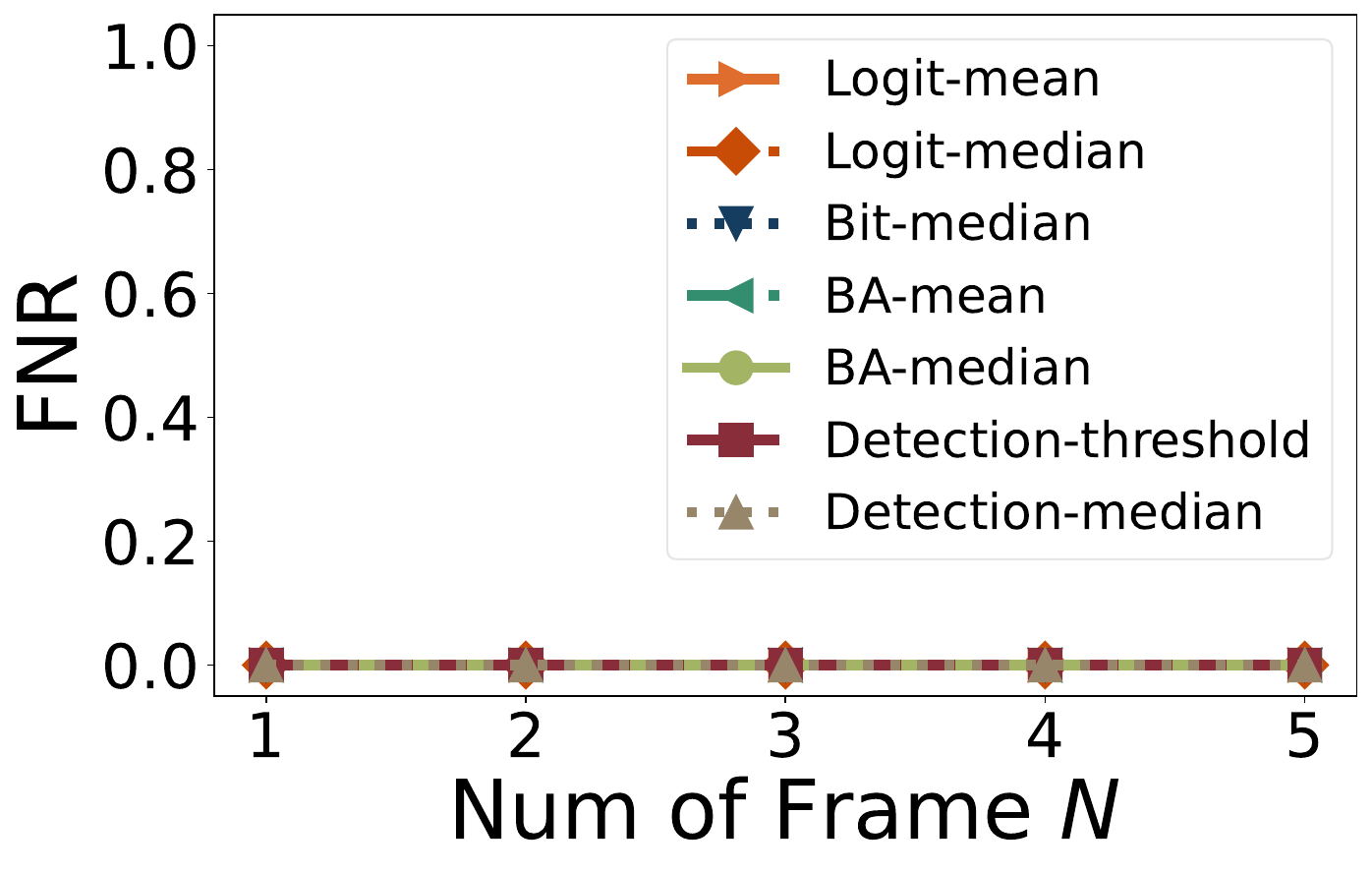}
        \caption{Frame Average}
    \end{subfigure}
    \begin{subfigure}{.23\linewidth}
        \centering
        \includegraphics[width=\linewidth]{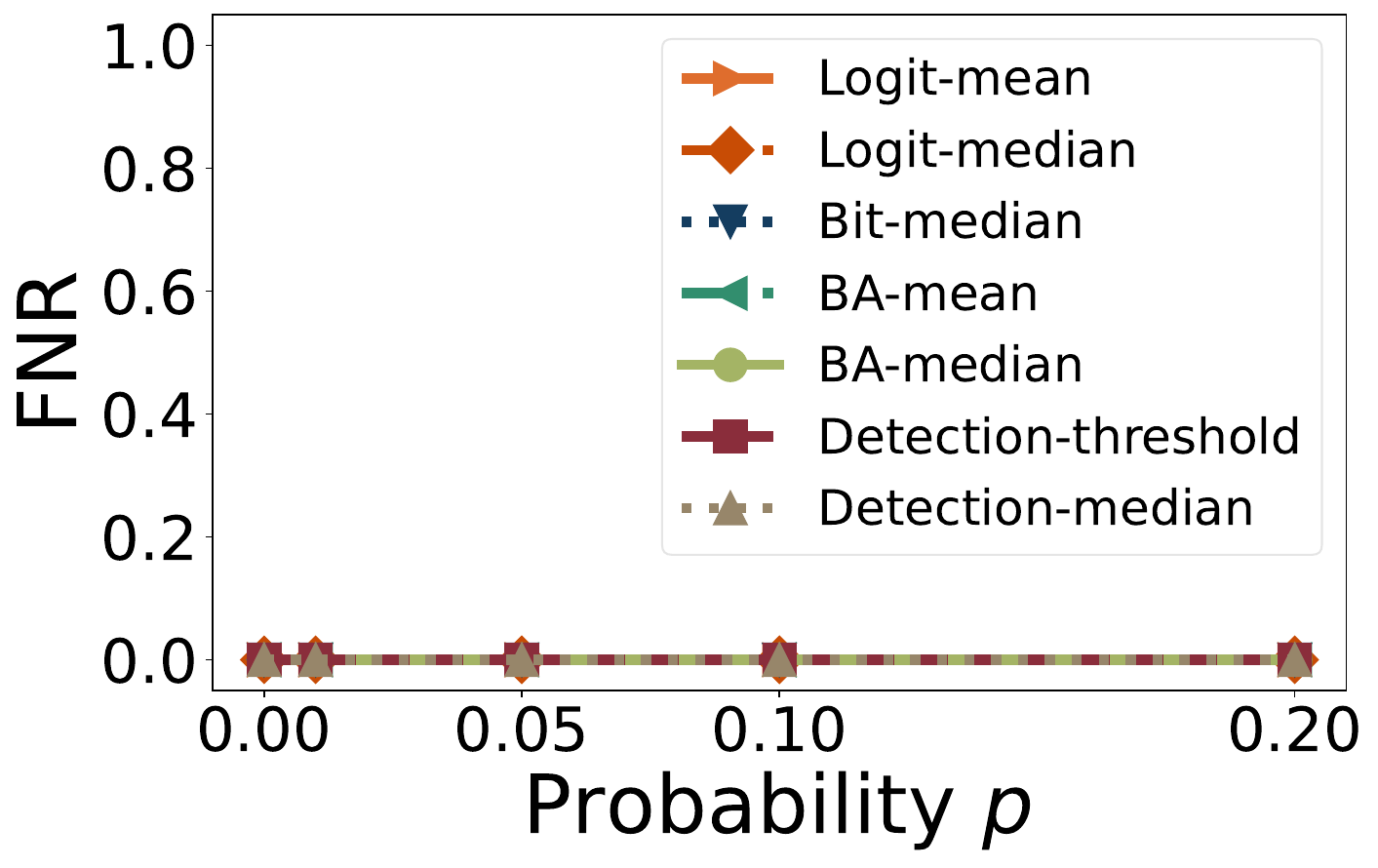}
        \caption{Frame Switch}
    \end{subfigure}
    \begin{subfigure}{.23\linewidth}
        \centering
        \includegraphics[width=\linewidth]{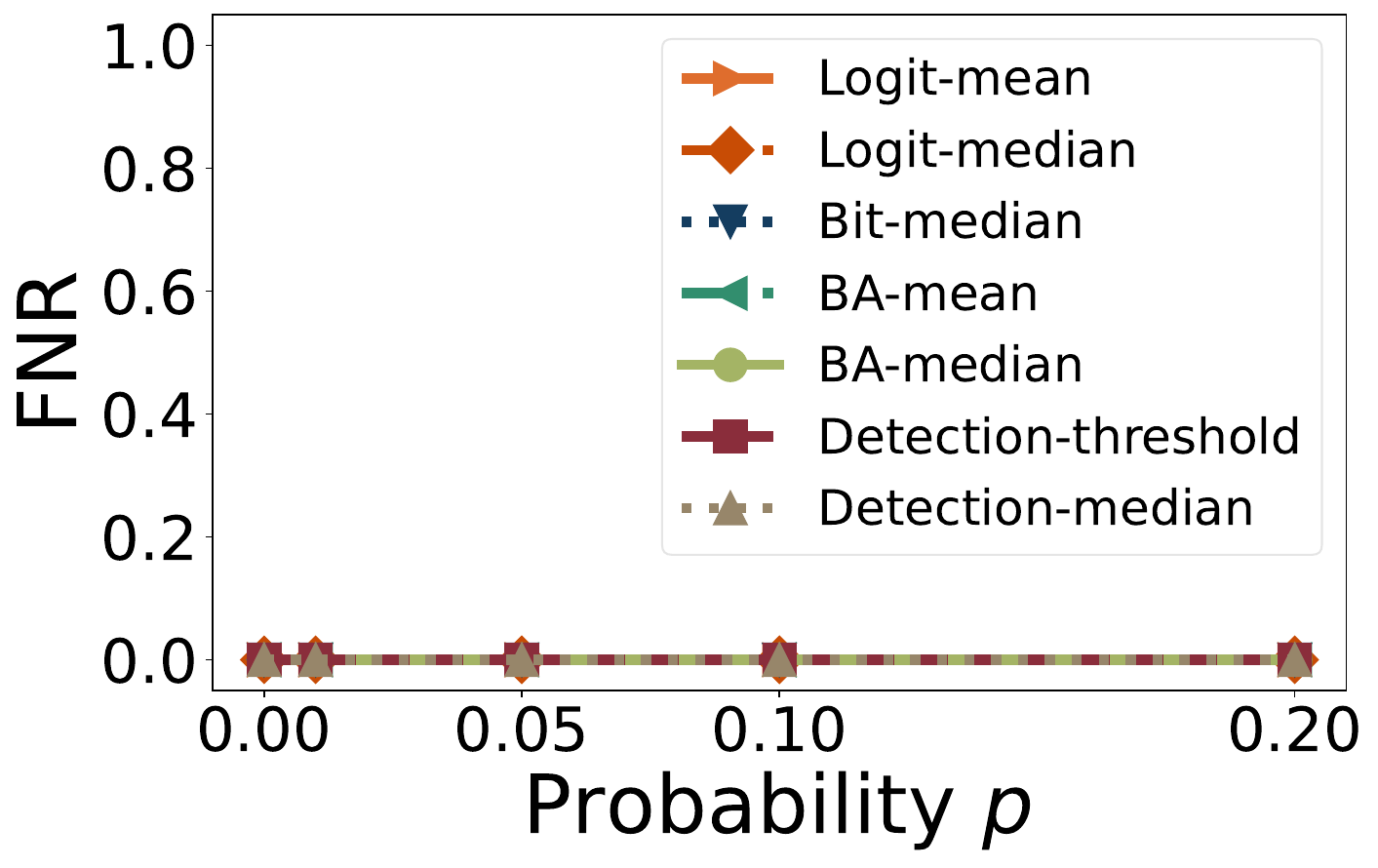}
        \caption{Frame Removal}
    \end{subfigure} \\

    \begin{subfigure}{.9\linewidth}
    \centering
    \caption*{Realistic video style}
    \end{subfigure}

    \begin{subfigure}{.23\linewidth}
        \centering
        \includegraphics[width=\linewidth]{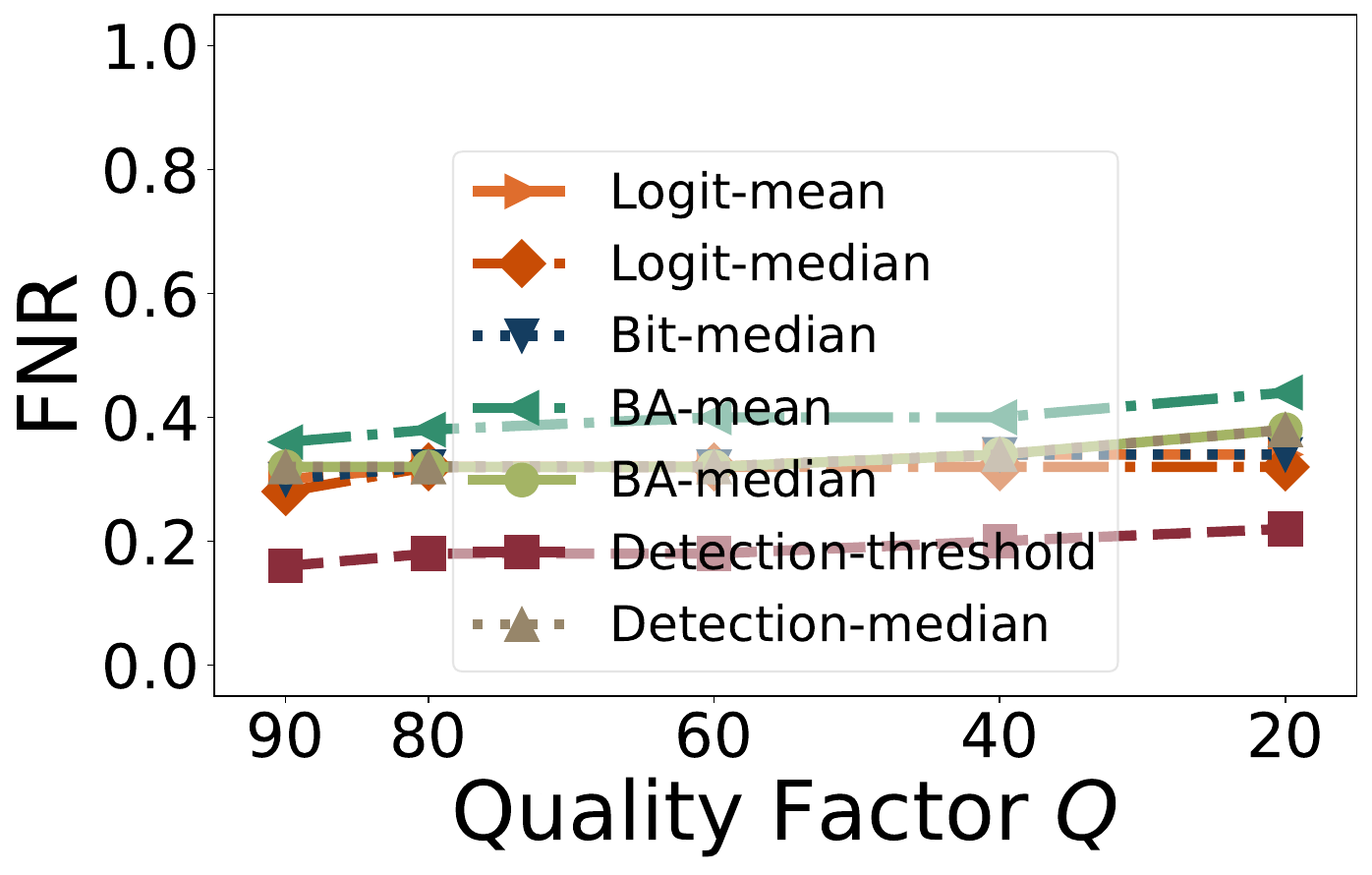}
        \caption{JPEG}
    \end{subfigure}
    \begin{subfigure}{.23\linewidth}
        \centering
        \includegraphics[width=\linewidth]{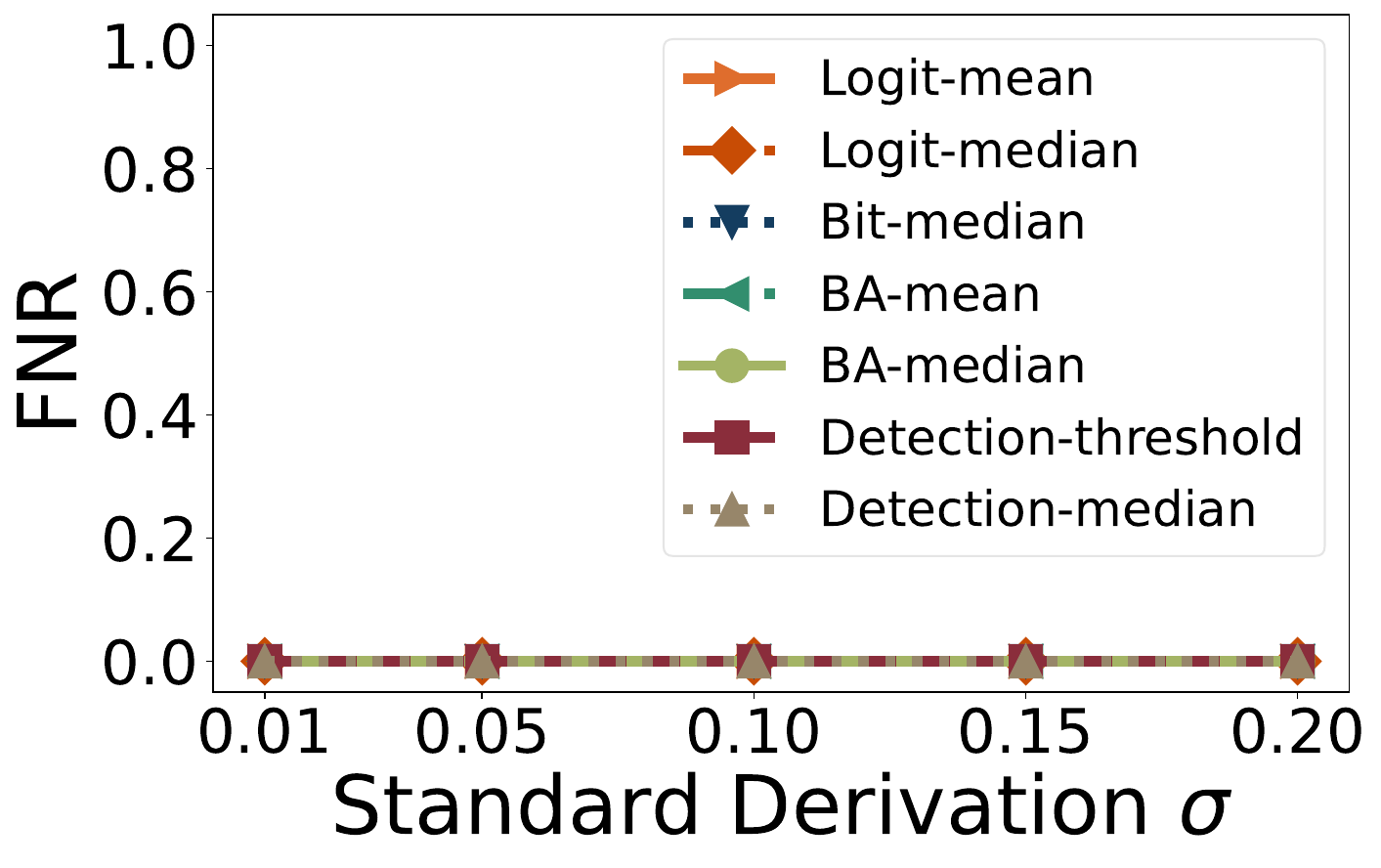}
        \caption{Gaussian Noise}
    \end{subfigure}
    \begin{subfigure}{.23\linewidth}
        \centering
        \includegraphics[width=\linewidth]{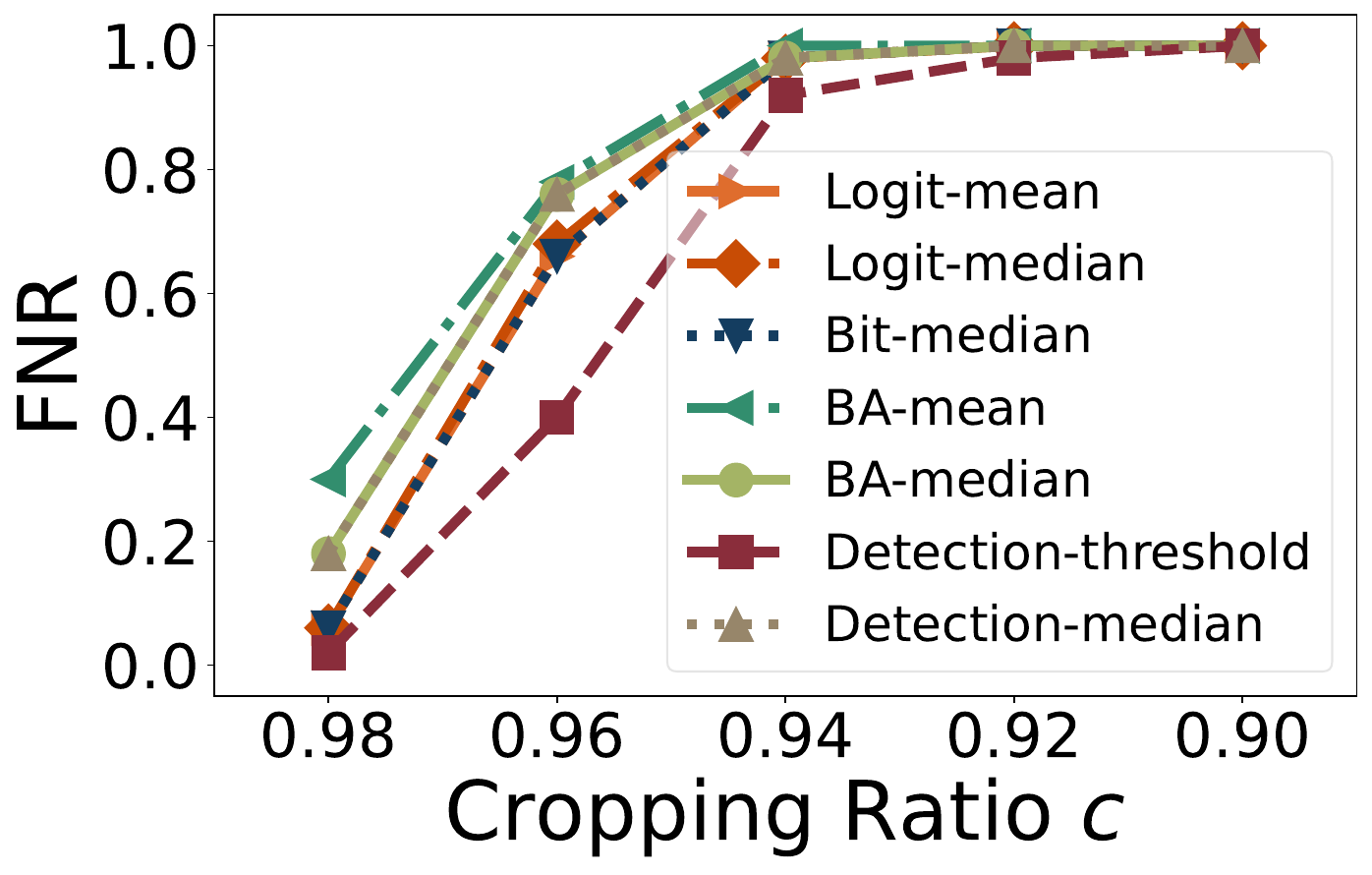}
        \caption{Cropping}
    \end{subfigure}
    \begin{subfigure}{.23\linewidth}
        \centering
        \includegraphics[width=\linewidth]{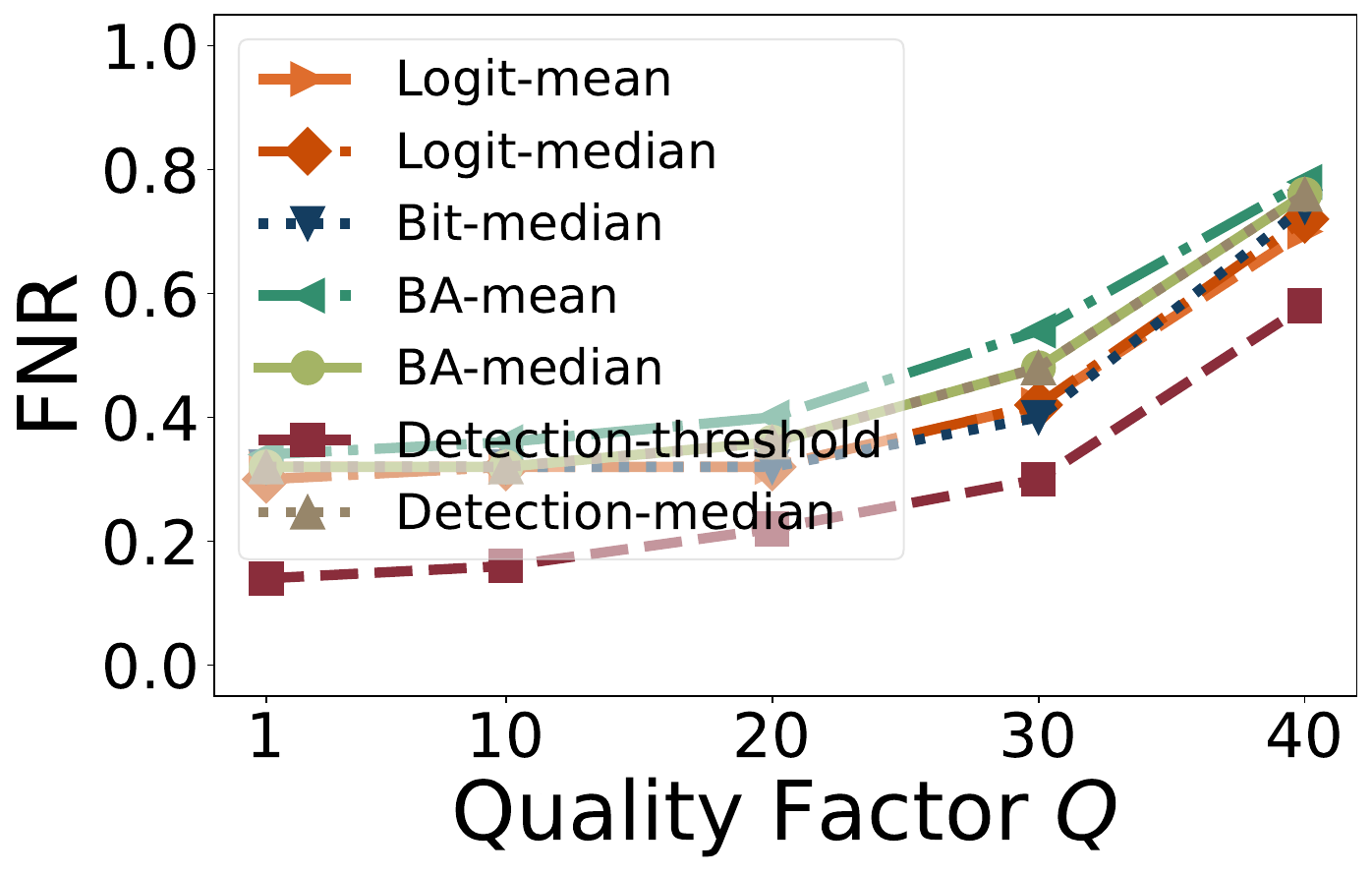}
        \caption{MPEG-4}
    \end{subfigure} \\
    
    \begin{subfigure}{.23\linewidth}
        \centering
        \includegraphics[width=\linewidth]{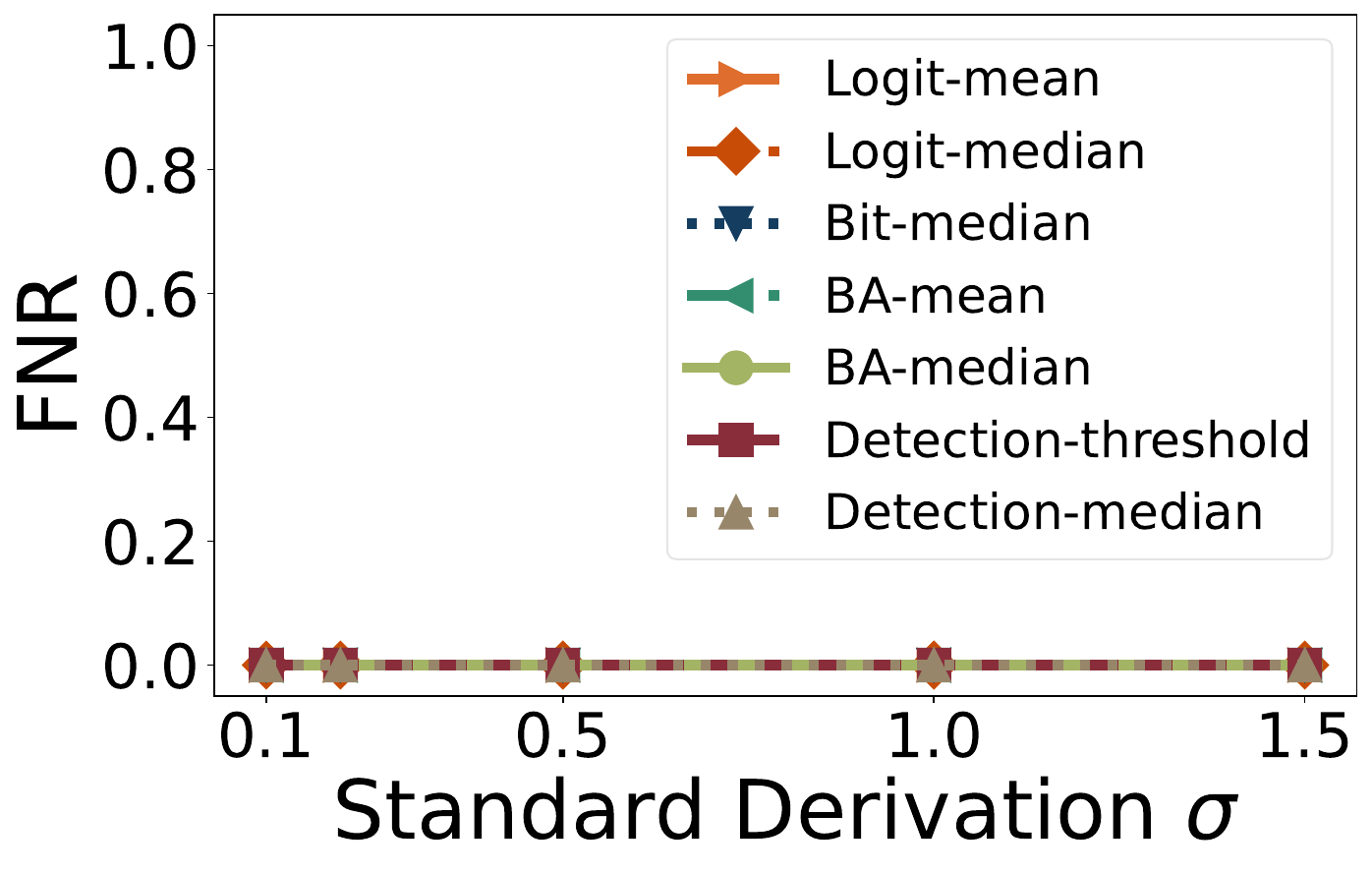}
        \caption{Gaussian Blur}
    \end{subfigure}
    \begin{subfigure}{.23\linewidth}
        \centering
        \includegraphics[width=\linewidth]{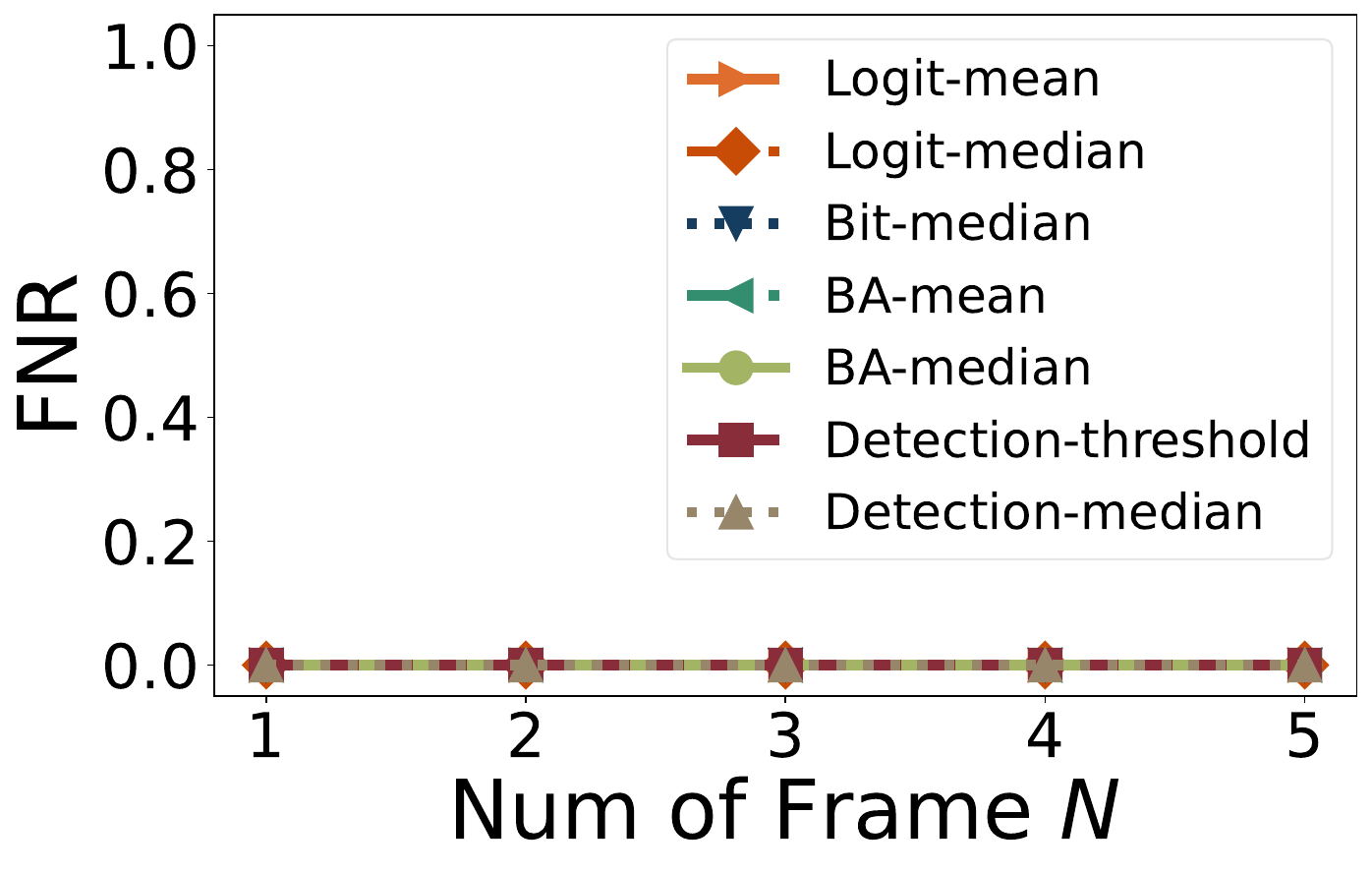}
        \caption{Frame Average}
    \end{subfigure}
    \begin{subfigure}{.23\linewidth}
        \centering
        \includegraphics[width=\linewidth]{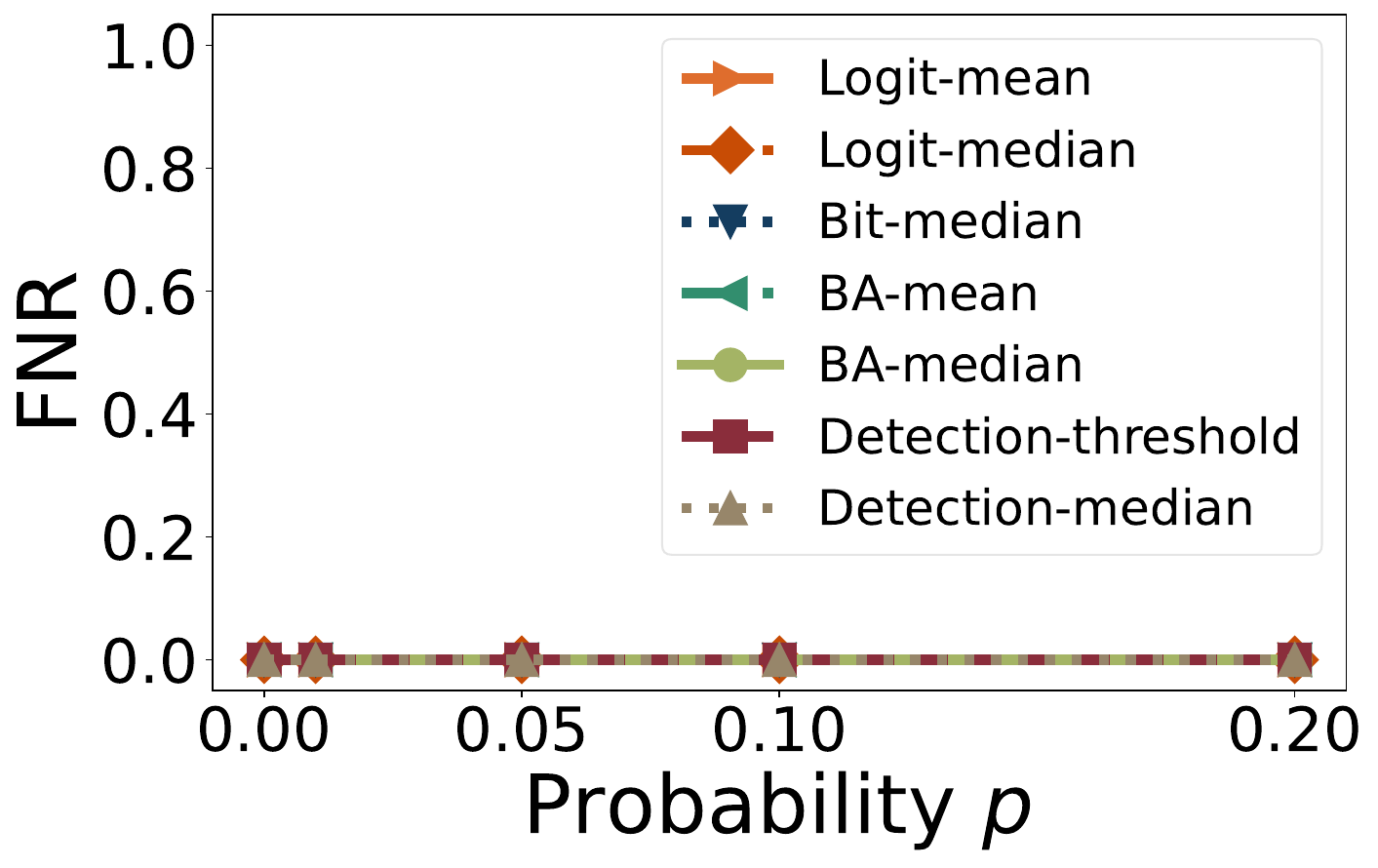}
        \caption{Frame Switch}
    \end{subfigure}
    \begin{subfigure}{.23\linewidth}
        \centering
        \includegraphics[width=\linewidth]{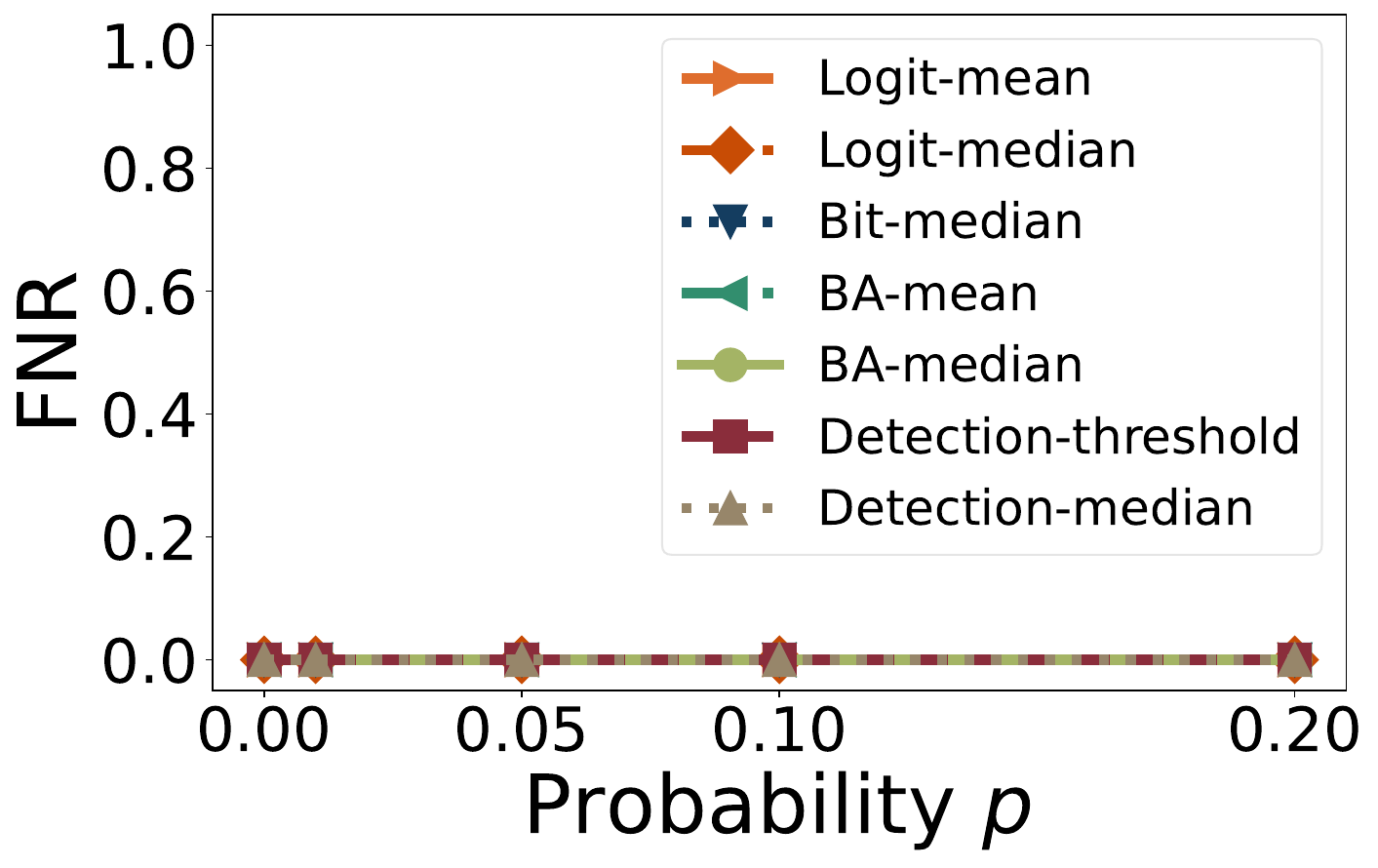}
        \caption{Frame Removal}
    \end{subfigure} \\

    \begin{subfigure}{.9\linewidth}
    \centering
    \caption*{Cartoon video style}
    \end{subfigure}

    \begin{subfigure}{.23\linewidth}
        \centering
        \includegraphics[width=\linewidth]{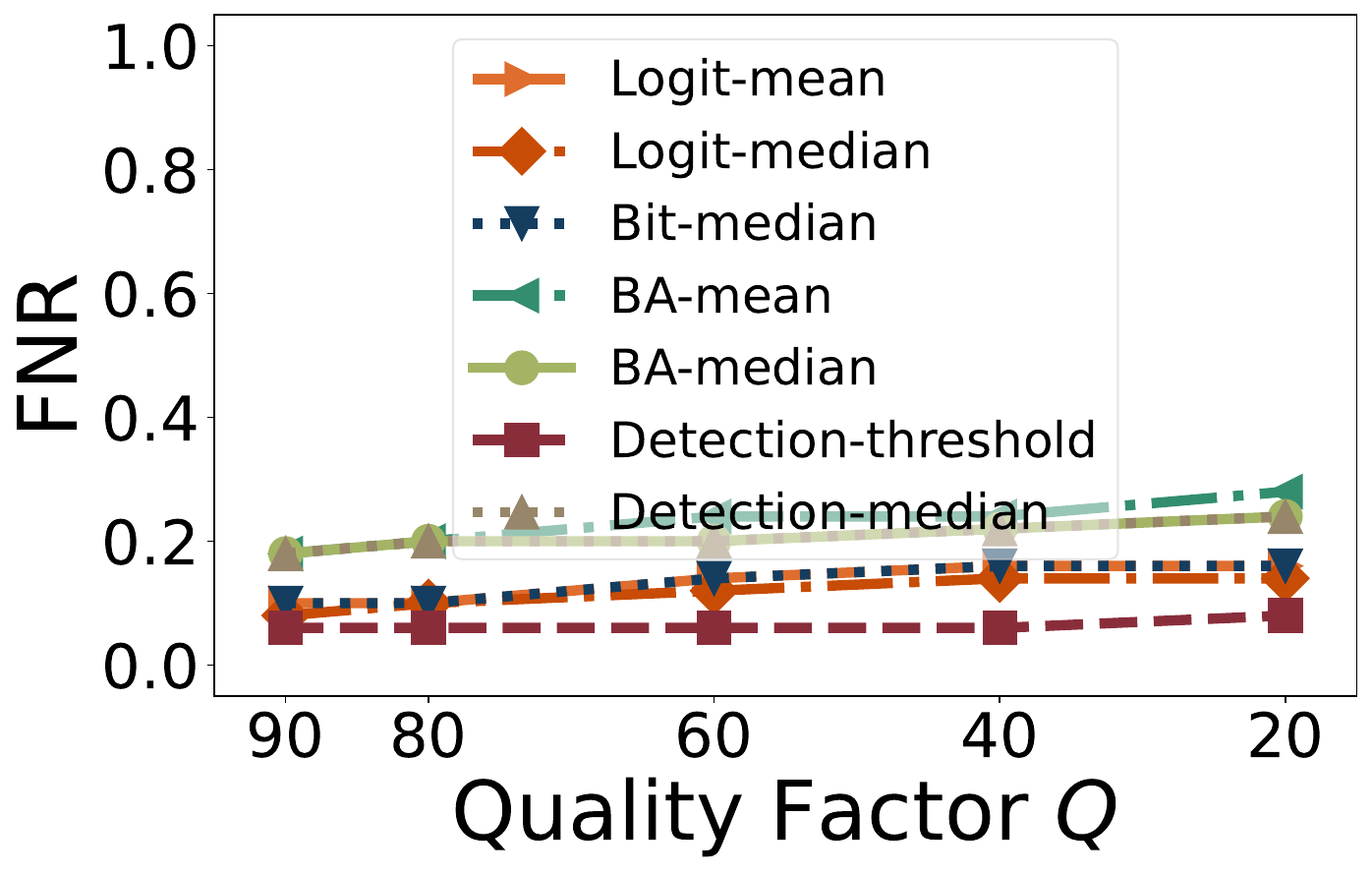}
        \caption{JPEG}
    \end{subfigure}
    \begin{subfigure}{.23\linewidth}
        \centering
        \includegraphics[width=\linewidth]{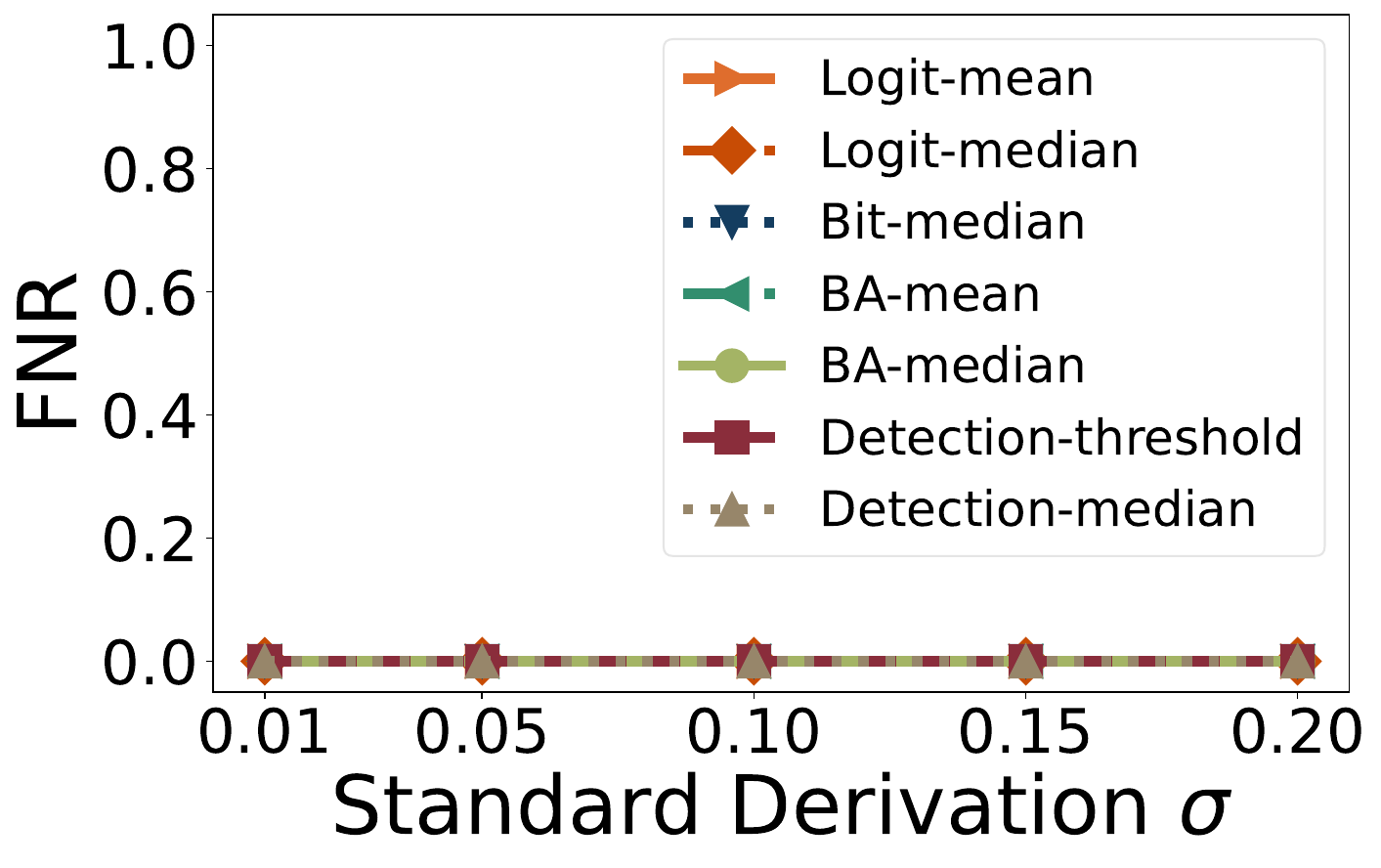}
        \caption{Gaussian Noise}
    \end{subfigure}
    \begin{subfigure}{.23\linewidth}
        \centering
        \includegraphics[width=\linewidth]{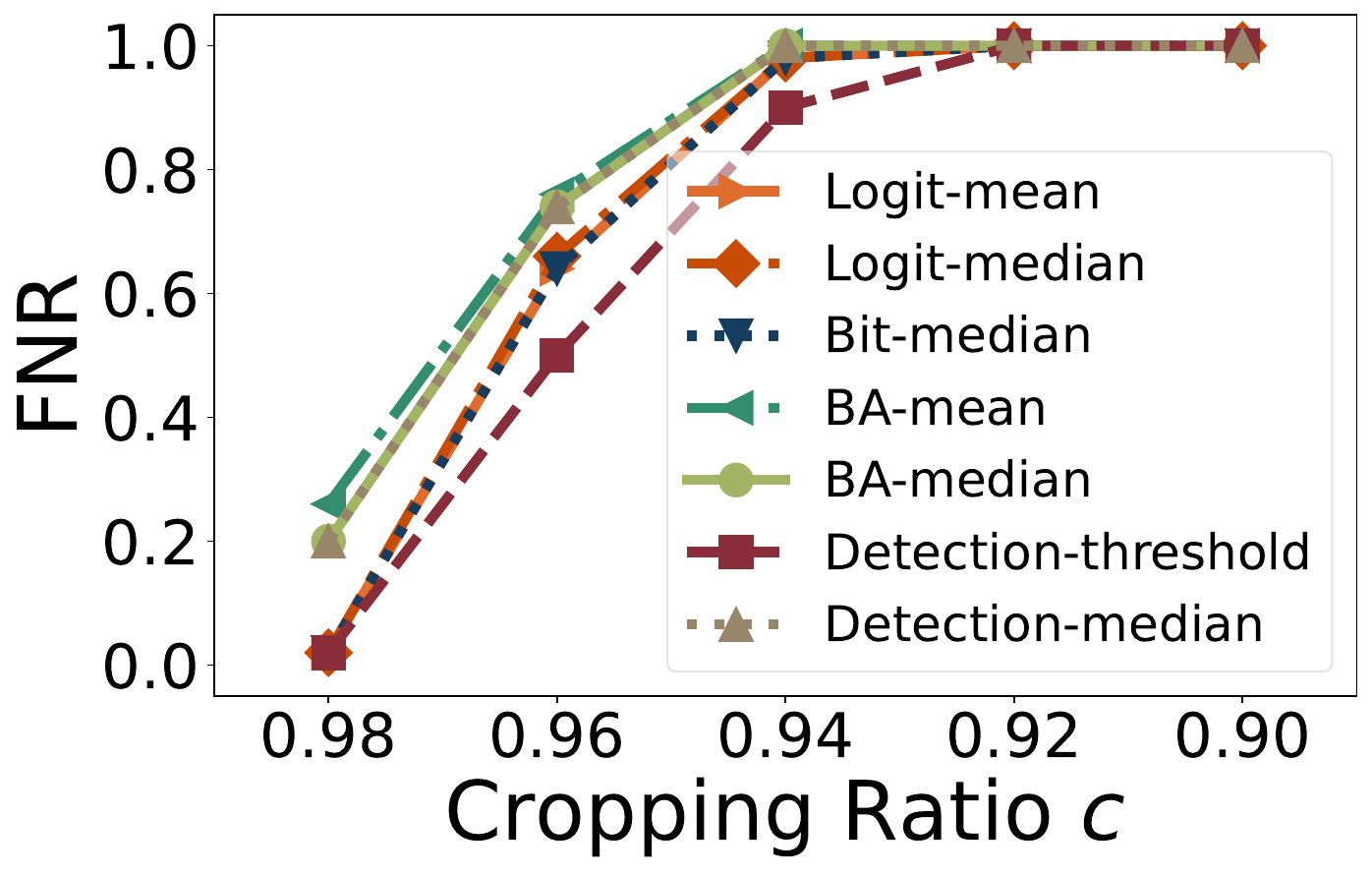}
        \caption{Cropping}
    \end{subfigure}
    \begin{subfigure}{.23\linewidth}
        \centering
        \includegraphics[width=\linewidth]{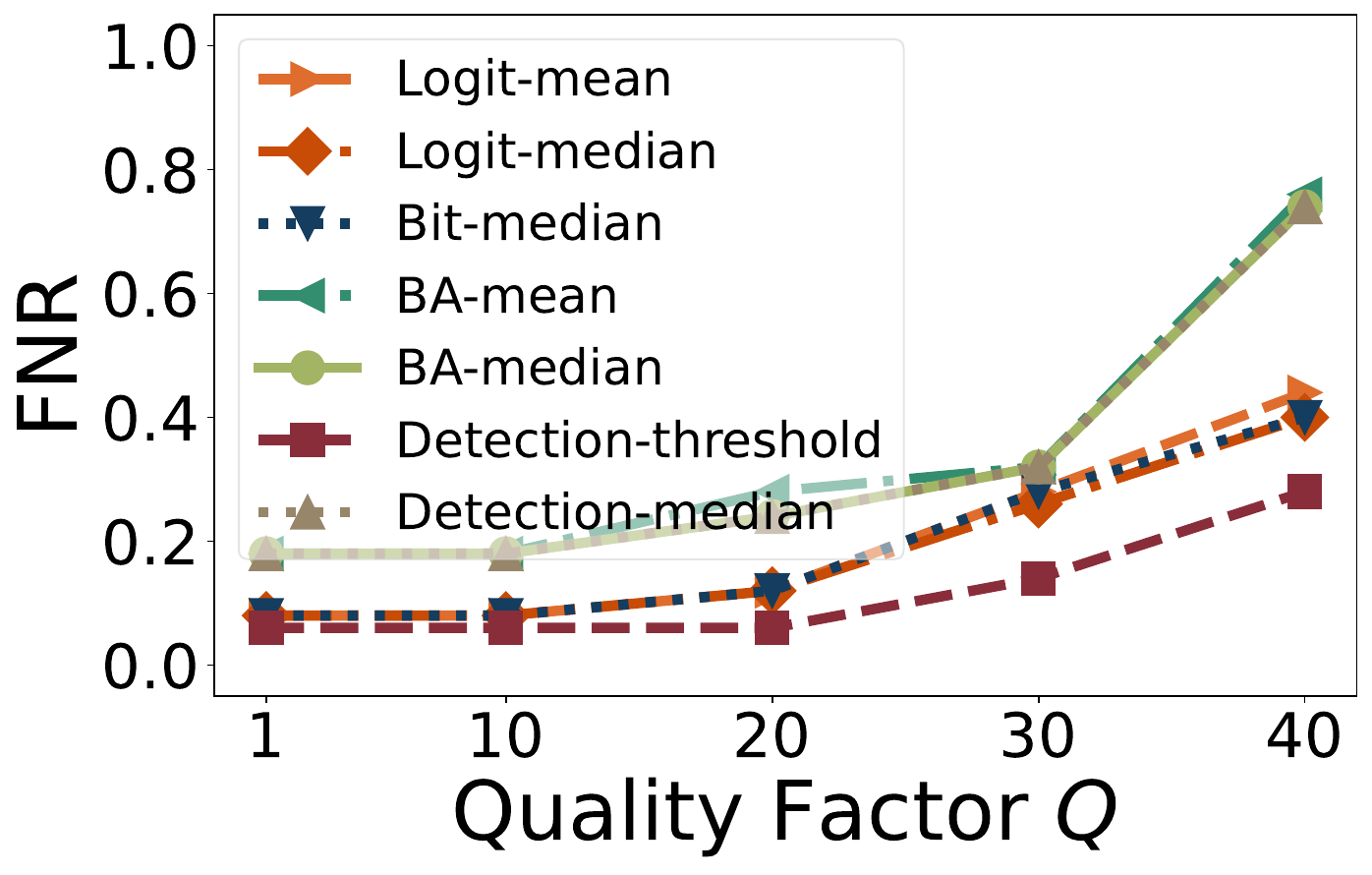}
        \caption{MPEG-4}
    \end{subfigure} \\
    
    \begin{subfigure}{.23\linewidth}
        \centering
        \includegraphics[width=\linewidth]{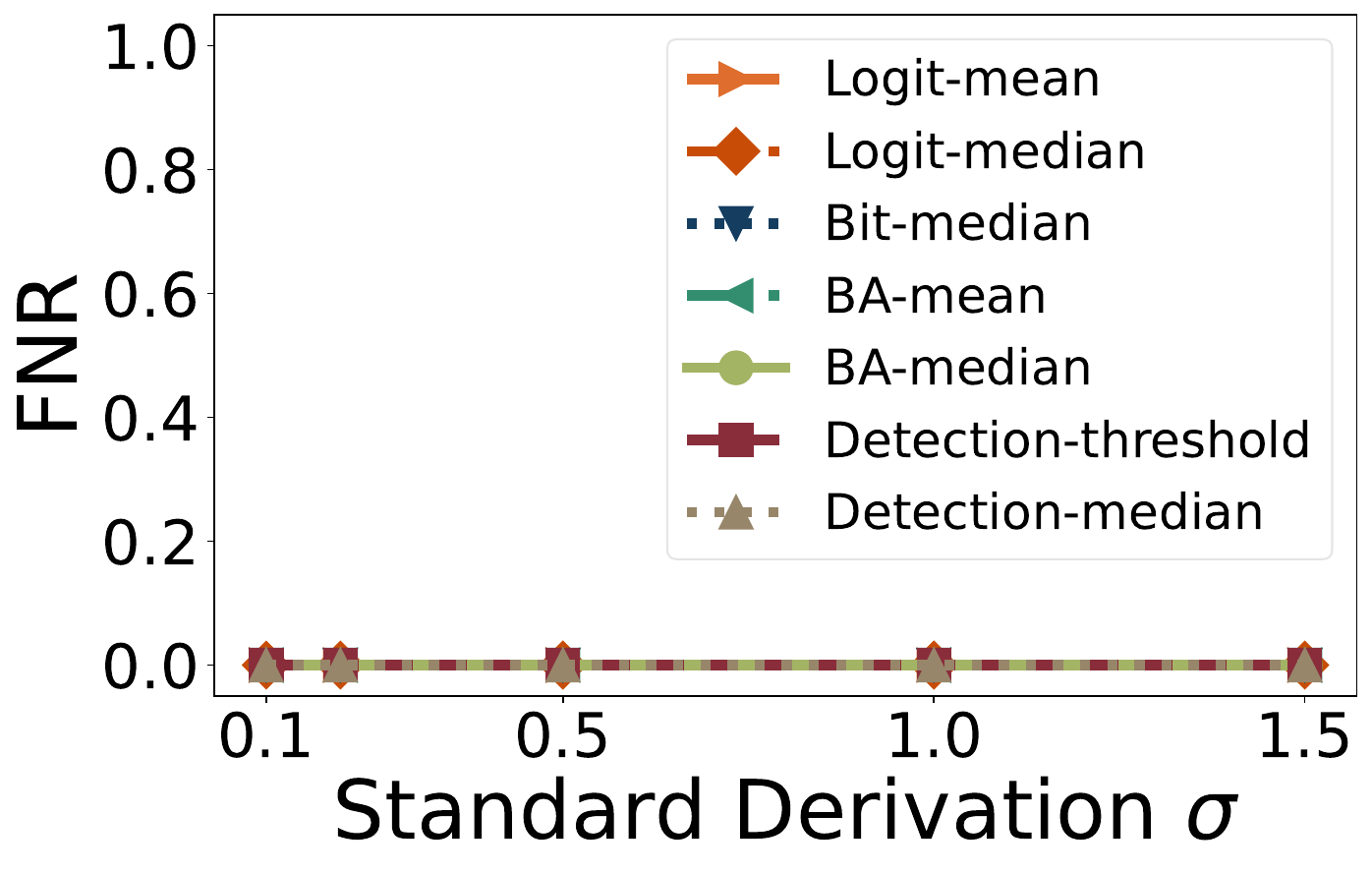}
        \caption{Gaussian Blur}
    \end{subfigure}
    \begin{subfigure}{.23\linewidth}
        \centering
        \includegraphics[width=\linewidth]{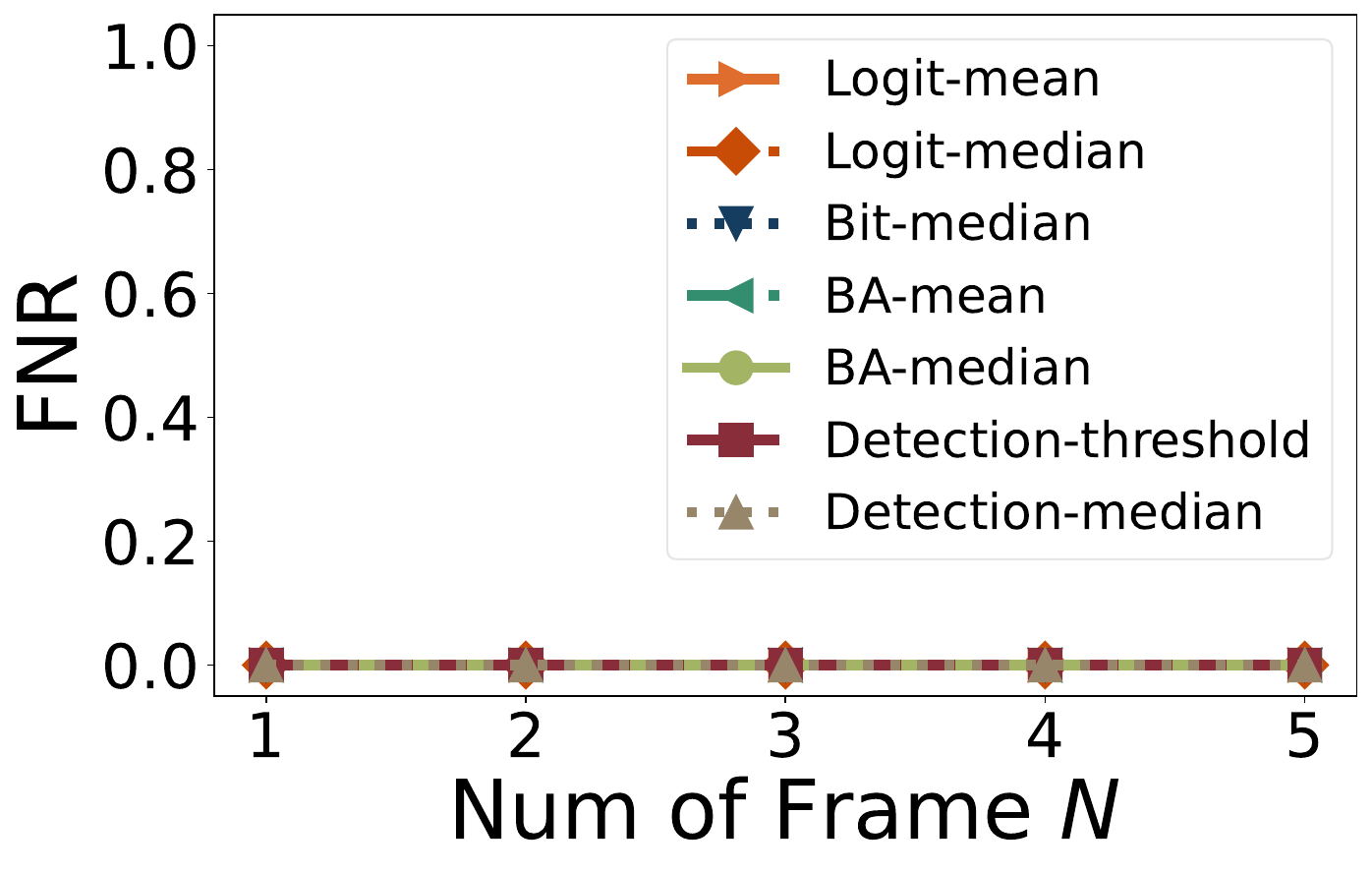}
        \caption{Frame Average}
    \end{subfigure}
    \begin{subfigure}{.23\linewidth}
        \centering
        \includegraphics[width=\linewidth]{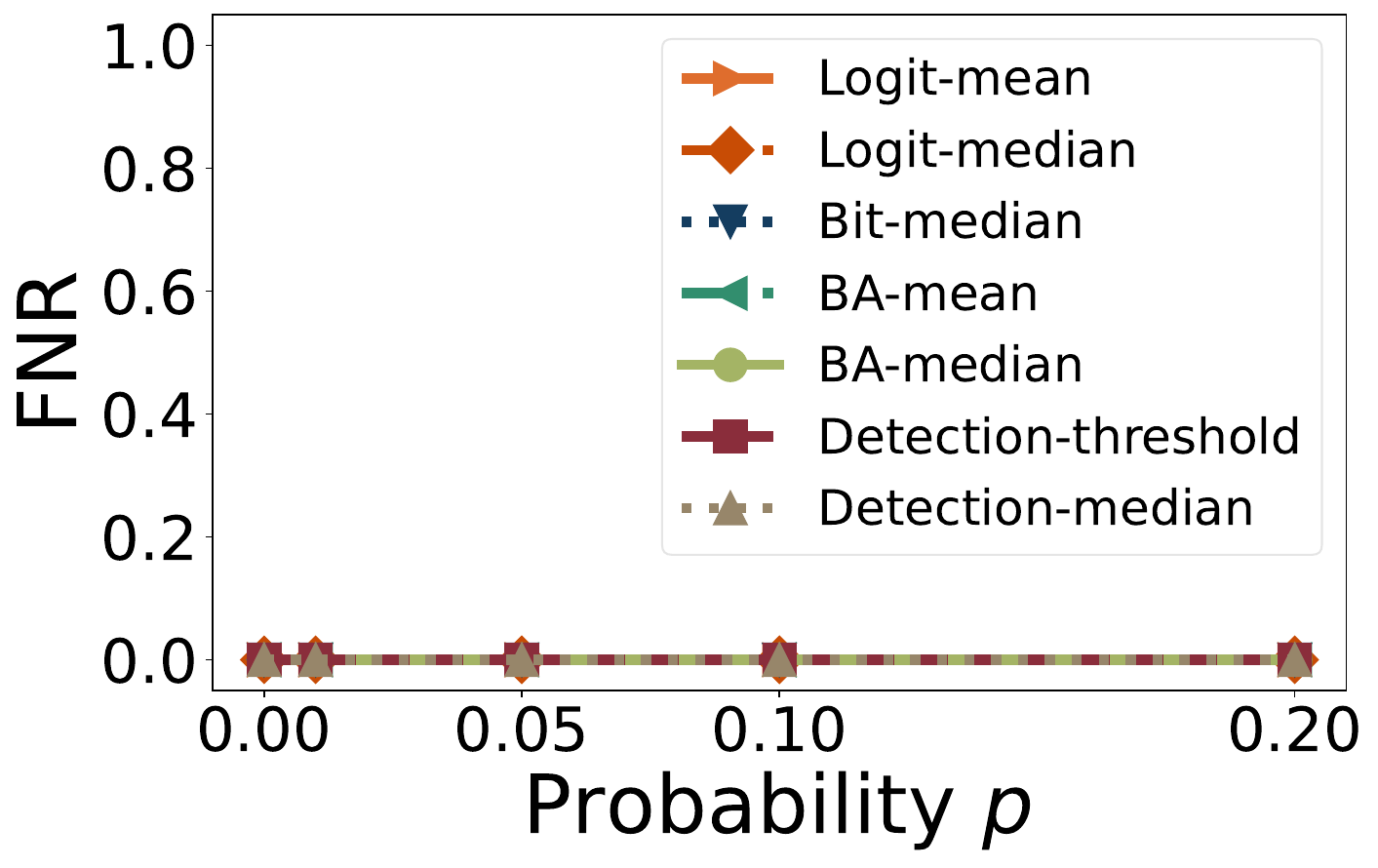}
        \caption{Frame Switch}
    \end{subfigure}
    \begin{subfigure}{.23\linewidth}
        \centering
        \includegraphics[width=\linewidth]{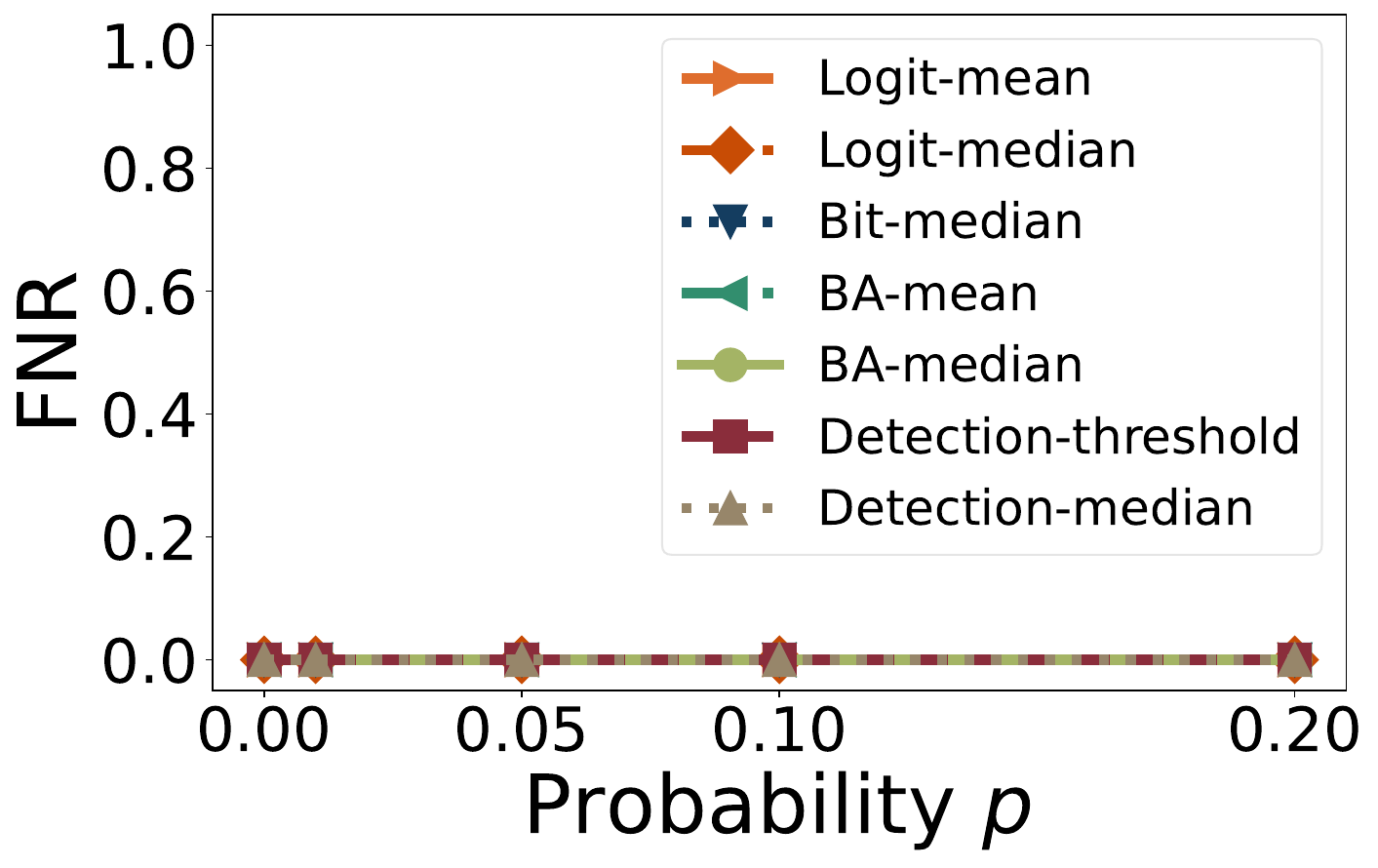}
        \caption{Frame Removal}
    \end{subfigure} \\

    \begin{subfigure}{.9\linewidth}
    \centering
    \caption*{Sci-fi video style}
    \end{subfigure}
    
    \caption{More fine-grained watermark removal results for StegaStamp on videos generated by Sora.}
\end{figure}

\begin{figure}[]
    \centering

    \begin{subfigure}{.23\linewidth}
        \centering
        \includegraphics[width=\linewidth]{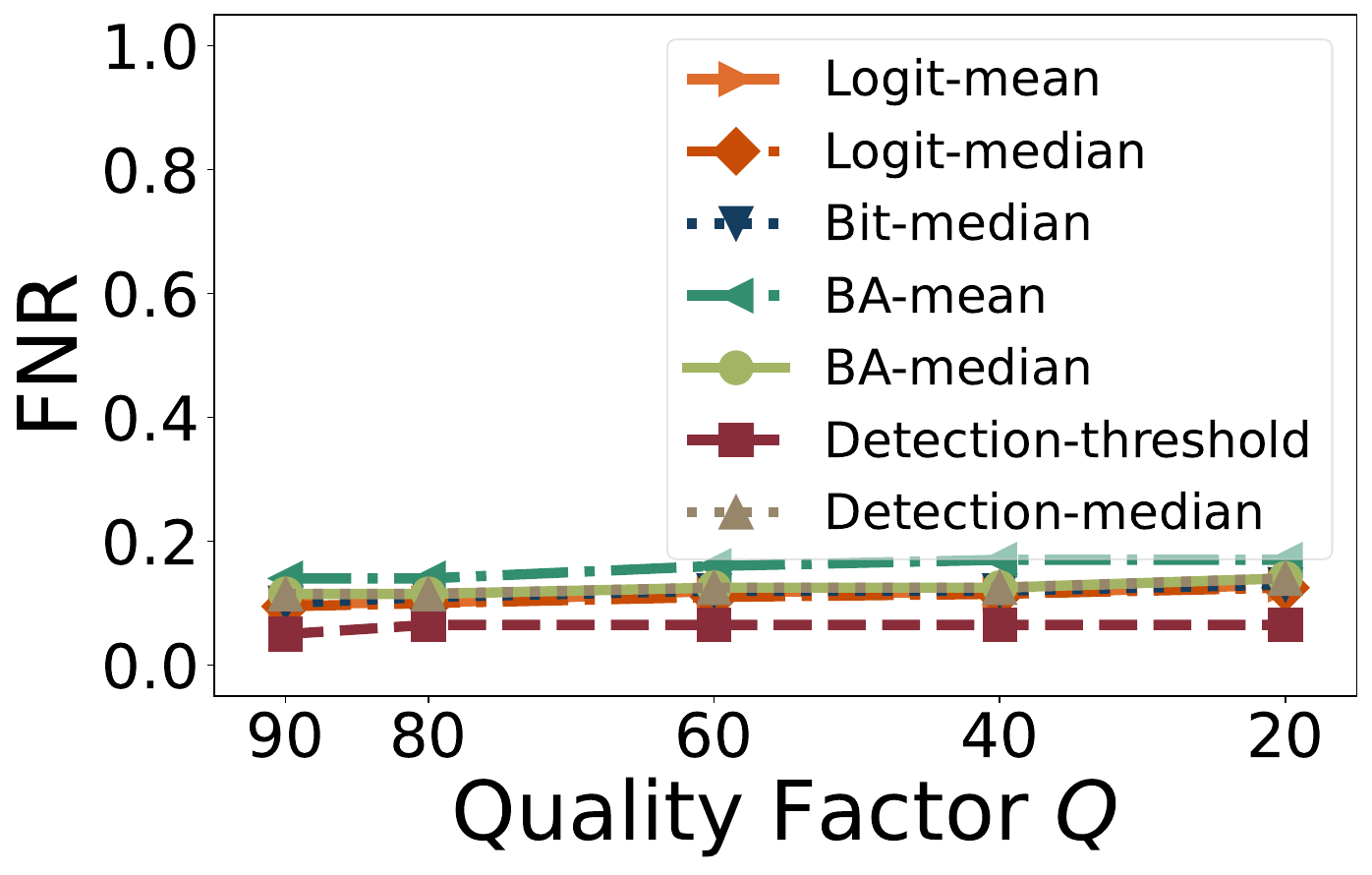}
        \caption{JPEG}
    \end{subfigure}
    \begin{subfigure}{.23\linewidth}
        \centering
        \includegraphics[width=\linewidth]{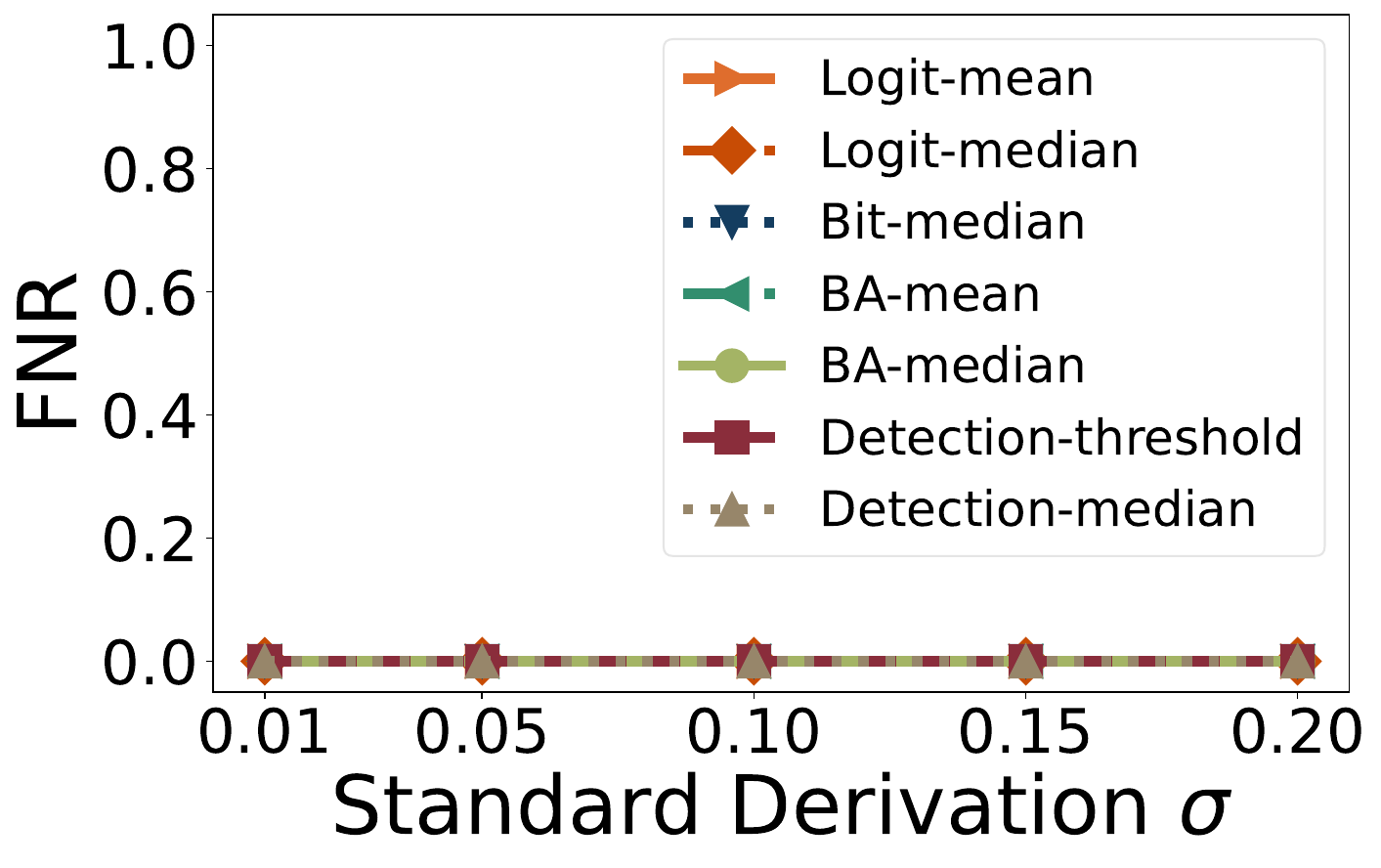}
        \caption{Gaussian Noise}
    \end{subfigure}
    \begin{subfigure}{.23\linewidth}
        \centering
        \includegraphics[width=\linewidth]{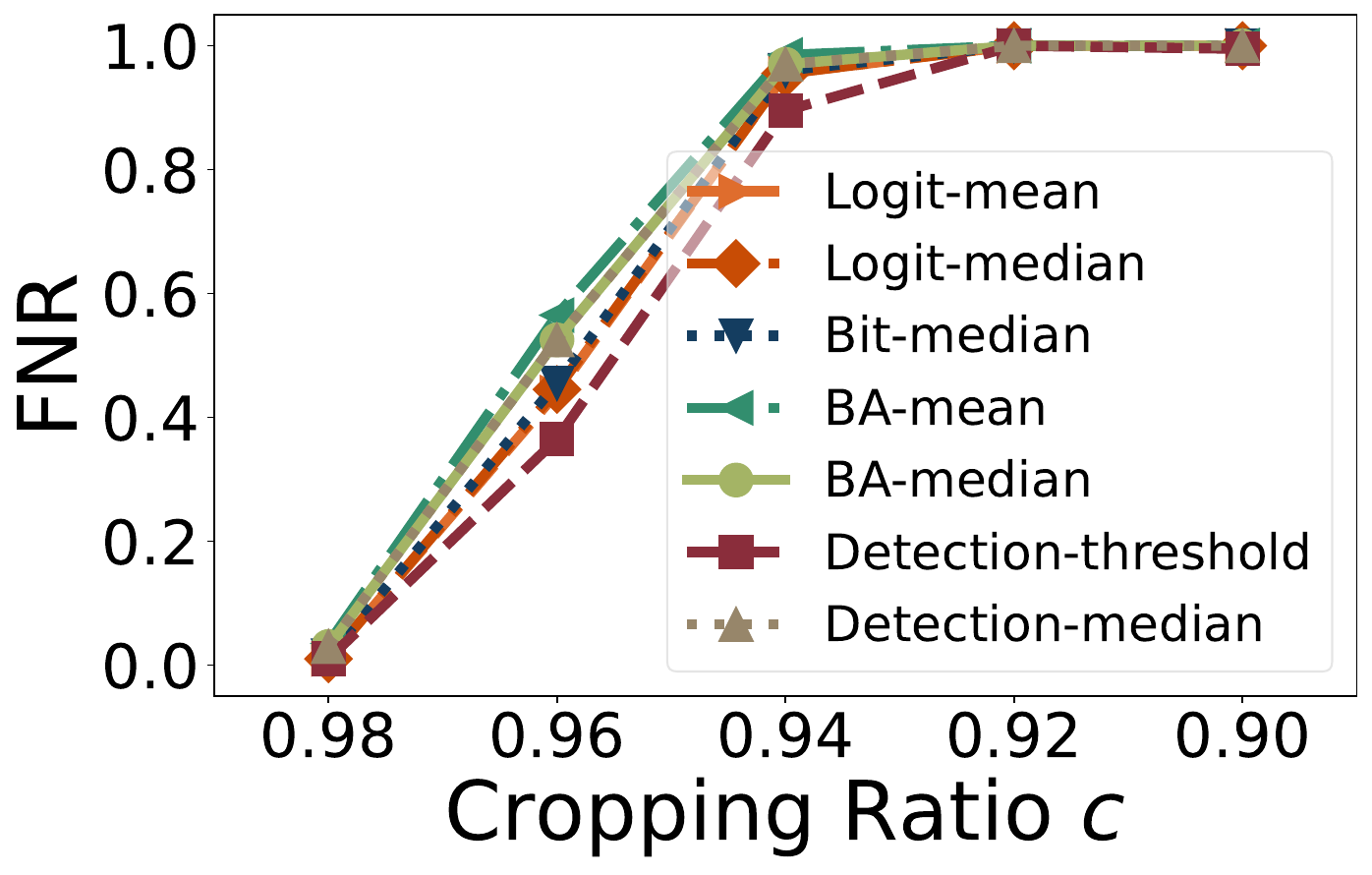}
        \caption{Cropping}
    \end{subfigure}
    \begin{subfigure}{.23\linewidth}
        \centering
        \includegraphics[width=\linewidth]{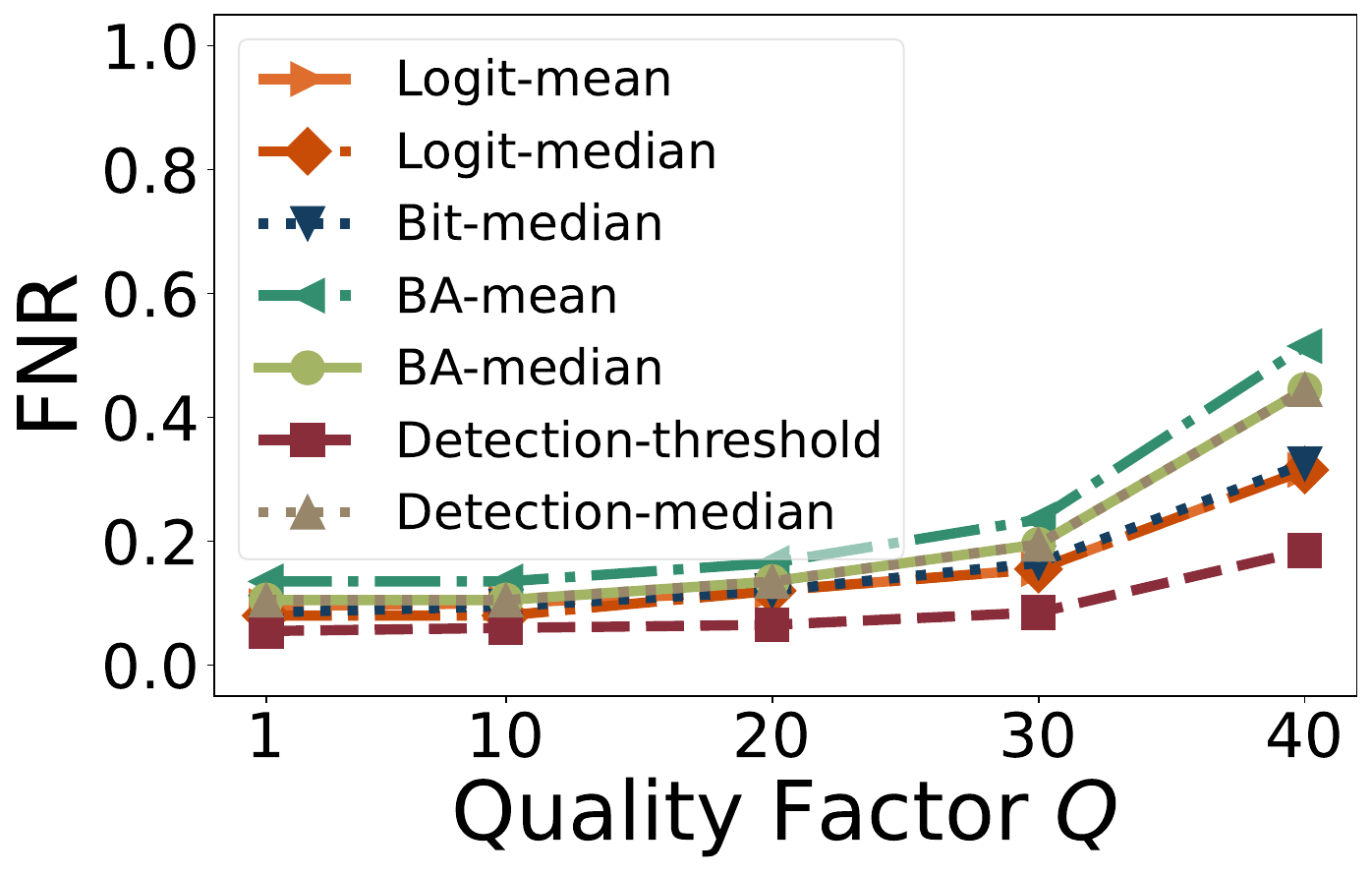}
        \caption{MPEG-4}
    \end{subfigure} \\
    
    \begin{subfigure}{.23\linewidth}
        \centering
        \includegraphics[width=\linewidth]{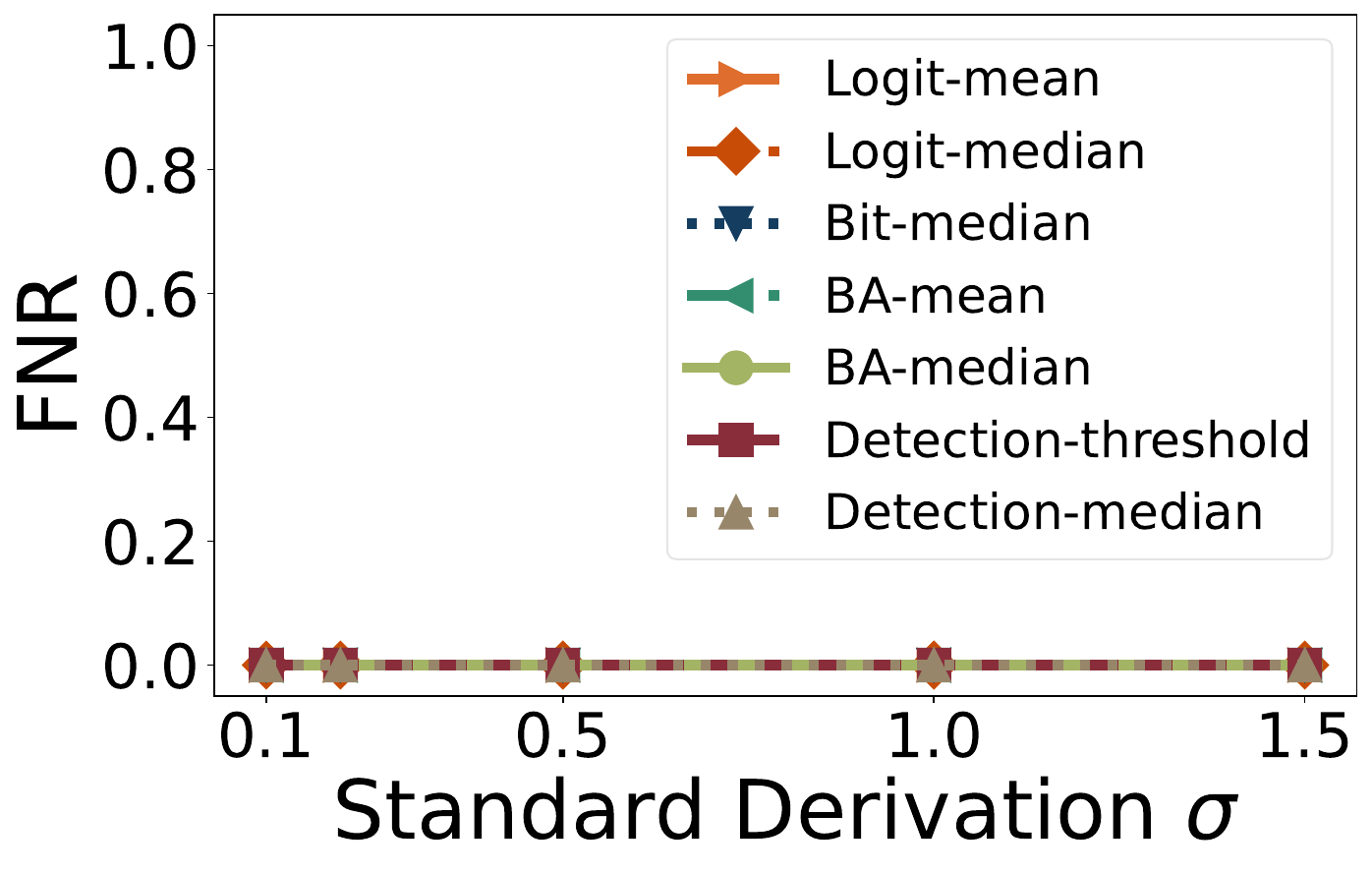}
        \caption{Gaussian Blur}
    \end{subfigure}
    \begin{subfigure}{.23\linewidth}
        \centering
        \includegraphics[width=\linewidth]{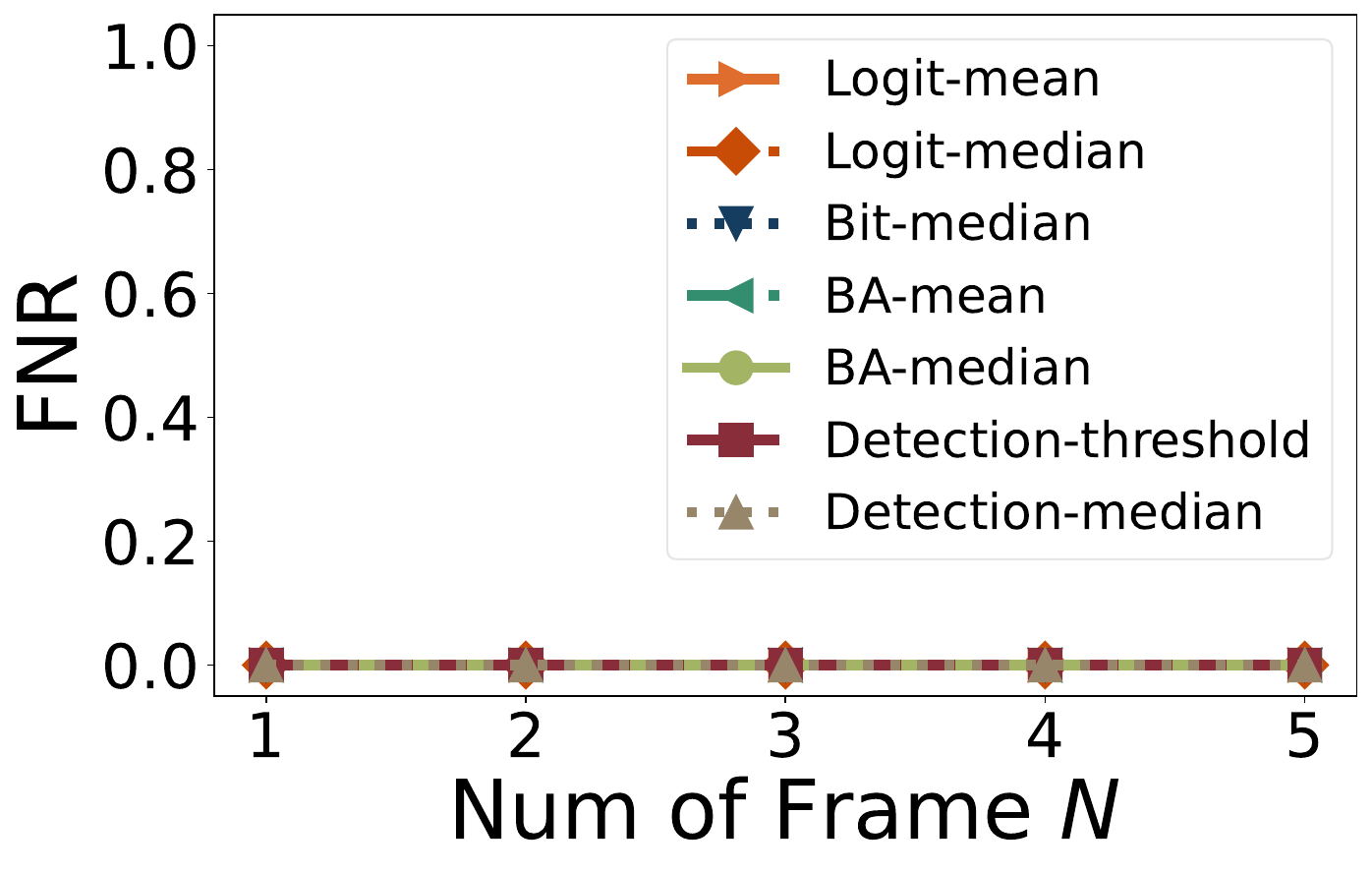}
        \caption{Frame Average}
    \end{subfigure}
    \begin{subfigure}{.23\linewidth}
        \centering
        \includegraphics[width=\linewidth]{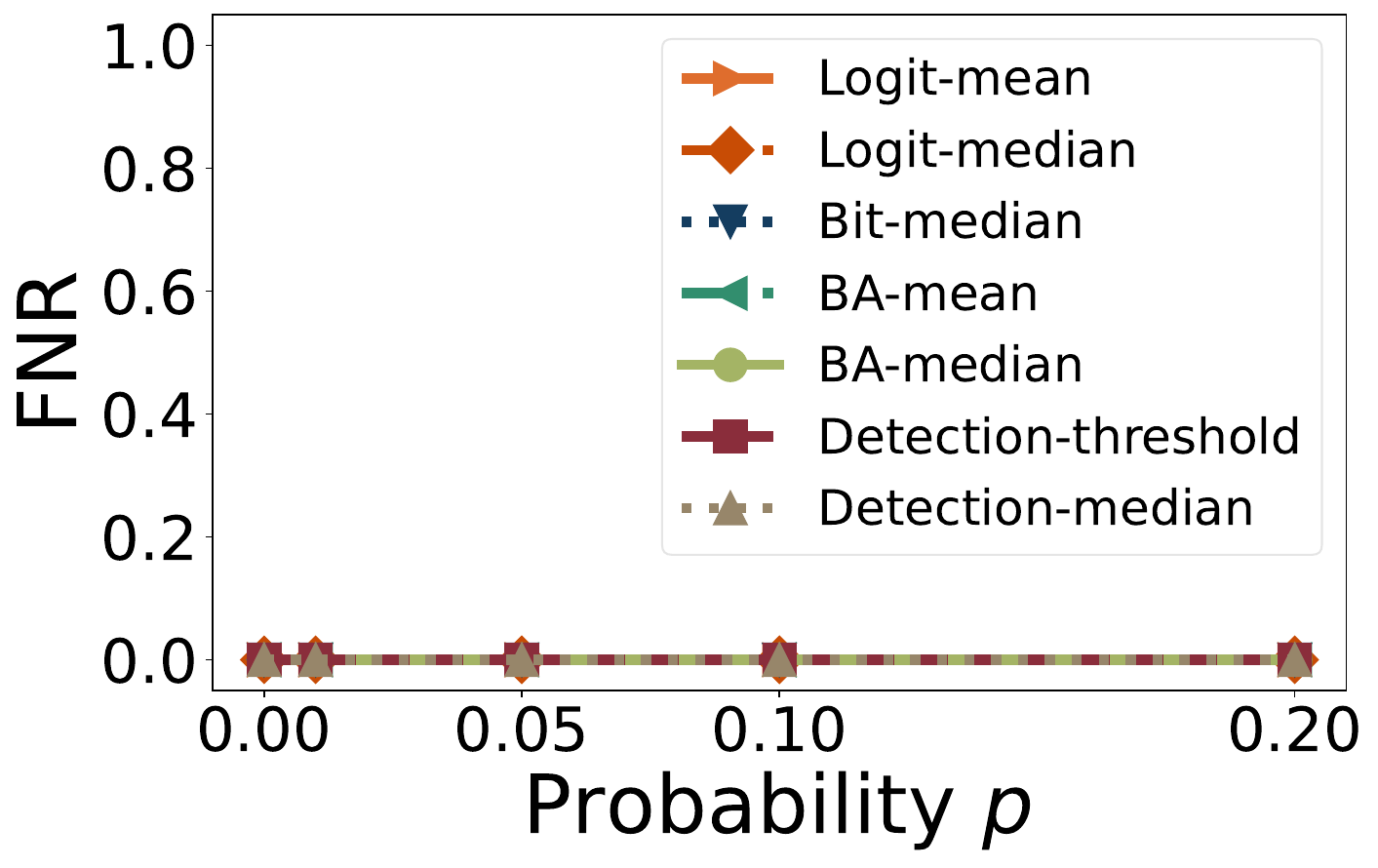}
        \caption{Frame Switch}
    \end{subfigure}
    \begin{subfigure}{.23\linewidth}
        \centering
        \includegraphics[width=\linewidth]{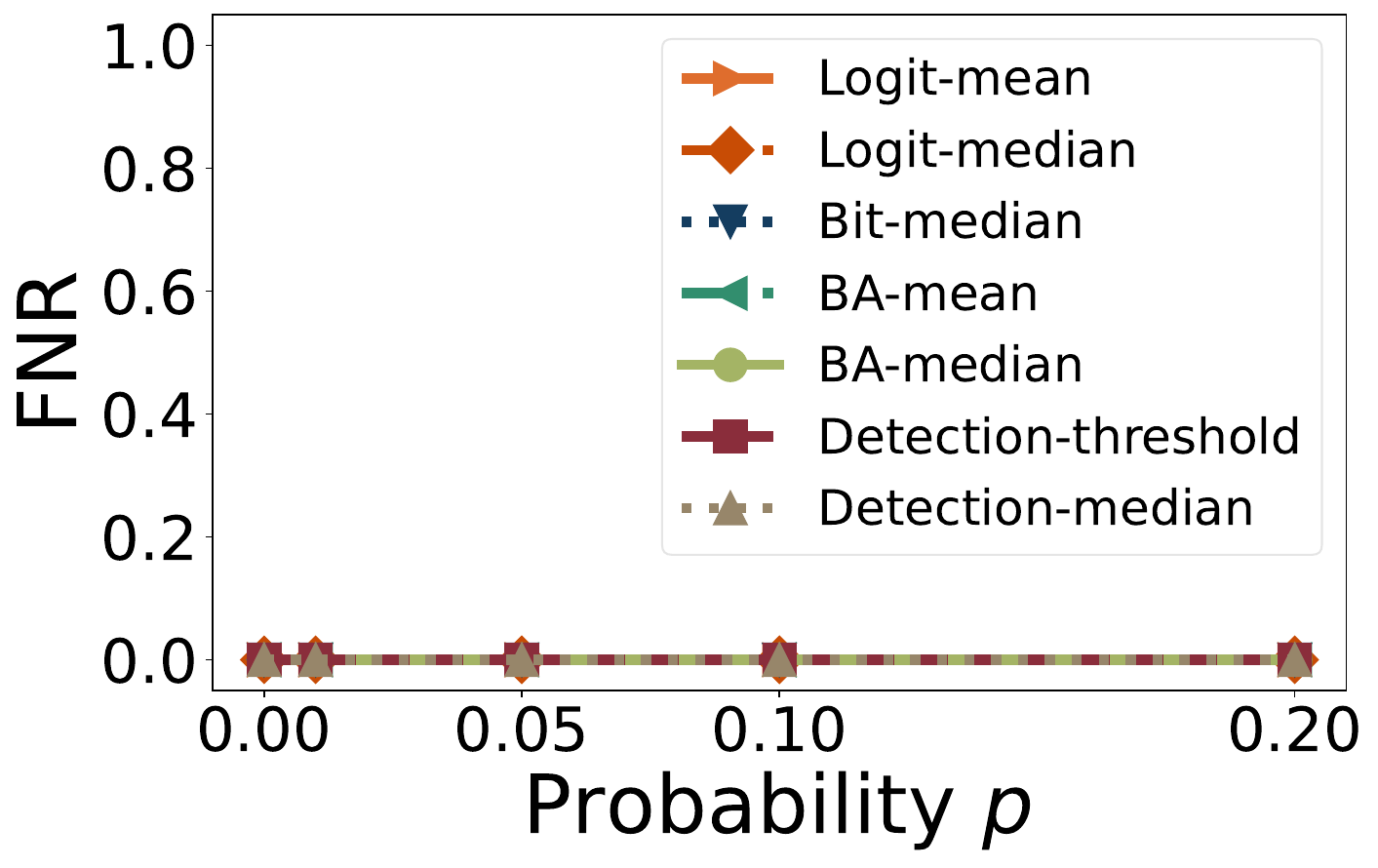}
        \caption{Frame Removal}
    \end{subfigure} \\

    \begin{subfigure}{.9\linewidth}
    \centering
    \caption*{Realistic video style}
    \end{subfigure}

    \begin{subfigure}{.23\linewidth}
        \centering
        \includegraphics[width=\linewidth]{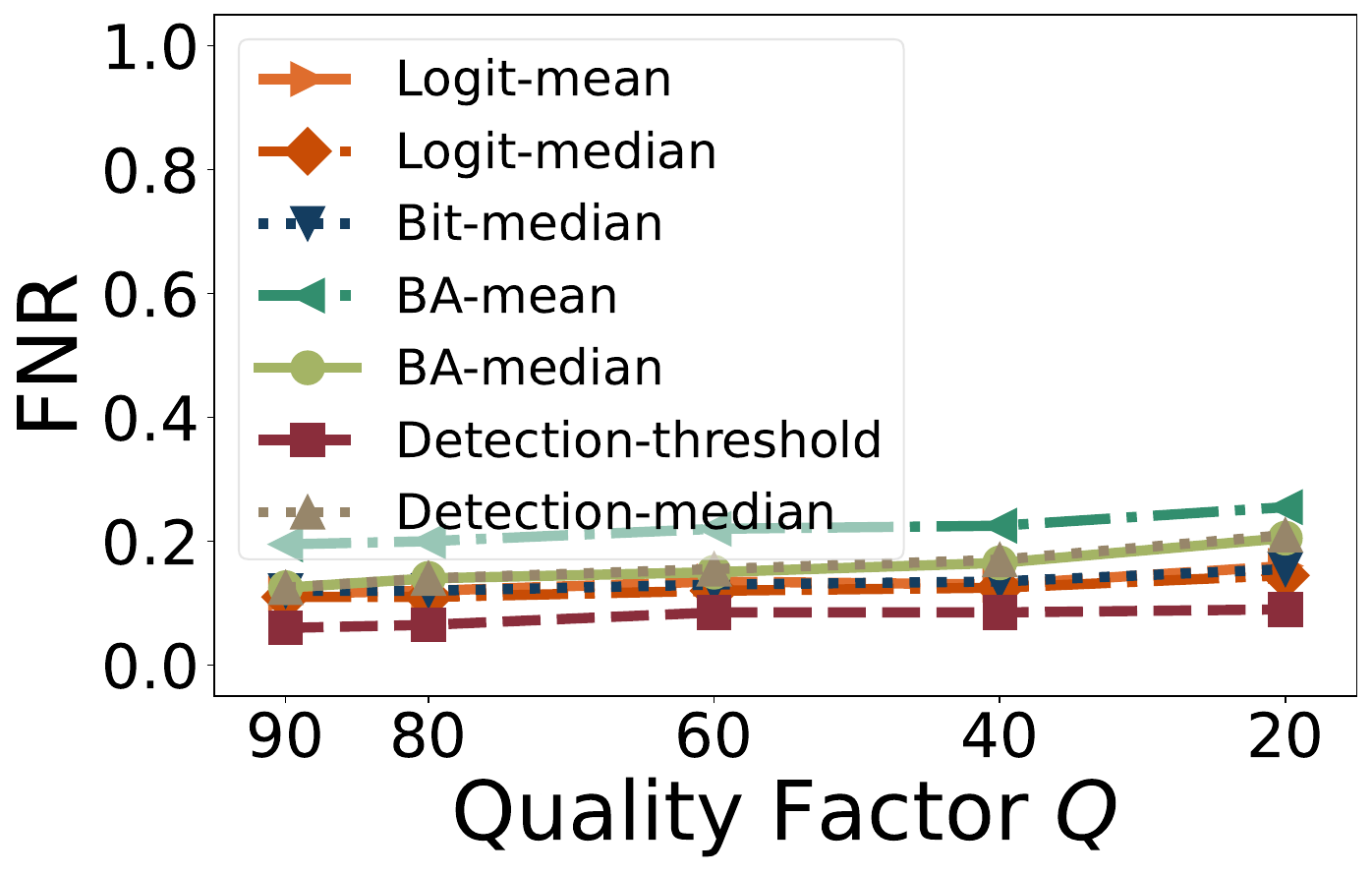}
        \caption{JPEG}
    \end{subfigure}
    \begin{subfigure}{.23\linewidth}
        \centering
        \includegraphics[width=\linewidth]{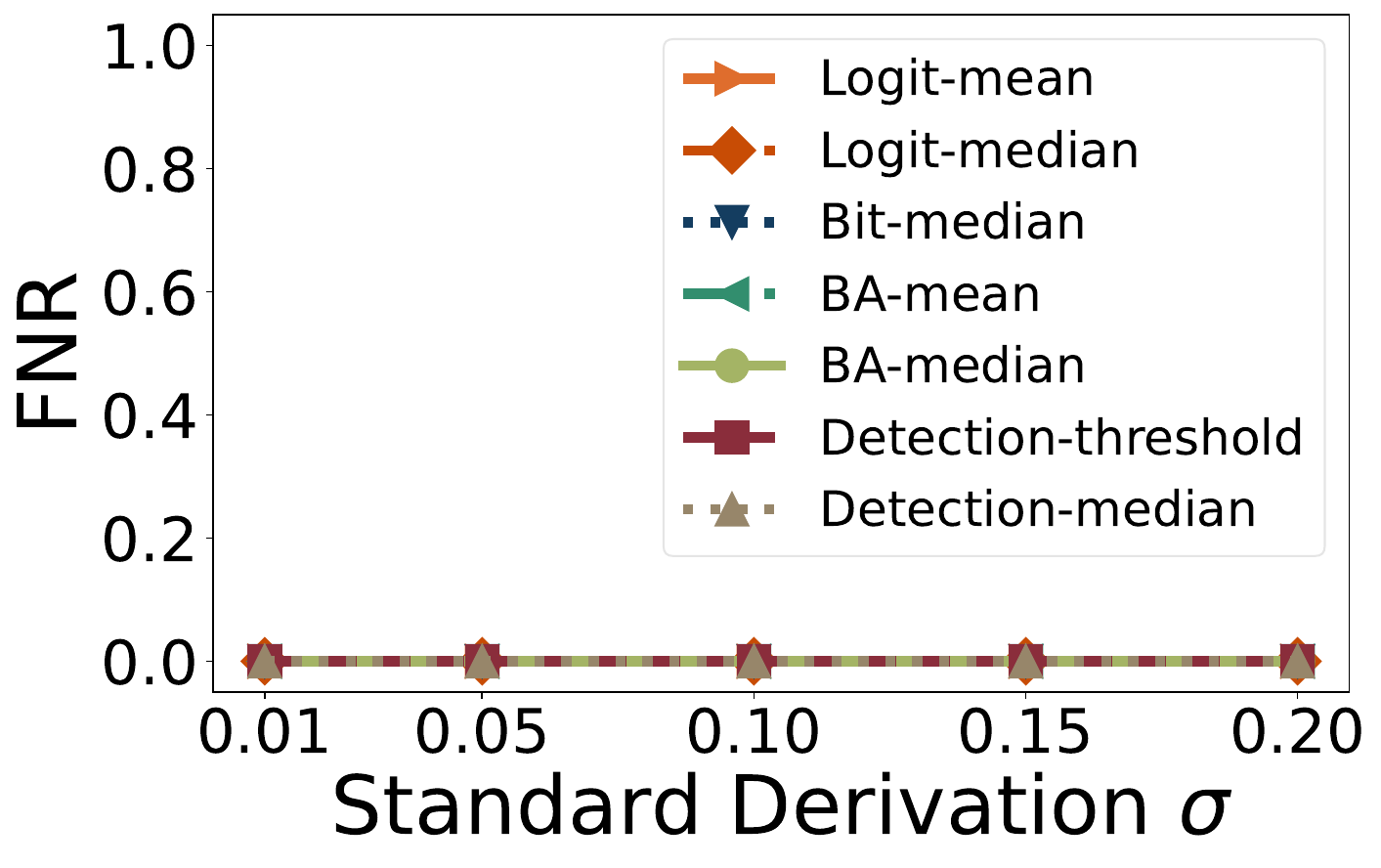}
        \caption{Gaussian Noise}
    \end{subfigure}
    \begin{subfigure}{.23\linewidth}
        \centering
        \includegraphics[width=\linewidth]{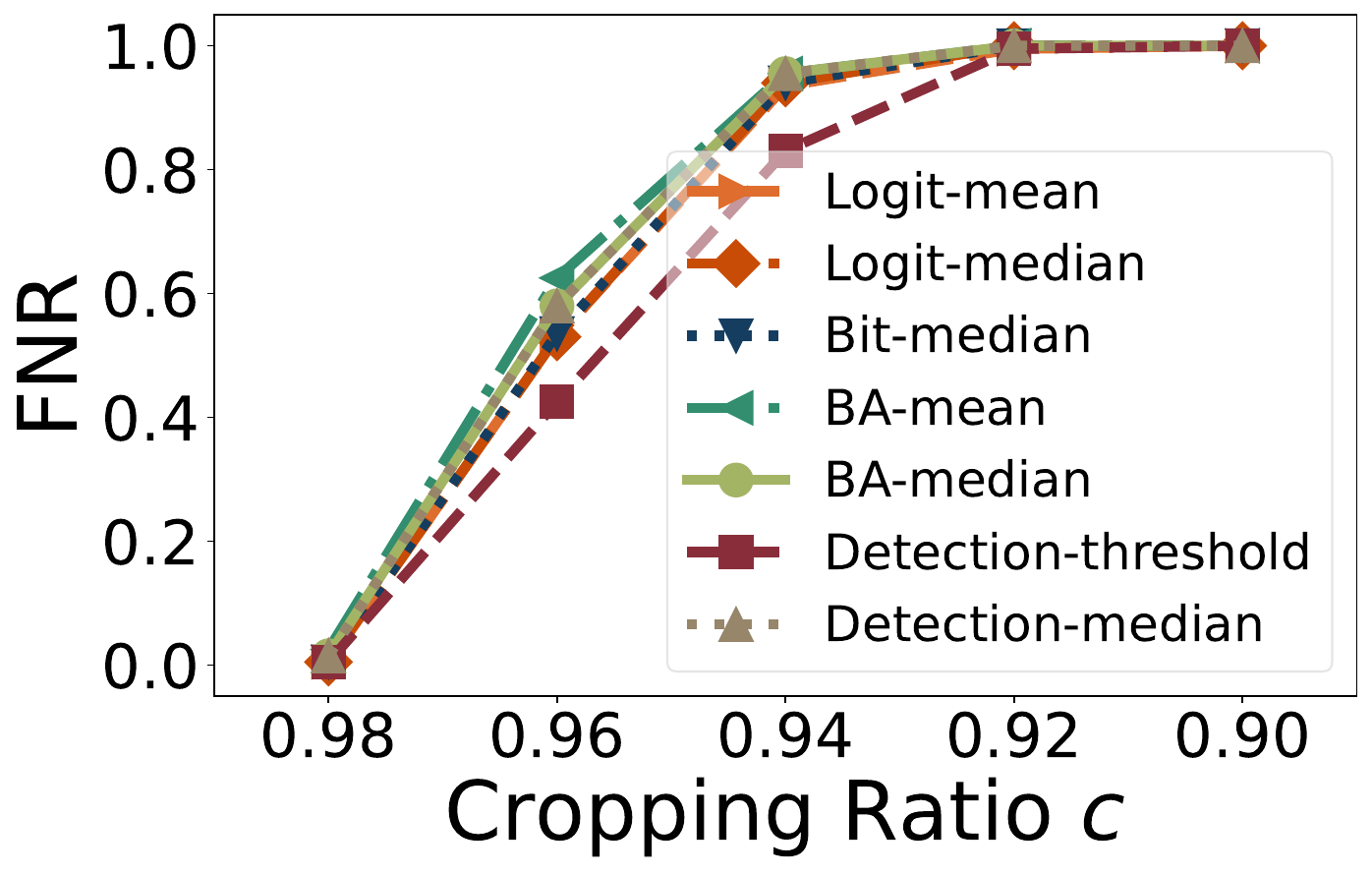}
        \caption{Cropping}
    \end{subfigure}
    \begin{subfigure}{.23\linewidth}
        \centering
        \includegraphics[width=\linewidth]{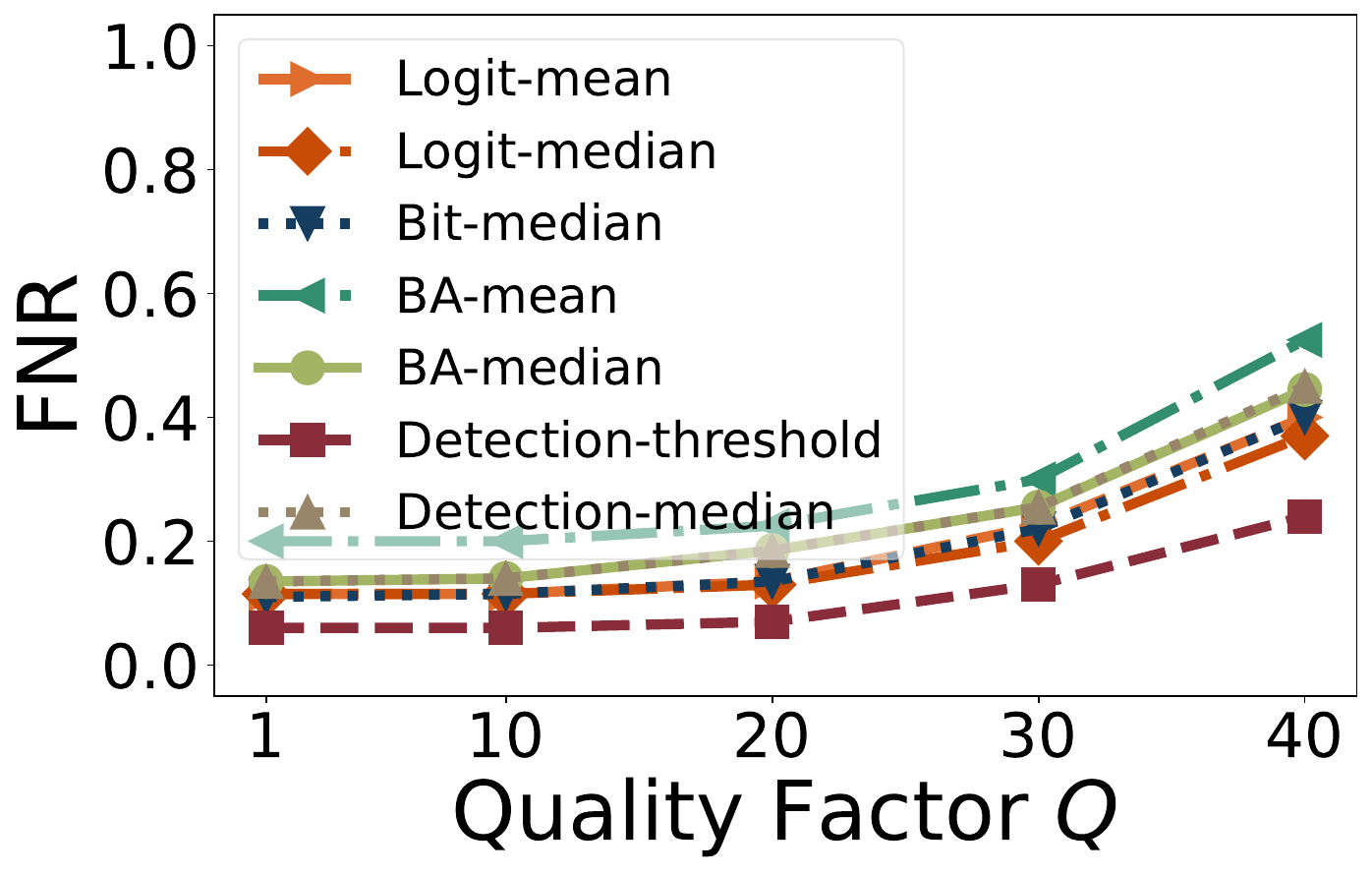}
        \caption{MPEG-4}
    \end{subfigure} \\
    
    \begin{subfigure}{.23\linewidth}
        \centering
        \includegraphics[width=\linewidth]{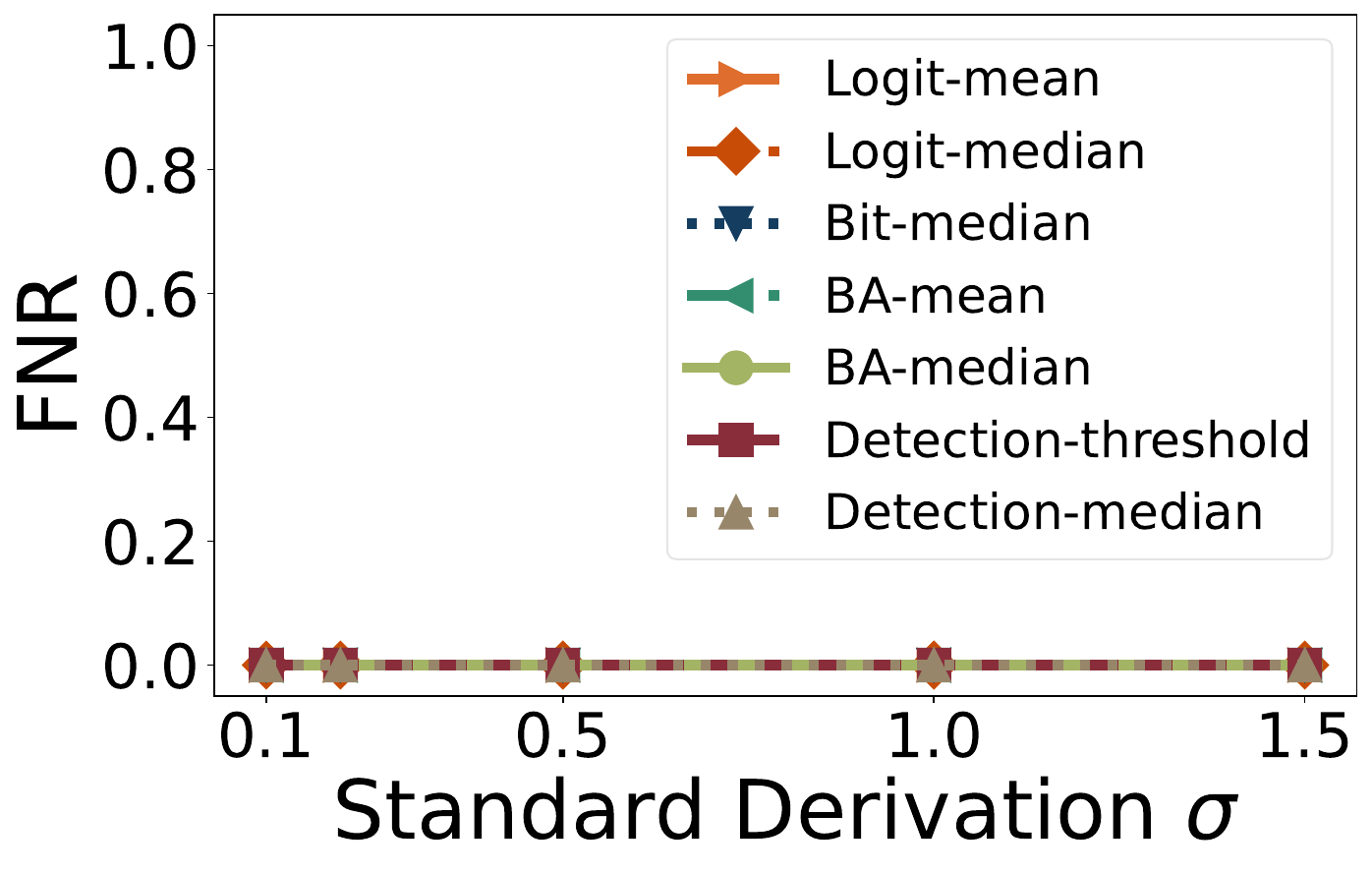}
        \caption{Gaussian Blur}
    \end{subfigure}
    \begin{subfigure}{.23\linewidth}
        \centering
        \includegraphics[width=\linewidth]{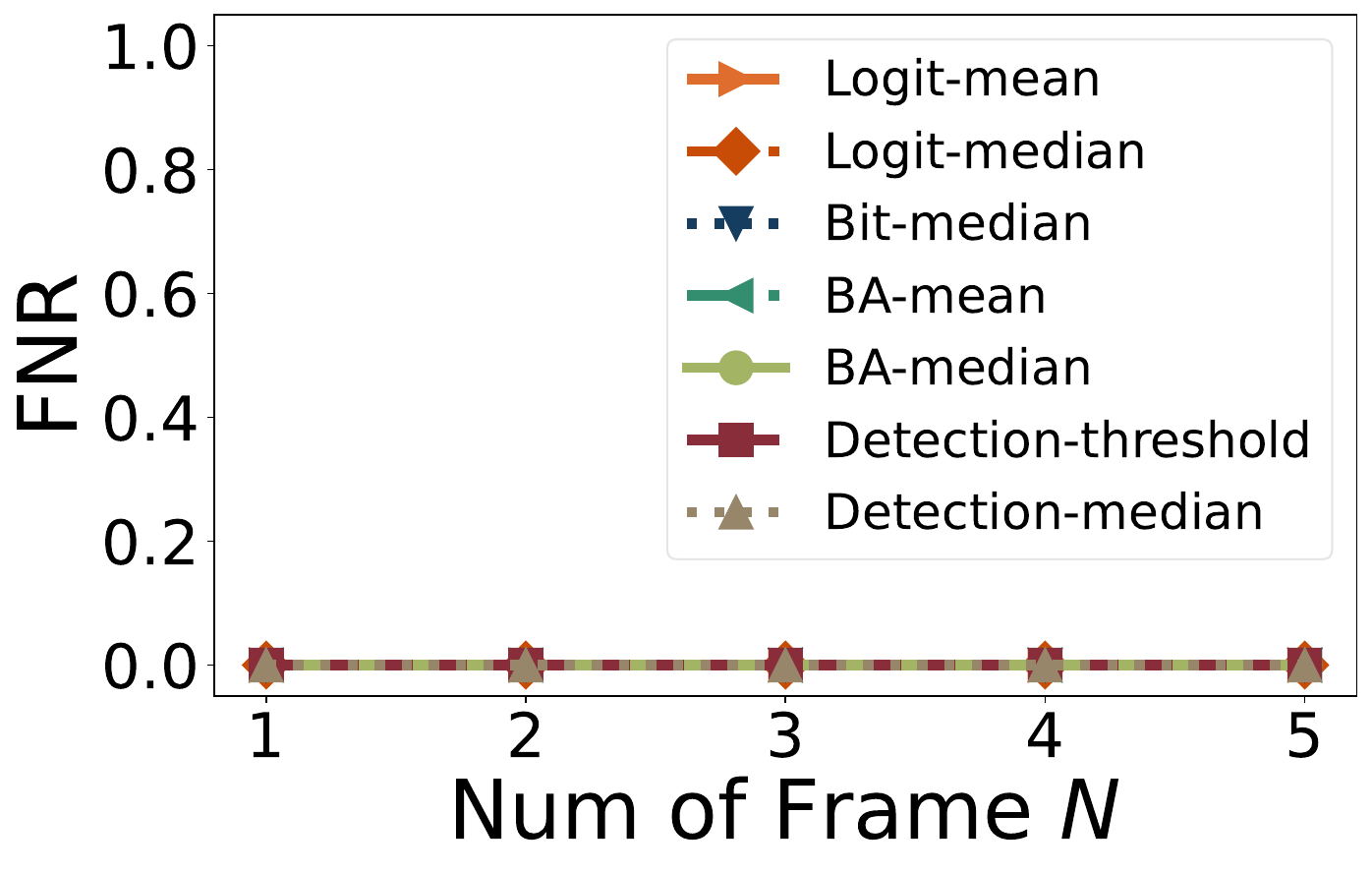}
        \caption{Frame Average}
    \end{subfigure}
    \begin{subfigure}{.23\linewidth}
        \centering
        \includegraphics[width=\linewidth]{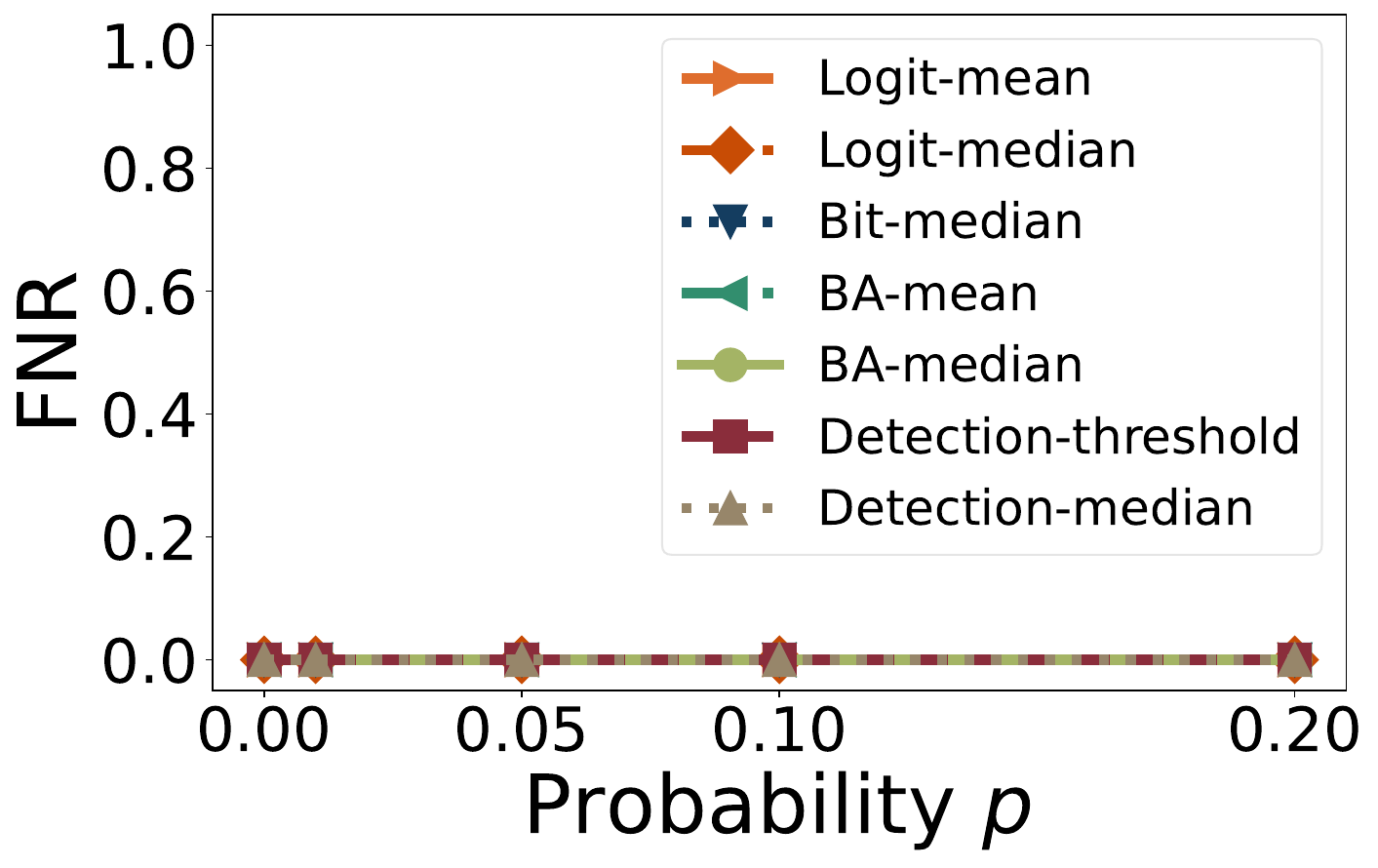}
        \caption{Frame Switch}
    \end{subfigure}
    \begin{subfigure}{.23\linewidth}
        \centering
        \includegraphics[width=\linewidth]{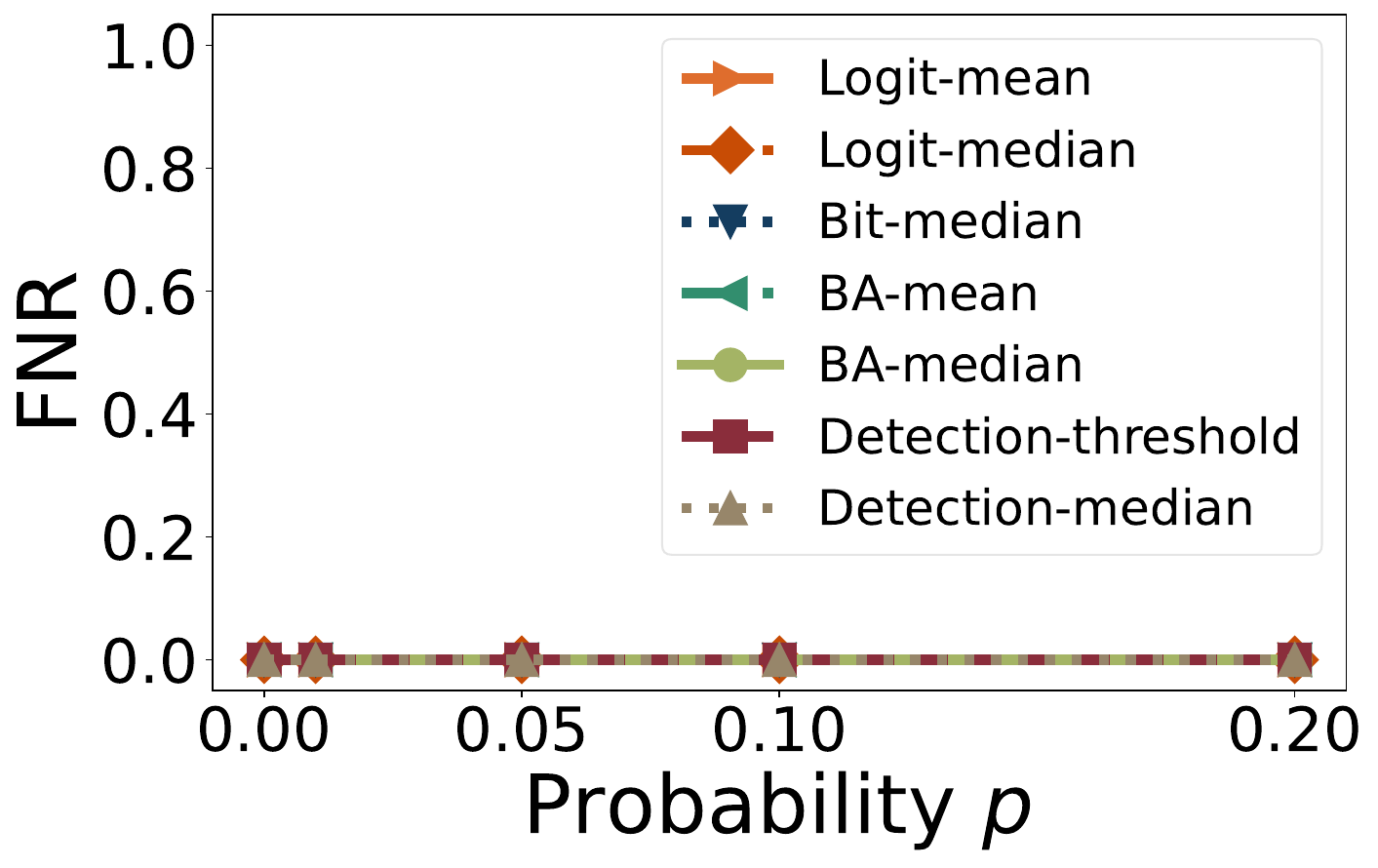}
        \caption{Frame Removal}
    \end{subfigure} \\

    \begin{subfigure}{.9\linewidth}
    \centering
    \caption*{Cartoon video style}
    \end{subfigure}

    \begin{subfigure}{.23\linewidth}
        \centering
        \includegraphics[width=\linewidth]{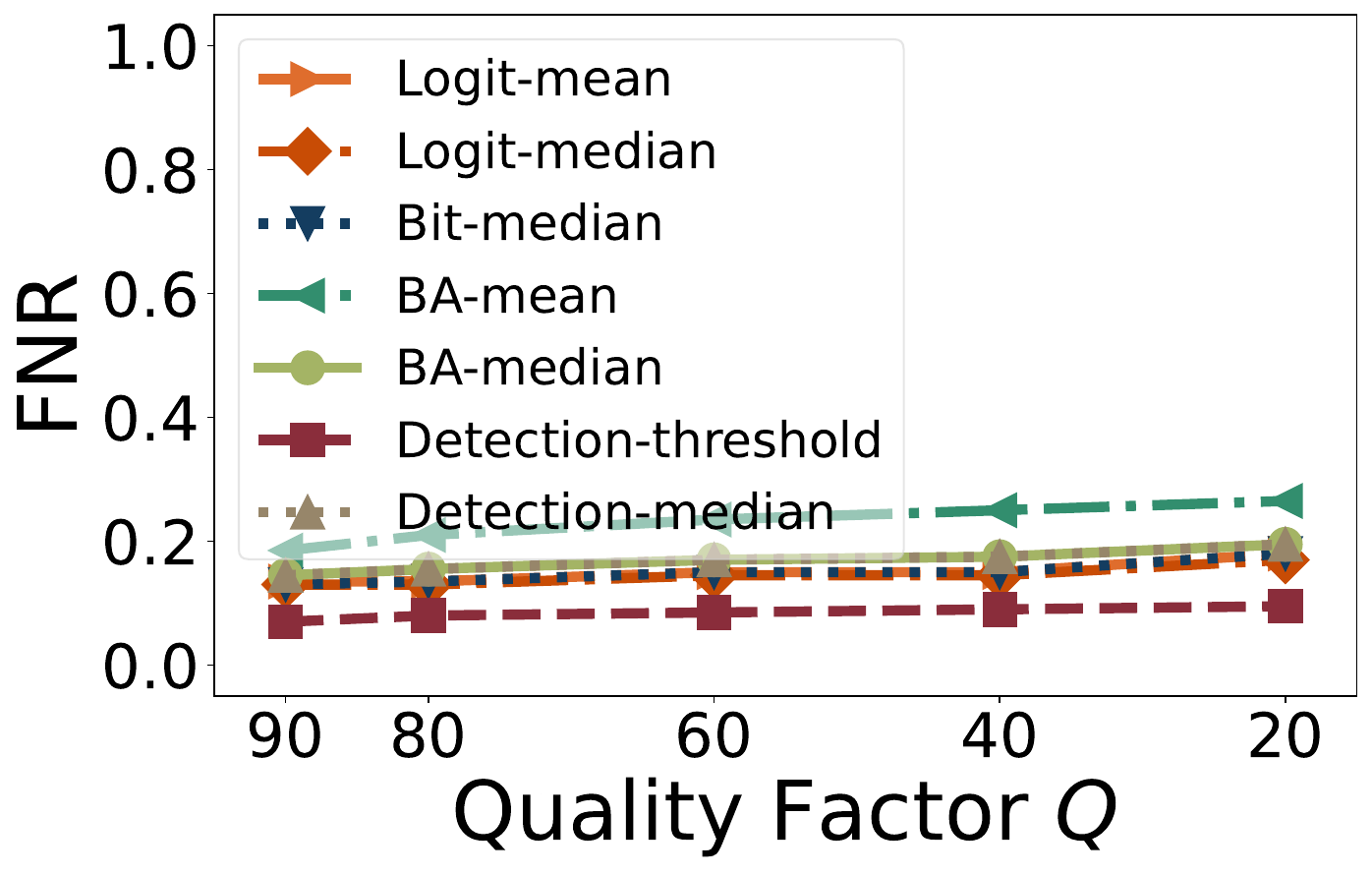}
        \caption{JPEG}
    \end{subfigure}
    \begin{subfigure}{.23\linewidth}
        \centering
        \includegraphics[width=\linewidth]{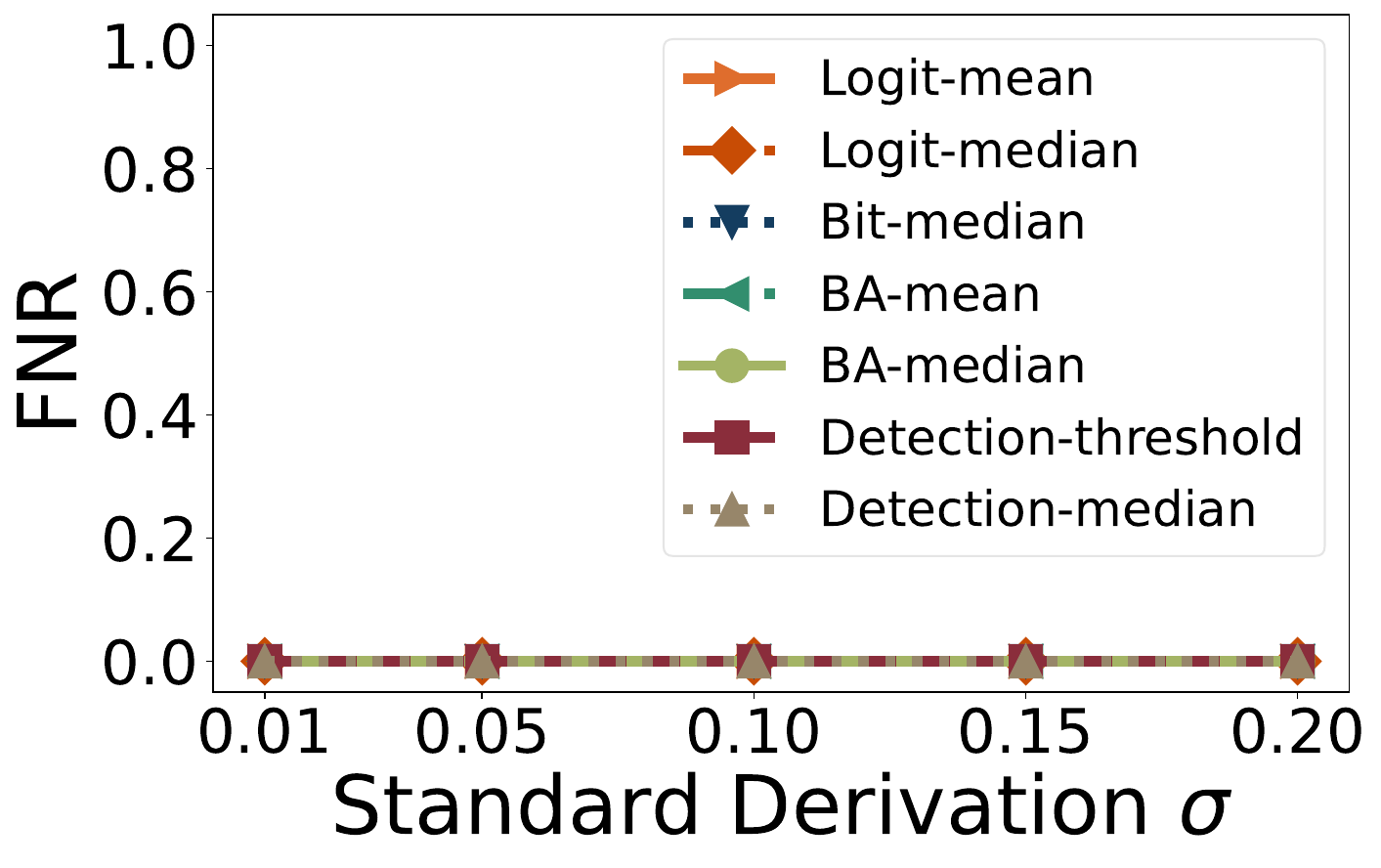}
        \caption{Gaussian Noise}
    \end{subfigure}
    \begin{subfigure}{.23\linewidth}
        \centering
        \includegraphics[width=\linewidth]{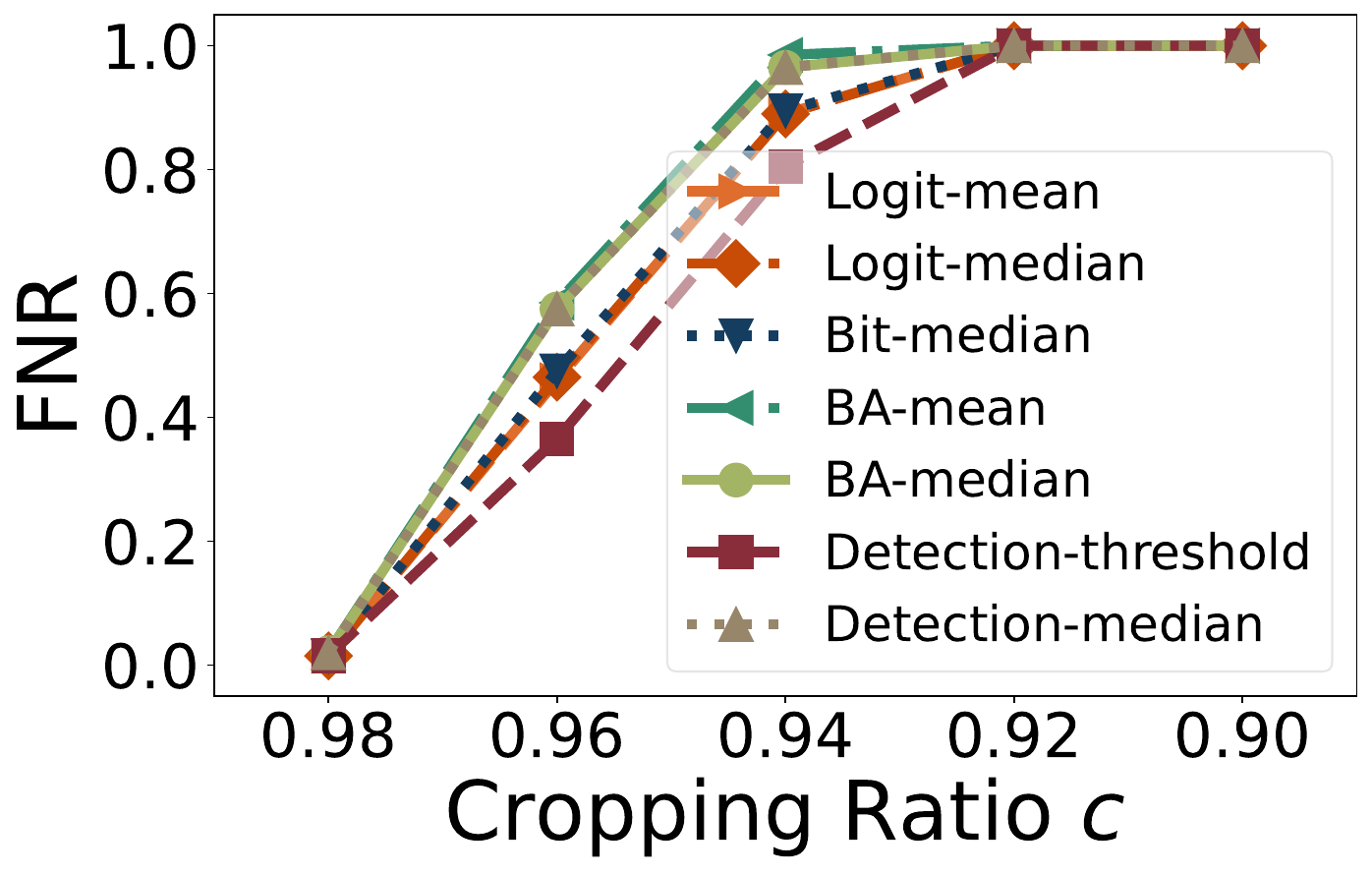}
        \caption{Cropping}
    \end{subfigure}
    \begin{subfigure}{.23\linewidth}
        \centering
        \includegraphics[width=\linewidth]{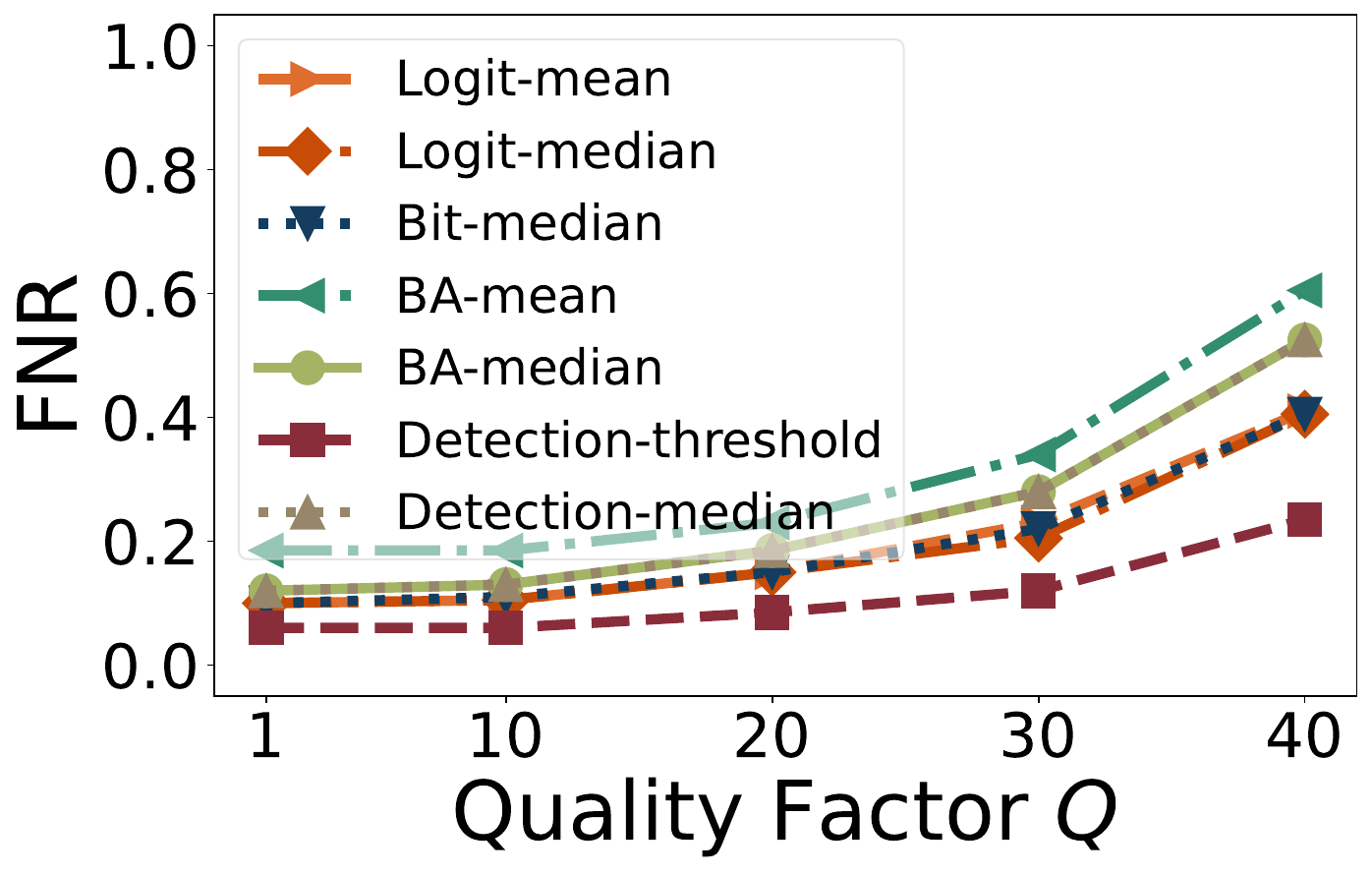}
        \caption{MPEG-4}
    \end{subfigure} \\
    
    \begin{subfigure}{.23\linewidth}
        \centering
        \includegraphics[width=\linewidth]{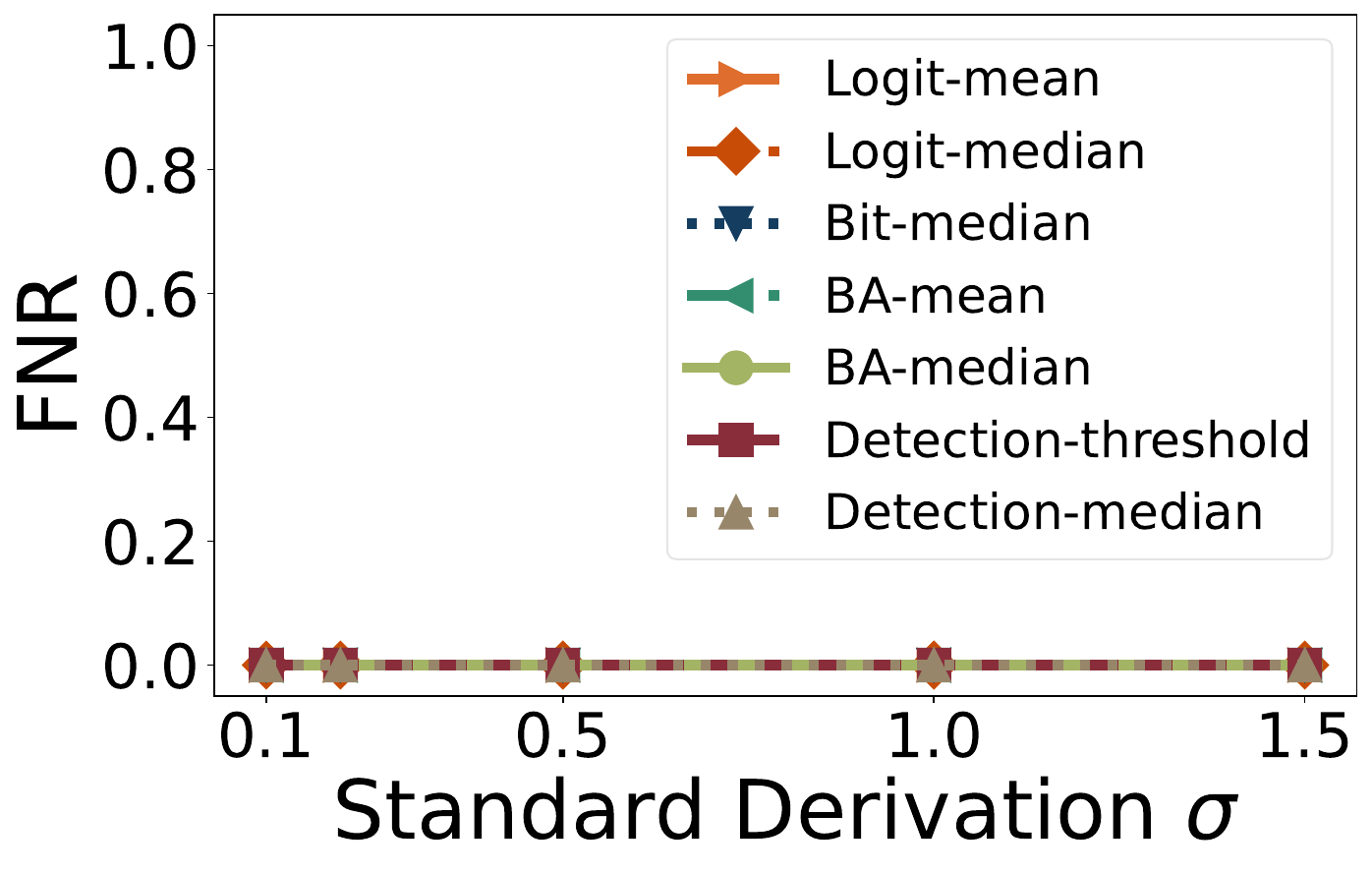}
        \caption{Gaussian Blur}
    \end{subfigure}
    \begin{subfigure}{.23\linewidth}
        \centering
        \includegraphics[width=\linewidth]{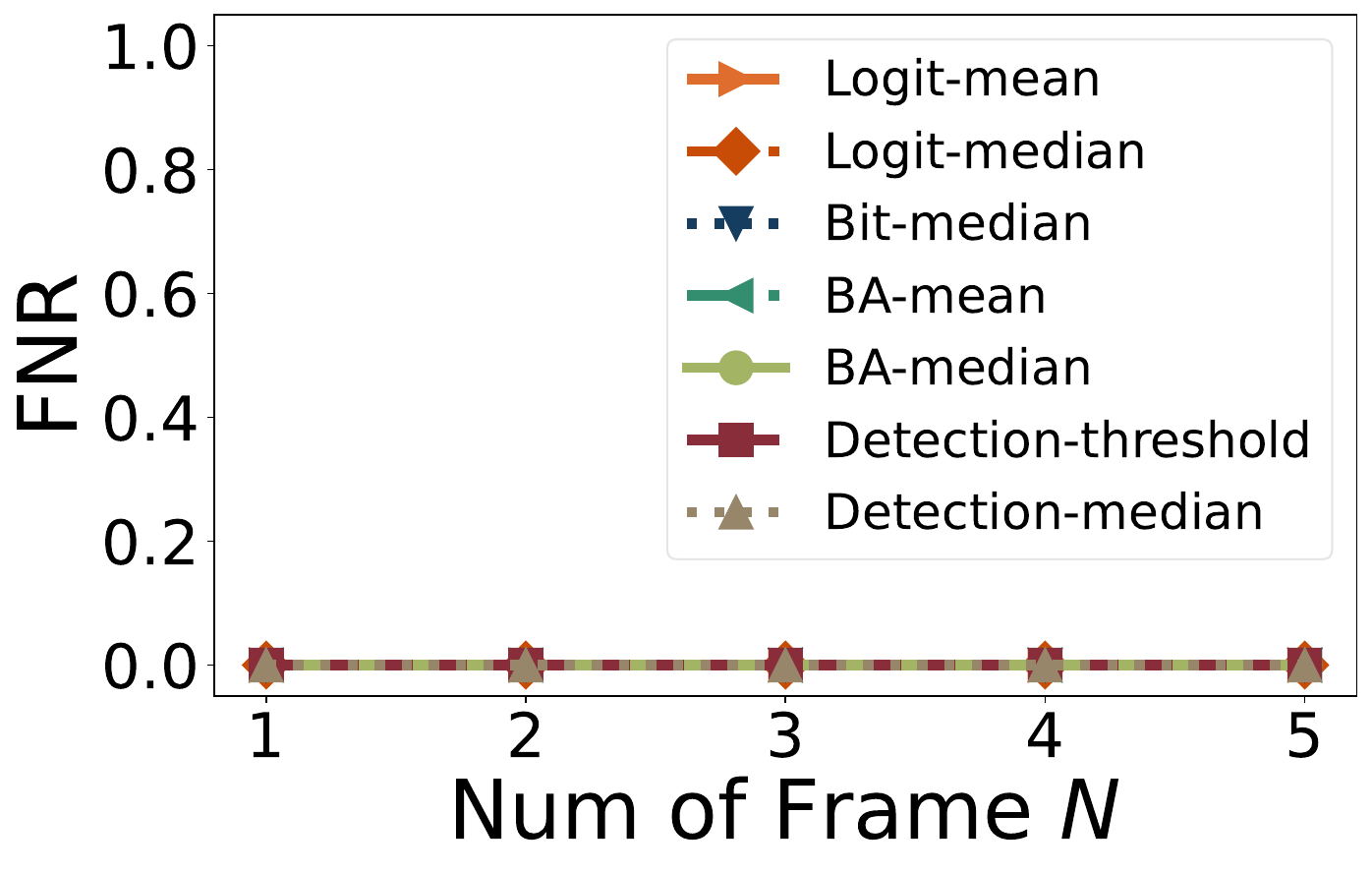}
        \caption{Frame Average}
    \end{subfigure}
    \begin{subfigure}{.23\linewidth}
        \centering
        \includegraphics[width=\linewidth]{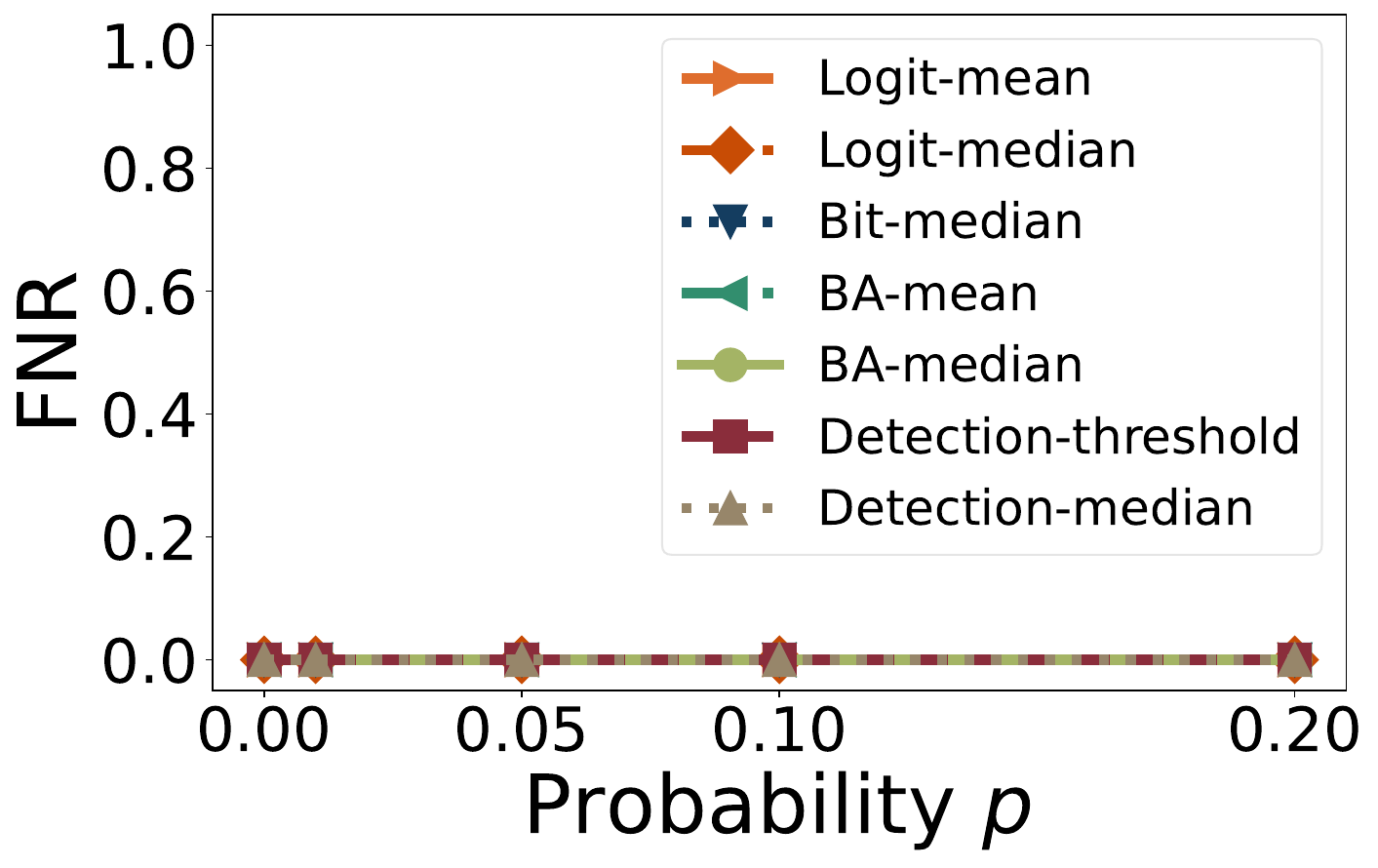}
        \caption{Frame Switch}
    \end{subfigure}
    \begin{subfigure}{.23\linewidth}
        \centering
        \includegraphics[width=\linewidth]{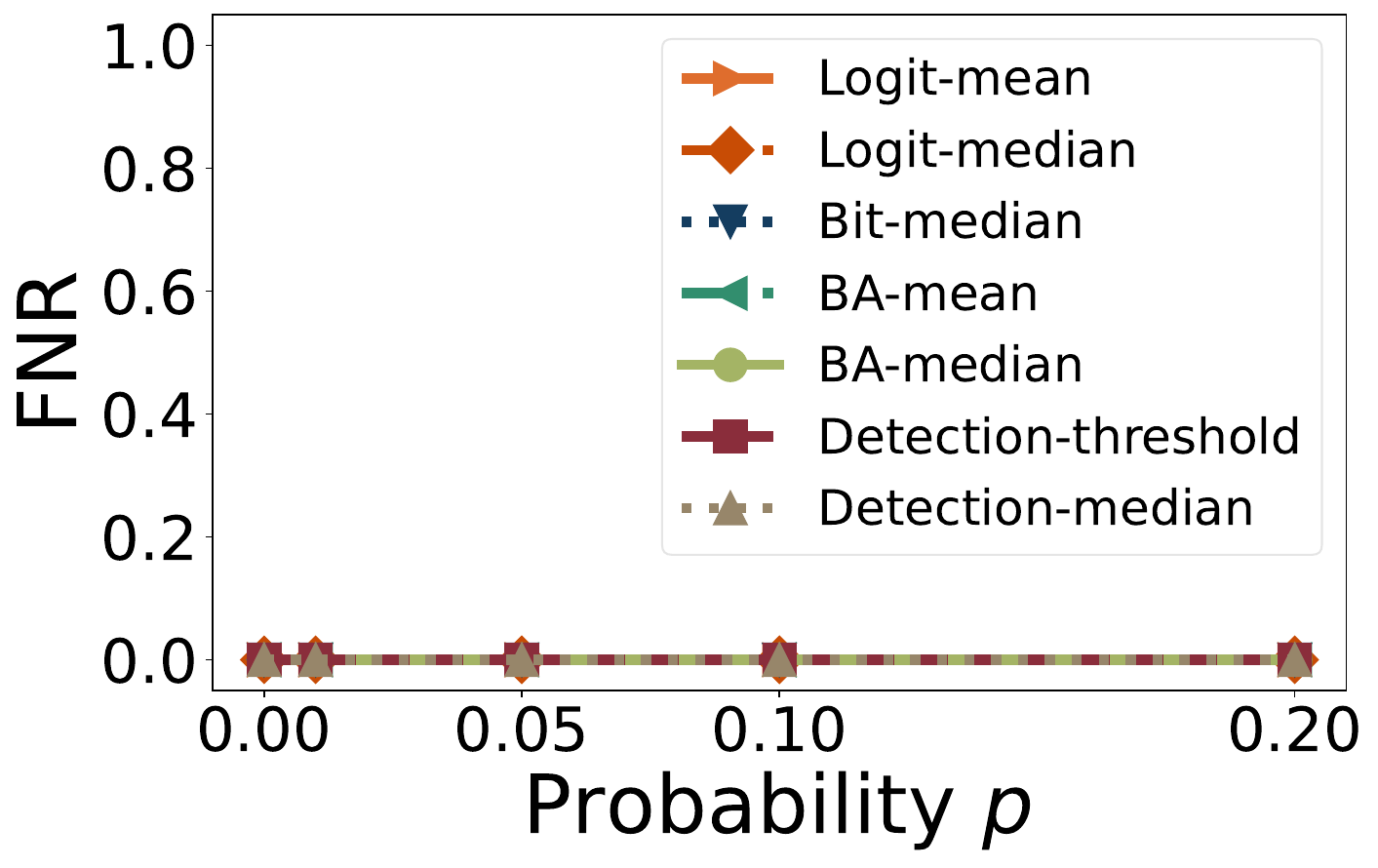}
        \caption{Frame Removal}
    \end{subfigure} \\

    \begin{subfigure}{.9\linewidth}
    \centering
    \caption*{Sci-fi video style}
    \end{subfigure}
    
    \caption{More fine-grained watermark removal results for StegaStamp on videos generated by Hunyuan Video.}
\end{figure}

\clearpage
\pagestyle{empty}
\begin{figure}[]
    \centering
    \begin{subfigure}{.23\linewidth}
        \centering
        \includegraphics[width=\linewidth]{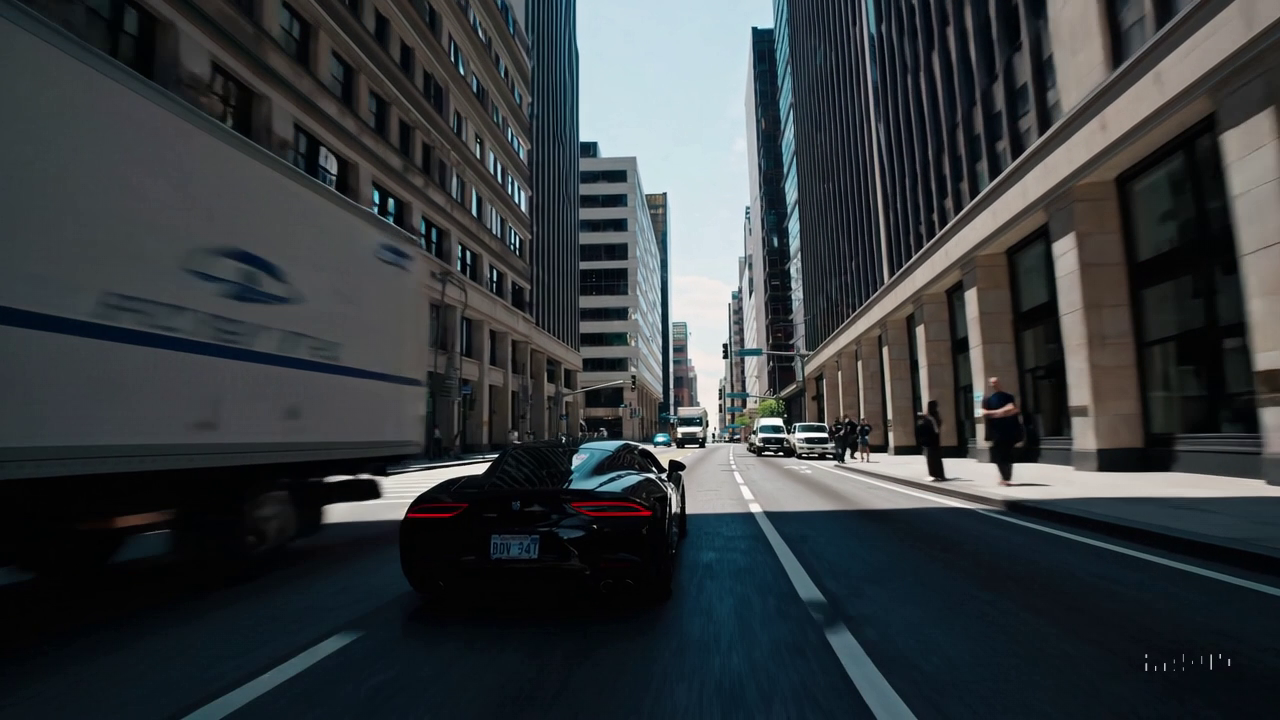}
    \end{subfigure}
    \begin{subfigure}{.23\linewidth}
        \centering
        \includegraphics[width=\linewidth]{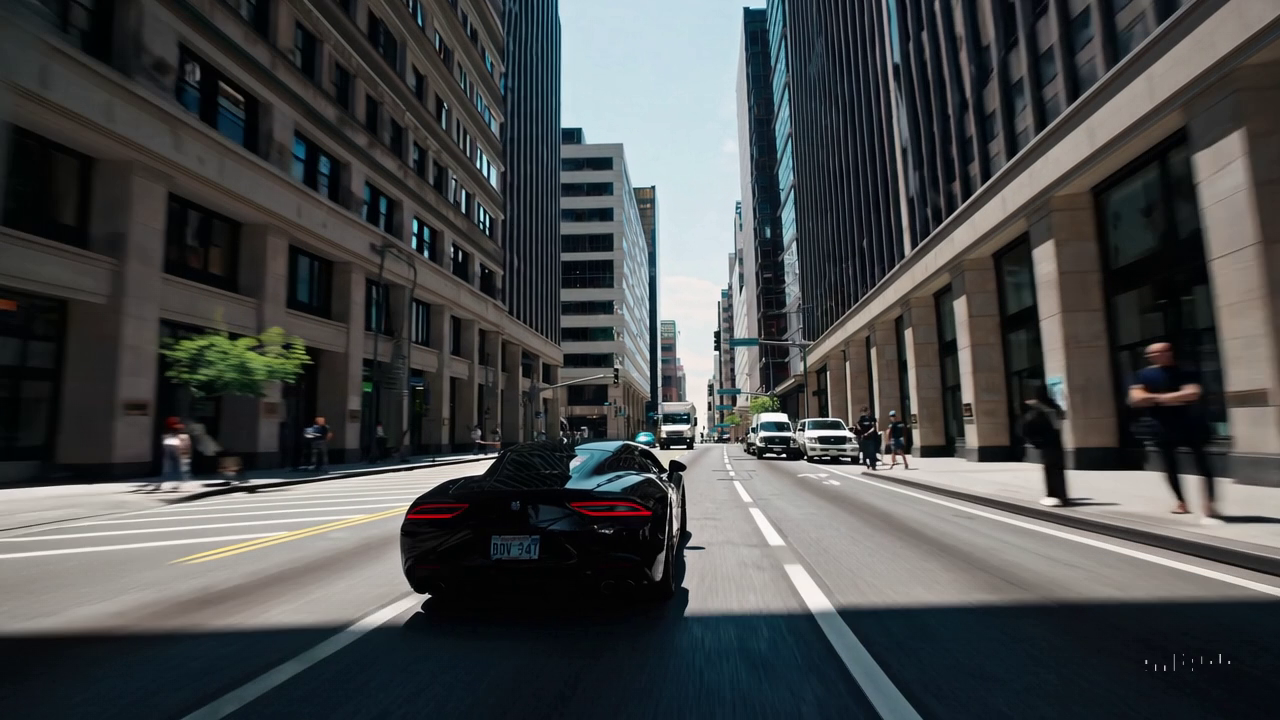}
    \end{subfigure}
    \begin{subfigure}{.23\linewidth}
        \centering
        \includegraphics[width=\linewidth]{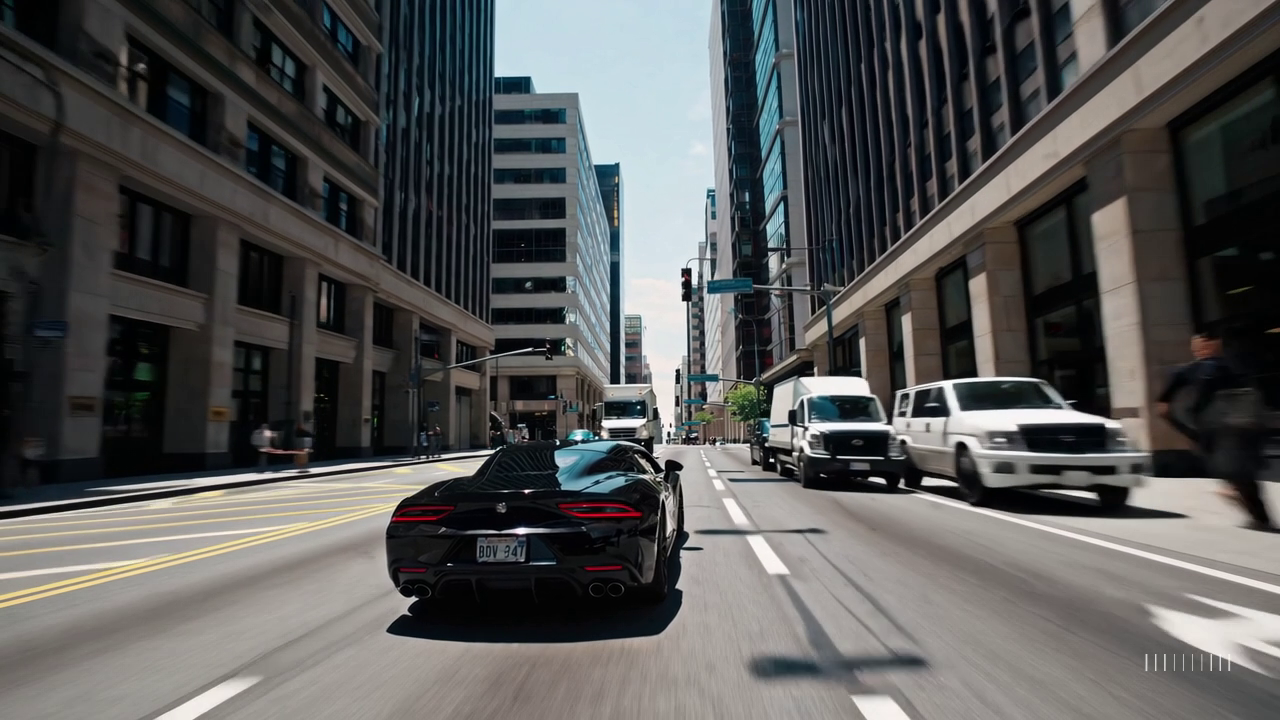}
    \end{subfigure}
    \begin{subfigure}{.23\linewidth}
        \centering
        \includegraphics[width=\linewidth]{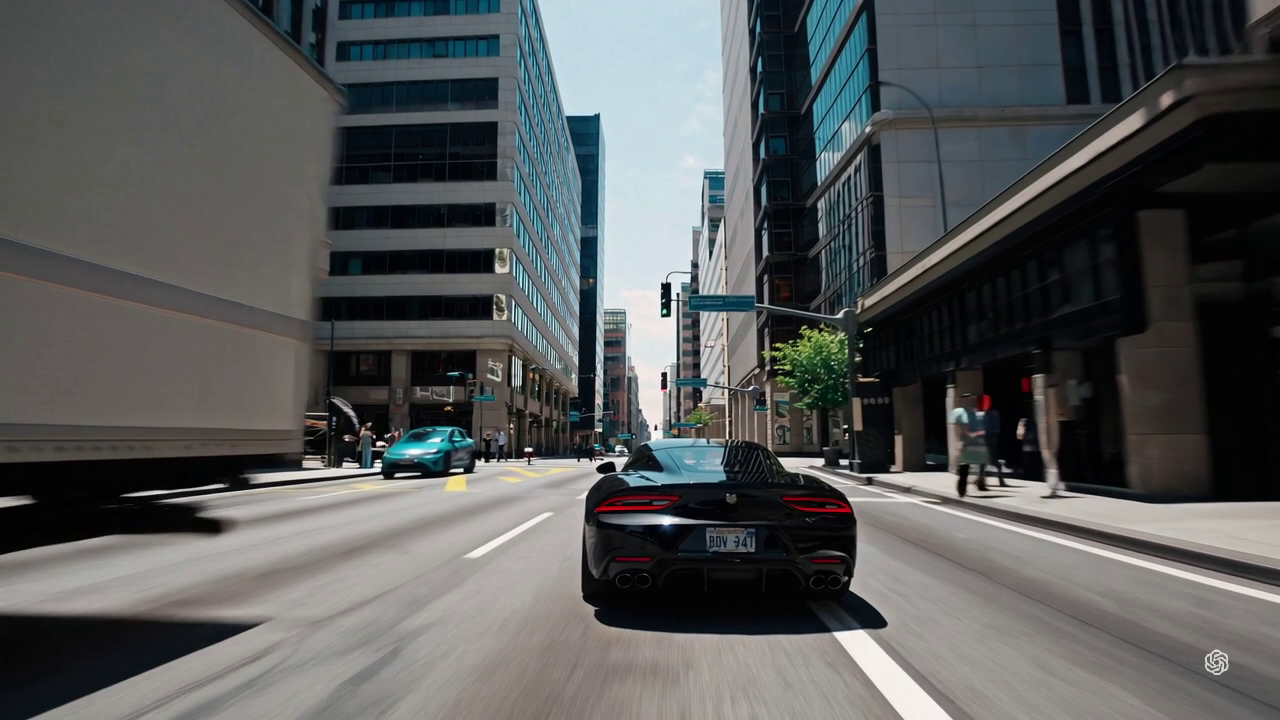}
    \end{subfigure} \\

    \begin{subfigure}{.23\linewidth}
        \centering
        \includegraphics[width=\linewidth]{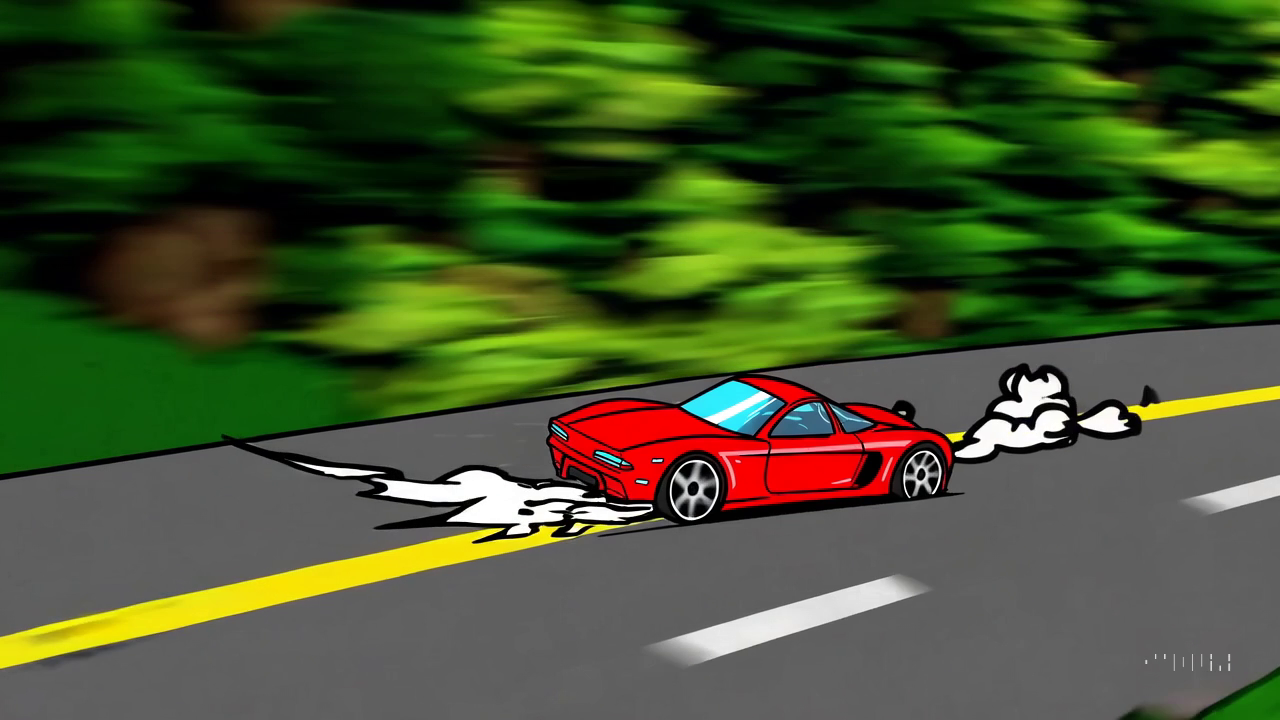}
    \end{subfigure}
    \begin{subfigure}{.23\linewidth}
        \centering
        \includegraphics[width=\linewidth]{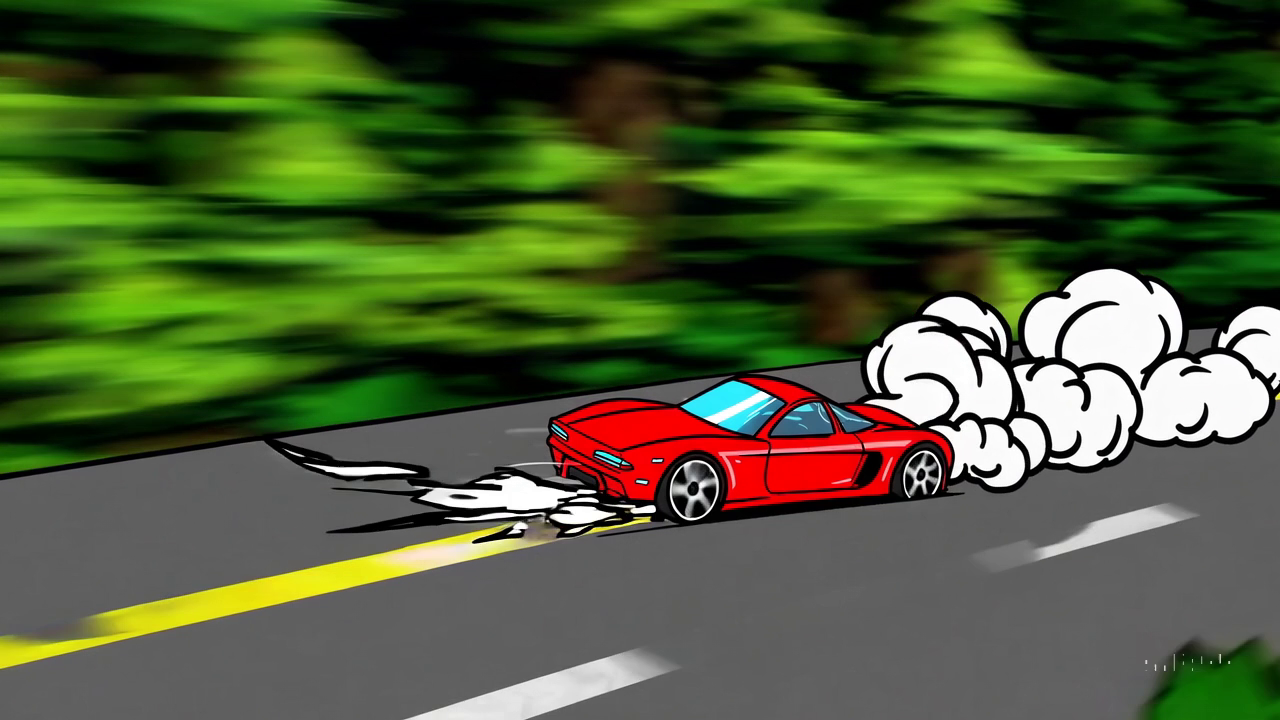}
    \end{subfigure}
    \begin{subfigure}{.23\linewidth}
        \centering
        \includegraphics[width=\linewidth]{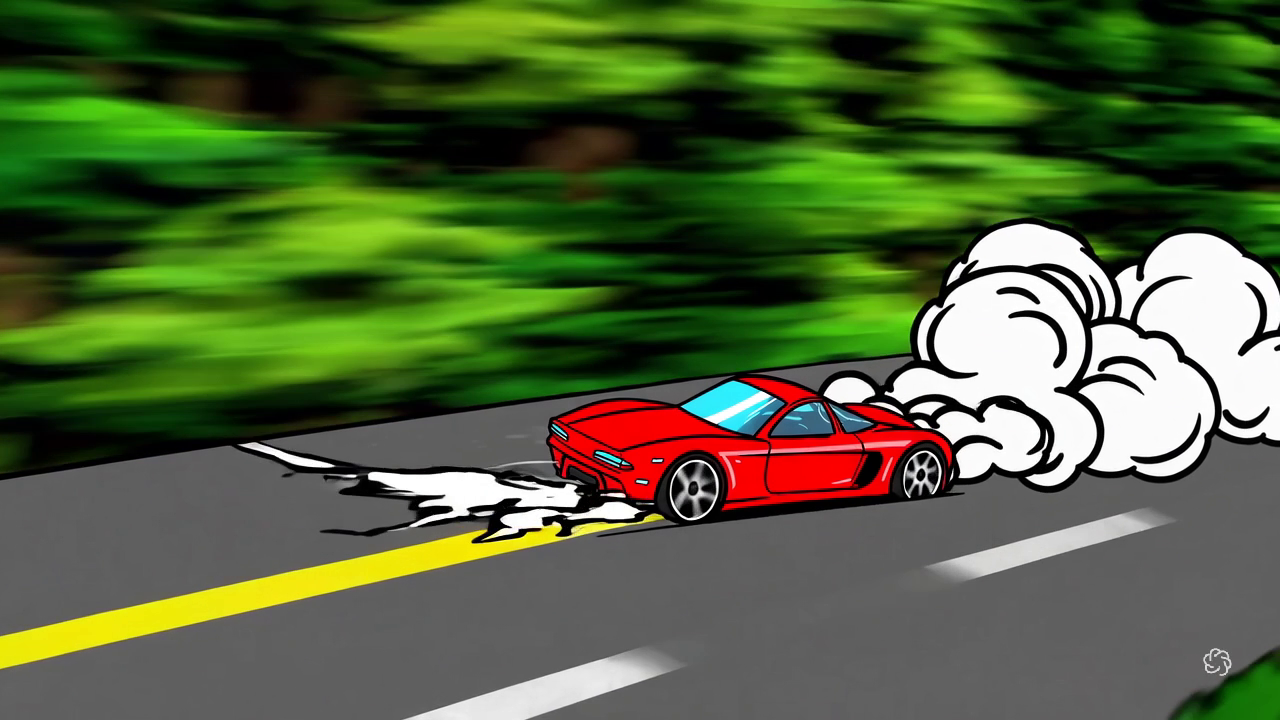}
    \end{subfigure}
    \begin{subfigure}{.23\linewidth}
        \centering
        \includegraphics[width=\linewidth]{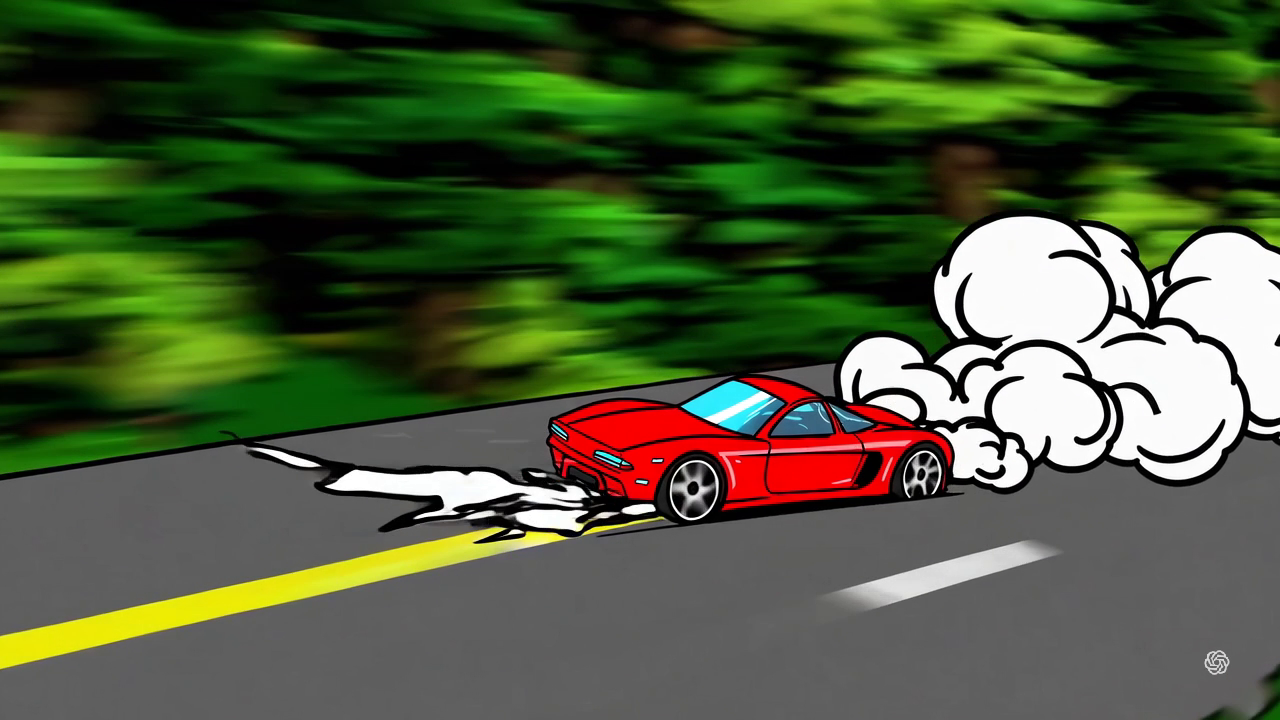}
    \end{subfigure} \\

    \begin{subfigure}{.23\linewidth}
        \centering
        \includegraphics[width=\linewidth]{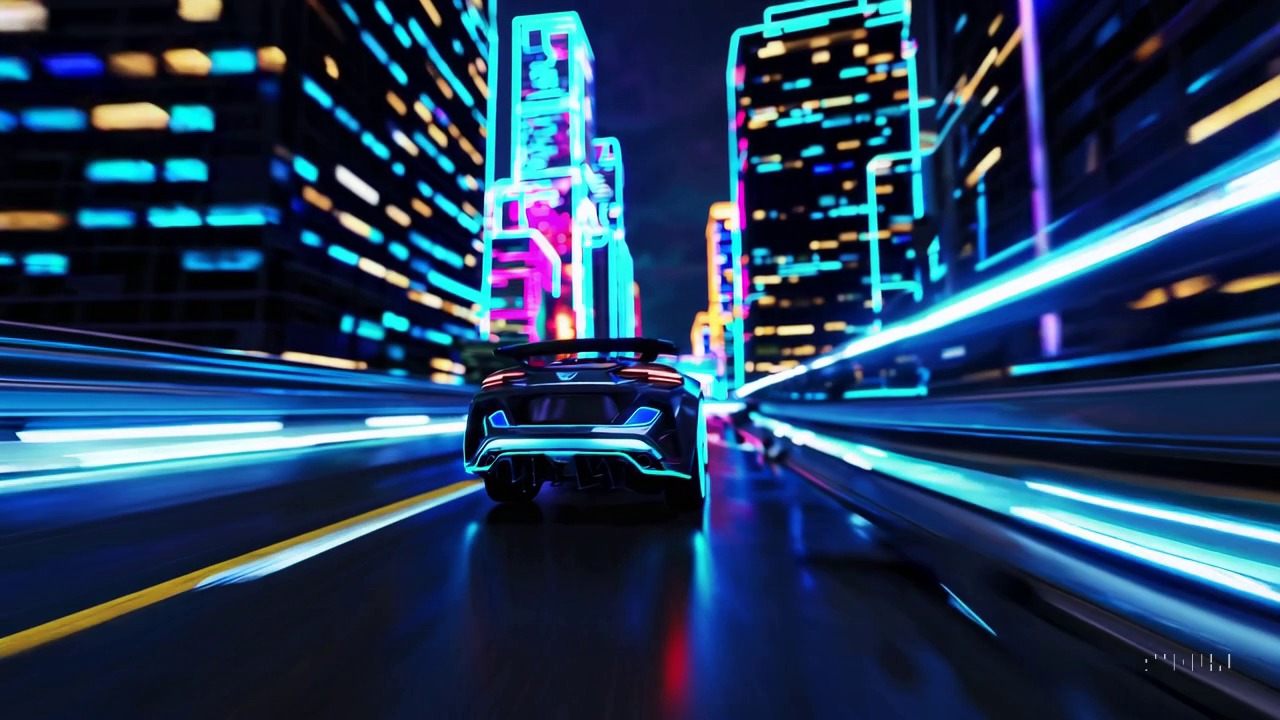}
    \end{subfigure}
    \begin{subfigure}{.23\linewidth}
        \centering
        \includegraphics[width=\linewidth]{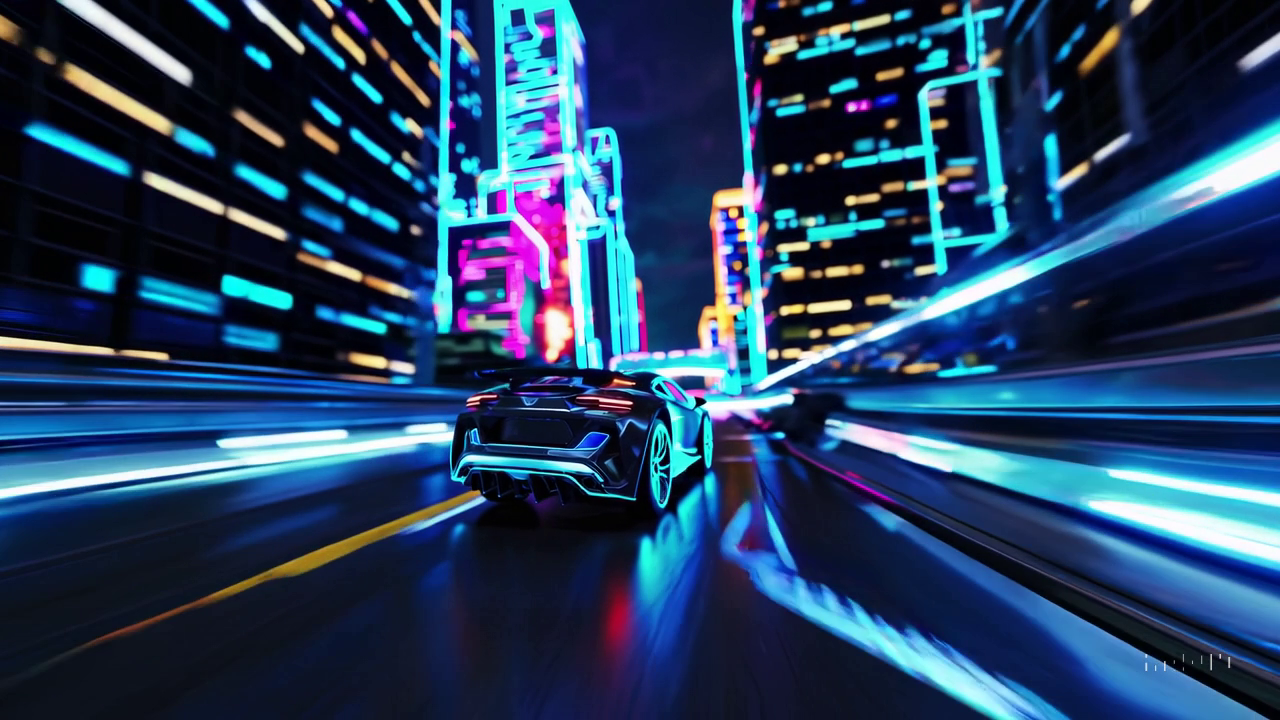}
    \end{subfigure}
    \begin{subfigure}{.23\linewidth}
        \centering
        \includegraphics[width=\linewidth]{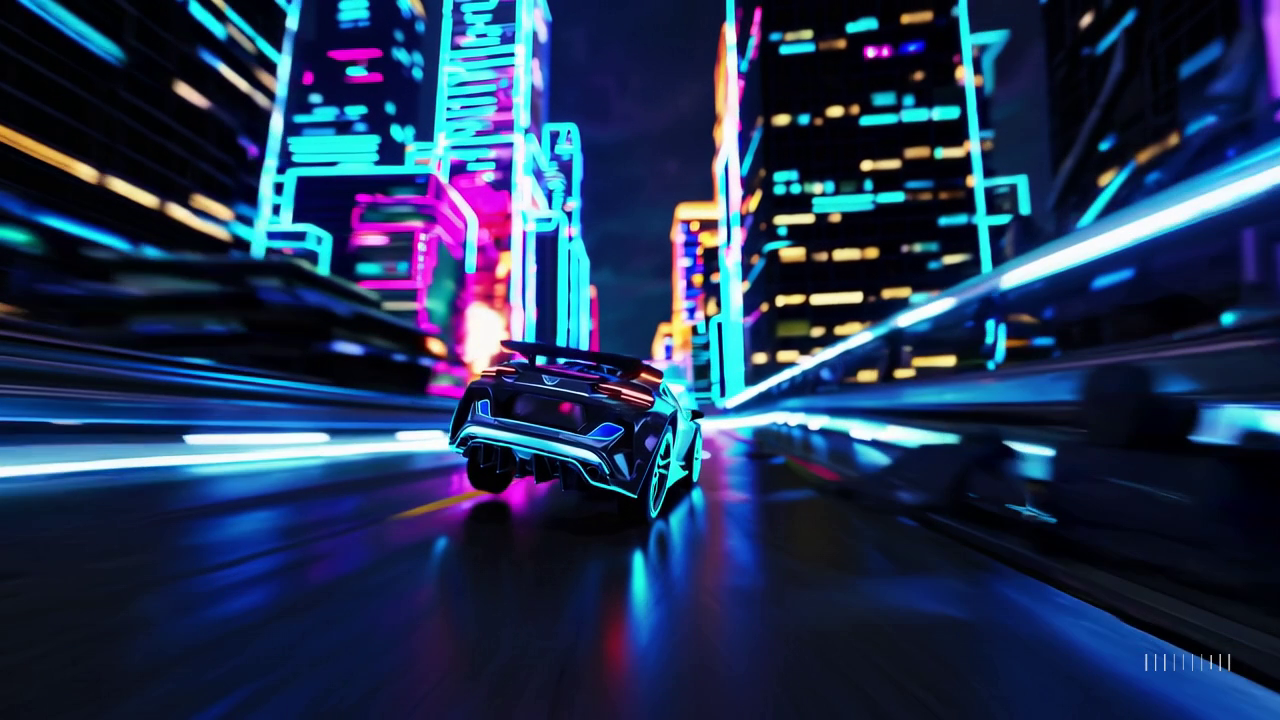}
    \end{subfigure}
    \begin{subfigure}{.23\linewidth}
        \centering
        \includegraphics[width=\linewidth]{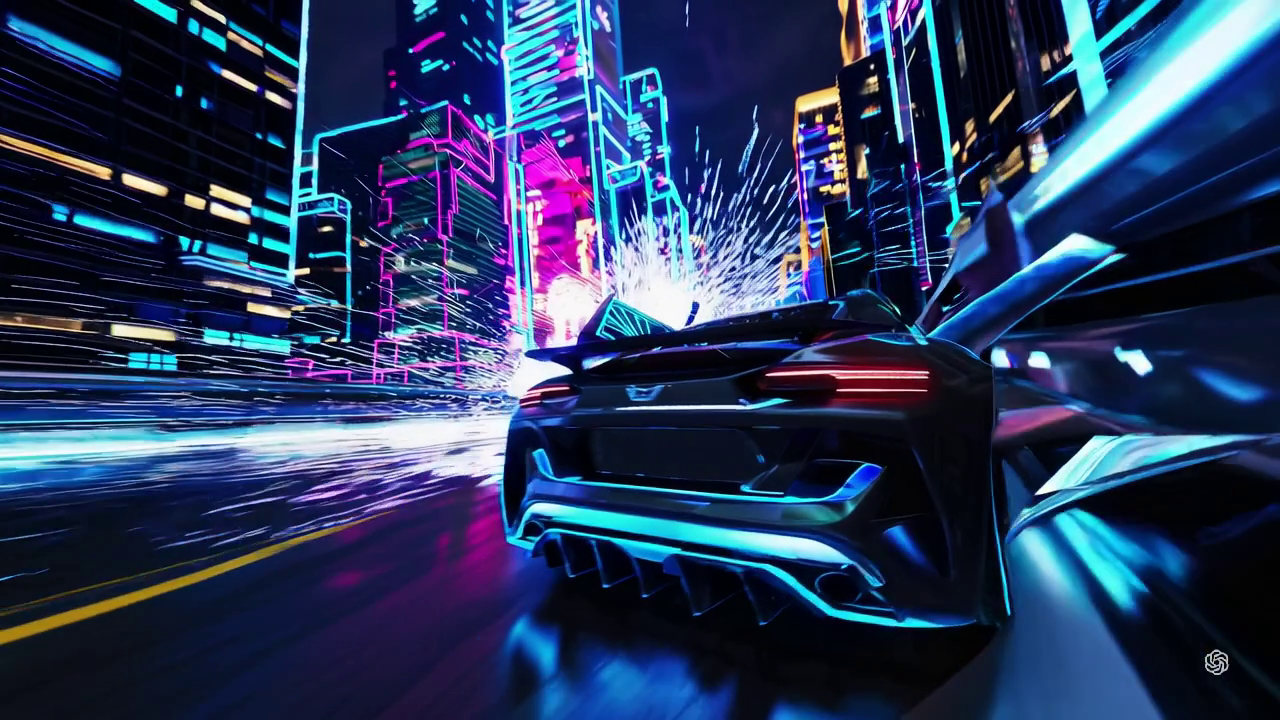}
    \end{subfigure} \\
    
    \begin{subfigure}{.9\linewidth}
    \centering
    \subcaption{Prompt: Generate a dynamic video with rapid frame changes featuring a high-speed car crash with flying debris and shattered glass.}
    \end{subfigure} 

    \begin{subfigure}{.23\linewidth}
        \centering
        \includegraphics[width=\linewidth]{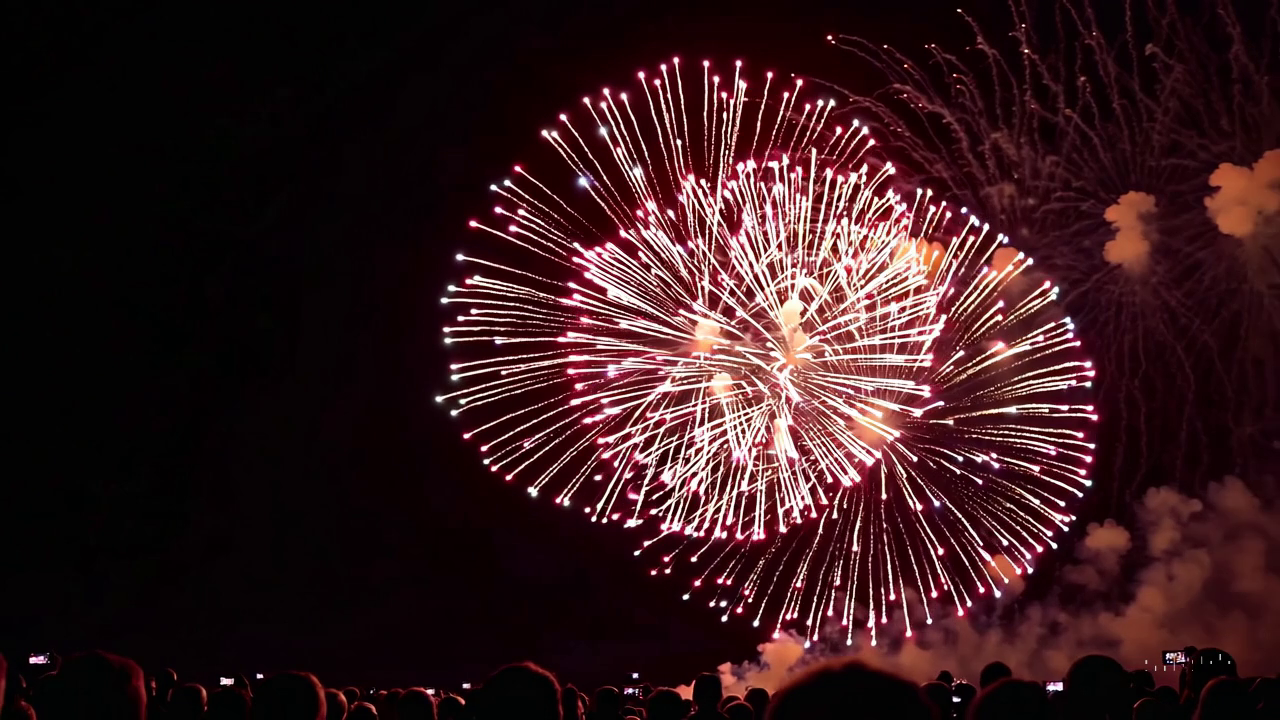}
    \end{subfigure}
    \begin{subfigure}{.23\linewidth}
        \centering
        \includegraphics[width=\linewidth]{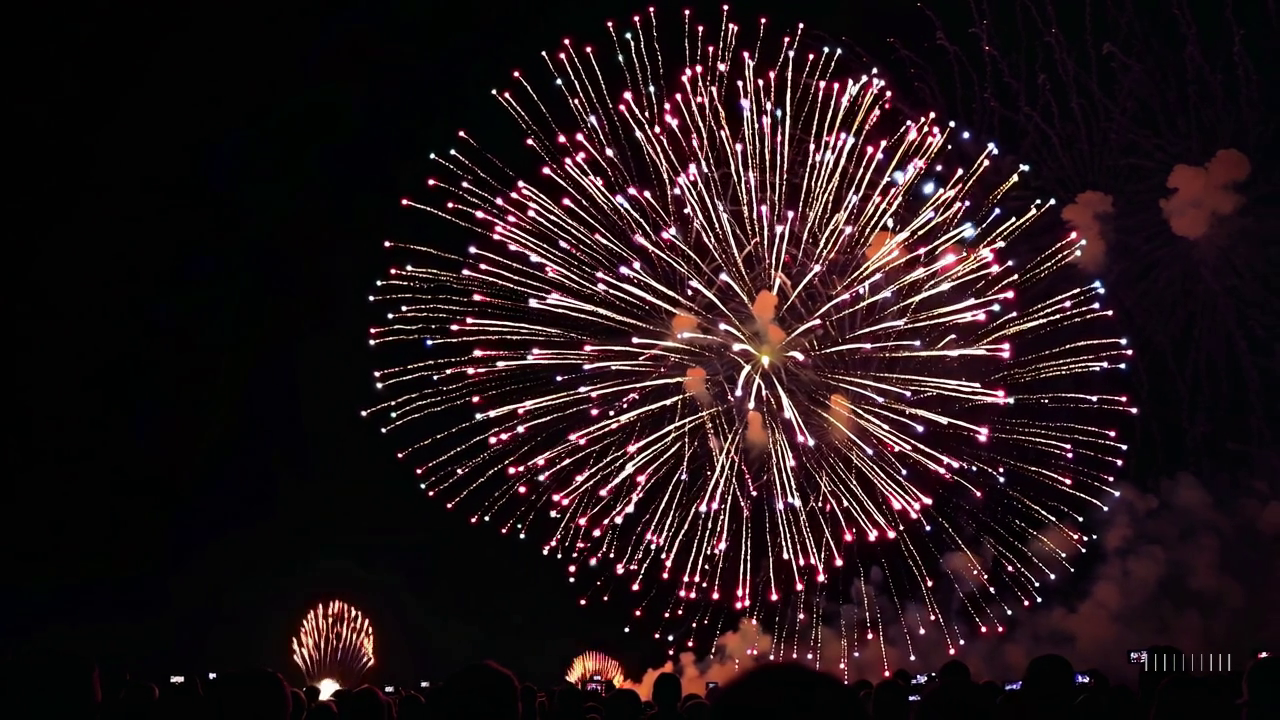}
    \end{subfigure}
    \begin{subfigure}{.23\linewidth}
        \centering
        \includegraphics[width=\linewidth]{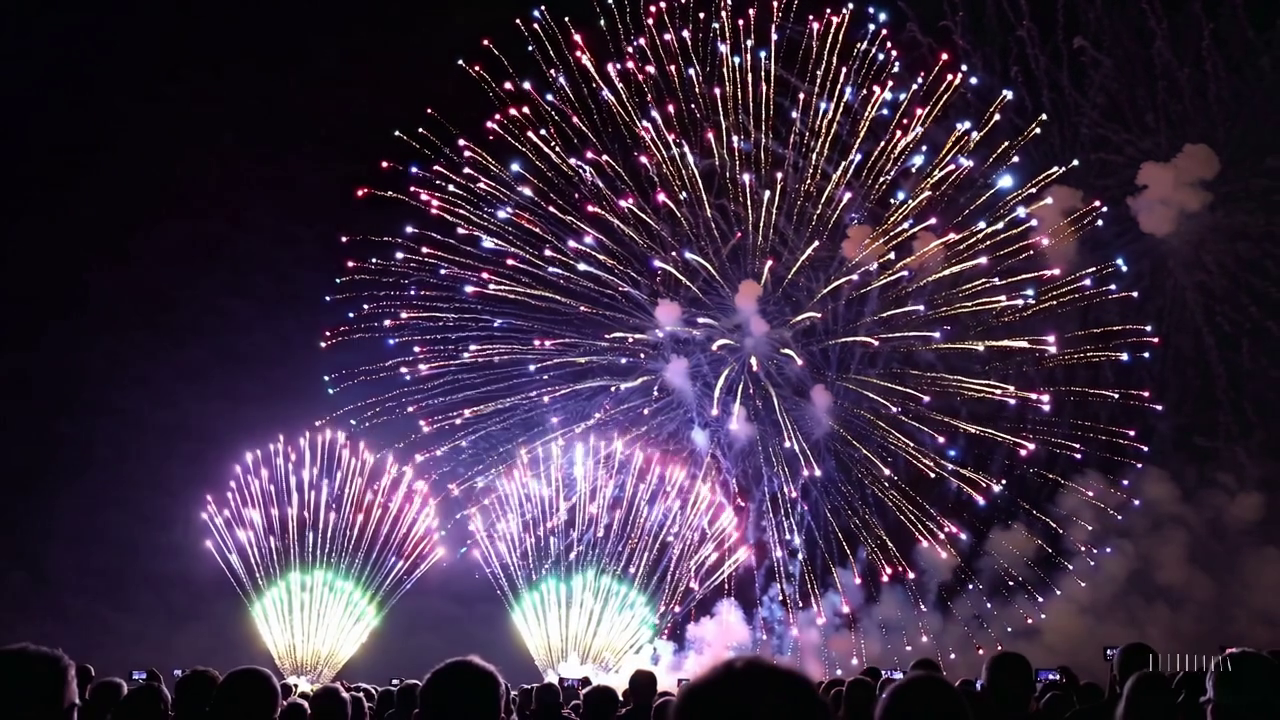}
    \end{subfigure}
    \begin{subfigure}{.23\linewidth}
        \centering
        \includegraphics[width=\linewidth]{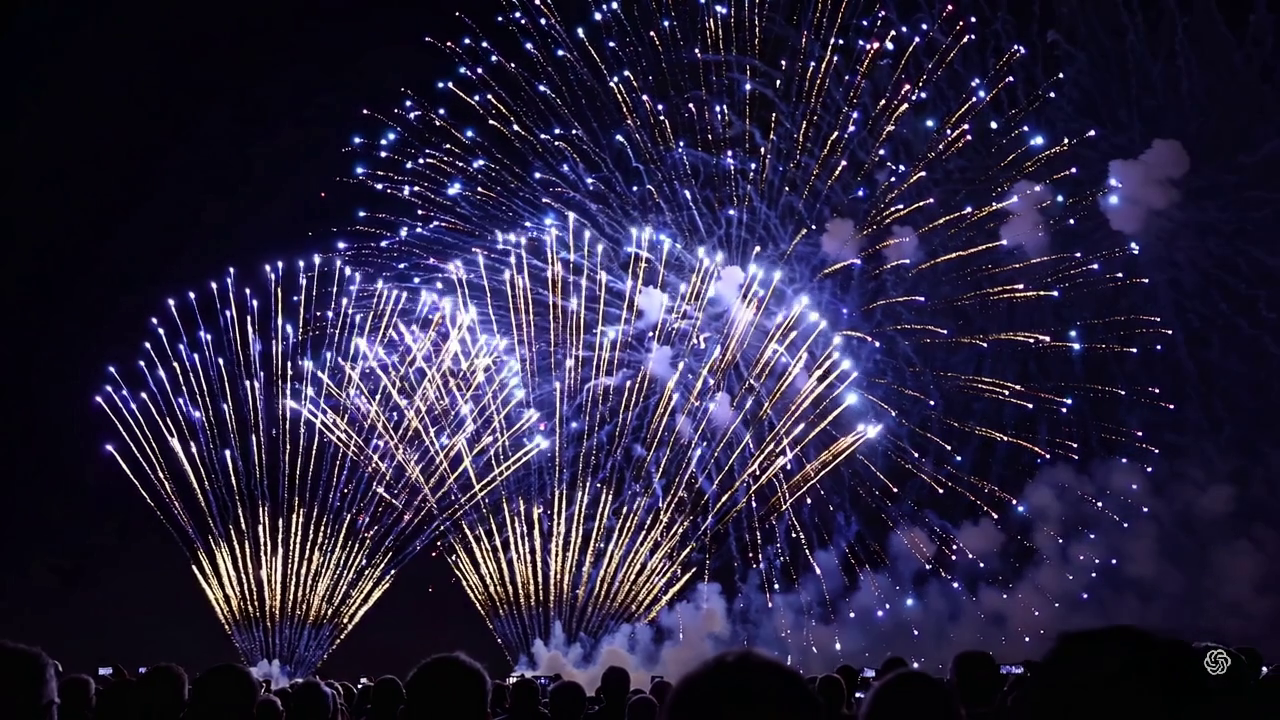}
    \end{subfigure} \\

    \begin{subfigure}{.23\linewidth}
        \centering
        \includegraphics[width=\linewidth]{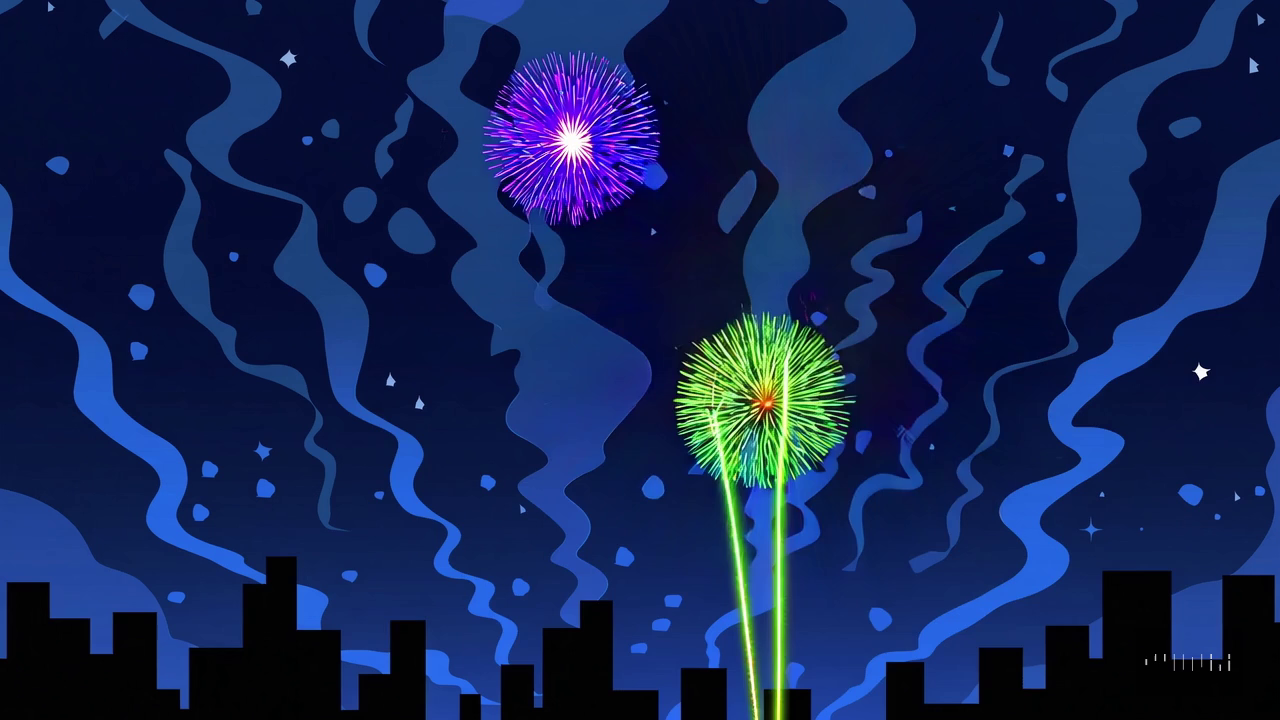}
    \end{subfigure}
    \begin{subfigure}{.23\linewidth}
        \centering
        \includegraphics[width=\linewidth]{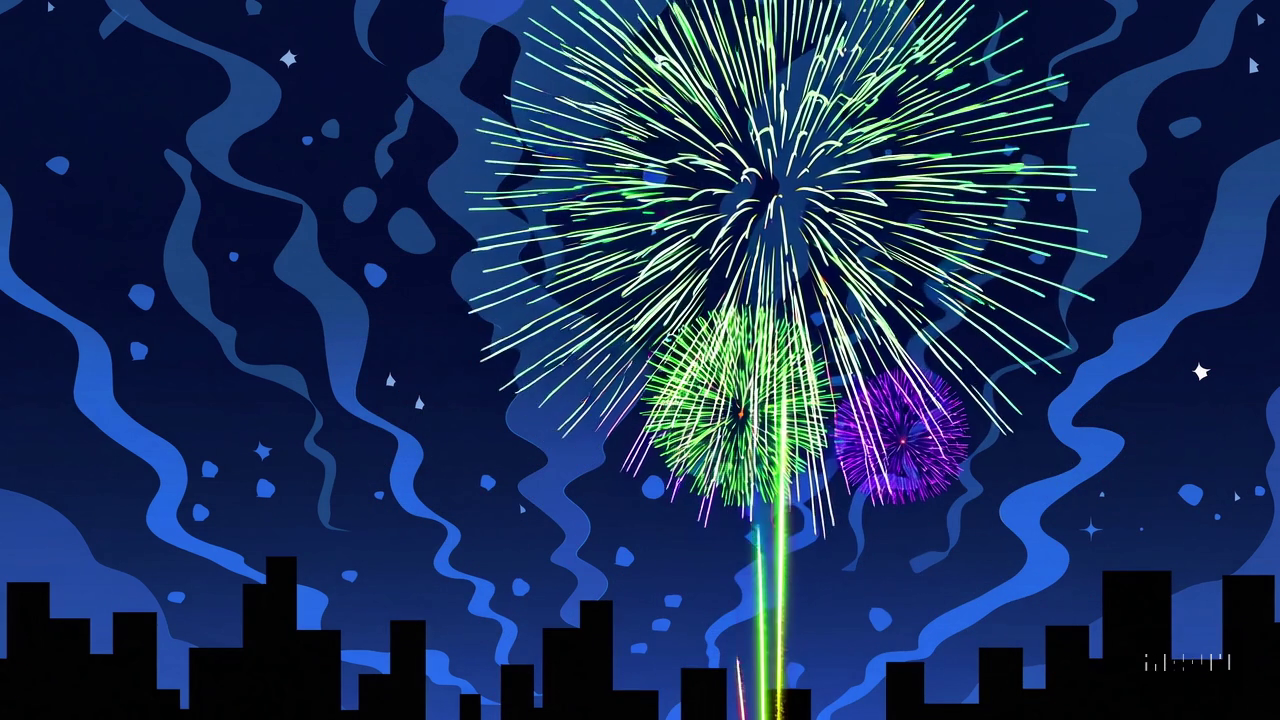}
    \end{subfigure}
    \begin{subfigure}{.23\linewidth}
        \centering
        \includegraphics[width=\linewidth]{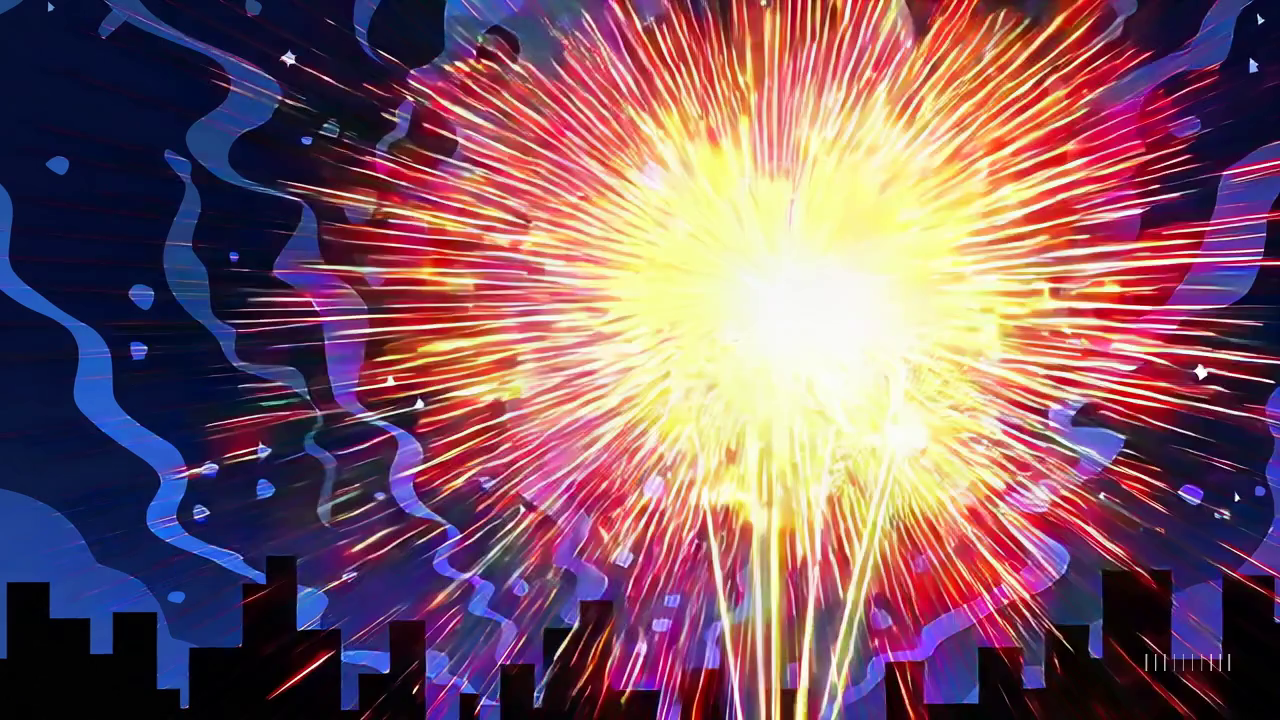}
    \end{subfigure}
    \begin{subfigure}{.23\linewidth}
        \centering
        \includegraphics[width=\linewidth]{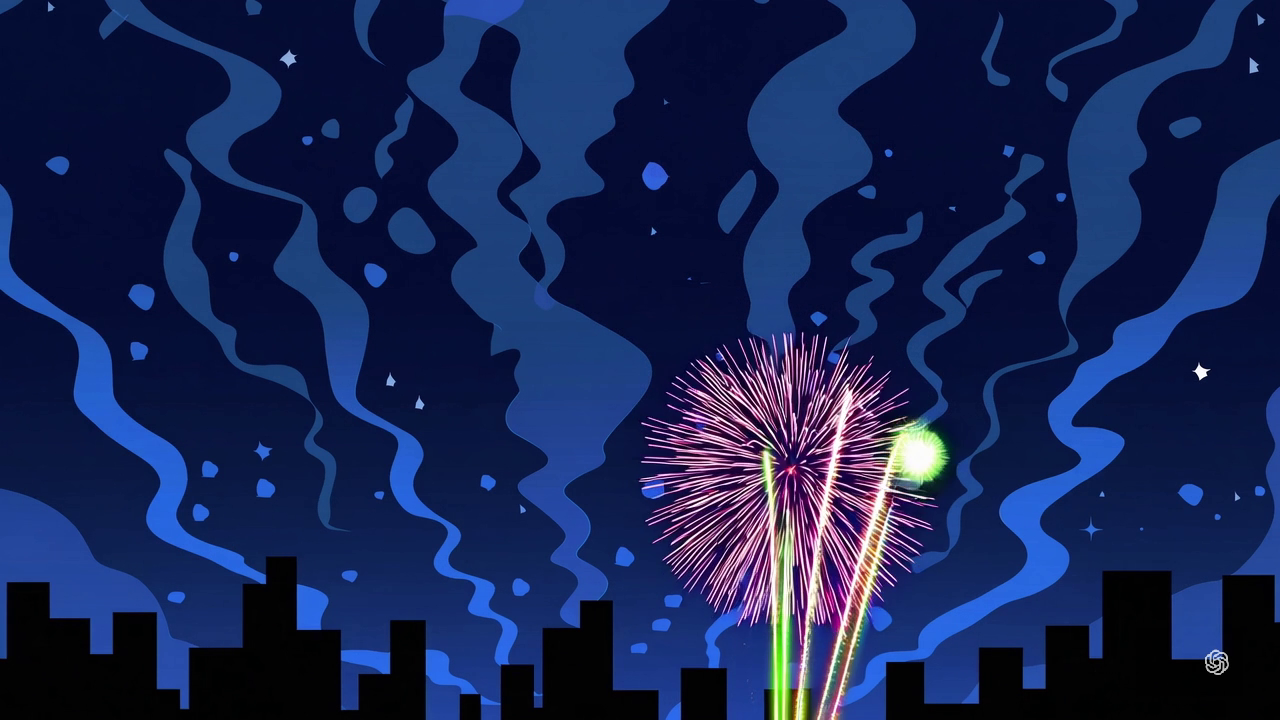}
    \end{subfigure} \\

    \begin{subfigure}{.23\linewidth}
        \centering
        \includegraphics[width=\linewidth]{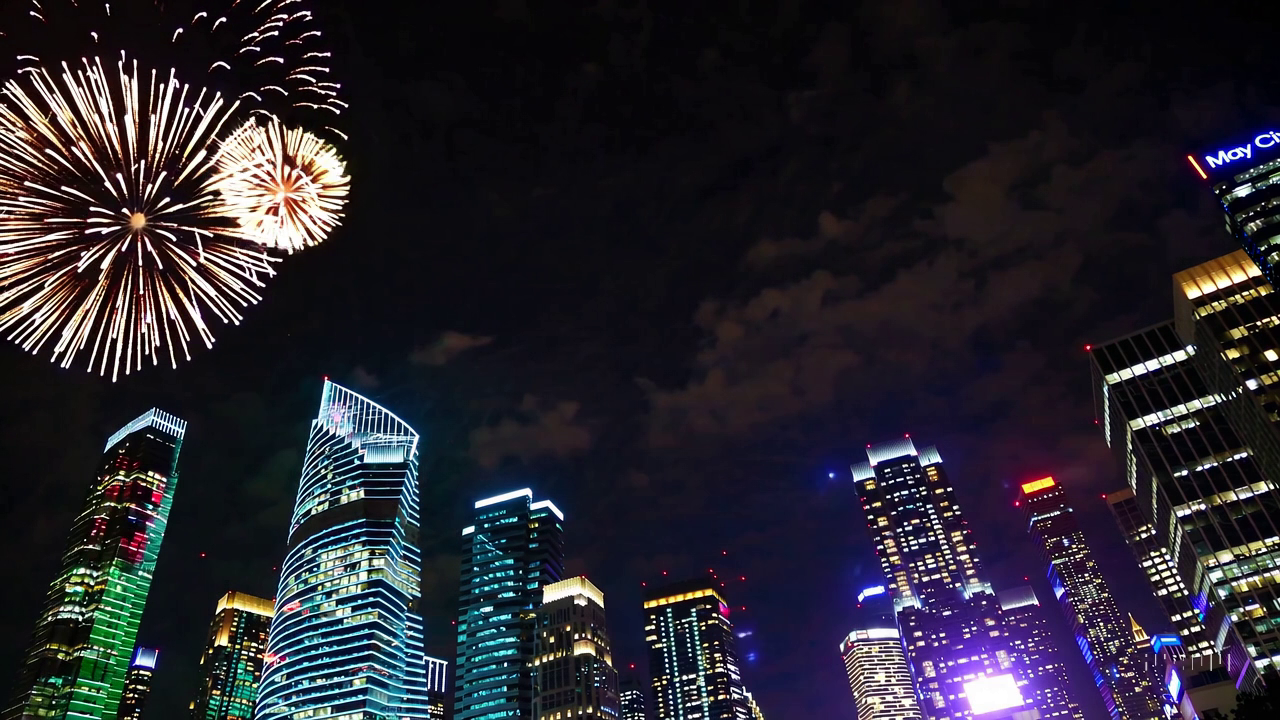}
    \end{subfigure}
    \begin{subfigure}{.23\linewidth}
        \centering
        \includegraphics[width=\linewidth]{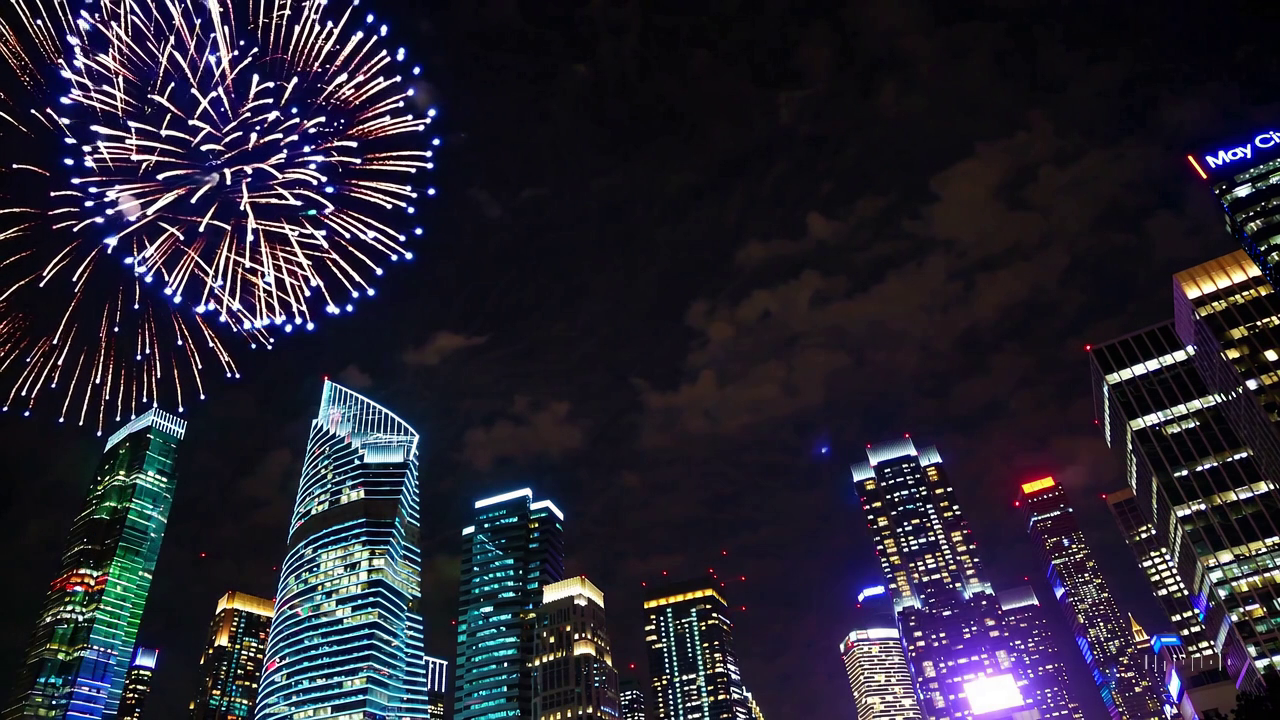}
    \end{subfigure}
    \begin{subfigure}{.23\linewidth}
        \centering
        \includegraphics[width=\linewidth]{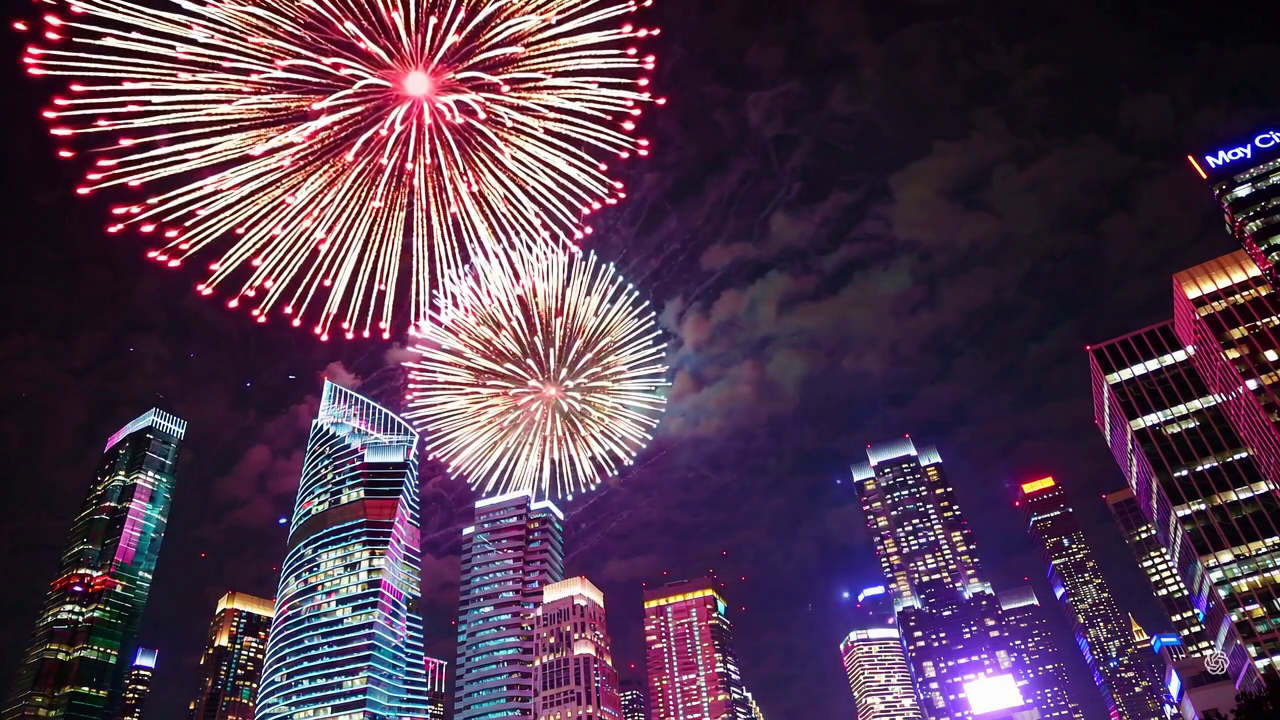}
    \end{subfigure}
    \begin{subfigure}{.23\linewidth}
        \centering
        \includegraphics[width=\linewidth]{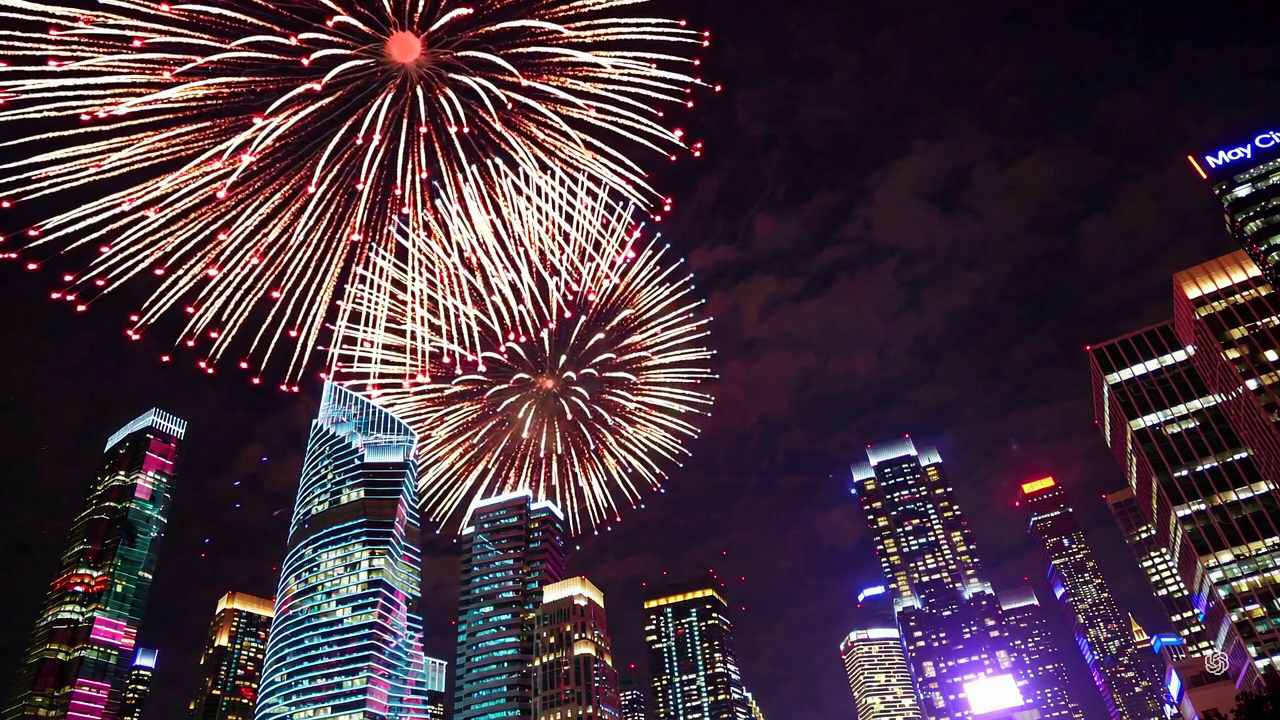}
    \end{subfigure} \\
    
    \begin{subfigure}{.9\linewidth}
    \centering
    \subcaption{Prompt: Generate a dynamic video with rapid frame changes featuring a dazzling fireworks display with vibrant explosions.}
    \end{subfigure}

    \begin{subfigure}{.23\linewidth}
        \centering
        \includegraphics[width=\linewidth]{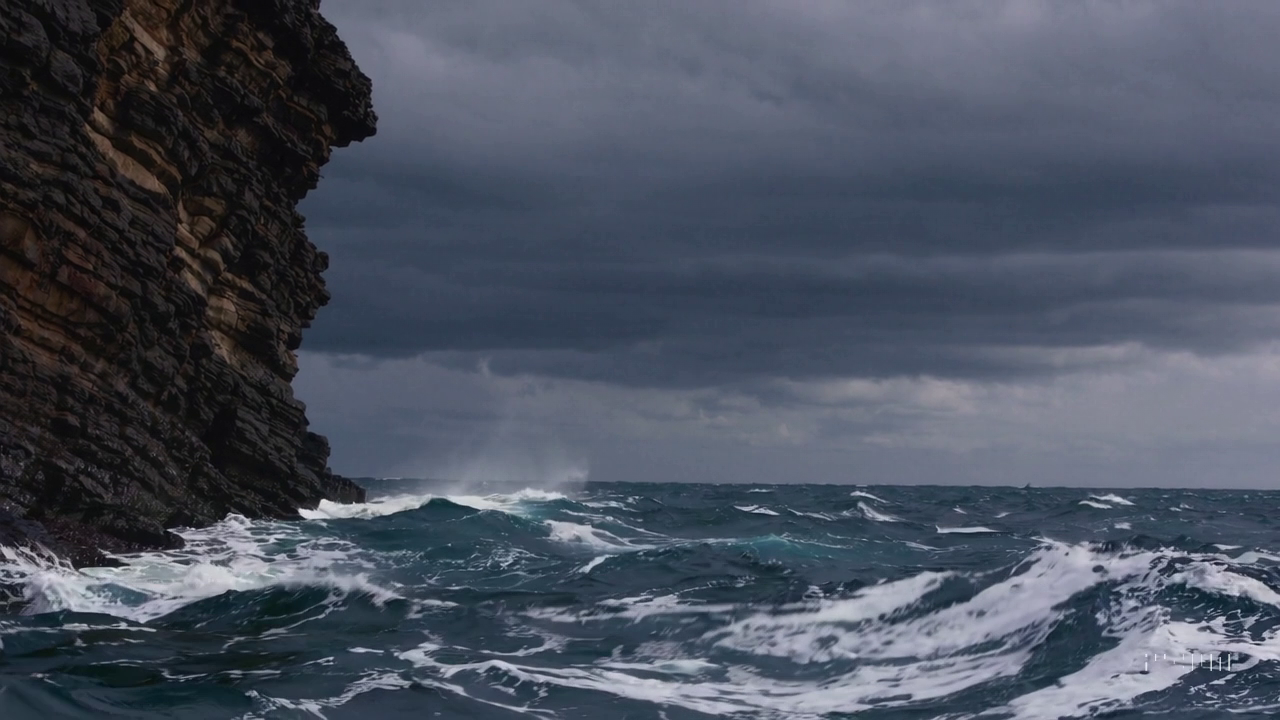}
    \end{subfigure}
    \begin{subfigure}{.23\linewidth}
        \centering
        \includegraphics[width=\linewidth]{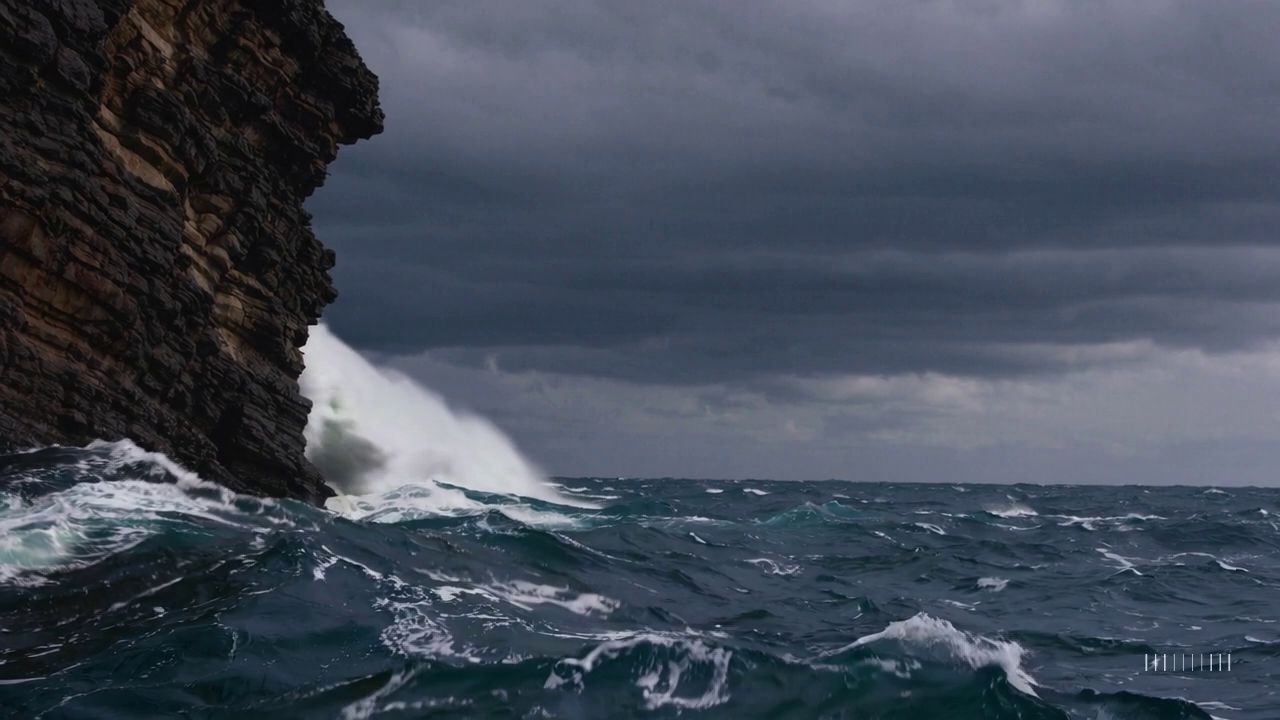}
    \end{subfigure}
    \begin{subfigure}{.23\linewidth}
        \centering
        \includegraphics[width=\linewidth]{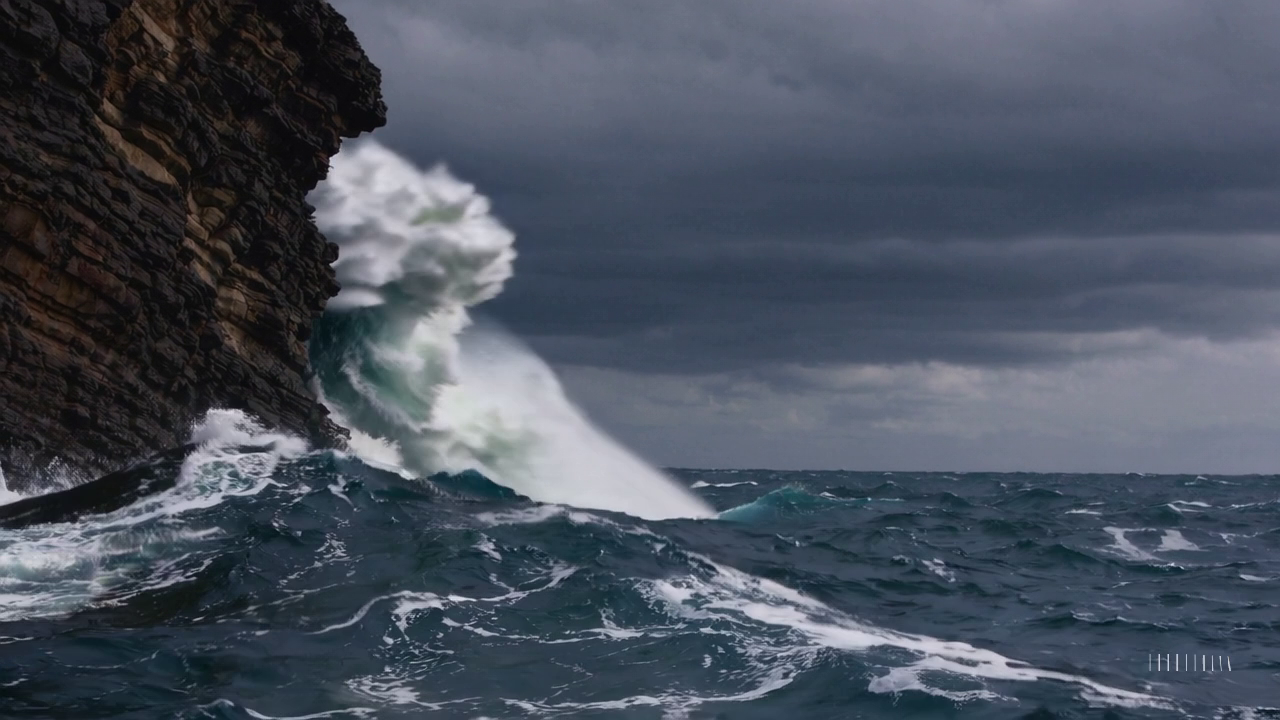}
    \end{subfigure}
    \begin{subfigure}{.23\linewidth}
        \centering
        \includegraphics[width=\linewidth]{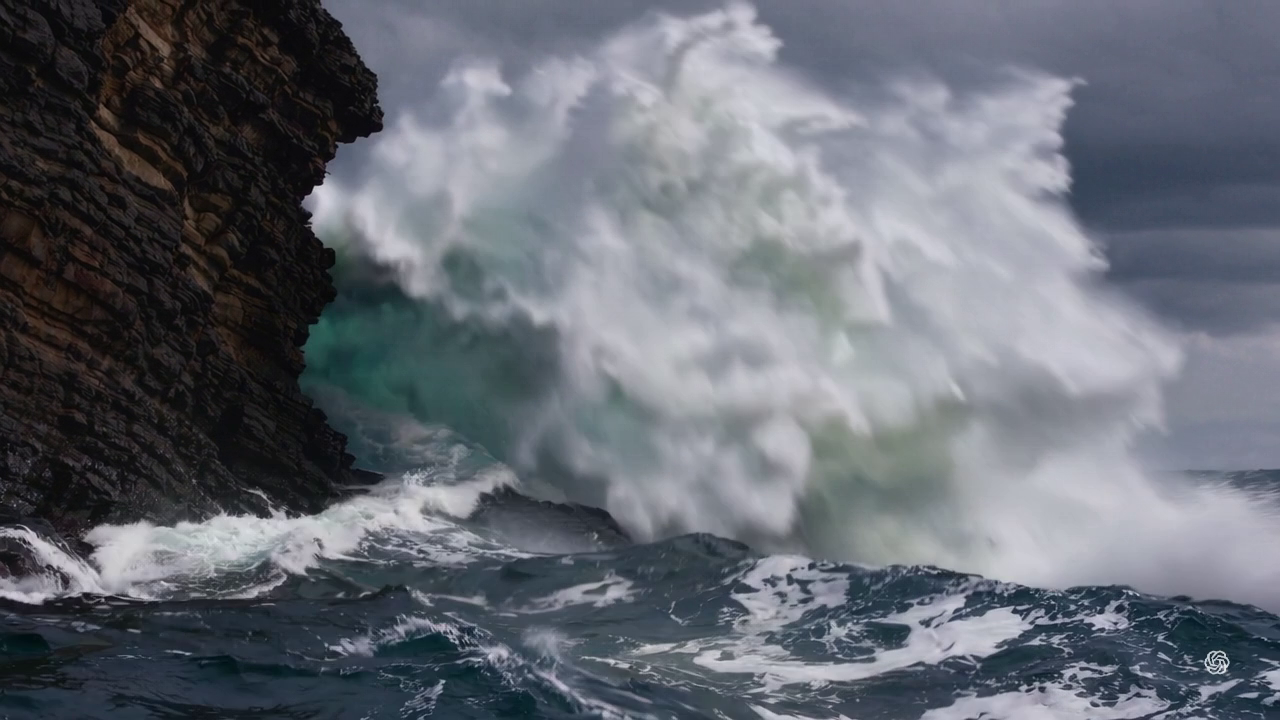}
    \end{subfigure} \\

    \begin{subfigure}{.23\linewidth}
        \centering
        \includegraphics[width=\linewidth]{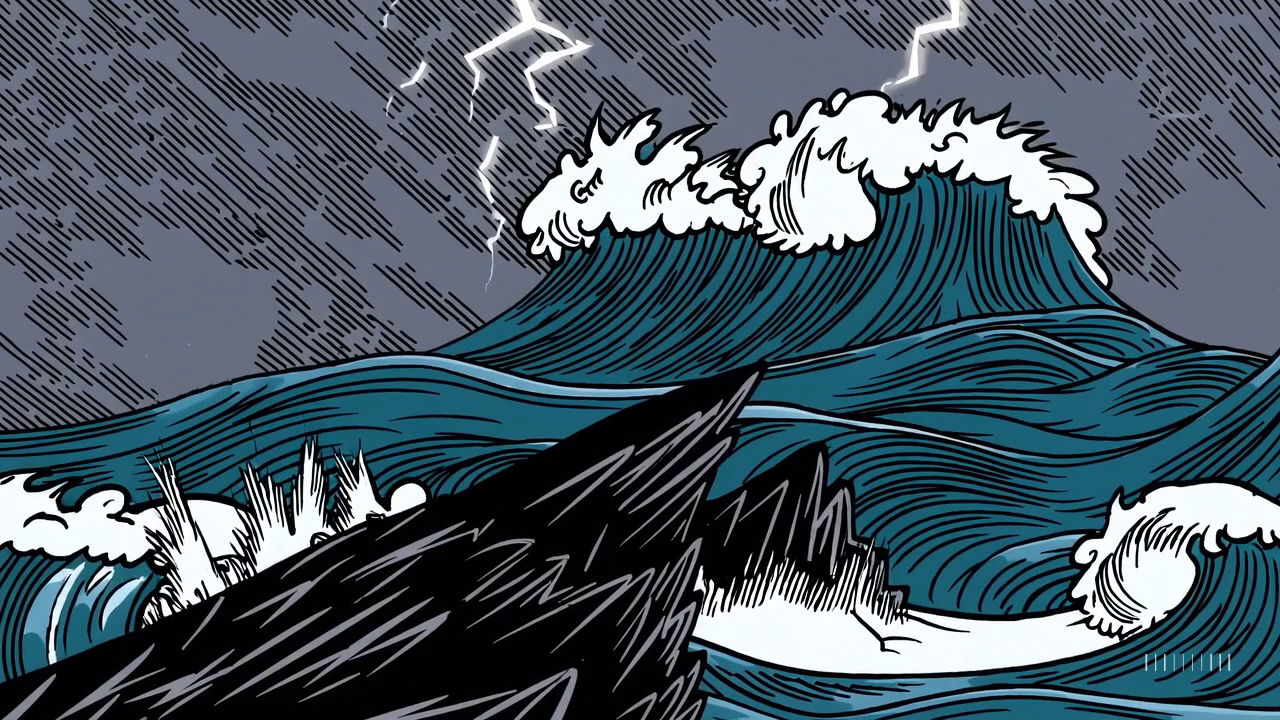}
    \end{subfigure}
    \begin{subfigure}{.23\linewidth}
        \centering
        \includegraphics[width=\linewidth]{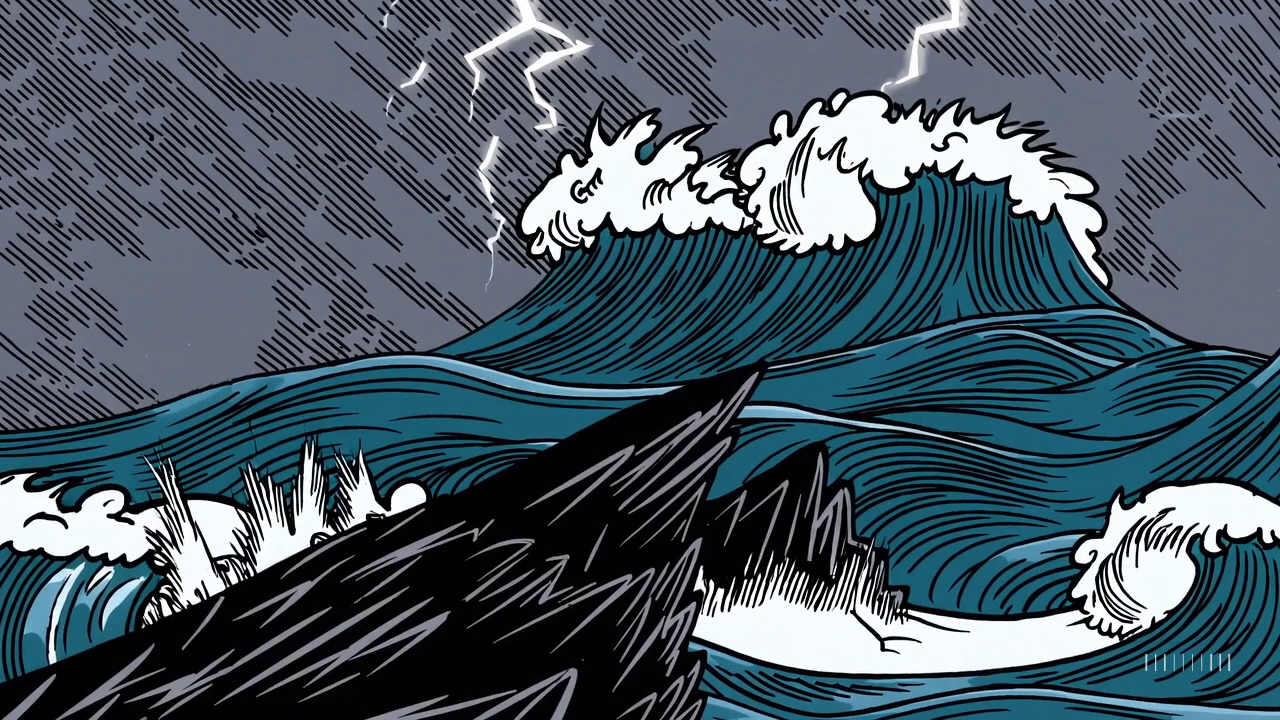}
    \end{subfigure}
    \begin{subfigure}{.23\linewidth}
        \centering
        \includegraphics[width=\linewidth]{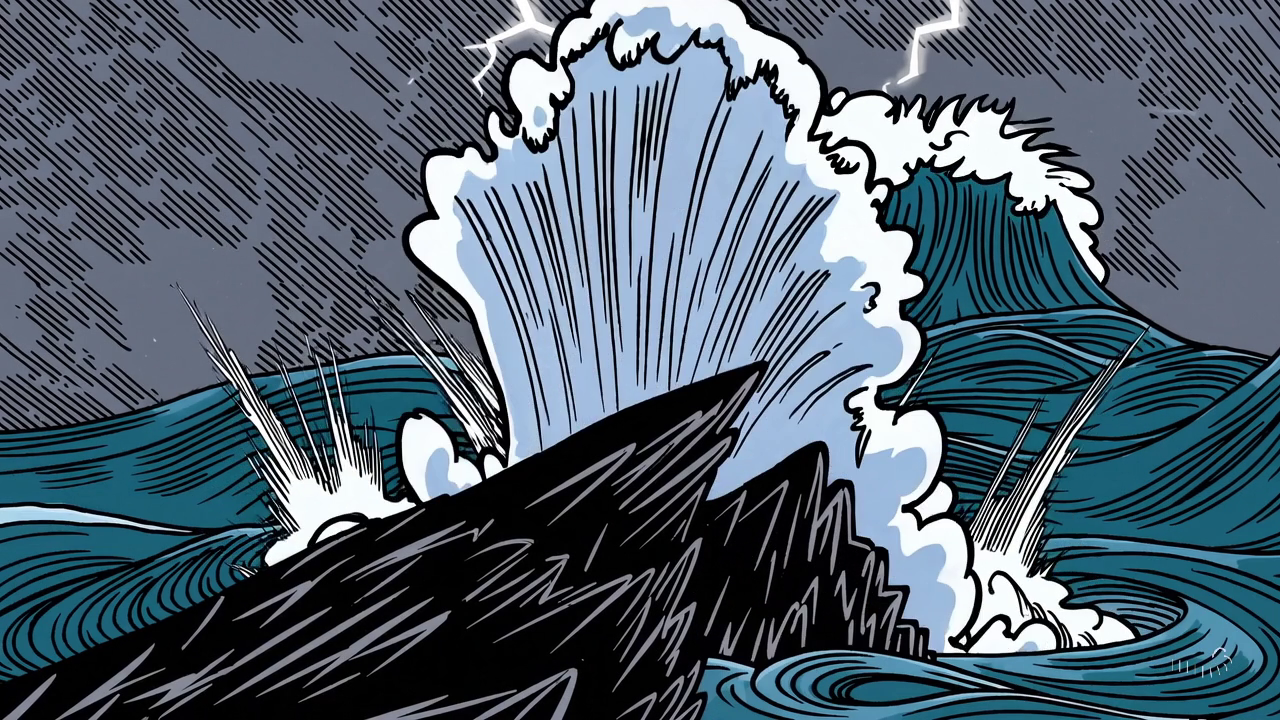}
    \end{subfigure}
    \begin{subfigure}{.23\linewidth}
        \centering
        \includegraphics[width=\linewidth]{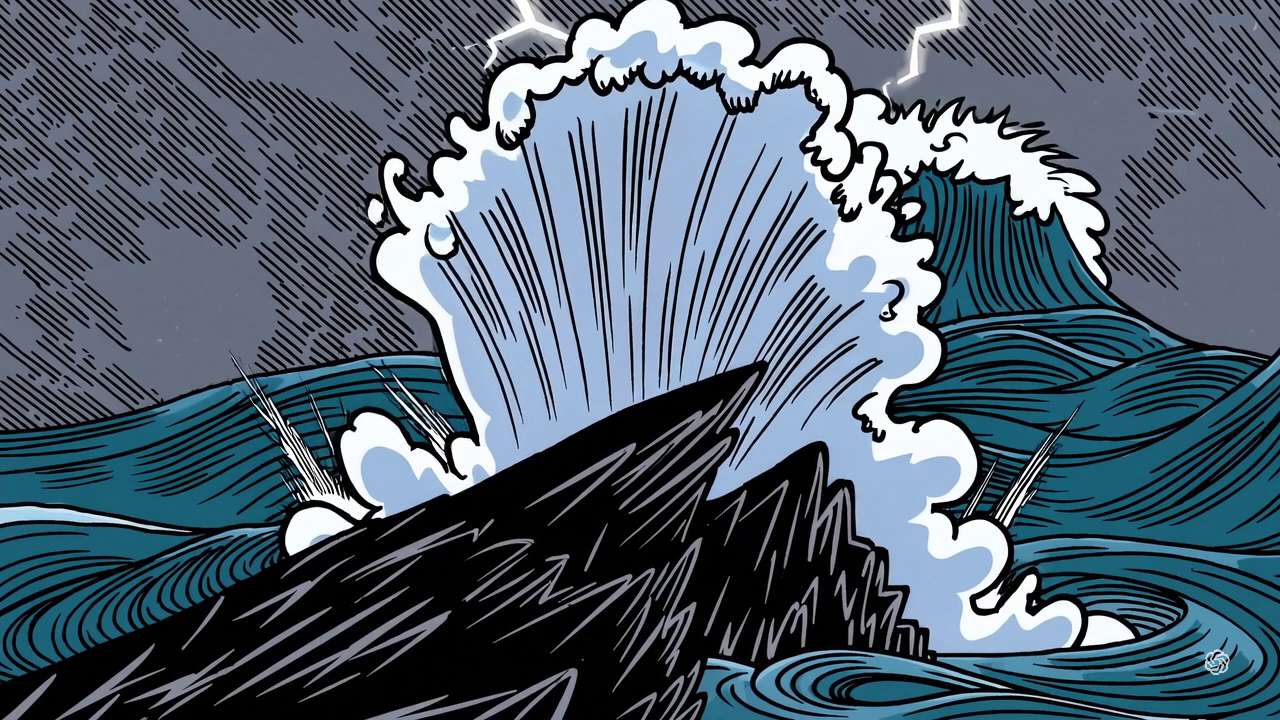}
    \end{subfigure} \\

    \begin{subfigure}{.23\linewidth}
        \centering
        \includegraphics[width=\linewidth]{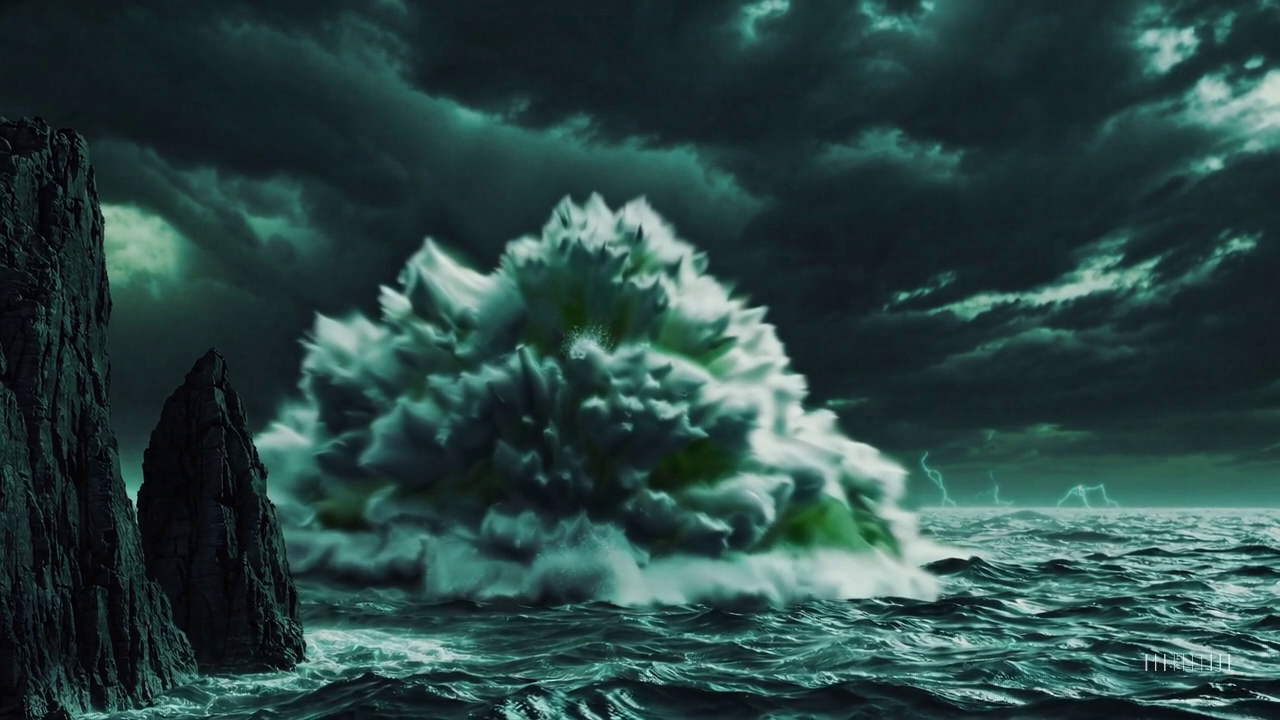}
    \end{subfigure}
    \begin{subfigure}{.23\linewidth}
        \centering
        \includegraphics[width=\linewidth]{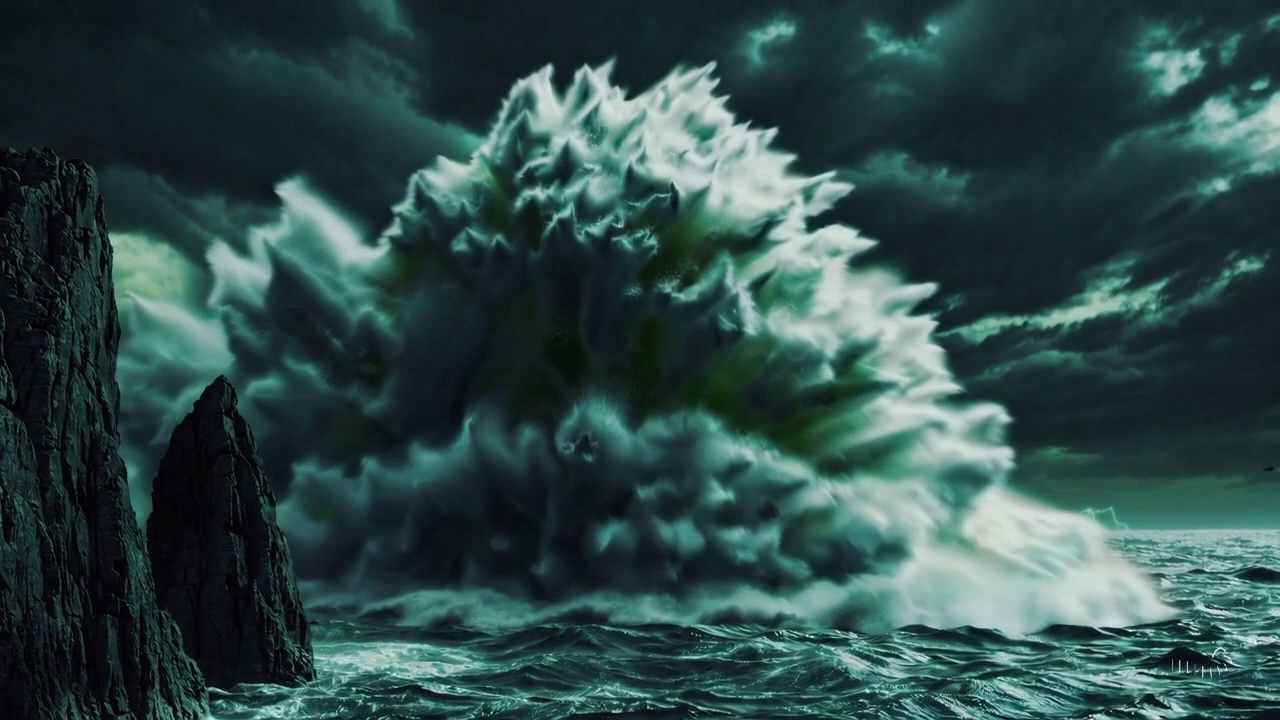}
    \end{subfigure}
    \begin{subfigure}{.23\linewidth}
        \centering
        \includegraphics[width=\linewidth]{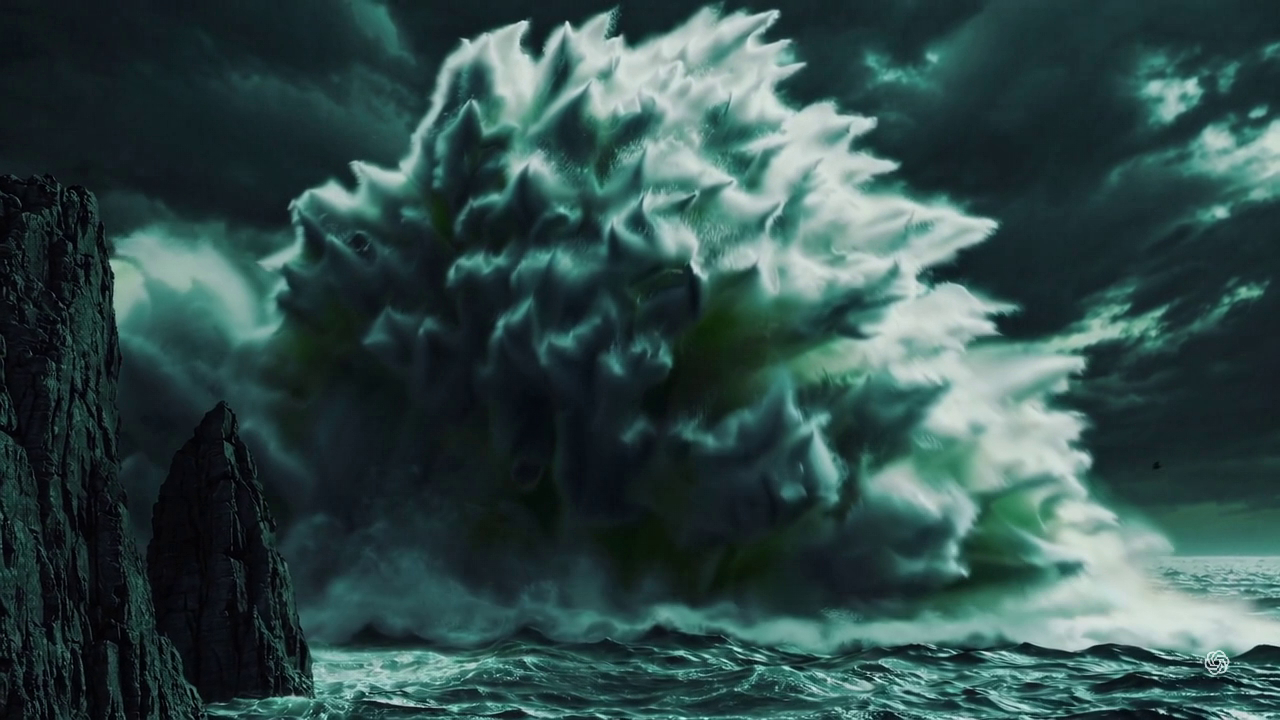}
    \end{subfigure}
    \begin{subfigure}{.23\linewidth}
        \centering
        \includegraphics[width=\linewidth]{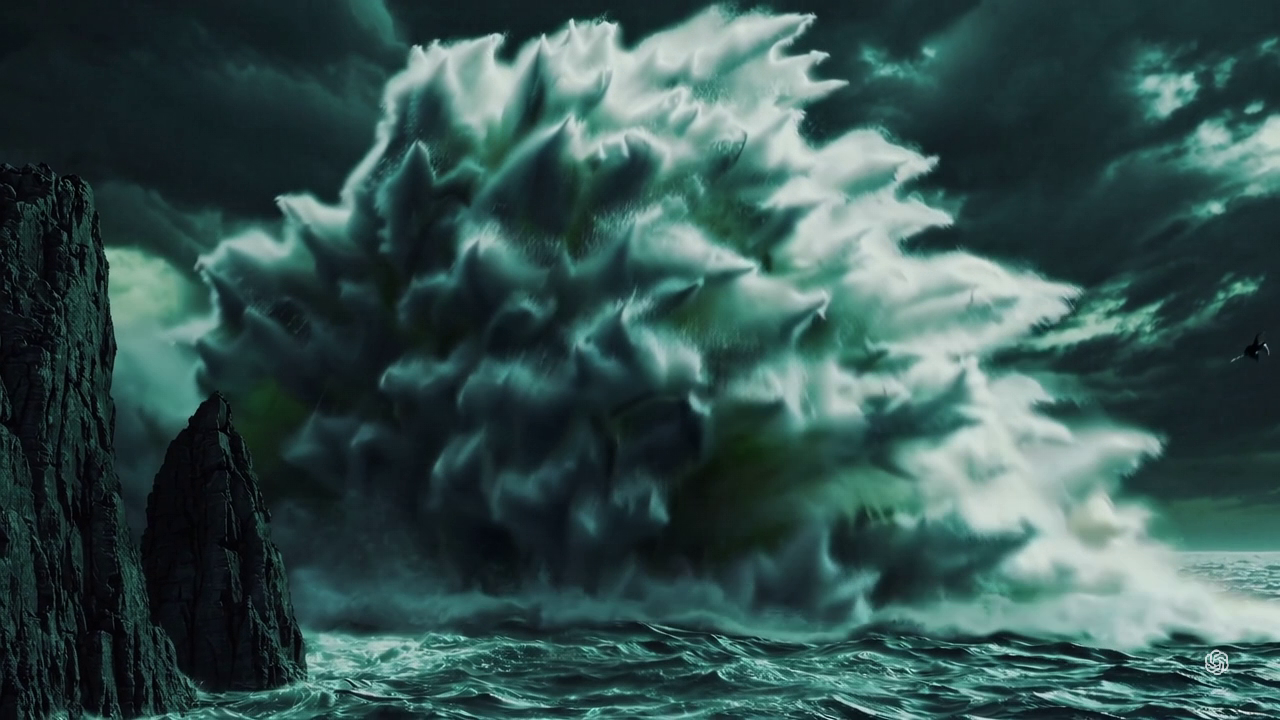}
    \end{subfigure} \\

    \begin{subfigure}{.9\linewidth}
    \centering
    \subcaption{Prompt: Generate a dynamic video with rapid frame changes featuring stormy ocean waves crashing against cliffs in a chaotic sequence.}
    \end{subfigure}
    
    \caption{Video examples generated by Sora. The first, second, and third rows correspond to the \emph{realistic}, \emph{cartoon}, and \emph{sci-fi} styles, respectively.}
\end{figure}

\end{document}